\documentclass[11pt]{amsart}
\usepackage[utf8]{inputenc}
\usepackage[T1]{fontenc}
\usepackage{mathptmx}
\usepackage{amsthm,amssymb}

\newtheorem{remark}{Remark}

\newcommand{\A}{\mathbb{A}}

\newcommand{\R}{\mathbb{R}}

\newcommand{\Degre}{\ensuremath{^\circ}}
\usepackage{url}
\usepackage[T1]{fontenc}
\usepackage[utf8]{inputenc} 

\def\di{\displaystyle}

\usepackage{float}
\usepackage{dsfont}
\usepackage{comment}
\usepackage{amsmath,amssymb,graphicx}
\usepackage{mathrsfs}
\usepackage{subcaption}
\usepackage{caption}

\usepackage{graphicx}

\usepackage[version=3]{mhchem}\usepackage{booktabs}
\usepackage[margin=2cm]{geometry}

\def\di{\displaystyle}

\usepackage{booktabs}
\usepackage{mathtools}
\usepackage{dsfont}
\usepackage{comment}
\usepackage{mathrsfs}
\usepackage{subcaption}
\usepackage{caption}
\usepackage{multicol}
\usepackage{lipsum}
\usepackage{color}
\usepackage{colortbl}
\usepackage{array}
\usepackage{commath}
\usepackage{arydshln}

\begin{document}
\title[Generation of Helical states - breaking of symmetries, Curie's principle, and excited states]{Generation of Helical states - breaking of symmetries, Curie's principle, and excited states}

\author{J. Sabalot-Cuzzubbo}
 
\author{D. Bégué}
    
\author{J. Cresson}

\date{\today}

\begin{abstract}
Following previous work of M.H. Garner, R. Hoffmann, S. Rettrup and G.C. Solomon, we discuss the generation of helical molecular orbitals (MOs) for linear chains of atoms. We first give a definition of helical MOs and we provide an index measuring how far a given helical states is from a perfect helical distribution. Structural properties of helical distribution for twisted $[n]$-cumulene and cumulene version of Möbius systems are given. We then give some simple structural assumptions as well as symmetry requirements ensuring the existence of helical MOs. Considering molecules which do not admit helical MOs, we provide a first way to induce helical states by the breaking of symmetries. We also explore an alternative way using excited conformations of given molecules as well as different electronic multiplicities. Several examples are given. 
\end{abstract}

\maketitle

\tableofcontents

\section{Introduction}

During the past few years, a huge interest has been developed for particular linear chains of atoms like cumulene or carbyne. One of the interests is to possess in some configurations, helical MOs along the chain opening new possibilities in nanoscale electronics. This phenomenon called {\bf electrohelicity} has been discussed in several different ways, in particular, using classical H\"uckel theory and symmetry group (see example \cite{garner}) or from a more physical point of view using properties of the Hamiltonian (see for example \cite{guna}). To our knowledge, these works are restricted to molecules in their ground state and no studies provide a discussion of the effect of excited states as well as electronic multiplicities on electrohelicity. This is precisely the subject of this paper.\\

First, we define helical MOs following the previous work of M.H. Garner et al. in \cite{garner} and S. Gunasekaran et al. in \cite{guna}. A helical MO is associated to a given distribution of vectors or angles associated to the $\pi$-system. The classical picture is to draw a helix representing the twist of the $\pi$-vectors along the structure. In \cite{bro}, W. Jeorgensen et al. study how far a given distribution is from a perfect helix, i.e. a distribution governed by a fixed pitch. We introduce an index based on the standard least square method giving the best perfect helix fitting to a given helical distribution. We take this opportunity to give general results about Hückel distributions of twisted [n]-cumulene and equivalent cumulene version of Möbius systems. \\

Second, we review previous work leading to a characterization of electrohelicity in term of basic structural assumptions on the linear chain and symmetry properties. This characterization can be used to precisely identify which kind of atoms can be used to construct molecule admitting electrohelicity. Chirality plays an important role in this setting and is supported by a general physical argument called the Curie's principle. \\

Third, taking molecules which do not support electrohelicity in their ground state, we explain how to obtain it by looking to excited states or electronic multiplicities. Such a possibility is only valid for molecules which in their ground state admit a subgroup with the basic ingredients of electrohelicity. In this case, using symmetry adapted MOs, we are able to predict the properties of the induced helical MOs. 

Finally, we illustrate our discussion with several examples.
\section{Helical states and some properties}

\subsection{Definition of helical states}

In this section, we follow the work of S. Gunasekaran and L. Venkataraman in \cite{guna}. We consider a chain of $N-1$ sp-atoms $A_i$ denoted by $C$, together with arbitrary left and right end groups denoted by $L$ and $R$, consisting of a connection with an $sp^2$-atom with a given structure (fig. \ref{fig 1}):

\begin{figure}[H]
	\centering
	\includegraphics[scale=0.6]{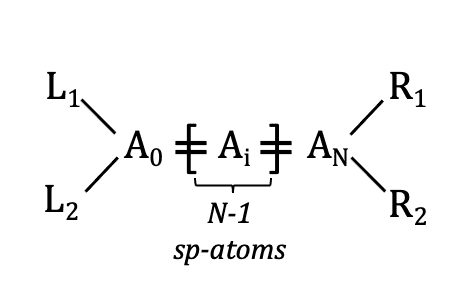}
	\caption{Representation of a standard chain}
	\label{fig 1}
\end{figure}

The chain $C$ of atoms is oriented along the $z$-axis. The $\pi$-system of the chain can be modelled with a basis set comprising a $p_x$ and $p_y$ orbital for each atom of the chain. The $p_z$-orbital is directed along the chain direction and is part of the $\sigma$-system. \\

The left and right groups can have arbitrary shapes, but the most usual situation encountered for this type of chain-compound is that of unsubstituted methylene (alkylidene) like cumulenes where $L_1=L_2=H$ and $R_1=R_2=H$. The cumulene molecule with N double bonds is denoted as $[N]$-cumulene. In their ground-states cumulenes with even numbers of carbon atoms and hence with odd numbers of double bonds are planar (eclipsed conformation). Cumulenes with odd numbers of carbon atoms are nonplanar, with orthogonal planes of the terminal methyl groups.\\

Our interest is on the {\bf orbital structure} restricted to the chain $C$, meaning that we will not discuss the particular orbital structure emerging in the left and right side groups.\\

In this work, we focus on {\bf helical orbitals} which were discussed by M.H. Garner and co-workers in \cite{garner} in connection with cumulenes and annulenes. Helical orbitals allow a $\pi$-electron delocalization twisting along the chain. \\

Formally, this can be viewed as follows:\\

Let $E$ be a given energy and $z$ the integer position of the atoms of the chain with $z=0$ corresponding to the left side and $z=N$ to the right side. Denoted by $\psi (z)$ the atomic molecular orbital associated to atoms $A (z)$, we have: 

\begin{equation}
\psi (z) =\di\left ( 
\begin{array}{c}
\psi_x (z) \\
\psi_y (z) 
\end{array}
\right )  = M R(z\omega ) v_0
\end{equation}
where $v_0$ is a fixed two-dimensional unit vector, $\omega \in \R$, $R(z\omega )$ is the rotation matrix of angle $z\omega$ and $M$ is a real symmetric matrix.\\

As the vector $R(z\omega ) v_0$ belongs to the unit circle $S^1$, the vector $\psi (z)$ belongs to the image of the unit circle by the linear map associated to $M$. Denoted by $v_a$ and $v_b$ the two orthogonal eigenvectors of $M$ associated to the eigenvalues $a$ and $b$ which are real, the image of $S^1$ is an ellipse whose major/minor axis are given by $a v_a$ and $b v_b$. The vector $\psi (z)$ rotates along the ellipse when $z$ increases. The orientation of the rotation when $z$ increases depends on $M$, which characterizes the elliptic polarization. If $\mid M \mid >0$ (resp. $\mid M \mid <0$) the rotation is clockwise (resp. counterclockwise). 
\subsection{Fitting a helix to helical MOs} 

We first introduce the notion of distribution of angles:\\

We call the {\bf distribution associated to $\psi$} and we denote by $\mathscr{D} (\psi )$ a finite family of angles $\phi (z_j )$, $j=0,\dots , m$ or equivalently a finite family of unit vectors $\psi (z_j ) \in S^1$, where $z_j \in \R$ is an increasing family of real values.

\begin{remark}
It must be noted that in principle a helical MO induces a continuous distribution of angles $\phi (z)$ but in practice we will have access only to a sampling of this distribution.
\end{remark}

We can represent a distribution by sections along a cylinder oriented along the $z$-axis. \\

A linear molecule being given, a distribution can be obtained in essentially two different ways:\\

\begin{itemize}
    \item First, we can use H\"uckel theory in order to determine an approximation of the wave function along the molecule, and a vector $\psi (z)$ corresponding to the decomposition of it in the basis given by $p_x (z)$ and $p_y (z)$ at the atom positioned at point $z$. Such a distribution will be denoted by $\mathscr{D}_{Huckel}$.\\
    
    \item We can use DFT (see Appendix 1) in order to determine the vector $\psi (z)$ using the technique developed by W. Jeorgensen, M.H. Garner and G. Solomon in \cite{bro}. Such a distribution will be denoted by $\mathscr{D}_{DFT}$. 
\end{itemize}

\begin{remark}
The previous distributions are not the only ones which can be defined. Observations deduced from DFT are sometimes subject to caution. As a consequence, one can imagine using more refined methods for example CASPT2 (Complete Active Space Perturbation Theory at the order 2) in order to check if the underlying phenomenon is method's dependant or not (see Appendix 1). In the following, we then also studied $\mathscr{D}_{CASPT2}$.
\end{remark}

We can associate to each distribution $\mathcal{D} (\psi )$ a continuous helix in the following sense: \\

For each set of real constants $b>0$, $\gamma \in \R$, $\epsilon =\pm 1$ and $r>0$, one can defined an helix $h_{r,b, \epsilon}$ whose parametrization is given by: 
\begin{equation}
    \left \{ 
    \begin{array}{lll}
    x(t) & = & r\cos (\gamma t),\\
    y(t) & = & \epsilon r\sin (\gamma t) ,\\
    z(t) & = & b \gamma t .
    \end{array}
    \right .
\end{equation}

Another way to write the parametrization is:
\begin{equation}
    h_{r,b,\gamma ,\epsilon} (t)=r \left ( \cos (\gamma t) e_x +\epsilon \sin (\gamma t) e_y \right ) +b\gamma t e_z ,
\end{equation}
where $(e_x,e_y,e_z)$ is the canonical basis defined by $e_x = (1,0,0)$, $e_y =(0,1,0)$, $e_z =(0,0,1)$.\\

In the following, we simply denote $h_{b,\gamma, \epsilon}$ for $h_{r,b,\gamma, \epsilon}$.

\begin{figure}[H]
	\centering
	\includegraphics[scale=0.4]{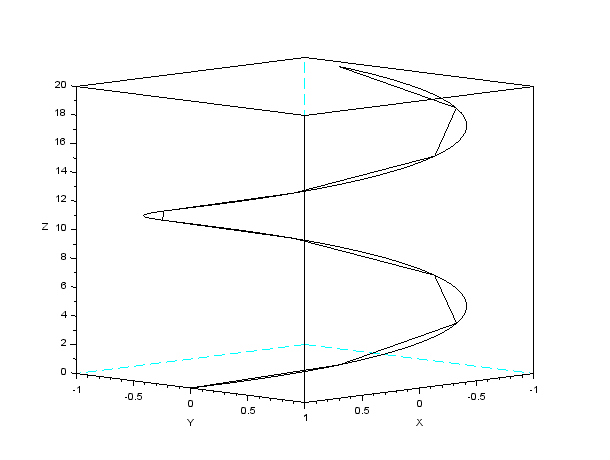}
	\caption{Perfect helix curve with $b=2$, $r=1$, $\gamma=1$, $\epsilon =1$.}
	\label{fig2-perfecthelix}
\end{figure}

If the parameter $\epsilon =1$ (resp. $-1$), then the helix is right-hand (resp. left-hand). \\

For each distribution $\mathcal{D} (\psi )$ one can look for the best helix $h_{b,\epsilon}$ which minimize the quantity: 
\begin{equation}
L(r, b,\gamma, \epsilon ) =\di\sum_{i=0}^m \parallel \psi (z_i) - h_{r, b,\gamma, \epsilon} (z_i) \parallel^2 .
\end{equation}

The sign of $\epsilon$ for a given distribution is easy to determine so that we are reduced to the minimization of the functional:
\begin{equation}
L(r, b,\gamma ) =\di\sum_{i=0}^m \parallel \psi (i) - h_{r, b,\gamma ,\epsilon} (i) \parallel^2 ,
\end{equation}
where $\epsilon$ is fixed.\\

Such a problem is solved using the {\bf non linear least squares method}. In general {\bf no unique solution is found} and {\bf no explicit formula} can be given from the data because the method is an iterative process. \\

Another possibility is to restrict our attention to the distribution angles instead of the specific geometry of the helix. This is done in the next section. 
\subsection{Fitting a helical distribution - cumulated angle}

Helix with a constant pitch $b$ induces a distribution of the form: 
\begin{equation}
    \mathscr{D}_{Helix} =\{ \phi (z_i ) = z_i /b, \ i=0,\dots ,N .\} . 
\end{equation}
As a consequence, if one considers the {\bf cumulated angle} defined by:
\begin{equation}
    \phi_c (i)=\di\sum_{1}^i \mid \phi (z_j) -\phi (z_{j-1} )\mid \ \ \mbox{\rm with}\  \ \phi_c (z_0 )=0 ,
\end{equation}
one obtains:
\begin{equation}
    \phi_c (z)=\di\frac{1}{b} z ,
\end{equation}
i.e. a straight line passing through $0$ with a constant slope $1/b$.\\

An idea is then to compare the set of points $(z_j ,\phi_c (z_j) )$, $j=0;\dots ,N$ of a given helical distribution with a set of helical points $(z_j , z_j /b ) $, by minimizing the quantity:
\begin{equation}
    \mathscr{L} (b)=\di\sum_{i=0}^n \left ( \phi_c (j) -f_j \right )^2 ,
\end{equation}
where $f_j =z_j /b$, i.e. to use the (linear) {\bf least squares method} or {\bf regression analysis} in order to obtain the best fitted {\bf straight line} of the form:
\begin{equation}
    \phi=\alpha \,(z-z_0) + \phi_0 ,
\end{equation}
where $\alpha \in \R$ has to be determined, to the set of data $\phi_c (z_j)$.\\

It must be noted that the special form of the straight line is due to the fact that we impose on the line to pass through the initial point $(z_0 ,\phi_c (0)$ which is the first value of the cumulated angles distribution. \\

As we usually choose $z_0 =0$ for the position of the first atom and $\phi_c (0))=0$, the optimization problem reduces to find the best linear line:
\begin{equation}
    \phi =\alpha z
\end{equation}
which fits the set of data $\phi_c (z_j)$.

\begin{remark} A similar idea is used by W. Bro-Jeorgensen, M.H. Garner, G. Solomon in \cite{bro}, where they define the {\bf MAD index} which corresponds to the {\bf mean of the absolute deviation} $\mid \phi_c (z_j ) -f_j \mid$. 
\end{remark}

The solution to the previous problem has an explicit analytical solution given by: 
\begin{equation}
    \alpha = \di\frac{\sigma (\phi_c (z) ,z) +\overline{\phi_c (z)} \bar{z}}{\sigma^2 (z)} ,
\end{equation}
where for two given series $x=(x_0 ,\dots ,x_N)$, $y=(y_0\dots ,y_N )$, we denote by:
\begin{equation}
\bar{x}=\di\frac{1}{N+1}\sum_{i=0}^N x_i,\ \ \sigma (x,y)=\di\sum_{i=0}^N \di\frac{(x_i -\bar{x} ) (y_i -\bar{y} ) }{N+1},\ \ \sigma^2 (x)=\di\sum_{i=0}^N\frac{(x_i -\bar{x} )^2}{N+1} ,
\end{equation}
the mean of $x$, covariance between $x$ and $y$ and variance of $x$.\\

Let us denote by $\rho (x,y)$ the {\bf correlation} factor defined by: 
\begin{equation}
    \rho (x,y)=\di\frac{\sigma (x,y)}{\sigma (x) \sigma (y)} .
\end{equation}
The quality of the approximation given by the regression line is measured by $\rho (\phi_c (z) , z)$. If $\rho^2 (\phi_c (z),z )$ is very close to one then the approximation is very good. 

\begin{remark}
As usual, the quality of the indicator is related to the quality of measure and it is subject to discussion. In general, for very good, measured quantities, a correlation $>0.95$ is assumed to represent a very good approximation by a straight line.
\end{remark}

Consequently, we propose as an indicator of helicity the quantity:
\begin{equation}
    HEL=\rho (\phi_c (z),z) ,
\end{equation}
instead of the MAD index proposed in \cite{bro}. Two advantages at least can be pointed out:\\

\begin{itemize}
    \item First, the correlation quantity is well-known and its interpretation, even if it is subject to discussion, is well documented.
    
    \item Second, a comparison is possible as $HEL \in [0,1]$. Contrarily, the MAD index can take arbitrary values and it is not clear what is the exact difference between a MAD index of $4$, $7$ or $10$ as no normalization is given. 
\end{itemize}

As an example, let us take the following angle distributions obtained for equivalent $4$-cumulene version of M\"obius systems (see section \ref{equivcum}) when $n=1$ given by: 
\begin{equation}
    \mathscr{D}_{4,1}=\left \{
    0,\, 32,\, 45,\, 58,\, 90
    \right \} .
\end{equation}
The associated cumulated distribution is given by:
\begin{equation}
    \phi_c =\left \{
    0,\, 32,\, 77,\, 135,\, 225
    \right \} .
\end{equation}

The "best" perfect helix fitting this set of data is given by: 
\begin{equation}
    \phi (z)=49.7 z ,
\end{equation}
and the correlation factor is given by: 
\begin{equation}
\rho (z,\phi_c(z))=0.9887211 ,    
\end{equation}
which shows very good agreement.

\begin{figure}[H]
	\centering
	\includegraphics[scale=0.35]{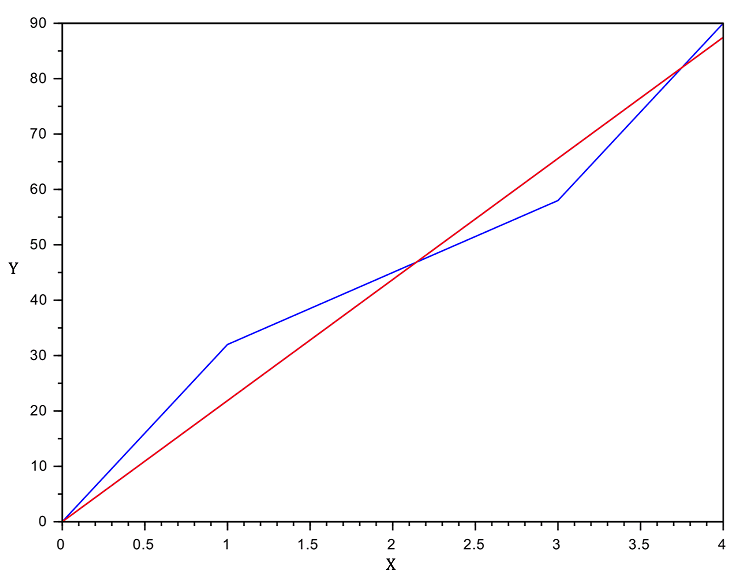}
	\caption{Best fitted helix for distribution of equivalent $4$-cumulene version of a M\"obius system and $n=1$}
	\label{fig3-best-fitted-helix-4-cum}
\end{figure}

The same can be done for the distribution of the equivalent $4$-cumulene version of a M\"obius system when $n=2$. In this case, the distribution is given by: 
\begin{equation}
    \mathscr{D}_{4,1}=\left \{
    0,\, 58,\, 135,\, 212,\, 270
    \right \} .
\end{equation}
The associated cumulated distribution is given by: 
\begin{equation}
    \phi_c =\left \{
    0,\, 58,\, 193,\, 405,\, 675
    \right \} .
\end{equation}

The "best" perfect helix fitting this set of data is given by: 
\begin{equation}
    \phi (z)=145.3 z ,
\end{equation}
and the correlation factor is given by: 
\begin{equation}
\rho (z,\phi_c(z))=0.9794173 ,    
\end{equation}
which again this shows a very good fit.

\begin{figure}[H]
	\centering
	\includegraphics[scale=0.4]{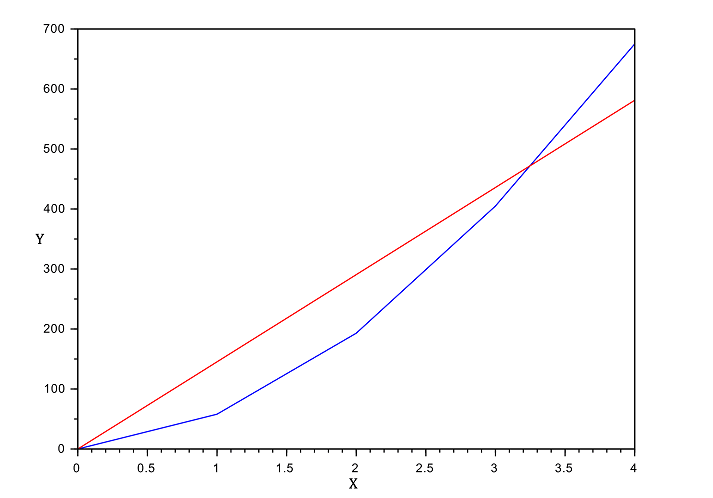}
	\caption{Best fitted helix for distribution of equivalent $4$-cumulene version of a Möbius system and $n=2$}
	\label{fig4-fitted-helix-42}
\end{figure}

Taking the solution of the minimisation problem, we can represent the $\pi$-system along the $C$ system like a helix:

\begin{figure}[H]
	\centering
	\includegraphics[scale=0.3]{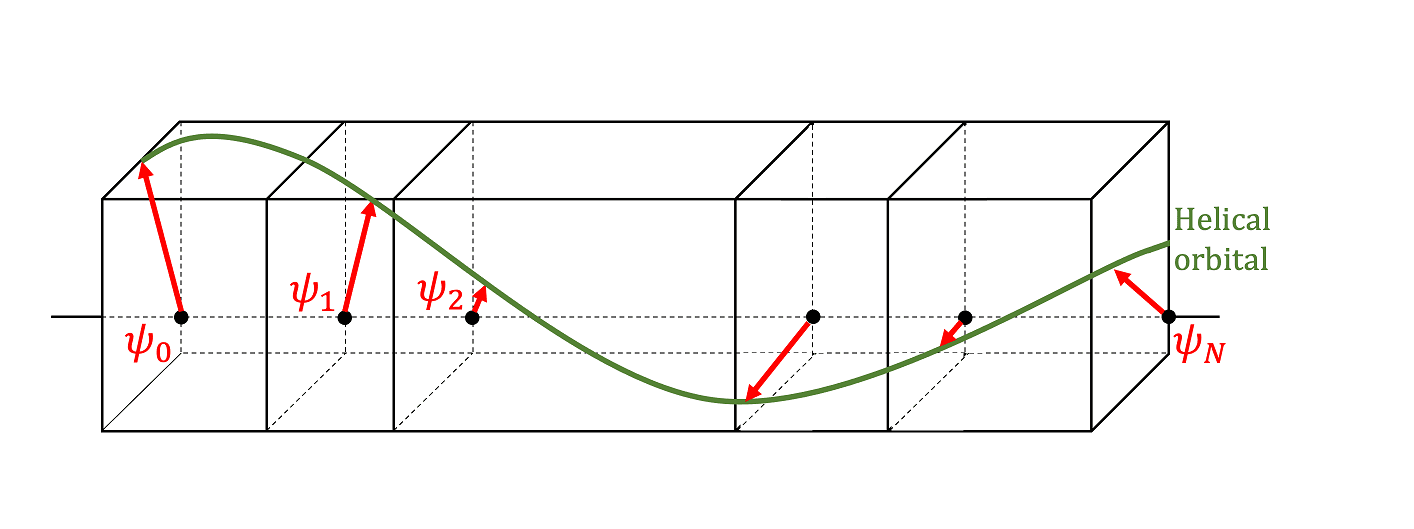}
	\caption{Representation of the $\pi$-system along the $C$ system}
	\label{fig5-pi-system}
\end{figure}
\subsection{Fitting a helix - general case}

Classical helixes are called {\bf perfect helixes} in \cite{bro}. However, perfect helixes are not the rule as many examples already proven (see the next sections and \cite{garner}). A more general class of helix is given by helix with a non-constant pitch, i.e. a helix of the form:
\begin{equation}
    \left \{ 
    \begin{array}{lll}
    x(t) & = & r\cos (t),\\
    y(t) & = & \epsilon r\sin (t) ,\\
    z(t) & = & P(t) ,
    \end{array}
    \right .
\end{equation}
where $P(t)$ is a smooth increasing function. 
\begin{figure}[H]
	\centering
	\includegraphics[scale=0.3]{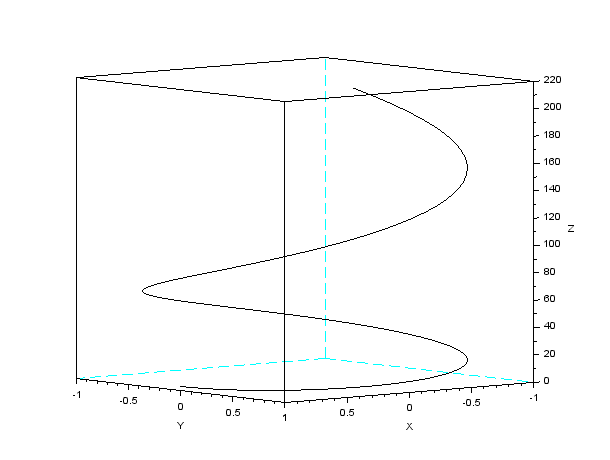}
	\caption{Imperfect helix with $P(t)=2t^2+3$.}
	\label{fig6-imperfect-helix}
\end{figure}
More generally, one can consider a helix with a non-constant radius of the form: 
\begin{equation}
    \left \{ 
    \begin{array}{lll}
    x(t) & = & r(t)\cos (t),\\
    y(t) & = & \epsilon r(t)\sin (t) ,\\
    z(t) & = & P(t) ,
    \end{array}
    \right .
\end{equation}
where $r:\R \mapsto \R^+$ is an arbitrary positive function. \\

A very good approximation of the function $P$ can be obtained from the set of data completing the best helix approximation just studied but in a nonlinear setting, for example using polynomial functions for $P(t)$ as well as for $r(t)$.\\

The previous generalization is not only a mathematical idea. As an example, the structure of the following linear molecule gives rise to a general helix structure:
\begin{figure}[H]
\centering
\includegraphics[width=0.12\linewidth]{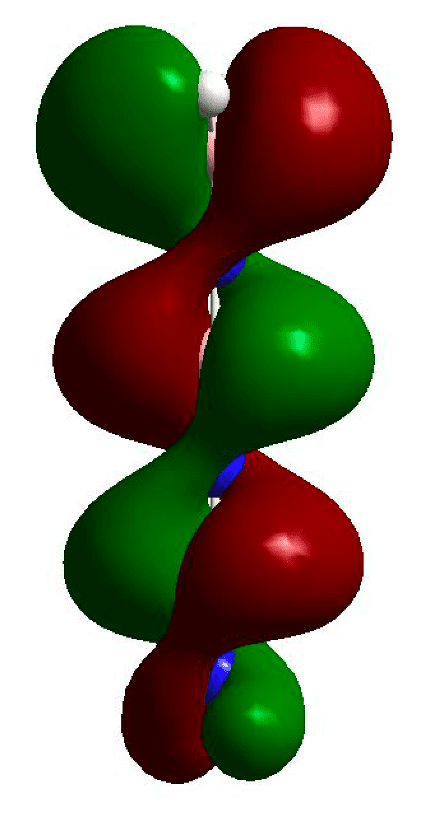}
\caption{Non standard helix}
\label{fig7-BN-helix}
\end{figure}
In fact, a very large class of morphology is possible, and the case of perfect helix seems to be not representative of helical state. In that respect, it seems unreasonable to take a perfect helix as a structure to make comparisons with.

Linear chains of boron nitride \cite{Cretu} could illustrate the existence of non standard helix. Like carbon nanostructures, boron nitride (BN) nanostructures present a wide variety of physical and chemical properties. They also present a wide variety of MOs according to both the number of atoms on the chains (even or odd), and to the nature of the $A_0$ and $A_N$ atoms (boron or nitrogen) as illustrated in the tables reported below.

\begin{table}[H]
    \caption{Molecular orbitals of a linear chain \textbf{B=N=B=N=B} obtained at the B3LYP/6-311G(d.p) level of theory - c = charge} 
    \label{table1-mo-BN}
    \centering
    \begin{tabular}{cccccc}
        \midrule
                     & $\alpha$-{HOMO-1} & $\alpha$-{HOMO} & $\alpha$-{LUMO} & $\alpha$-{LUMO+1} \\
   $c = 0$ - doublet &
   {\includegraphics[height=1.30cm]{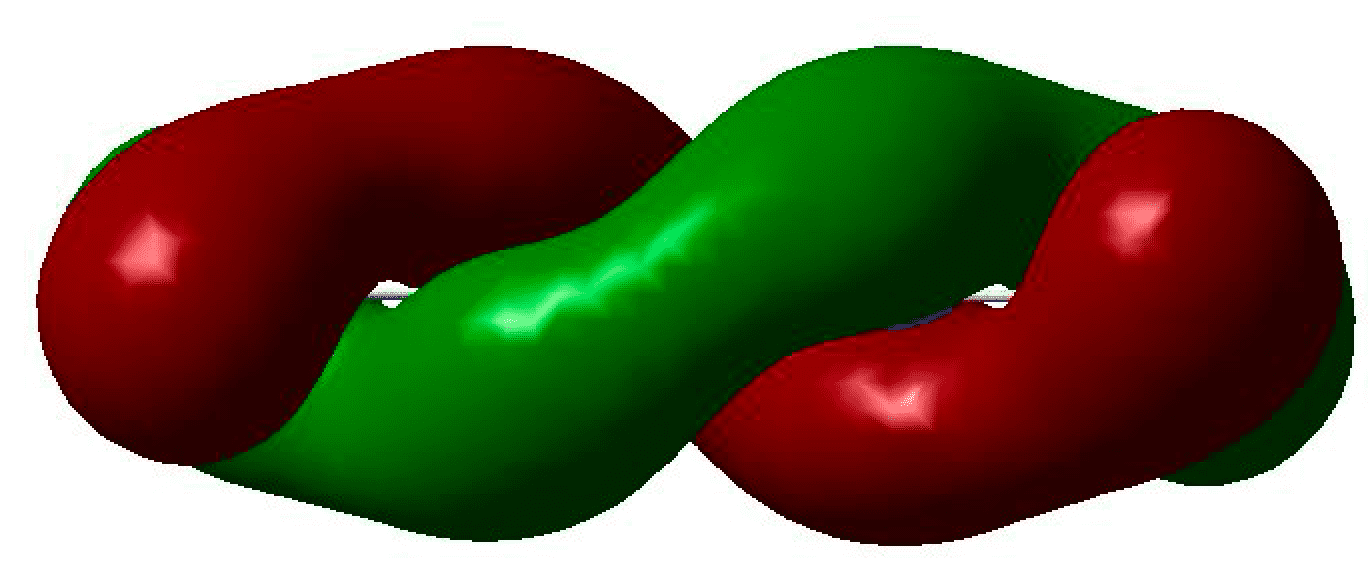}} &
   {\includegraphics[height=1.40cm]{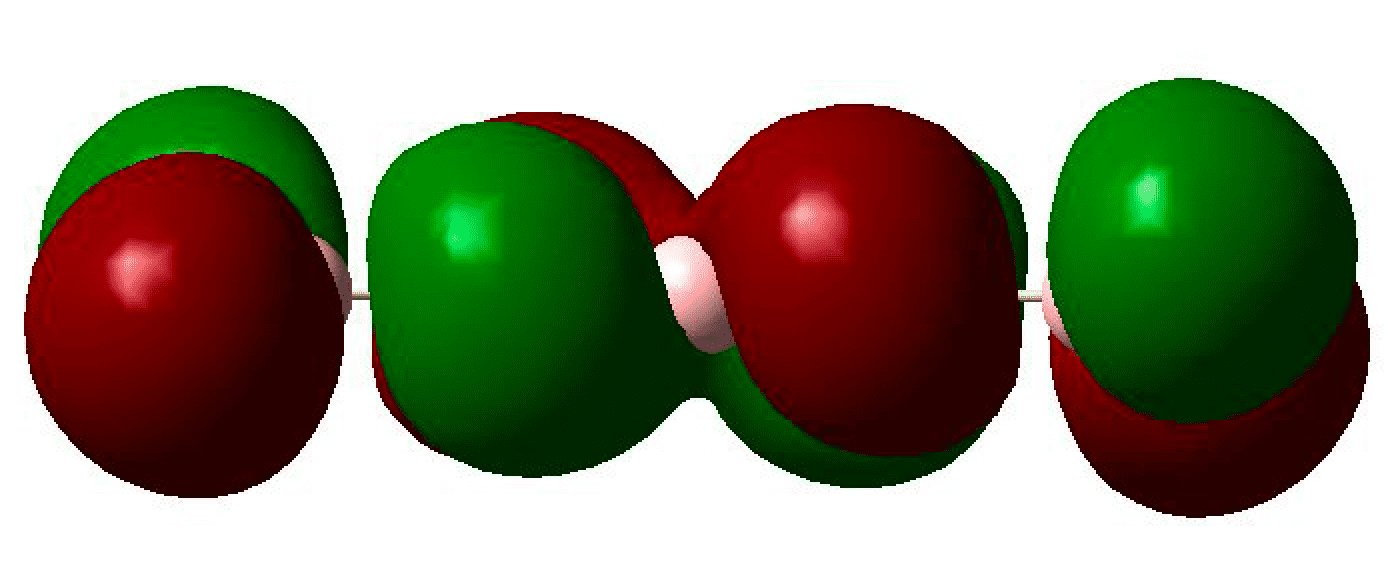}} &
   {\includegraphics[height=1.40 cm]{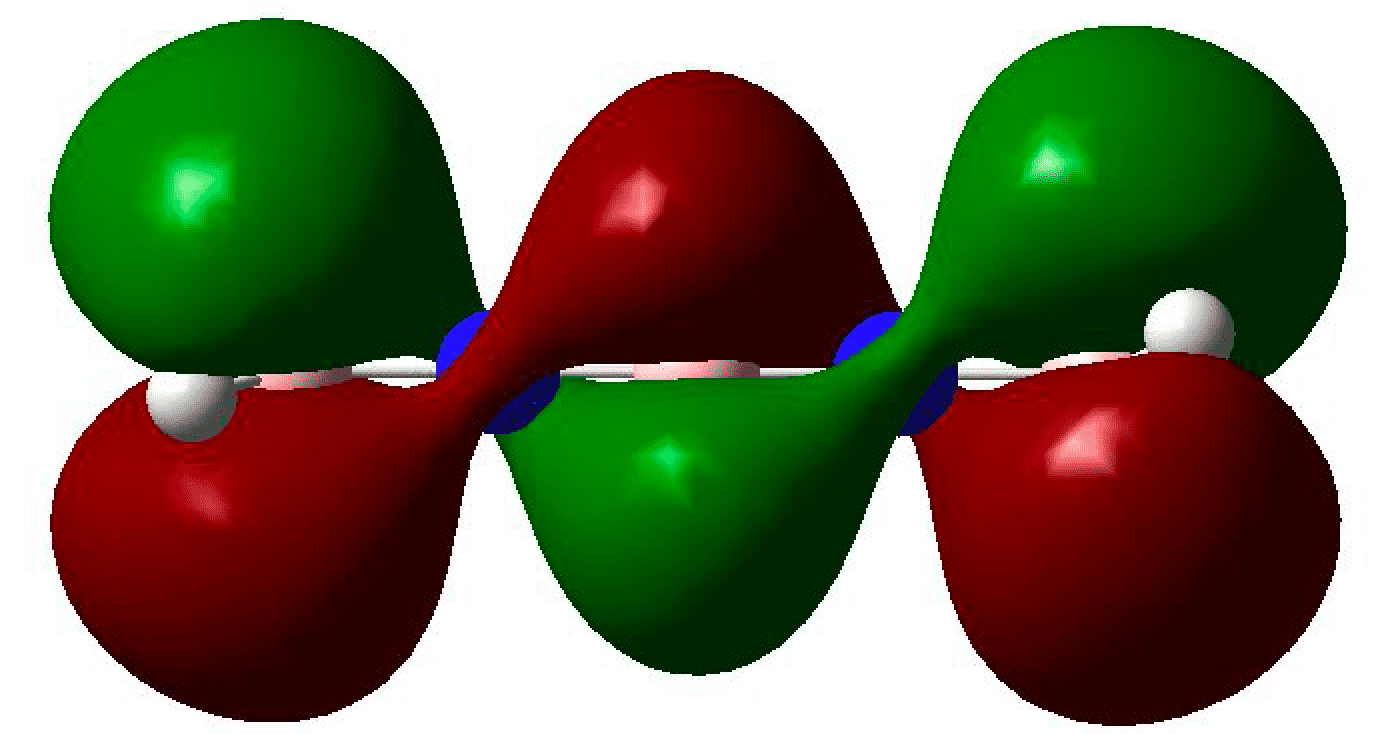}} & 
   {\includegraphics[height=1.40cm]{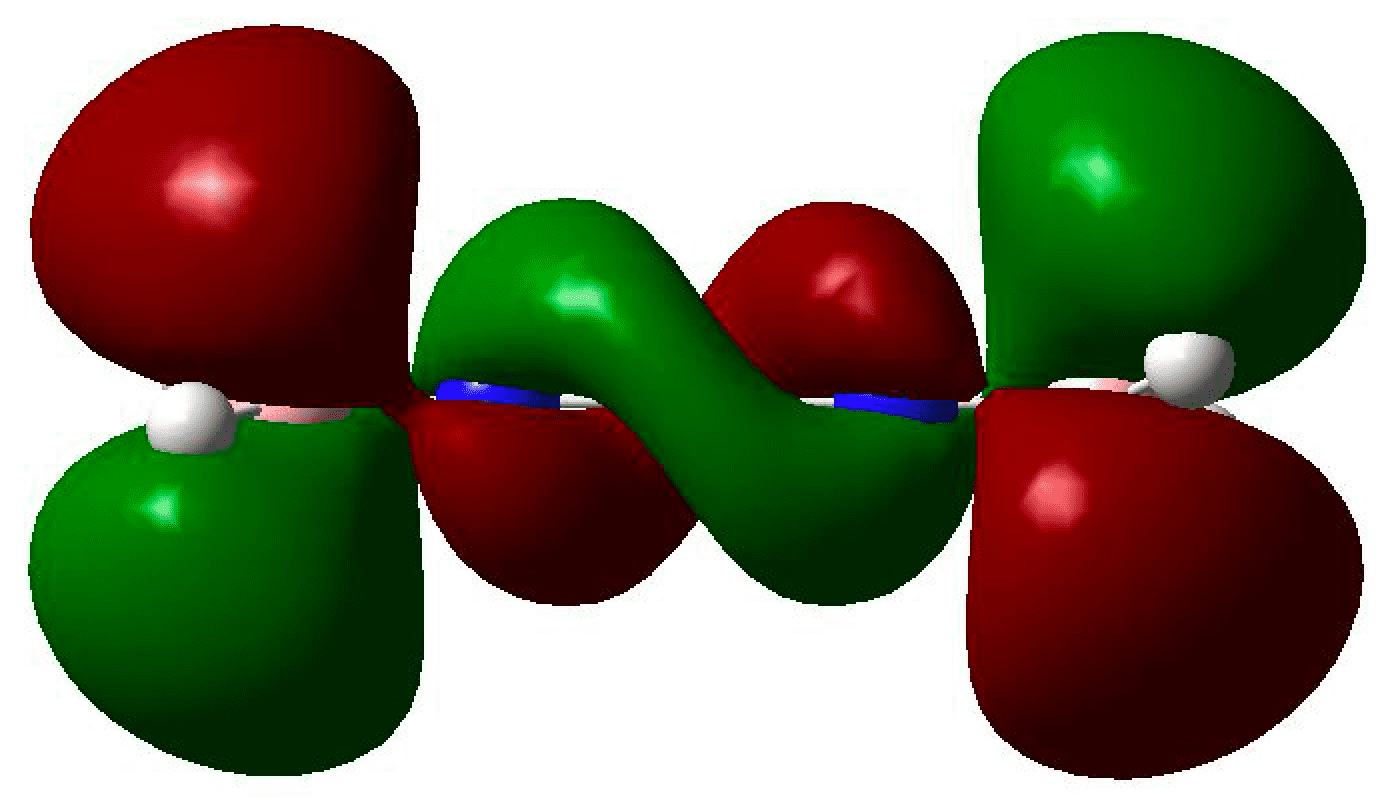}} \\ 
    & $\beta$-{HOMO-1} & $\beta$-{HOMO} & $\beta$-{LUMO} & $\beta$-{LUMO+1} \\
  & {\includegraphics[height=1.40cm]{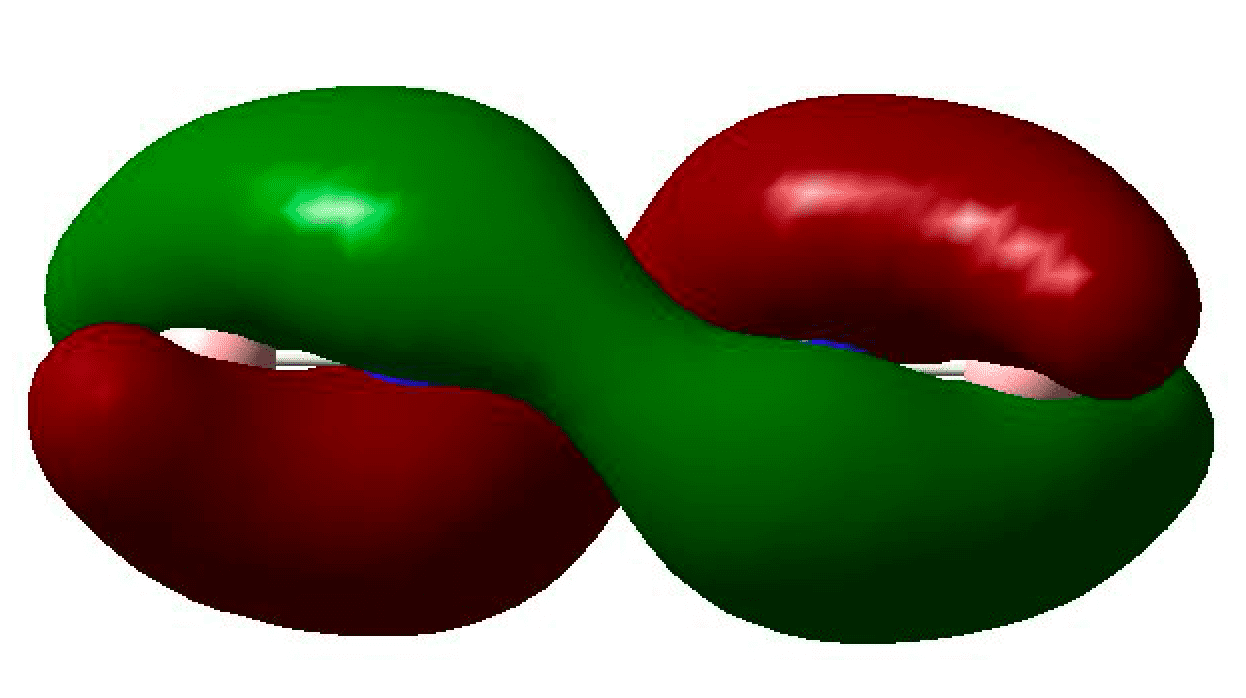}} &
   {\includegraphics[height=1.40cm]{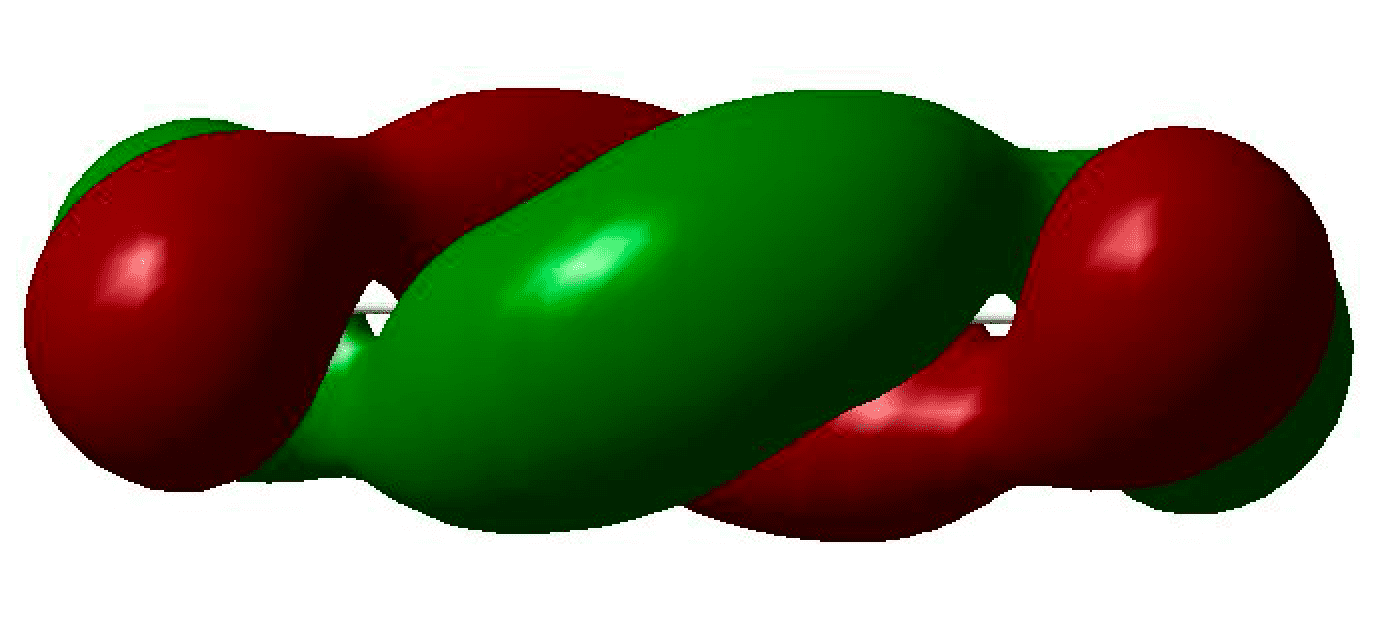}} &
   {\includegraphics[height=1.40 cm]{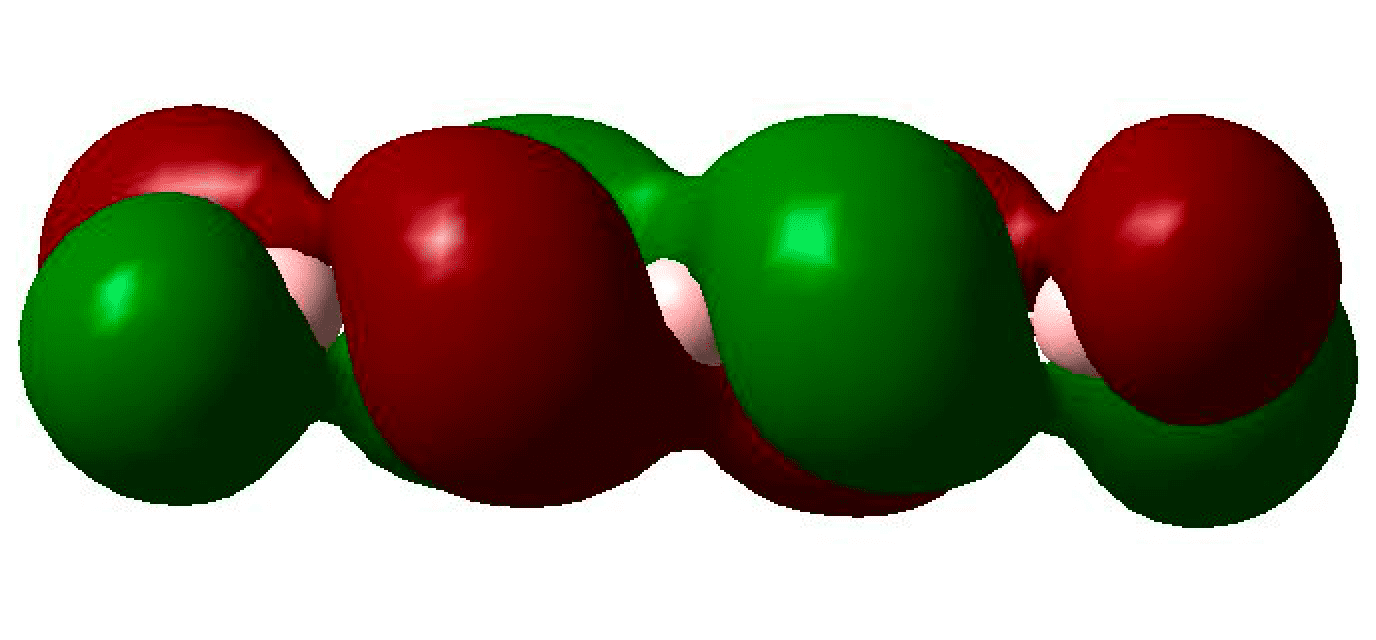}} & 
   {\includegraphics[height=1.40cm]{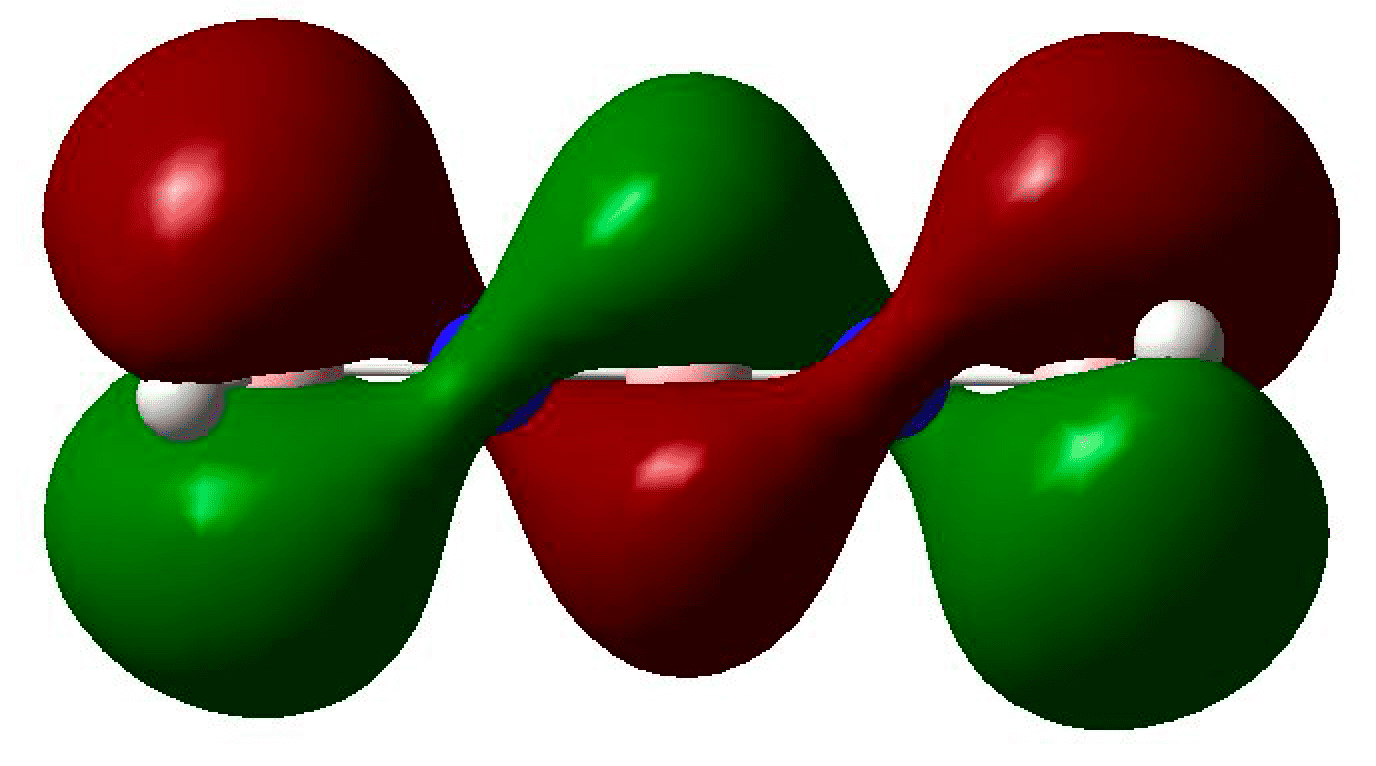}} \\ 
     \hdashline[1pt/1pt]
     
       & $\alpha$-{HOMO-1} & $\alpha$-{HOMO} & $\alpha$-{LUMO} & $\alpha$-{LUMO+1} \\
    $c = 0$ - quadruplet &
    {\includegraphics[height=1.30cm]{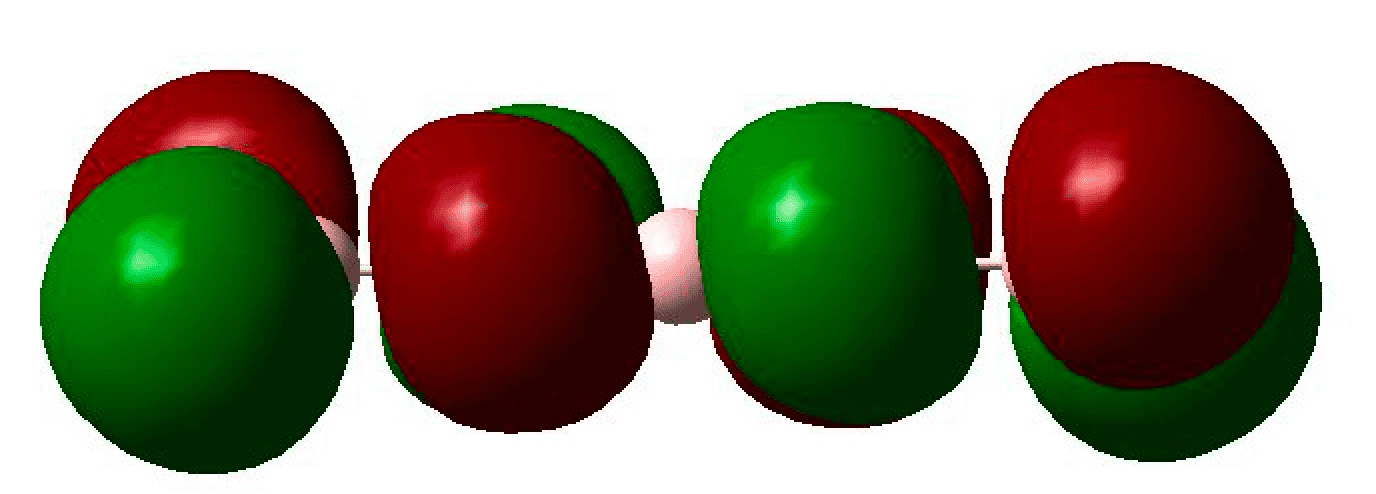}} &
    {\includegraphics[height=1.40cm]{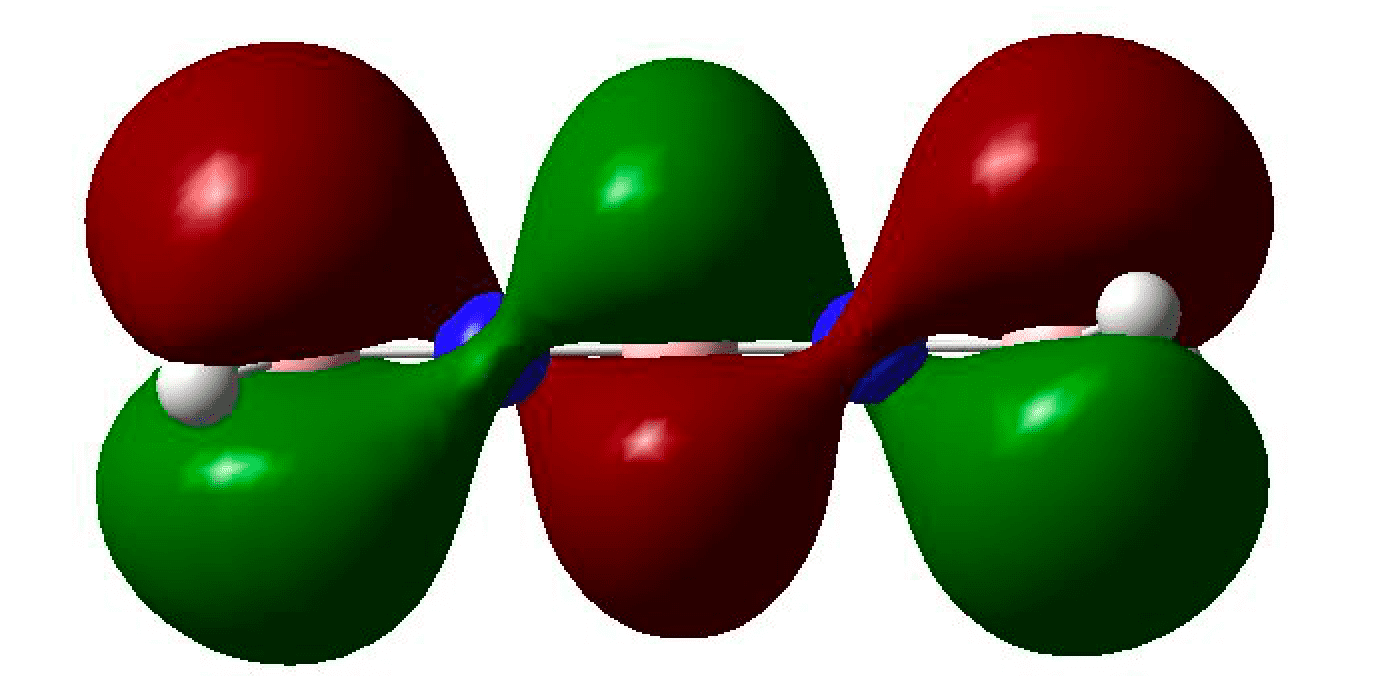}} &
    {\includegraphics[height=1.40 cm]{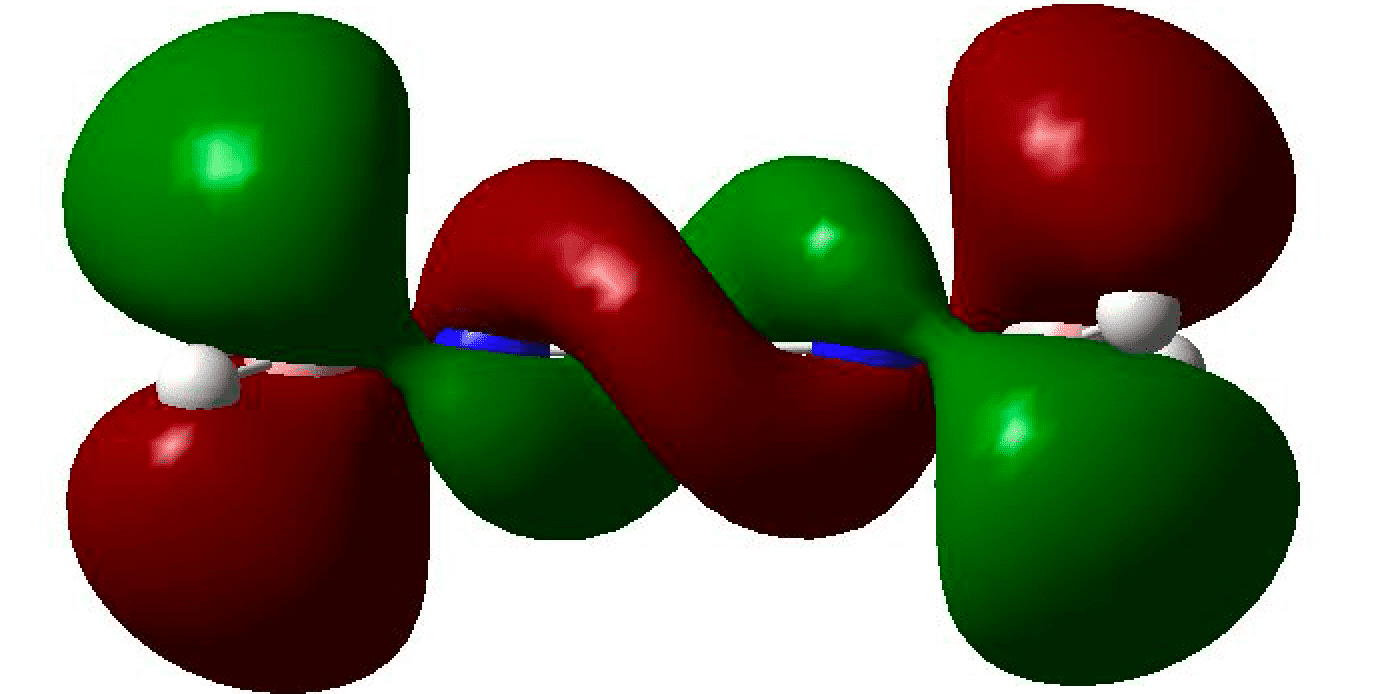}} & 
    {\includegraphics[height=1.40cm]{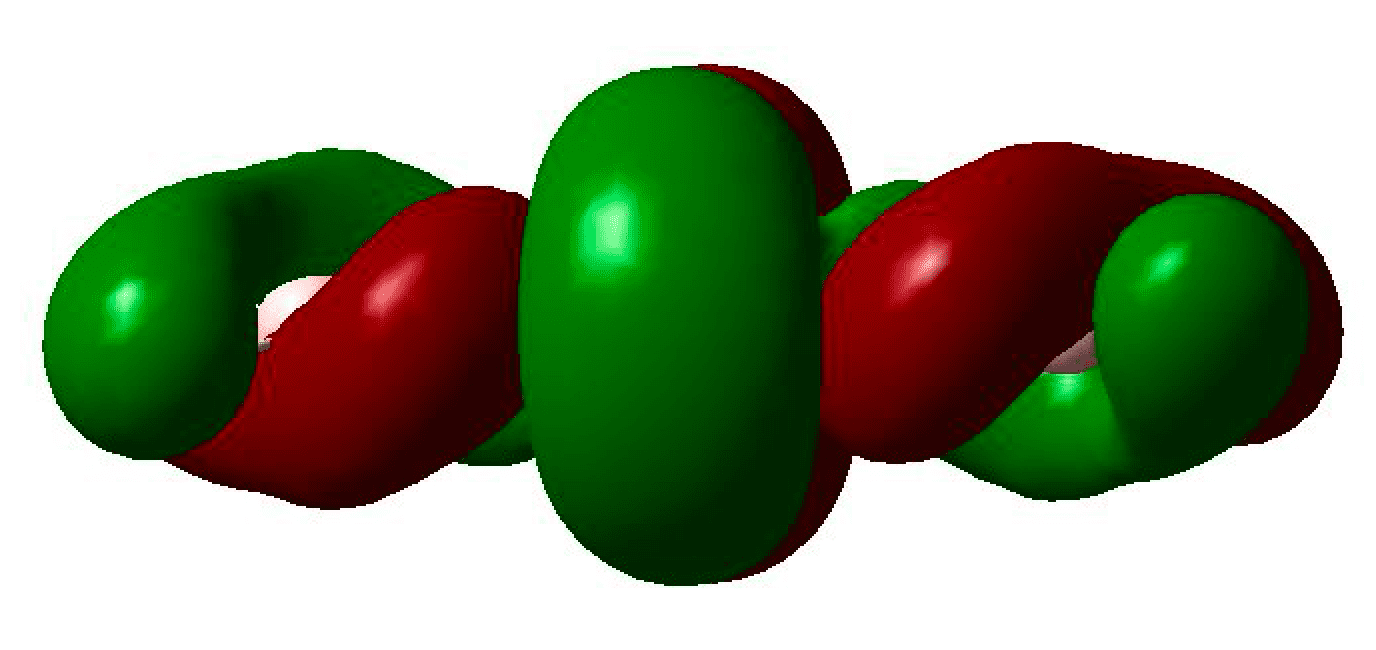}} \\ 
      & $\beta$-{HOMO-1} & $\beta$-{HOMO} & $\beta$-{LUMO} & $\beta$-{LUMO+1} \\
    & {\includegraphics[height=1.40cm]{B-N-5-m2-b-homo-1.png}} &
    {\includegraphics[height=1.40cm]{B-N-5-m2-b-homo.png}} &
    {\includegraphics[height=1.40 cm]{B-N-5-m2-b-lumo.png}} & 
    {\includegraphics[height=1.40cm]{B-N-5-m2-b-lumo+1.png}} \\ 
     \bottomrule
     \end{tabular}
      \end{table}
      
     \begin{table}[H]
    \caption{Molecular orbitals of a linear chain \textbf{N=B=N=B=N} obtained at the B3LYP/6-311G(d.p) level of theory. c = charge} 
    \label{table2-mo-BN}
    \centering
    \begin{tabular}{cccccc}
        \midrule
       & HOMO-1 & HOMO & LUMO & LUMO+1 \\
      c = 1 - singlet  & 
    {\includegraphics[height=1.30cm]{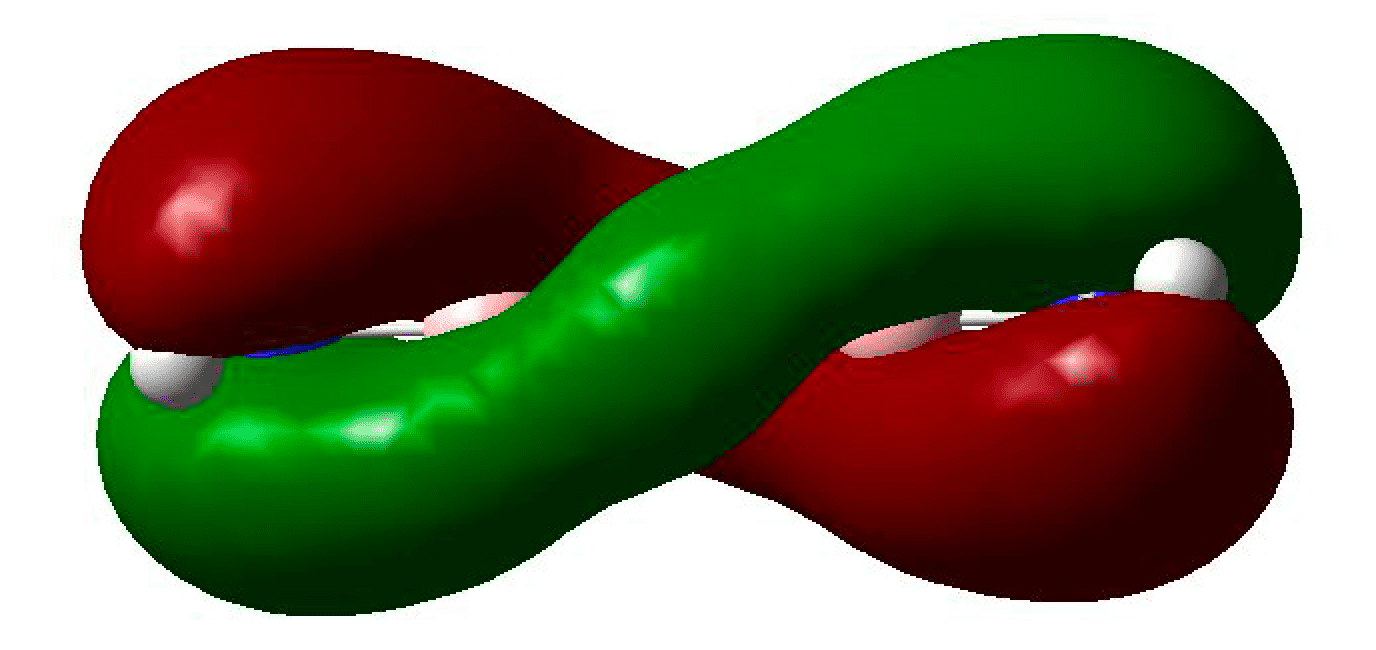}} &
    {\includegraphics[height=1.30cm]{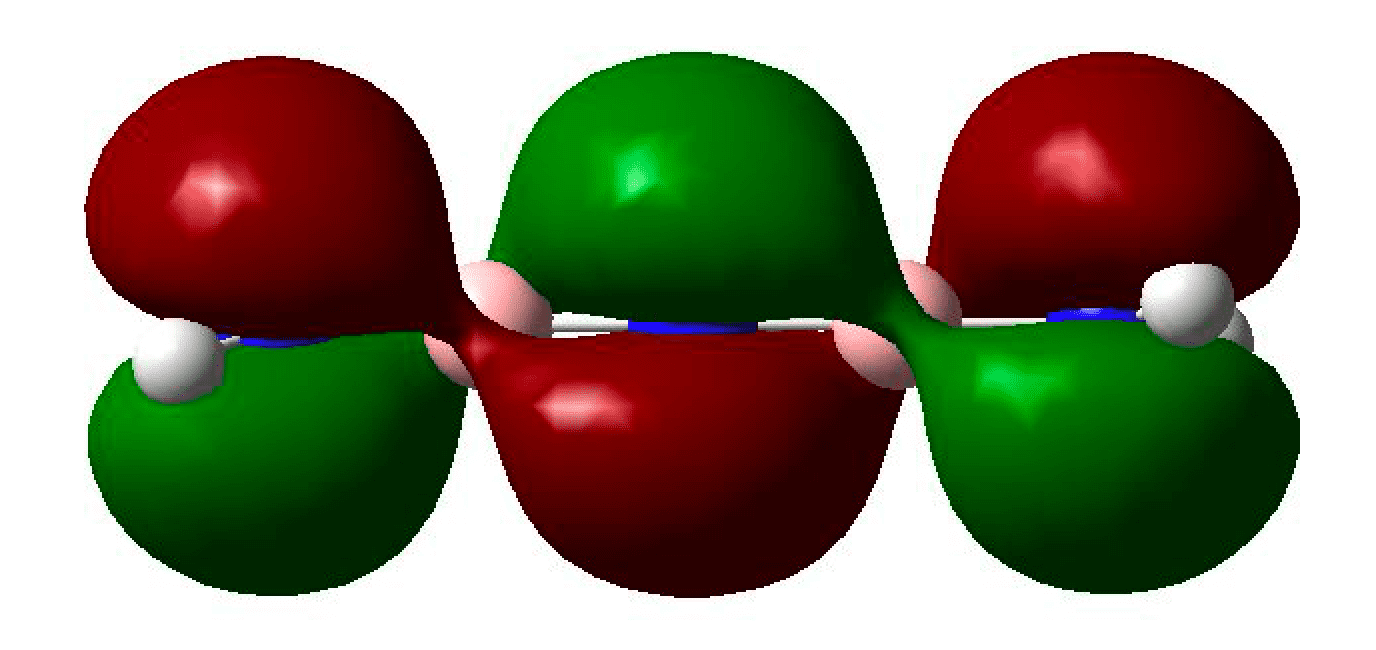}} &
    {\includegraphics[height=1.30 cm]{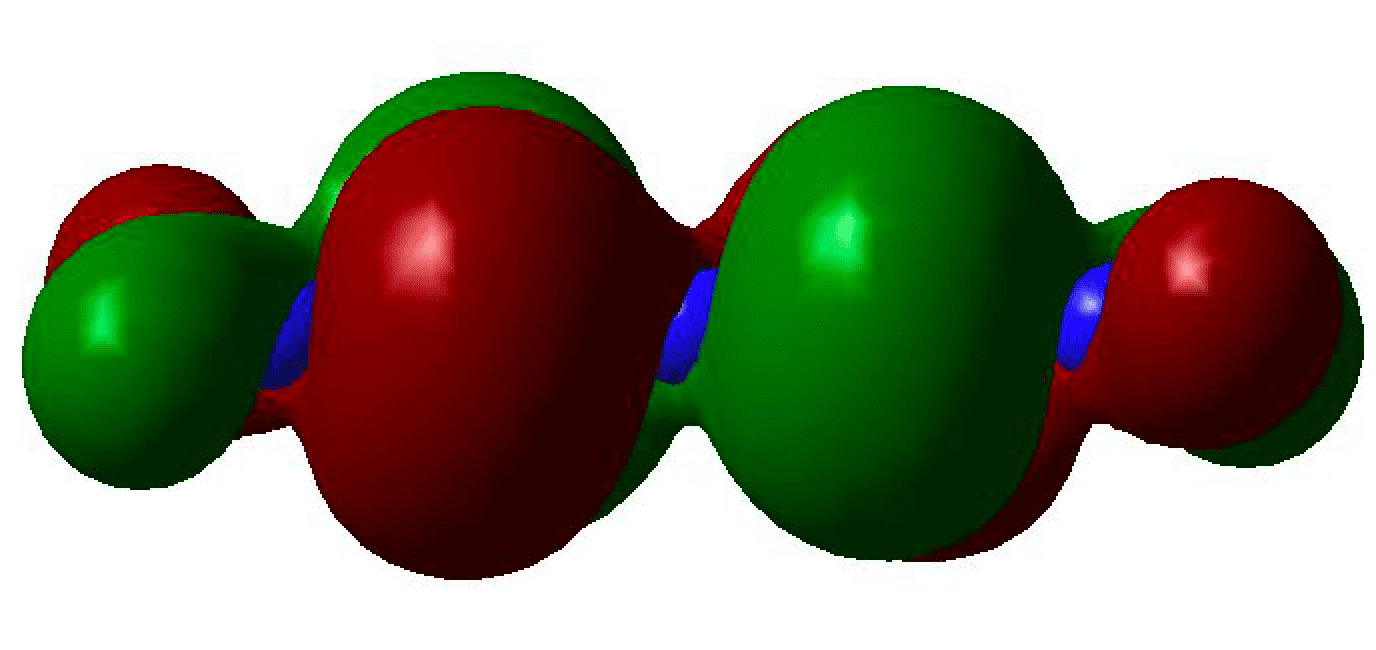}} & 
    {\includegraphics[height=1.50cm]{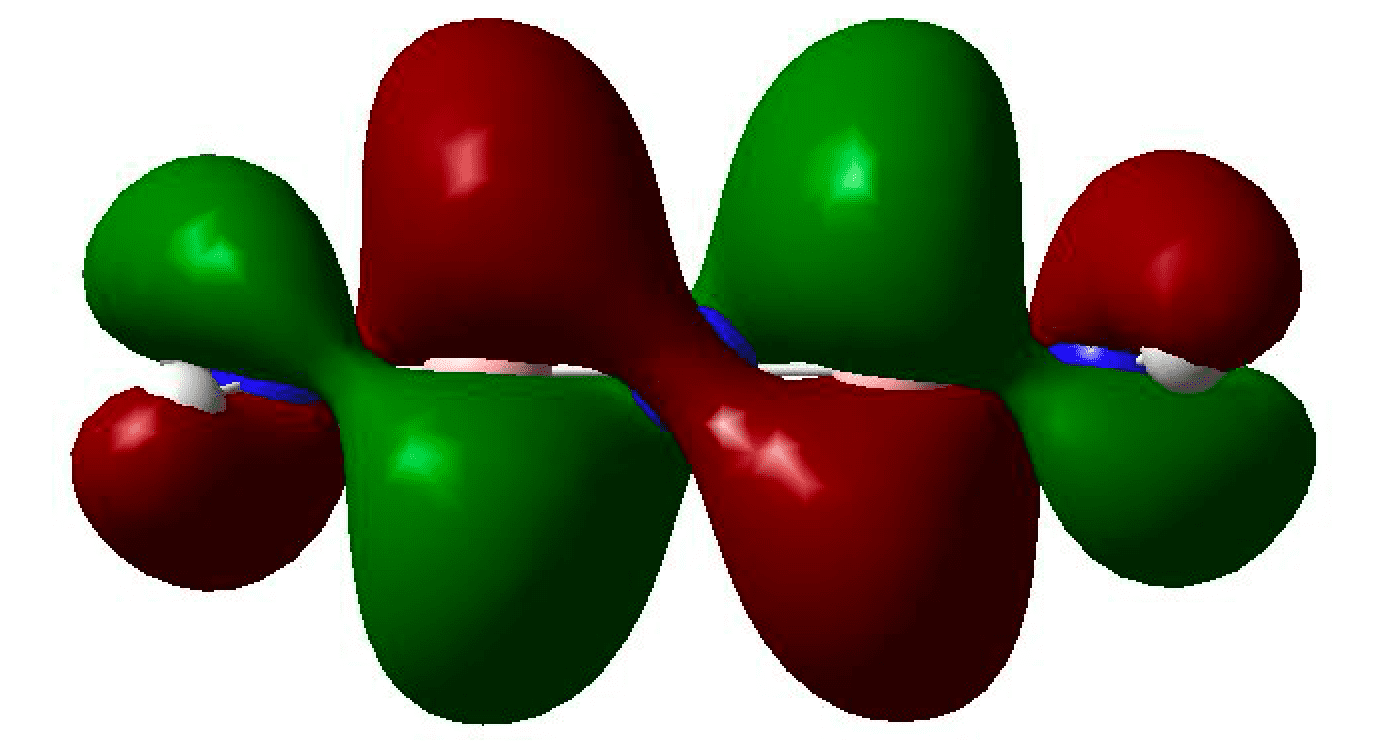}} \\ 
      \hdashline[1pt/1pt]
     & $\alpha$-HOMO-1 & $\alpha$-HOMO & $\alpha$-LUMO & $\alpha$-LUMO+1 \\
     c = 1 - triplet &
    {\includegraphics[height=1.40cm]{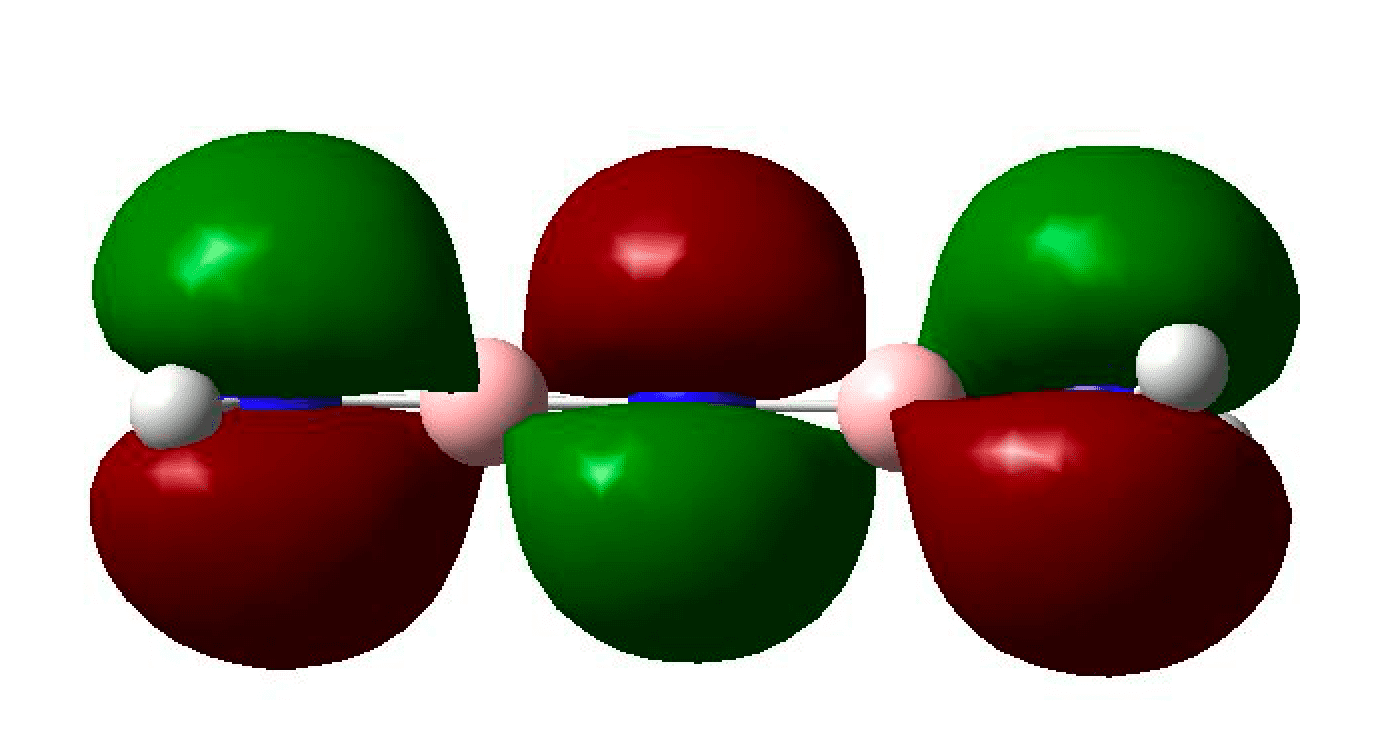}} &
    {\includegraphics[height=1.50cm]{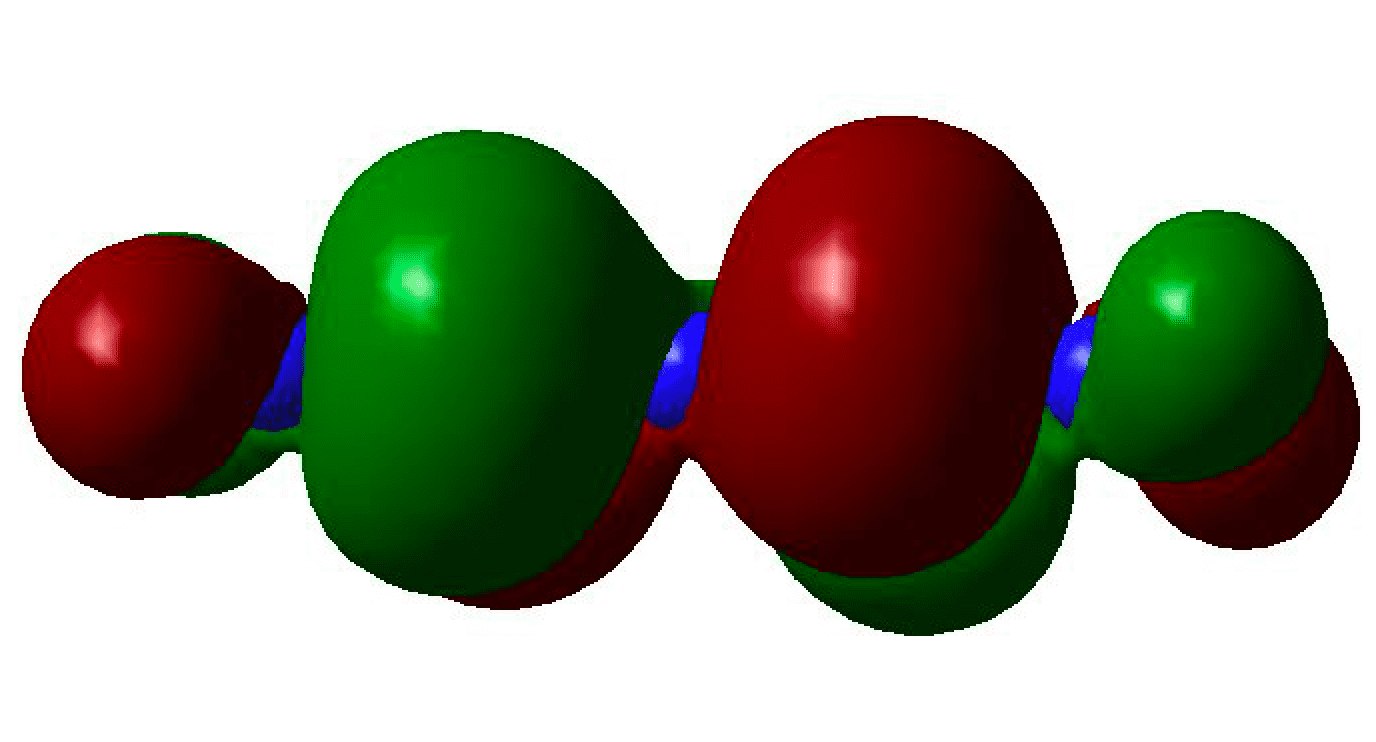}} &
    {\includegraphics[height=1.50cm]{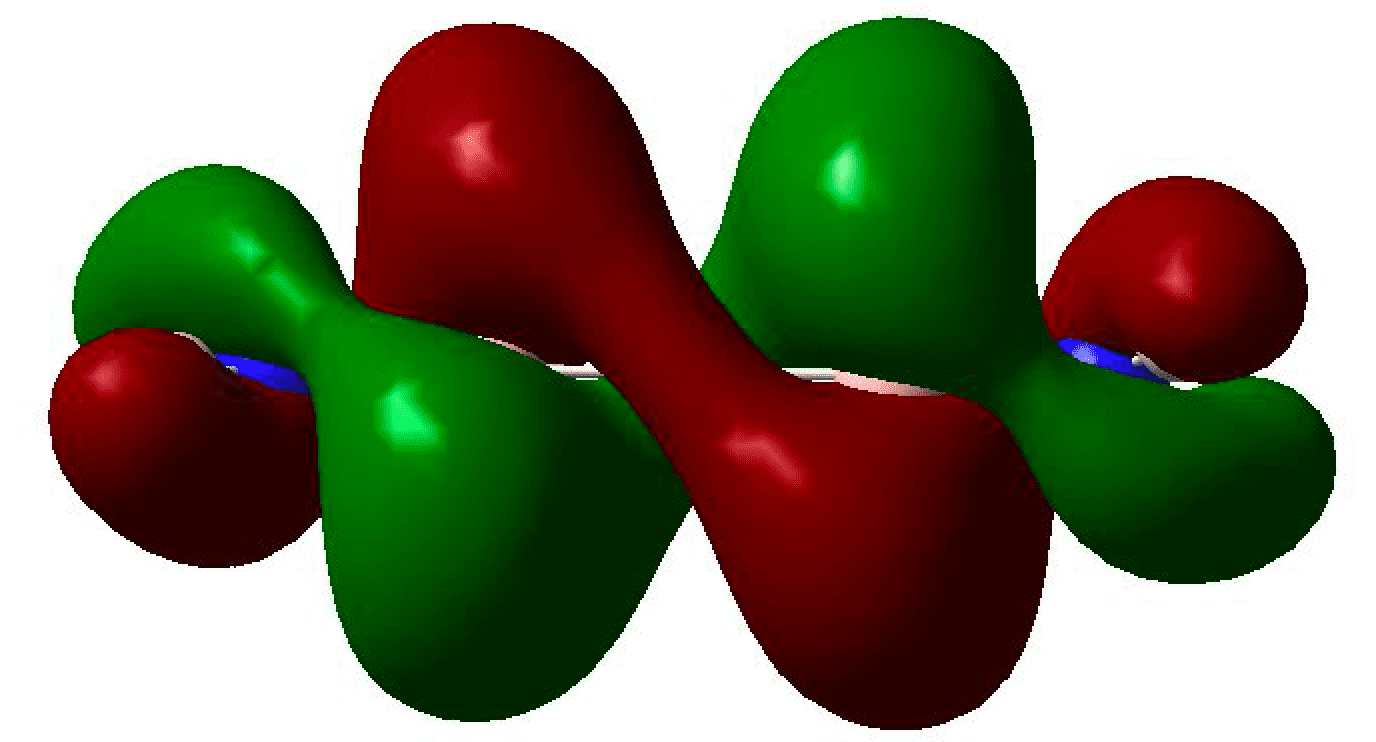}} & 
    {\includegraphics[height=1.50cm]{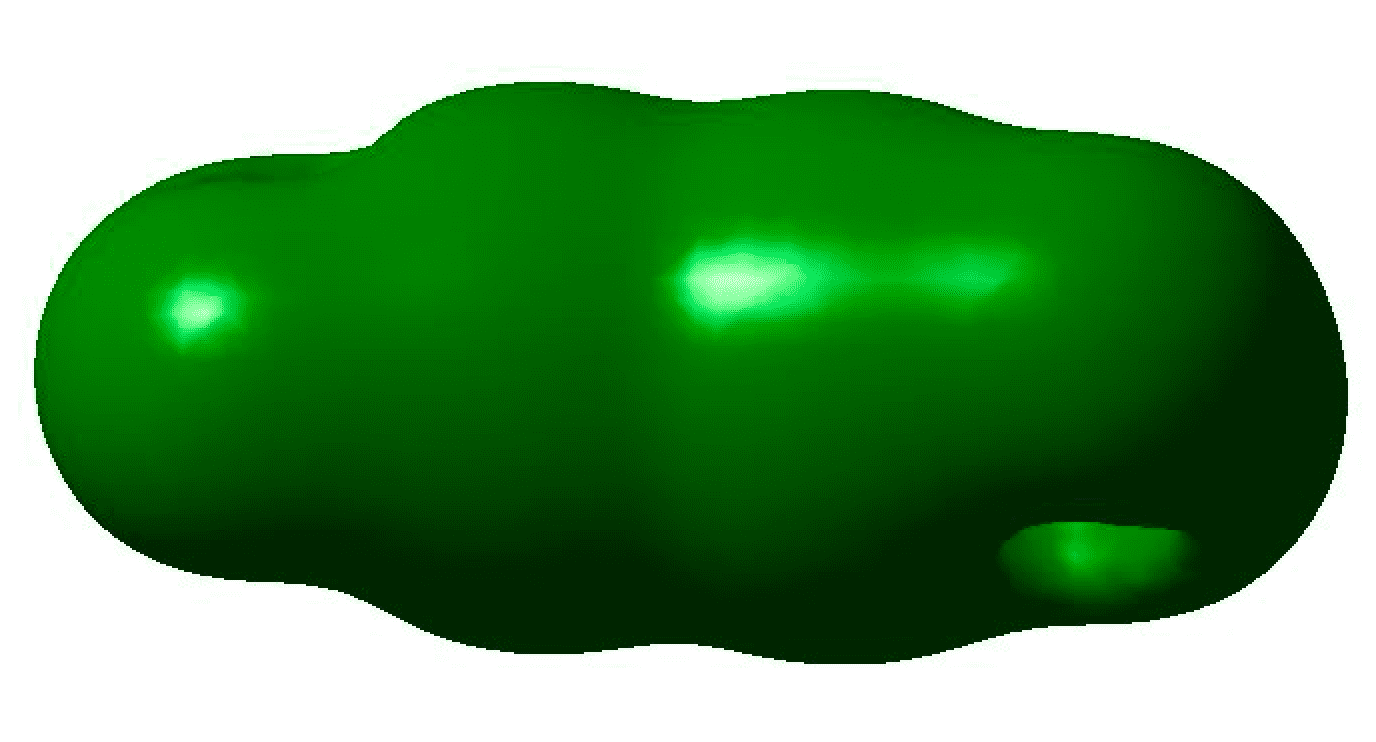}} \\ 
    & $\beta$-HOMO-1 & $\beta$-HOMO & $\beta$-LUMO & $\beta$-LUMO+1 \\
     &
    {\includegraphics[height=1.50cm]{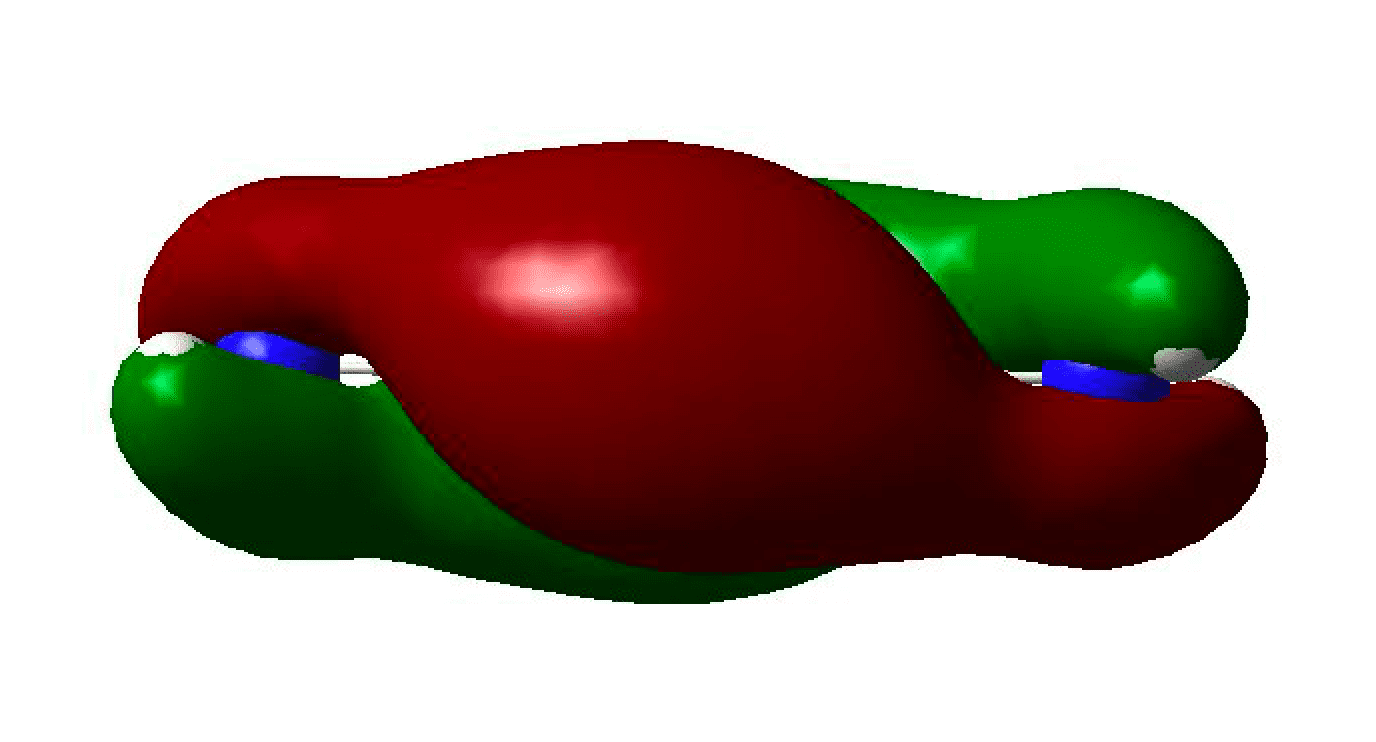}} &
    {\includegraphics[height=1.50cm]{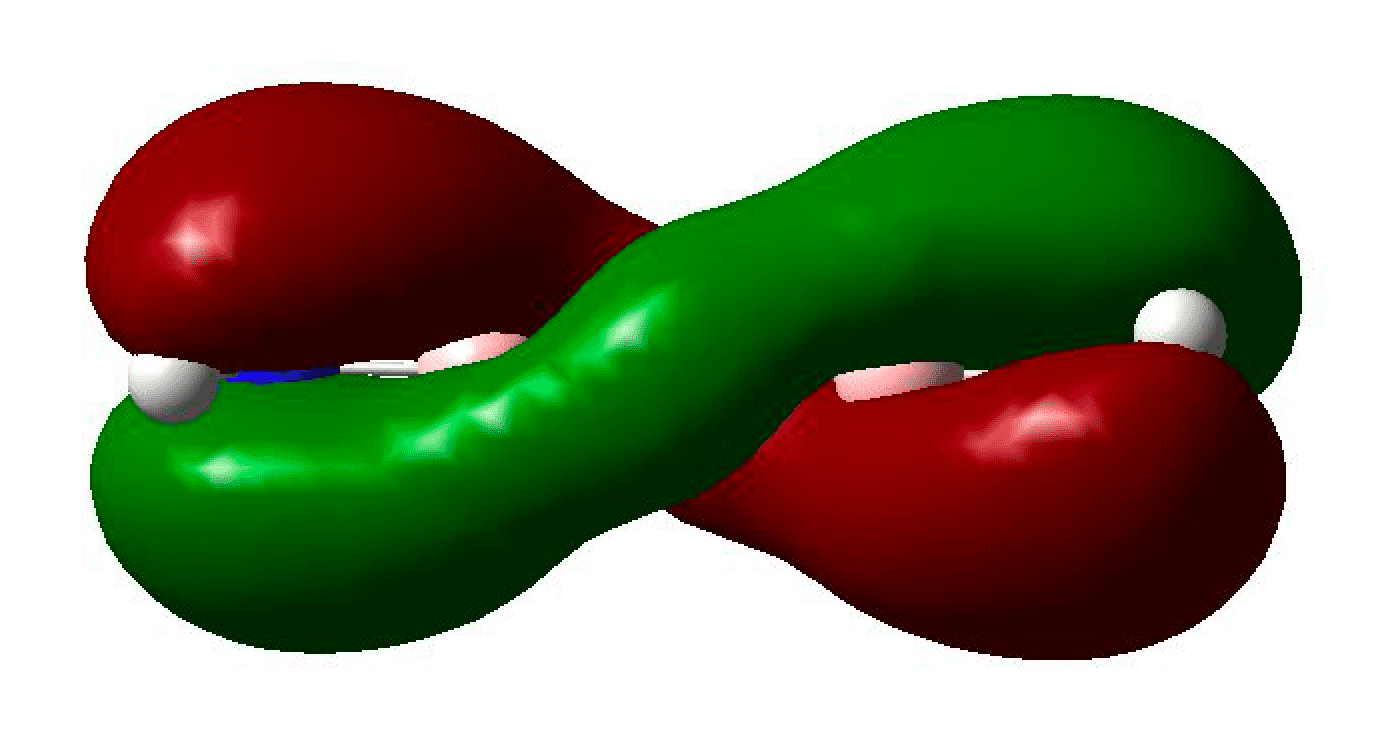}} &
    {\includegraphics[height=1.50cm]{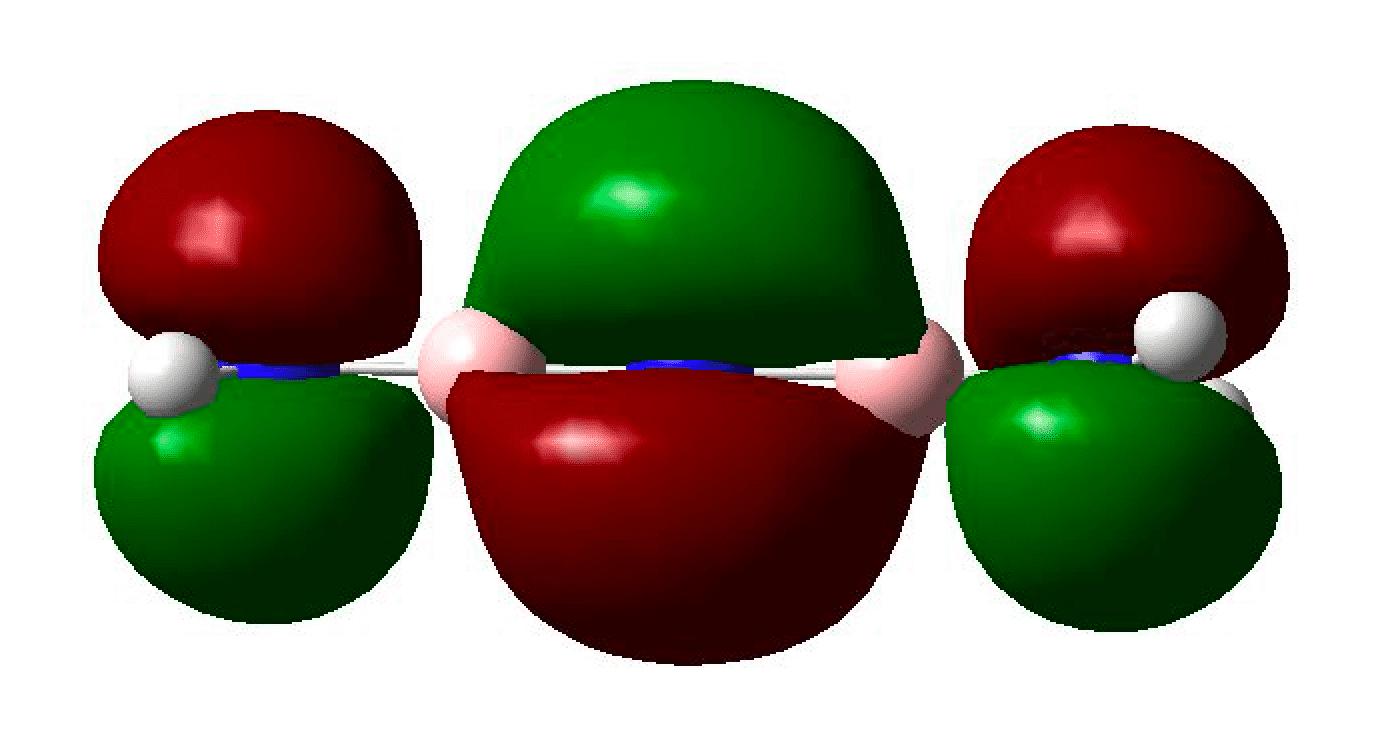}} & 
    {\includegraphics[height=1.50cm]{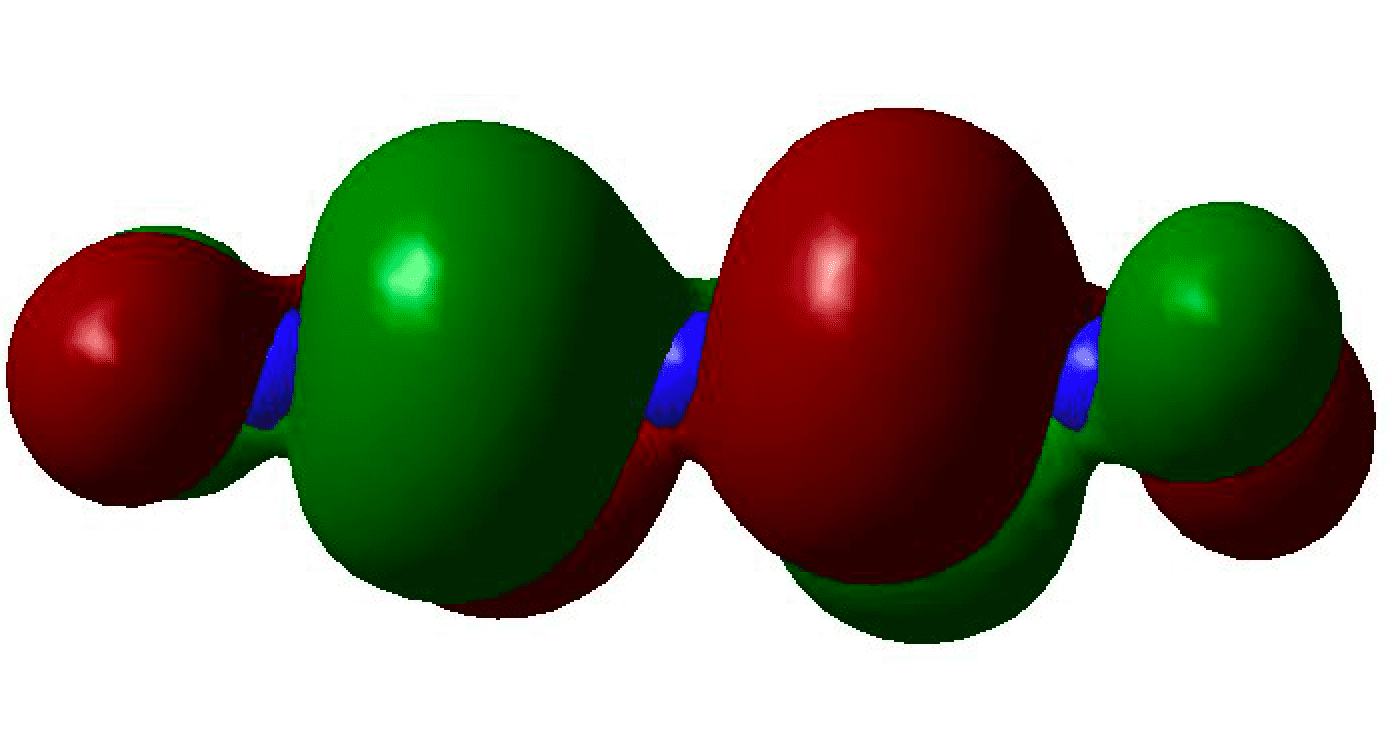}} \\
    \bottomrule
     \end{tabular}
      \end{table}
      
      \begin{table}[H]
    \caption{Molecular orbitals of a linear chain \textbf{B=N=B=N=B=N} obtained at the B3LYP/6-311G(d.p) level of theory. c = charge} 
    \label{table3-mo-BN}
    \centering
    \begin{tabular}{cccccc}
        \midrule 
     & HOMO-1 & HOMO & LUMO & LUMO+1 \\
   c = 0 - singlet &
    {\includegraphics[height=1.30cm]{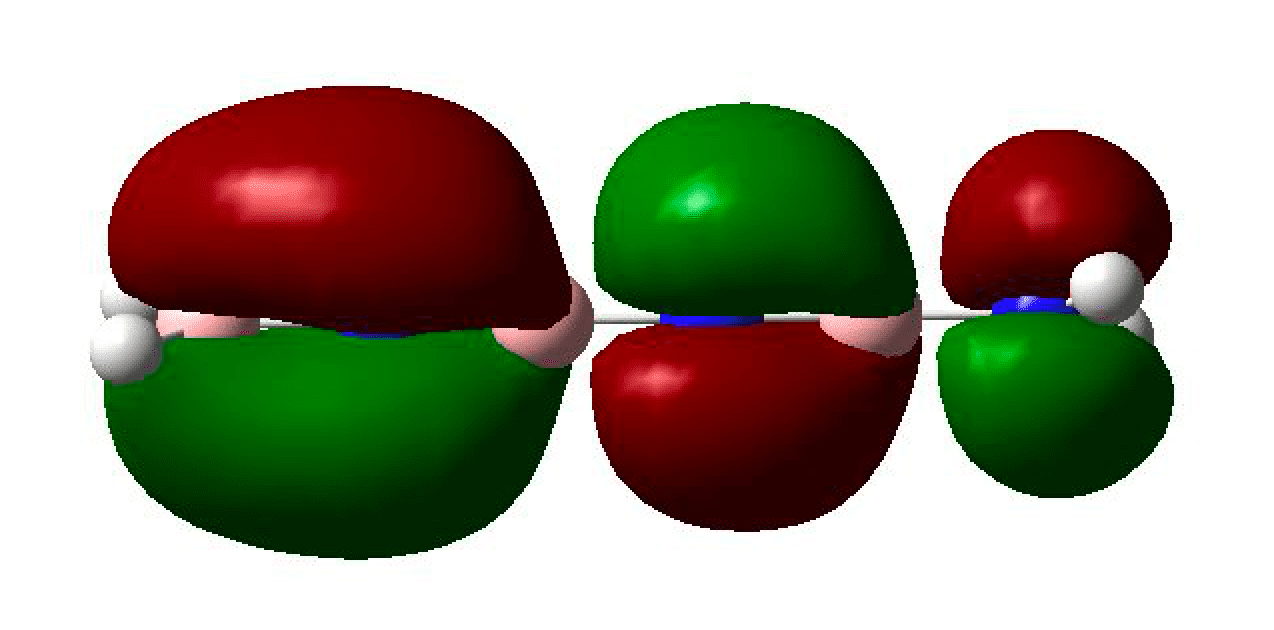}} &
    {\includegraphics[height=1.40cm]{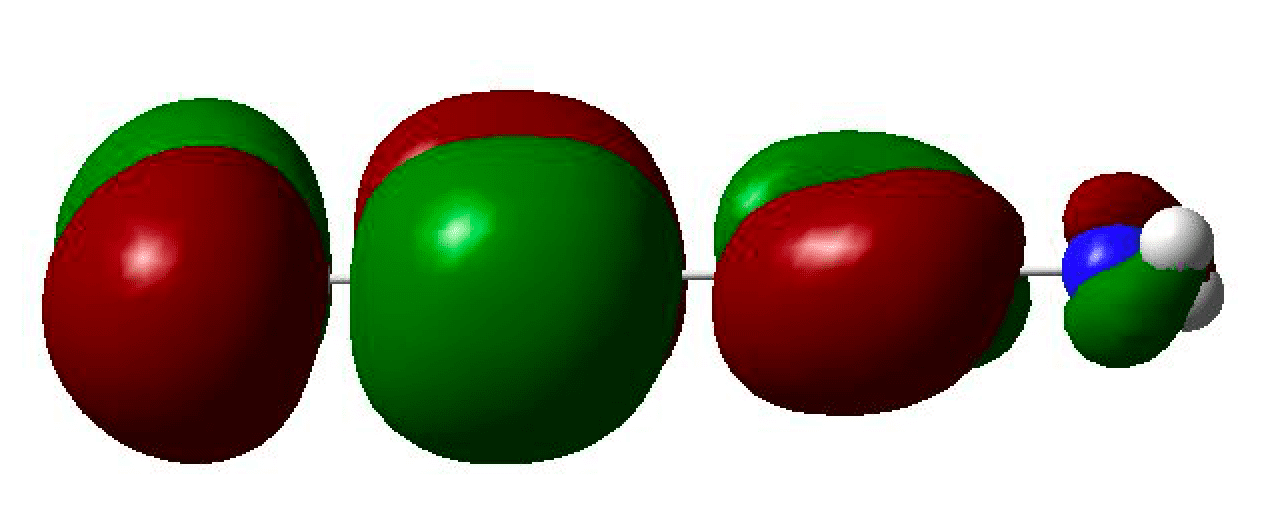}} &
    {\includegraphics[height=1.40 cm]{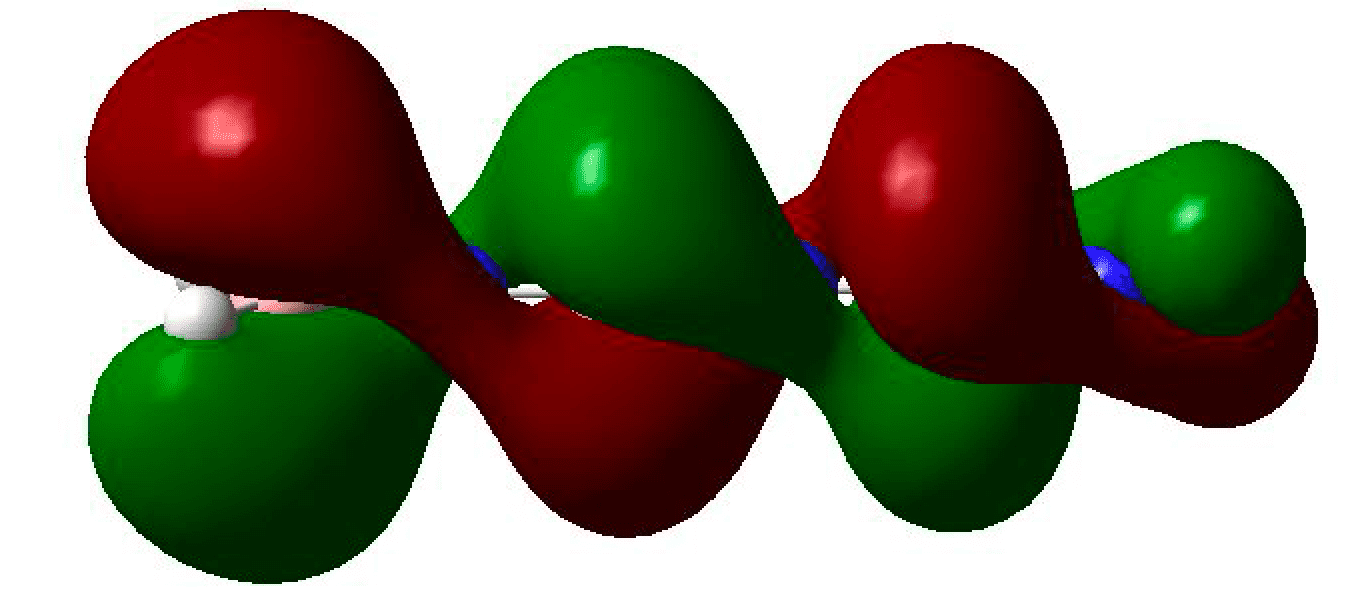}} & 
    {\includegraphics[height=1.40cm]{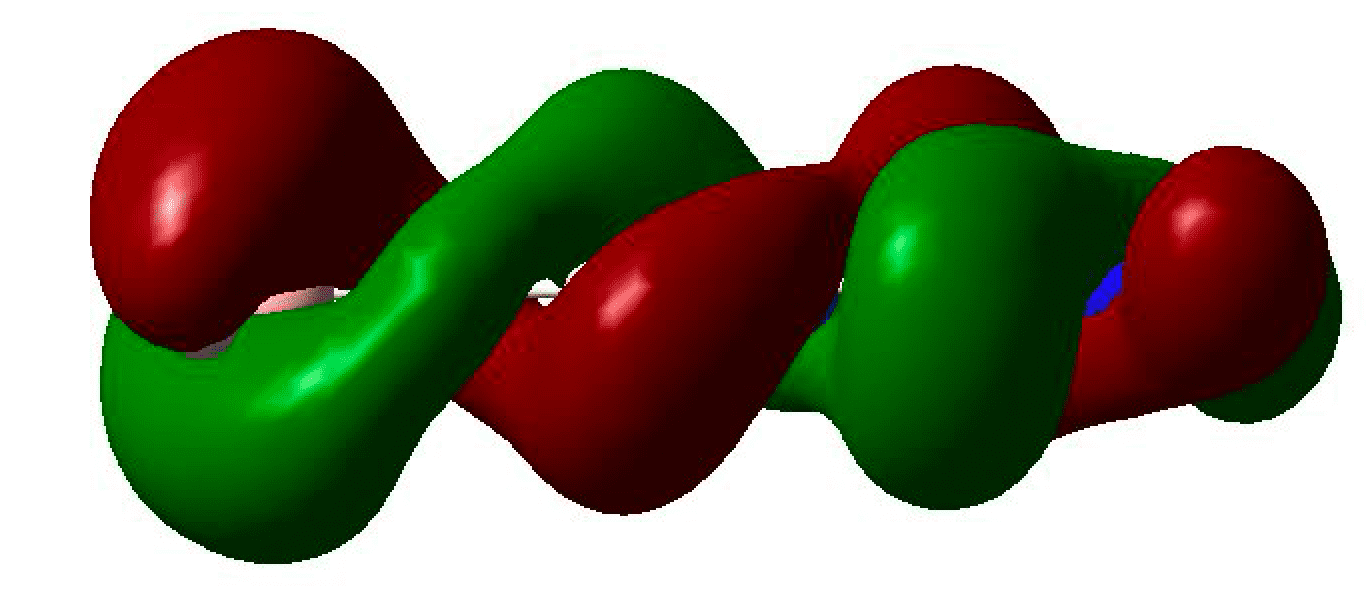}} \\ 
    \hdashline[1pt/1pt]
     & $\alpha$-HOMO-1 & $\alpha$-HOMO & $\alpha$-LUMO & $\alpha$-LUMO+1 \\
     c = 1 - doublet &
    {\includegraphics[height=1.30cm]{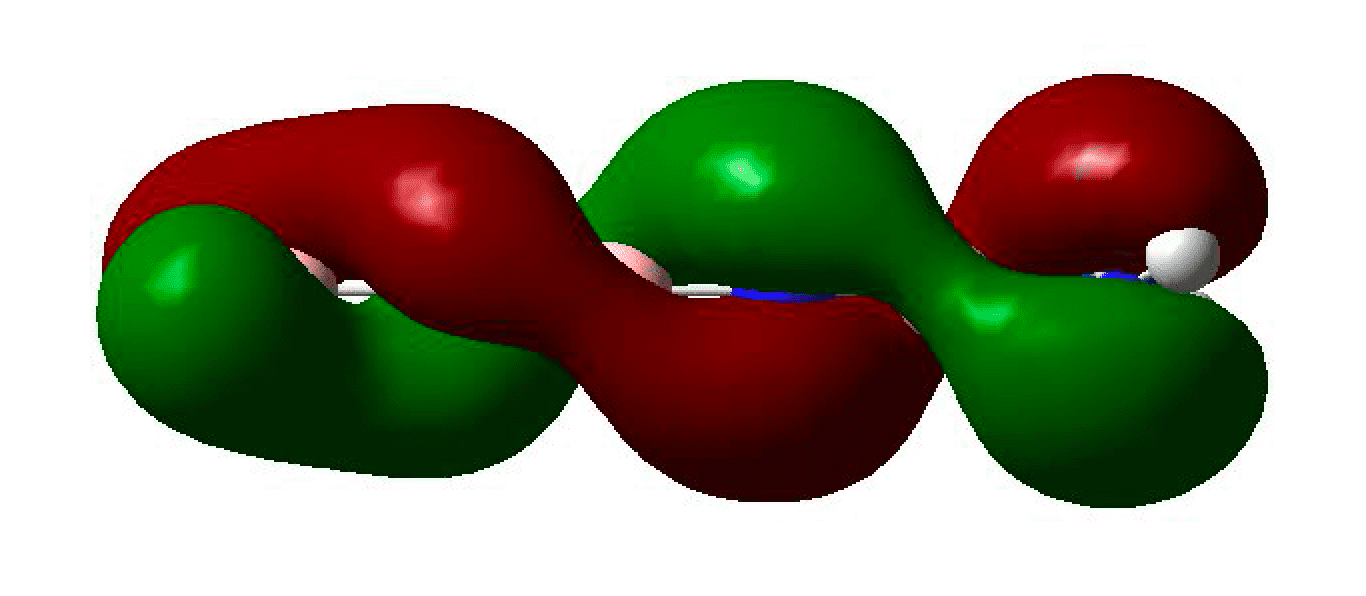}} &
    {\includegraphics[height=1.40cm]{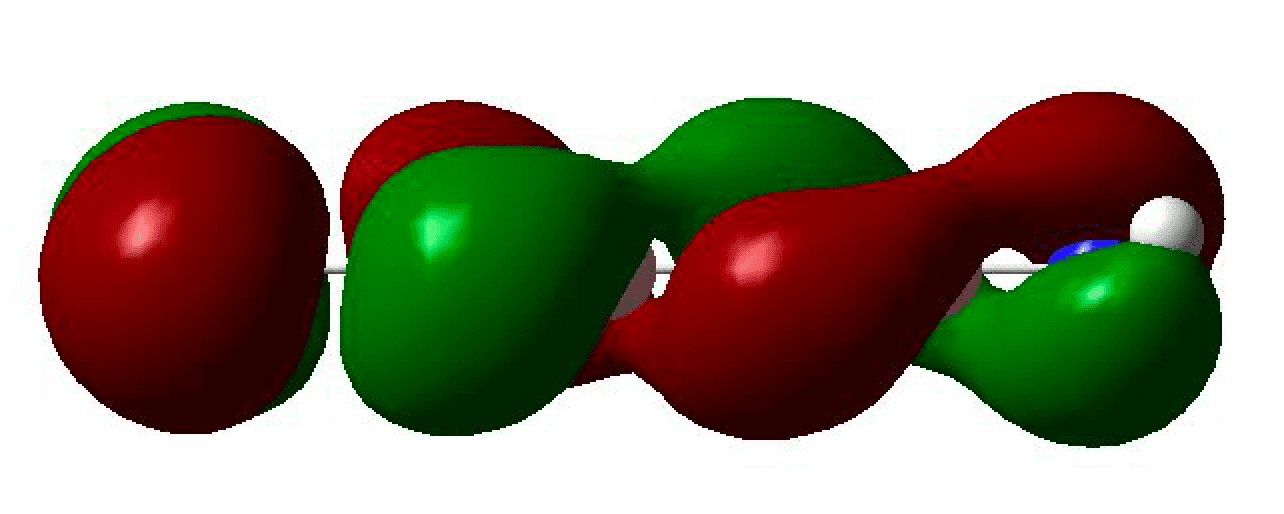}} &
    {\includegraphics[height=1.40cm]{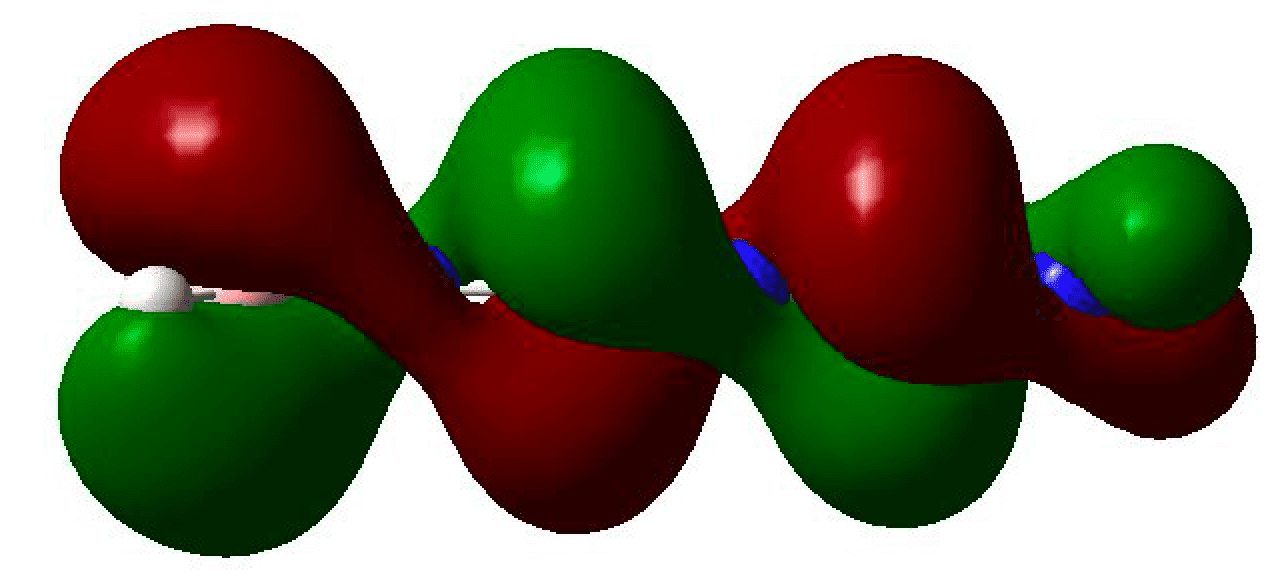}} & 
    {\includegraphics[height=1.40cm]{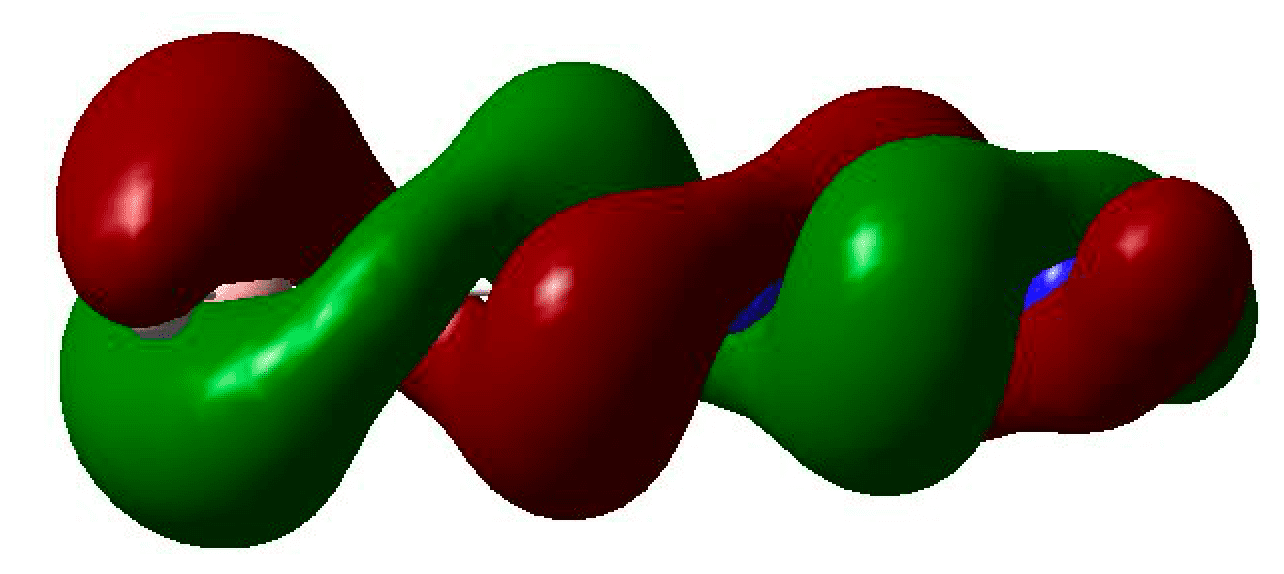}} \\ 

    & $\beta$-HOMO-1 & $\beta$-HOMO & $\beta$-LUMO & $\beta$-LUMO+1 \\
     &
    {\includegraphics[height=1.30cm]{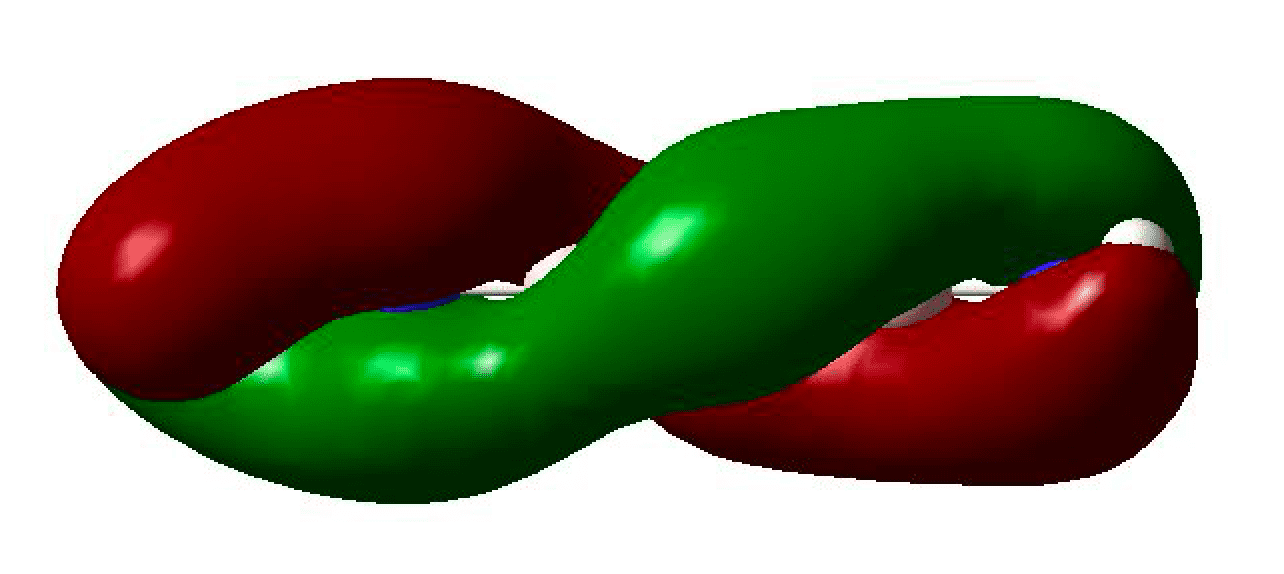}} &
    {\includegraphics[height=1.40cm]{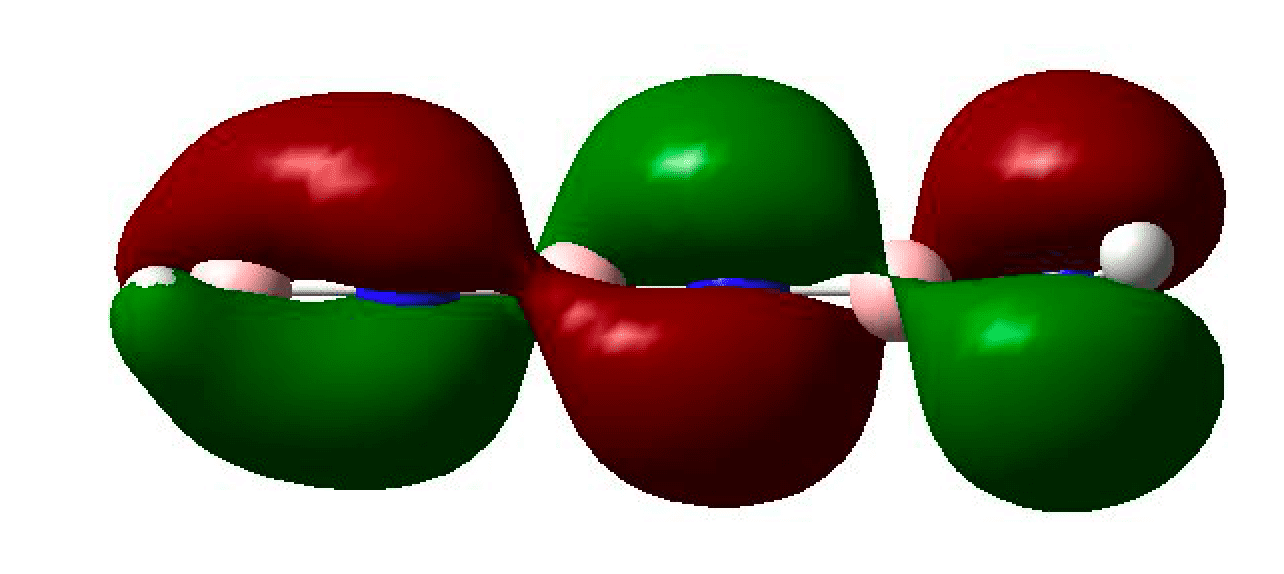}} &
    {\includegraphics[height=1.40cm]{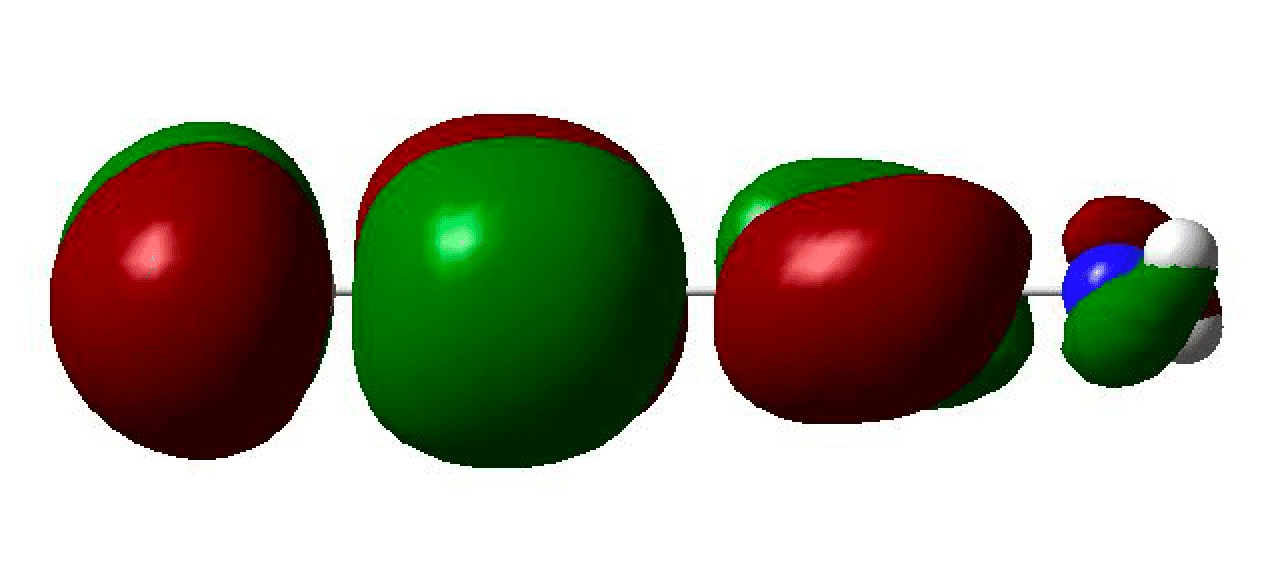}} & 
    {\includegraphics[height=1.40cm]{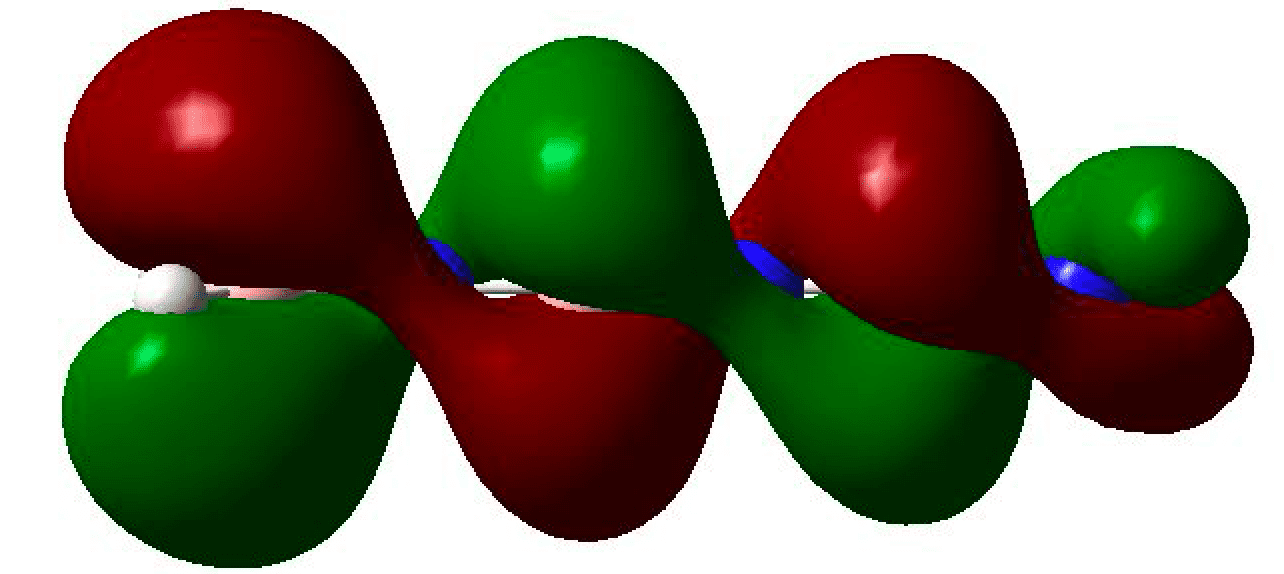}} \\
     \hdashline[1pt/1pt]
    
     & $\alpha$-HOMO-1 & $\alpha$-HOMO & $\alpha$-LUMO & $\alpha$-LUMO+1 \\
     c = 2 - triplet &
     {\includegraphics[height=1.30cm]{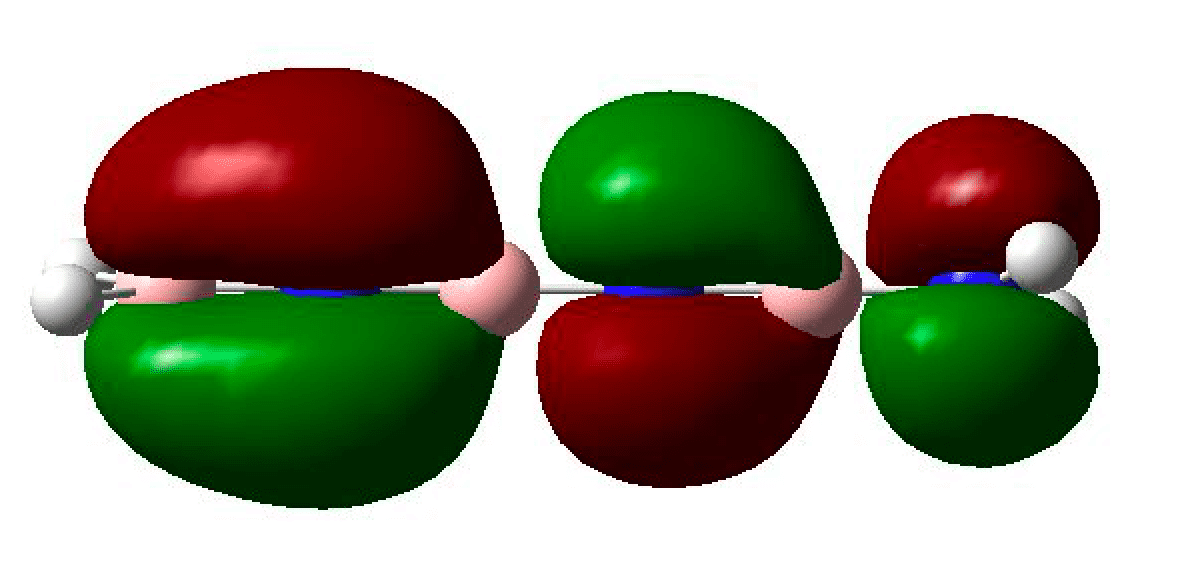}} &
    {\includegraphics[height=1.40cm]{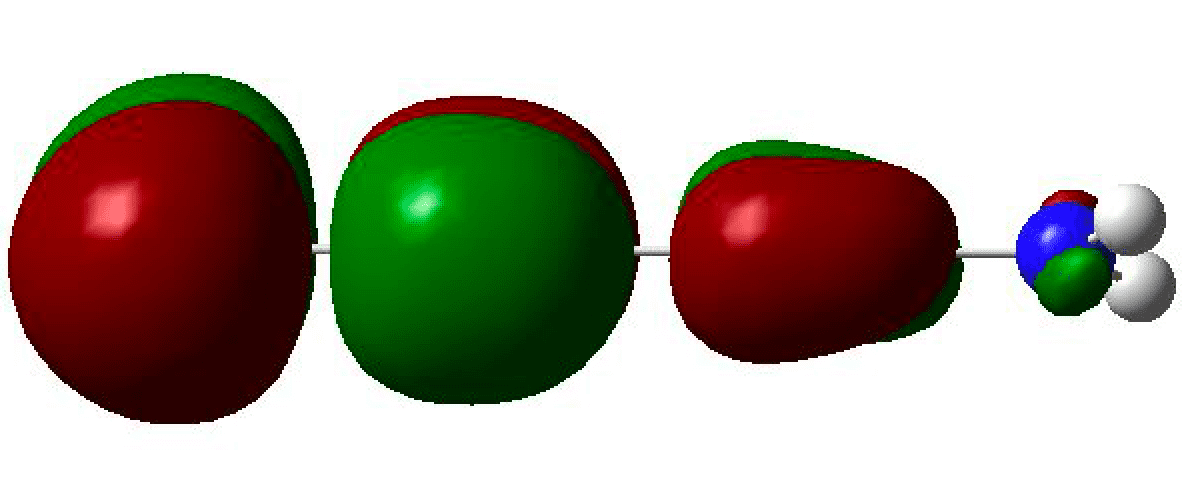}} &
    {\includegraphics[height=1.40cm]{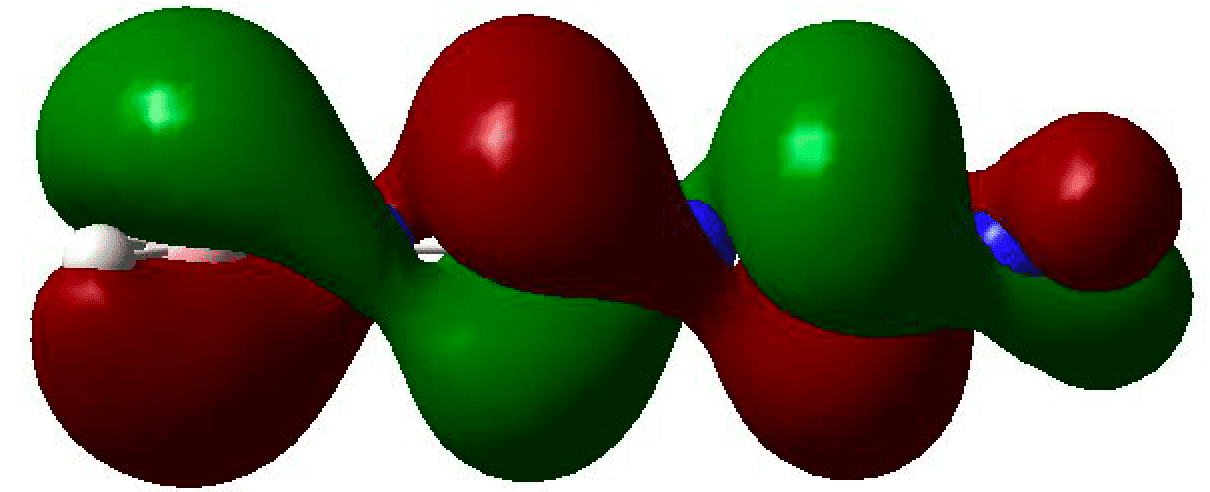}} & 
    {\includegraphics[height=1.40cm]{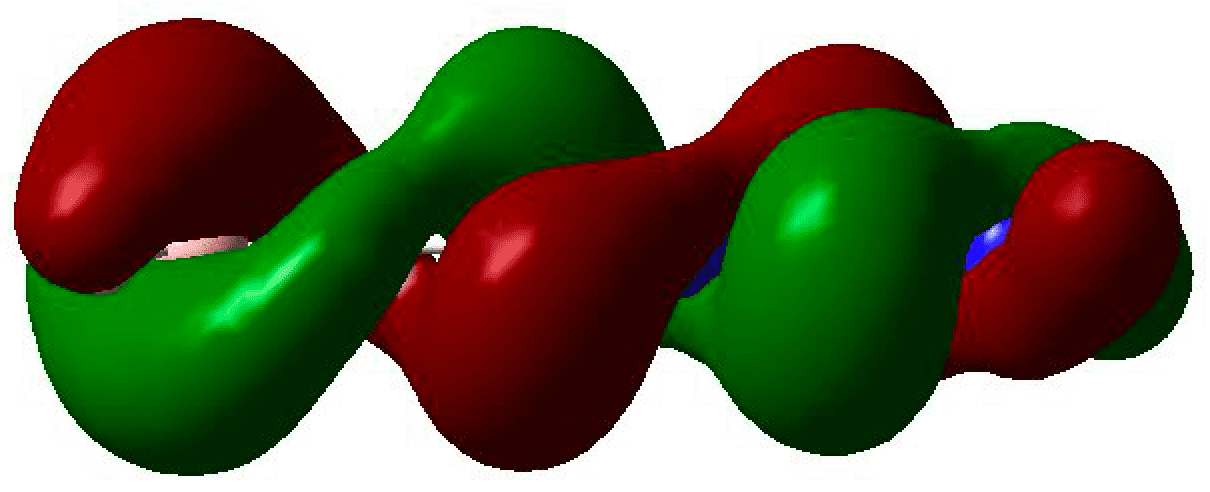}} \\ 
    & $\beta$-HOMO-1 & $\beta$-HOMO & $\beta$-LUMO & $\beta$-LUMO+1 \\
     &
    {\includegraphics[height=1.30cm]{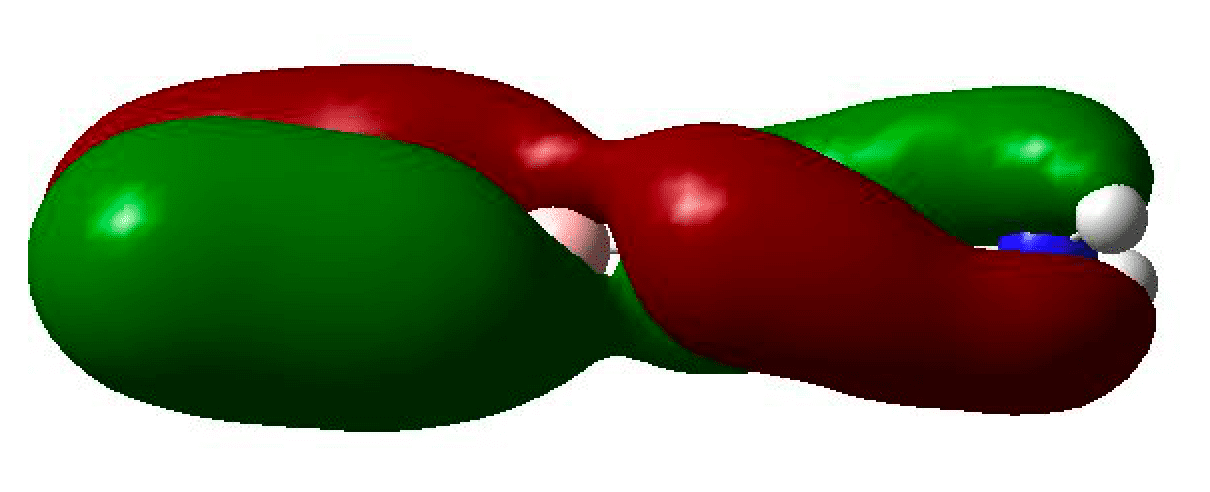}} &
    {\includegraphics[height=1.30cm]{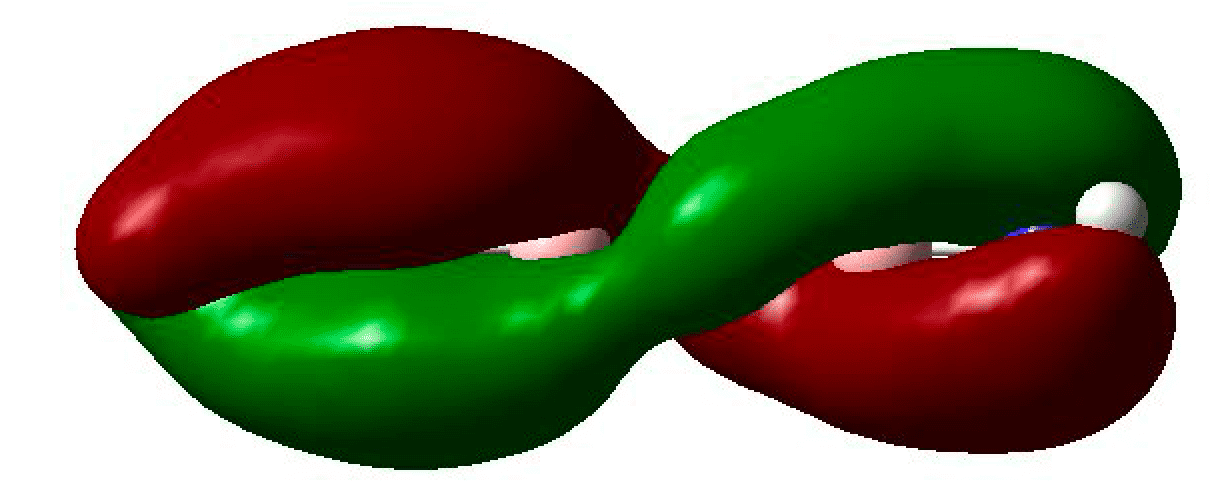}} &
    {\includegraphics[height=1.30cm]{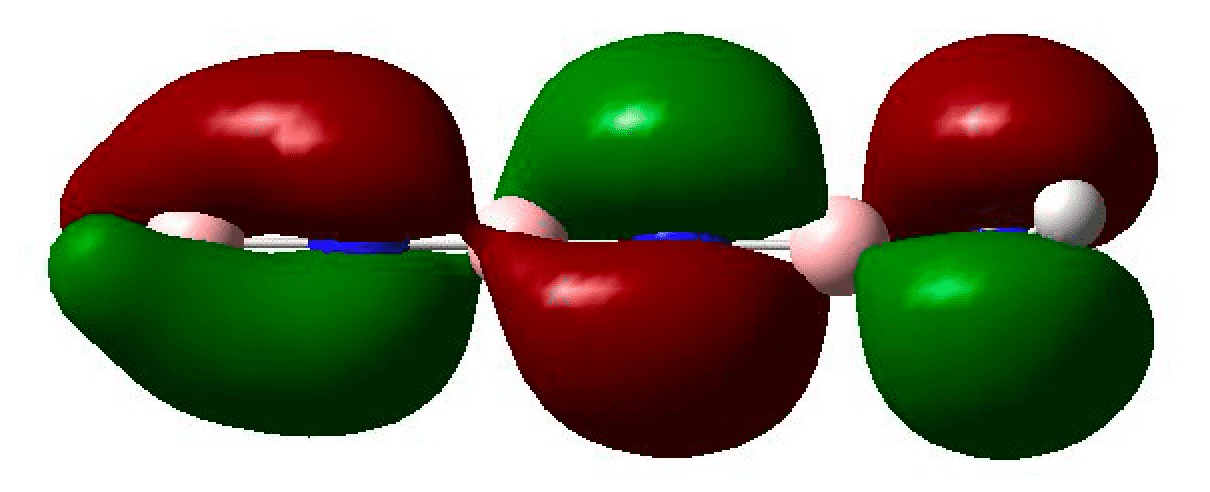}} & 
    {\includegraphics[height=1.30cm]{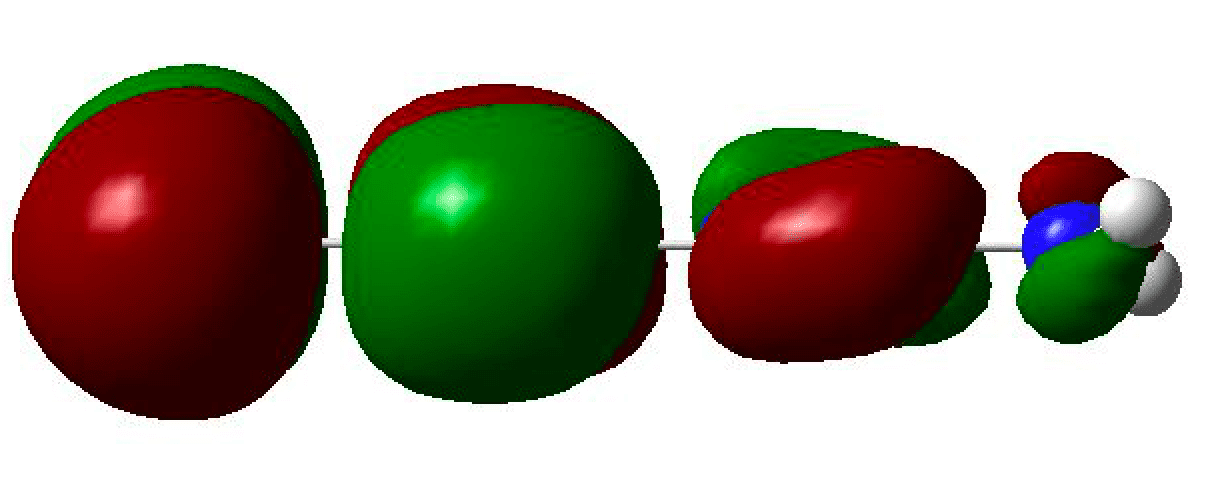}} \\
     \bottomrule
     \end{tabular}
     \end{table}

\subsection{Some properties of H\"uckel distributions for $0<\theta <\pi/2$ twisted $[N]$-cumulene}

From a theoretical point of view, helical H\"uckel distributions are the only ones that can be dealt with. A complete treatment can be done for twisted  $[N]$-cumulene. In this case the H\"uckel distribution $\mathscr{D}_{Huckel} (\theta )$ is obtained through the set of vectors given by:
\begin{equation}
    \psi_n (z)=a_n R ({\theta /2}) D_{b_n /a_n} R (k_n z+\delta_n ) ,
\end{equation}
where $D_{\lambda} :\R^2 \mapsto \R^2$ denotes a dilatation map of weight $\lambda$ defined by $(x,y)\mapsto (x,\lambda y)$, $\delta_n$ satisfies: 
\begin{equation}
    \tan (\delta_n )=-a_n /b_n \tan (\theta /2) ,
\end{equation}
and $k_n$ satisfies the equation:
\begin{equation}
    \sin^2 ((N+1)k_n ) =\cos^2 (\theta ) \, \sin^2 (k_n ),\ \ \mbox{\rm for}\ \ 1\leq n\leq 2N .
\end{equation}

In general, it is not possible to obtain explicit values for the distribution angles unless the case $\theta=0$ or $\theta=\pi/2$ which are discussed in the next section. We can nevertheless give some hints on its structure. Indeed, for each twist angle $\theta$, the H\"uckel distribution $\mathscr{D}_{Huckel} (N,\theta )$ of twisted $[N]$-cumulene by an angle $\theta$ possesses the following structure.\\

The distribution $\mathscr{D}_{Huckel} (N,\theta )$ is such that: 
\begin{equation}
\label{symdistribution}
    \phi (N-z)-\phi (N/2)=\phi (N/2)-\phi (z),\ \ z=0,\dots ,N.
\end{equation}
The distribution possesses an axis of symmetry directed along the vector $\phi (N/2)$. As $\phi (0)=0$ and $\phi (N)=\theta$, this axis is at $\theta/2$ or $\theta/2 +\pi/2$ depending on the number of turns and if the helix is right handed or left handed.\\

\newpage
It must be noted that this property is not as trivial as it looks due to the fact that we have a dilatation map which sends the components of $\psi (z)$ on an ellipse. It can be seen as follows:

\begin{figure}[H]
\centering
\includegraphics[width=1.0\linewidth]{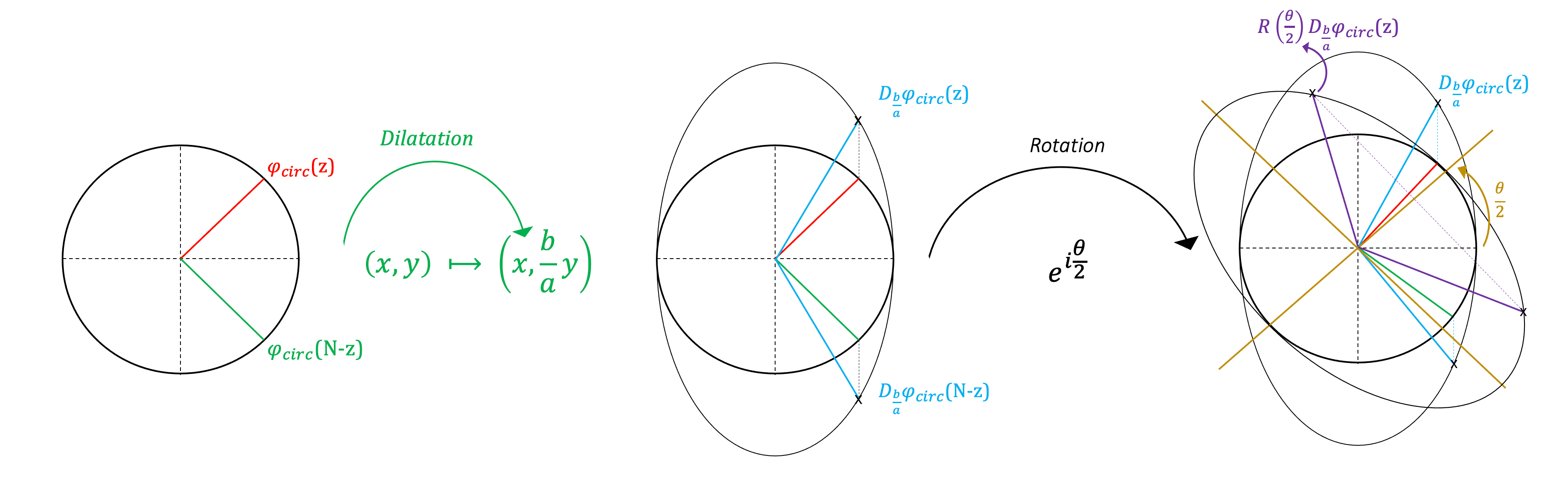}
\caption{Construction of the helical distribution from the circular distribution.}
\label{fig8-ellipse}
\end{figure}

The distribution is obtained from the circular distribution $\mathscr{D}_{circ} (\theta )$ defined by:
\begin{equation}
    \phi_n (z)=k_n z +\delta_n .
\end{equation}
This distribution is symmetric with respect to $0$, meaning that we have: 
\begin{equation}
    \phi_n (N-z)=-\phi_n (z) ,
\end{equation}
for $z=0,\dots ,N$. 

The distribution $\mathscr{D}_{Huckel} (\theta )$ is then obtained from $\mathscr{D}_{circ} (\theta )$ by a dilatation map $D_{b/a}$ from the vectors $v_n (z)=a e^{i\phi_n (z)}$, $i^2 =-1$ followed by a rotation $e^{i\theta /2}$. Rotations are isometric transformations so that they preserve angles. Dilatations are not preserving angles, but they preserve the structure of the circular distribution as can be seen by a simple computation. \\

A {\bf  formal proof} of the previous result can be obtained {\bf using the circularly polarized MOs} version of the previous case studied by S. Gunasekaran et al. in (\cite{guna}, p.4-5). They consider a modified H\"uckel matrix by increasing the coupling between the terminal lone $p$ orbitals from $t$ to $\sqrt{2} t$ instead of a constant coupling $t$ between the atoms of the chain (see \cite{guna}, fig. 5 (a) p.5).  

This simple change made the mathematical analysis greatly simplified and explicit expressions for the distribution given by $\phi_n (0)=0$, $\phi_n (N)=\theta$ can be obtained:
\begin{equation}
    \phi_n (z)=k_n z ,\ \ z=1,\dots ,N-1,\ \ Nk_n = \left \{
    \begin{array}{cl}
    \theta+m\pi,\ \ & n=2m+1 \\
    m\pi-\theta, & n=2m 
    \end{array}
\right .
\end{equation}

For a given $z\in \{ 1,\dots , N-1\}$, we can rewrite $\phi_n (z)$ as:
\begin{equation}
    \phi_n (z)=Nk_n \di\frac{z}{N} = \left \{
    \begin{array}{ll}
         (\theta+m\pi ) \di\frac{z}{N},\ \ & n=2m+1 \\
          (m\pi-\theta)\di\frac{z}{N}, & n=2m 
    \end{array}
\right .
\end{equation}

The quantity $\phi_n (z)$ is well defined for $z=N/2$ and gives:
\begin{equation}
    \phi_n (N/2)=\left \{
    \begin{array}{ll}
         (\theta+m\pi ) \di\frac{1}{2},\ \ & n=2m+1 \\
          (m\pi-\theta)\di\frac{1}{2}, & n=2m 
    \end{array}
\right .
\end{equation}

We can verify directly the previous symmetry results on the distributions $\mathscr{D}_{Huckel} (\theta )$, i.e. that (\ref{symdistribution}) is satisfied. Indeed, we have:

\begin{equation}
    \phi_n (N/2) -\phi_n (z)=\left \{
    \begin{array}{ll}
         (\theta+m\pi ) (\di\frac{1}{2} -\di\frac{z}{N}),\ \ & n=2m+1 \\
          (m\pi-\theta) (\di\frac{1}{2} -\di\frac{z}{N}), & n=2m 
    \end{array}
\right .
\end{equation}
and 
\begin{equation}
    \phi_n (N-z) -\phi_n (N/2)=\left \{
    \begin{array}{ll}
         (\theta+m\pi ) (1-\di\frac{z}{N} -\di\frac{1}{2} ),\ \ & n=2m+1 \\
          (m\pi-\theta) (1-\di\frac{z}{N}-\di\frac{1}{2} ), & n=2m 
    \end{array}
\right .
= \phi_n (N/2) -\phi_n (z)
\end{equation}
\subsection{Properties of distributions for $C_2$-adapted MOs of $\theta=0$ or $\theta=\pi/2$ twisted cumulene}
\label{equivcum}

For $\theta=0$ or $\theta=\pi/2$, cumulenes do not possess helical MOs. This is due to the fact that the $\pi_x$-system and the $\pi_y$-system are decoupled in this case (see section \ref{basic} for more details). However, we can look for $C_2$-adapted MOs, where $C_2$ mixes the $\pi_x$ and $\pi_y$ system (see section \ref{}). 

\begin{remark}
Looking for $C_2$-adapted MOs is natural when cumulenes are considered as linear versions of cyclic chains under the M\"obius topology. This can be understood as {\bf replacing the electronic complexity by a geometric complexity keeping symmetry properties} here given by a $C_2$ symmetry group. We refer to \cite{garner} for more details. In particular, one can note that the angular distribution for the M\"obius cyclic system is perfect but this property is lost when looking for a cumulene analogue. We refer to \cite{garner} where this problem is studied in details and to the following sections.
\end{remark}
\subsubsection{The case $\theta=\pi/2$} 

For $\theta=\pi/2$, the $\pi$-system of cumulenes can be obtained by a $N$-chain of atoms from $z=0$ to $z=N-1$ governing the $p_y$ part and a $N$-chain for $p_x$ from $z=1$ to $z=N$.\\

For a $N$-linear chain, the coefficients $c_{N,n} (z)$ of the wave function $\psi_n (z)$ for $z=1$ to $N$ are of the form:
\begin{equation}
\label{general}
    c_{N,n} (z)= \di\sqrt{\di\frac{2}{N+1}} a_{N,n} (z)
\end{equation}
where:
\begin{equation}
\label{a_n(z)}
    a_{N,n} (z)=\sin (k_n z),\ \ z=1,\dots ,N,\\
\end{equation}
with $k_n$ satisfying the equation:
\begin{equation}
    k_n =\di\frac{n\pi}{N+1},\ \mbox{\rm for}\ 1\leq n\leq N
\end{equation}

The special form of the H\"uckel matrix in this case implies that (see Appendix \ref{proofcoefpisur2}): 
\begin{equation}
c_y (z)=\di\sqrt{\di\frac{2}{N+1}} a_{N,n} (z+1),\ \ \ z=0,\dots ,N-1,\ \  c_x (z)=\di\sqrt{\di\frac{2}{N+1}} a_{N,n} (z) ,\ \ \ z=1,\dots ,N ,
\end{equation}
with $c_y (N)=0$ and $c_x (0)=0$.\\

Looking for $C_2$-adapted MOs, we obtain: 
\begin{equation}
    \psi_{+,n} =\di\frac{1}{\sqrt{2}} \left ( c_y (z) p_y +c_x (z) p_x \right ) ,\ \ \mbox{\rm and}\ \ \ 
    \psi_{-,n} =\di\frac{1}{\sqrt{2}} \left ( c_y (z) p_y -c_x (z) p_x \right )
\end{equation}
Choosing $\psi_{+,n}$ or $\psi_{-,n}$ give rise to a right (resp. left) hand helix.\\

In the following, we focus on $\psi_{+,n}$. We then have: 
\begin{equation}
\label{psi_n(z)}
    \psi_n (z)= \di\sqrt{\di\frac{2}{N+1}} 
\left ( 
\begin{array}{l}
    a_{N,n} (z) \\
    a_{N,n} (z+1)
\end{array}
\right ) .
\end{equation}

For $z=0,\dots ,N$, we want to compare $\phi_{+,n} (z)$ and $\phi_{+,n} (N-z)$. As the coefficients $a_n$ satisfies:
\begin{equation}
\label{demo1}
    a_{N,n} (N-z+1)=(-1)^{n+1} a_{N,n} (z) ,
\end{equation}
(see Appendix \ref{techni}) we deduce that $\psi_n (N-z)$ can be written as: 
\begin{equation}
    \psi_n (z)= \di\sqrt{\di\frac{2}{N+1}} 
\left ( 
\begin{array}{l}
    a_{N,n} (N-z) \\
    a_{N,n} (N-z+1)
\end{array}
\right ) 
=(-1)^{n+1}
\di\sqrt{\di\frac{2}{N+1}} 
\left ( 
\begin{array}{l}
    a_{N,n} (z+1) \\
    a_{N,n} (z)
\end{array}
\right ) 
.
\end{equation}

As a consequence, $\psi_{+,n} (N-z)$ is obtained from $\psi_n (z)$ by axial symmetry along the line $y=x$ (diagonal symmetry) and multiplication by a factor $(-1)^{n+1}$. Denoting by $S:(x,y)\rightarrow (y,x)$ the diagonal symmetry map, we have: 
\begin{equation}
\label{symorb}
    \psi_{+,n} (N-z) =(-1)^{n+1} S \left (\psi_{+,n} (z) \right ).
\end{equation}

We directly deduce the relationship on the distribution of the angles \eqref{symdistribution}.

In this case (see Appendix \ref{proofangle}), an explicit expression can be obtained for the angle $\mathscr{A}_{N,n,+,0,z}$ between $\psi_{+,n} (0)$ and $\psi_{+,n} (z)$ and is given by: 
\begin{equation}
    \mathscr{A}_{N,n,+,0,z} = cos^{-1} 
    \left ( \epsilon (a_{N,n} (1)) \epsilon (a_{N,n} (z+1))
\frac{1}{\sqrt{1 +(a_{N,n} (z)/a_{N,n} (z+1))^2}}
    \right ) ,\ \ z=0,\dots ,N-1, 
\label{distrib}
\end{equation}
and 
\begin{equation}
\label{pisur2prop}
    \mathscr{A}_{N,n,+,0,N} = (-1)^{n+1} \pi/2 ,
\end{equation}
where $\epsilon (x)$ denotes the sign of $x$.\\

As an example, we obtain for $N=3$ and $N=4$ the following distributions: 

\begin{table}[H]
    \caption{Distribution for $\theta=\pi/2$, $N=3$, $n=1$, $n=2$ and $n=3$ - Corresponding MOs obtained by the CASPT2(8:10)/6-311G(d.p) level of theory for the $[3]$-cumulene in its ground triplet state (see Table S2 for details).}
    \label{}
    \centering
    \begin{tabular}{c}
    \includegraphics[width=0.97\linewidth]{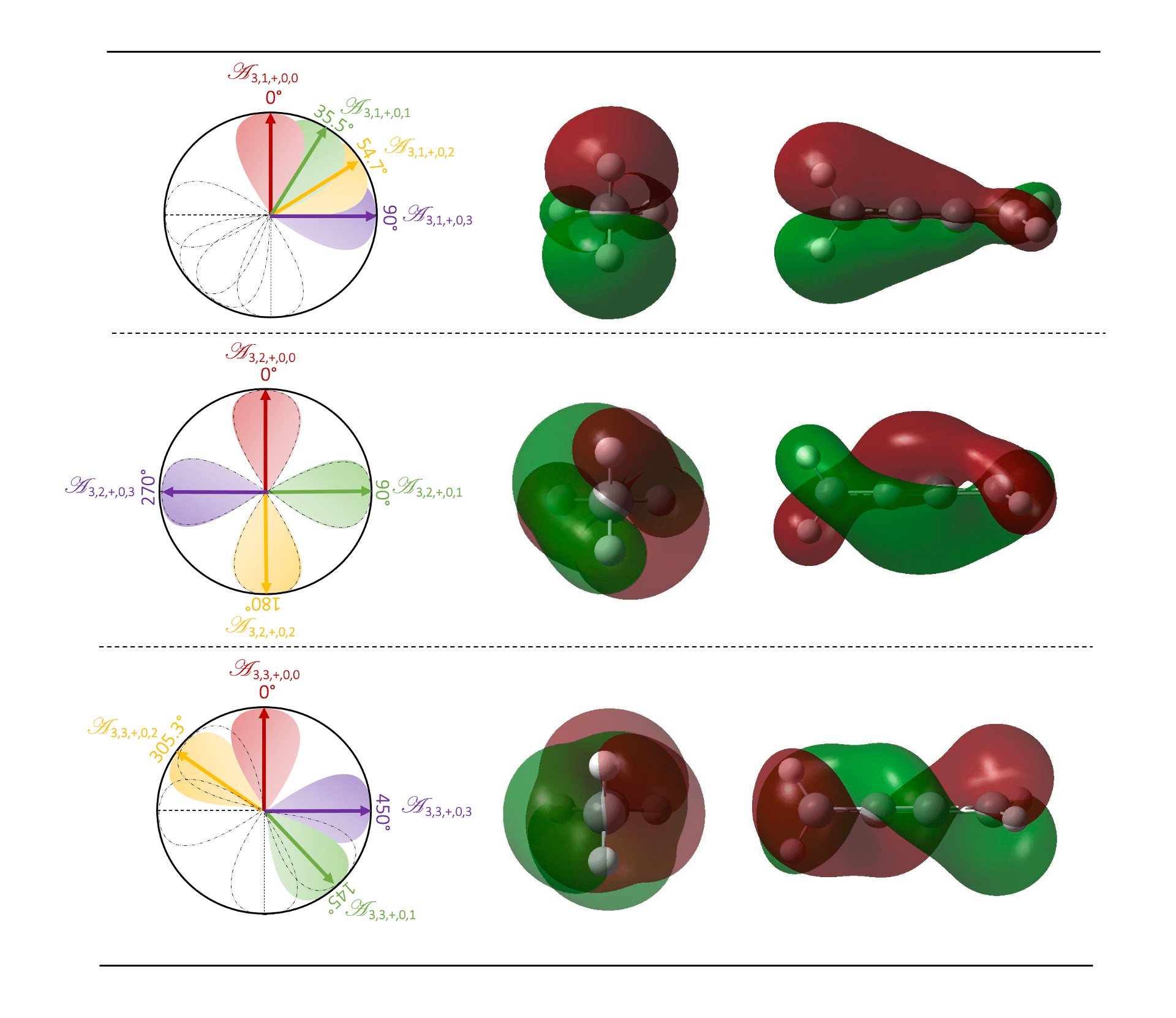} 
    \label{table4-distribution-mo-3-cum}
\end{tabular}
\end{table}

\begin{figure}[H]
\centering
\includegraphics[width=0.65\linewidth]{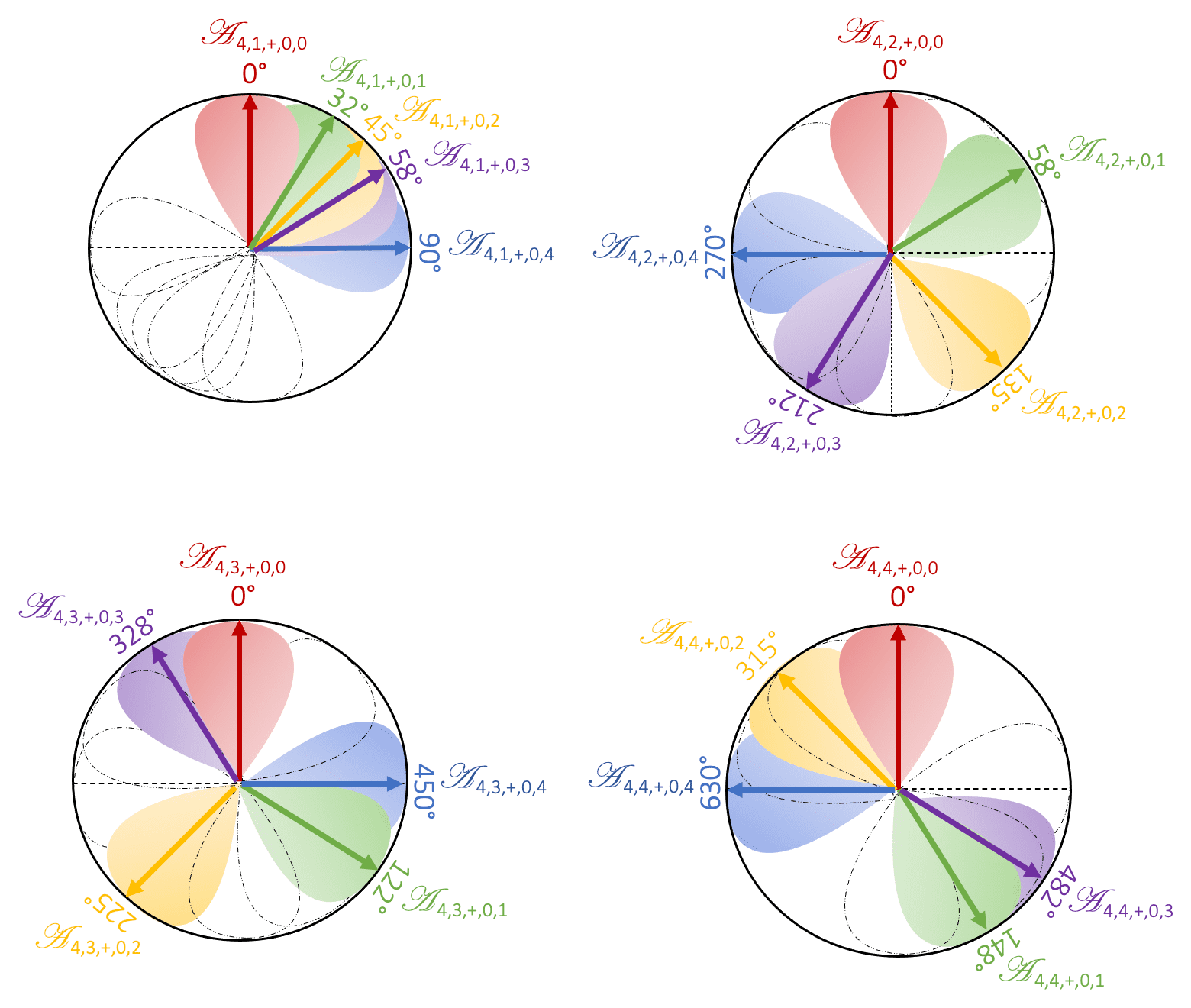}
\caption{Distribution for $\theta=\pi/2$, $N=4$, $n=1$, $n=2$, $n=3$ and $n=4$.}
\label{fig9-distribution-theta-pi-2}
\end{figure}

We can check some consequences of our previous results using the simulations, namely that:
\begin{itemize}
\item By \eqref{pisur2prop}, the angle between $\psi_n (0)$ and $\psi_n (N)$ is equal to $\pi/2$ when $n$ is odd and $-\pi/2$ when $n$ is even. \\

\item For all $n$, the family of vectors $\psi_n (z)$, $z=1,\dots, N$ are making a turn of an angle of $(2n-1)/4$ with respect to its initial position $\psi_n (0)$. As an example, for $N=3$, $n=1$ (a quarter turn) , $n=2$ (three quarter turn) and $n=3$ (five quarter turn). In the same way, for $N=4$, $n=1$ (a quarter turn) , $n=2$ (three quarter turn) , $n=3$ (five quarter turn) and $n=4$ (seven quarter turn) etc.
\end{itemize}

\subsubsection{The case $\theta=0$}

For $\theta=0$, the $\pi$-system of cumulenes can be obtained by a $N+1$-chain of atoms from $z=0$ to $z=N$ governing the $p_y$ part and a $N-1$ chain for $p_x$ for $z=1$ to $z=N-1$.\\

The special form of the H\"uckel matrix in this case implies that (see Appendix \ref{proofcoefzero}) : \\
\begin{equation}
\left .
\begin{array}{l}
c_y (z)=\di\sqrt{\di\frac{2}{N+2}} a_{N+1,n} (z+1) =\di\sqrt{\di\frac{2}{N+2}} \sin \left (\di\frac{n\pi (1+z)}{N+2} \right ),\ \ \ z=0,\dots ,N\\  
c_x (z)=\di\sqrt{\di\frac{2}{N}} a_{N-1,n-1} (z)=\di\sqrt{\di\frac{2}{N}} \sin \left ( \di\frac{(n-1) \pi z}{N} \right )  ,\ \ \ z=1,\dots ,N-1 ,
\end{array}
\right .
\end{equation} \\
with $c_x (0)=0$ and $c_x (N)=0$.\\

Looking for $C_2$-adapted MOs, we obtain, focusing on $\psi_{+,n}$:
\begin{equation}
    \psi_n (z)=  
\left ( 
\begin{array}{l}
\di\sqrt{\di\frac{2}{N}}    \sin \left (\di\frac{(n-1) \pi z}{N} \right ) \\
    \di\sqrt{\di\frac{2}{N+2}} \sin \left ( \di\frac{n\pi (z+1)}{N+2} \right )
\end{array}
\right ) 
\end{equation}
for $z=1,\dots ,N-1$.

Here again, one can compute explicitly the distribution of angles for particular values of $n$. In particular, using the same notations as before, we have (see Appendix \ref{proofangle0}):
\begin{equation}
    \mathscr{A}_{N,n,+,0,N} = \di\frac{(1-(-1)^{n+1})}{2} \pi =\left \{ 
    \begin{array}{cc}
        0  \ \ \ \ & \mbox{\rm if $n$ is odd,}  \\
        \pi & \mbox{\rm otherwise}. 
    \end{array}.
\right .
\end{equation}
As an example, we obtain for $\theta=0$, $N=3$ and $N=4$ the following distributions:

\begin{figure}[H]
\centering
\includegraphics[width=0.35\linewidth]{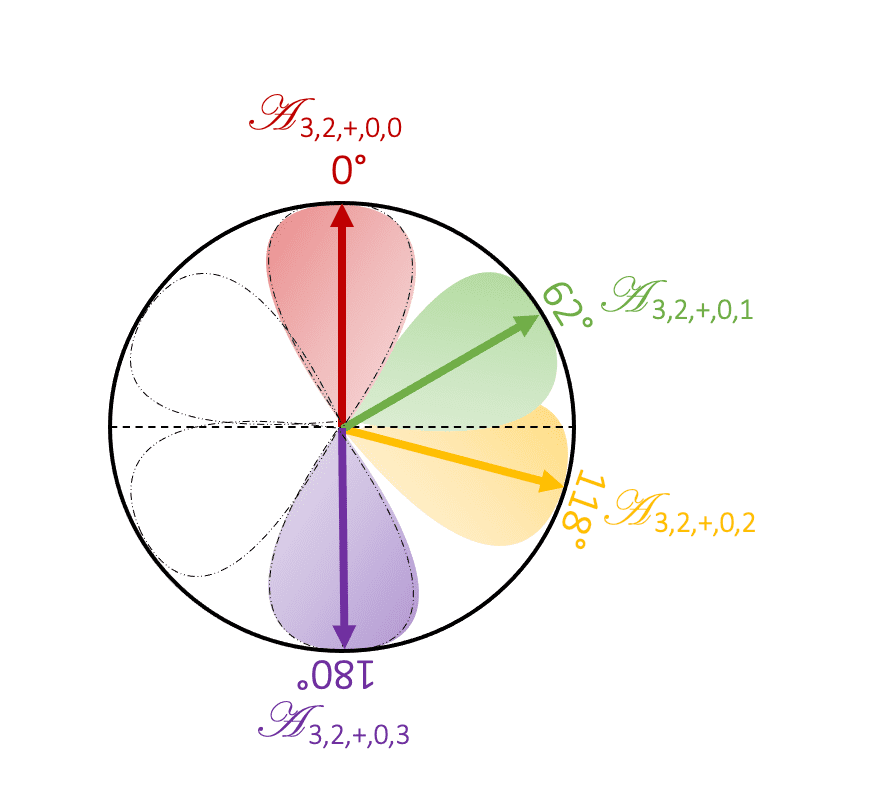}
\includegraphics[width=0.36\linewidth]{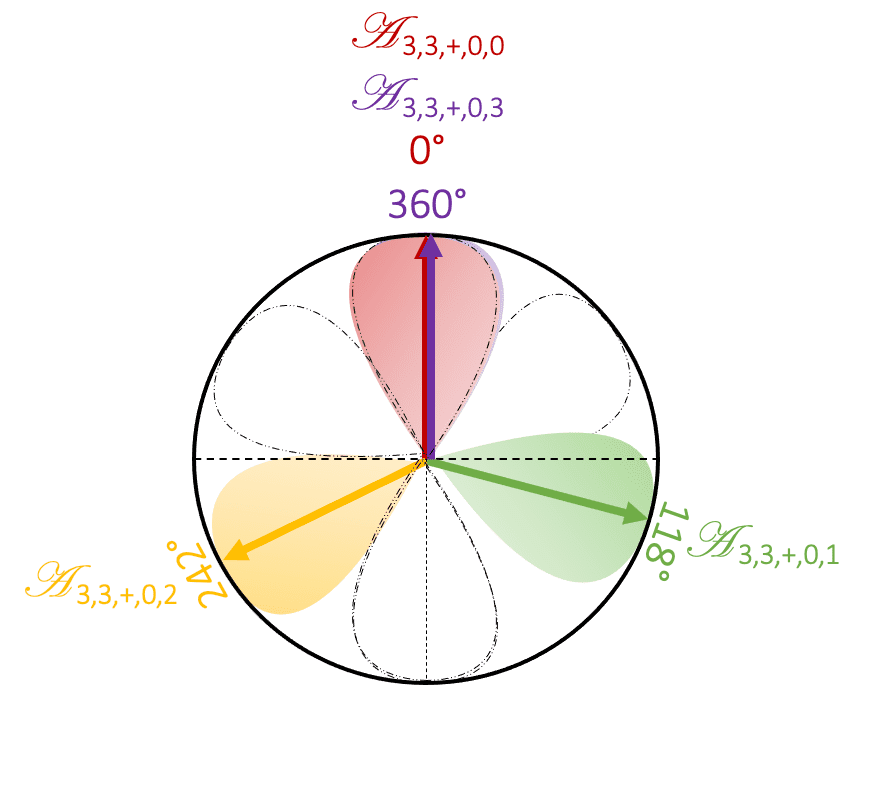}
\caption{Distribution for $\theta=0$, $N=3$, $n=2$, $n=3$.}
\label{fig10-distribution-theta-0}
\end{figure}

\begin{table}[H]
    \caption{Distribution for $\theta=0$, $N=4$, $n=2$, $n=3$ and $n=4$ - Corresponding MOs obtained by B3LYP/6-311G(d.p) level of theory for the $[4]$-cumulene in its ground singlet state (see Table S3 for details).}
    \label{}
    \centering
    \begin{tabular}{c}
    \includegraphics[width=0.75\linewidth]{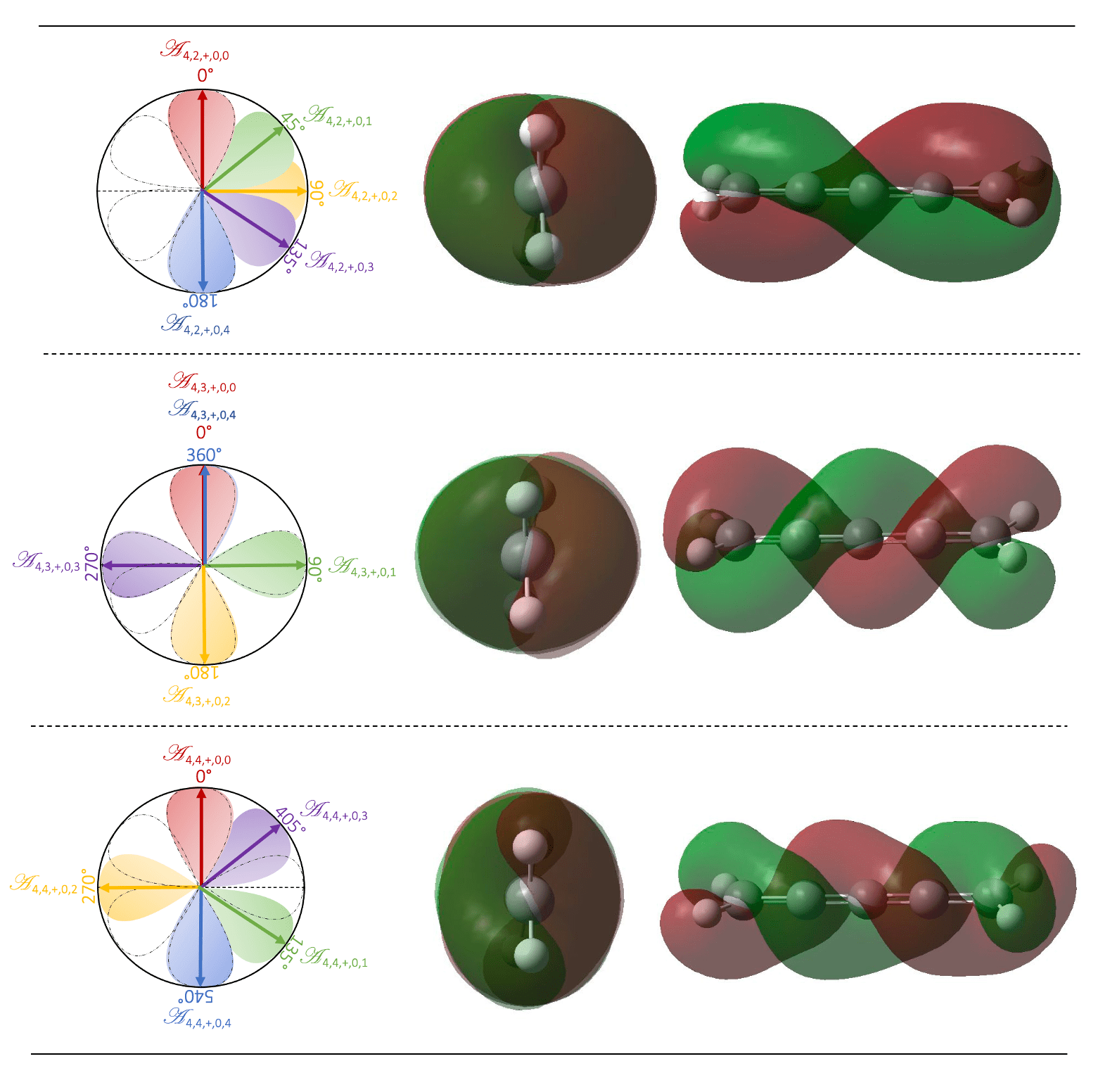} 
    \label{table5-distribution-theta-0-4-cum}
\end{tabular}
\end{table}

The simulations illustrate our previous results:\\

\begin{itemize}
    \item The angle between $\psi_n (0)$ and $\psi_n (N))$ is $0$ when $n$ is  odd and $180$ when $n$ is even.\\
    
    \item For all $n$, the family of vectors $\psi_n (z)$ are making $(n-1)/2$ turns with respect to its initial position $\psi_n (0)$.In particular, for $N=3$, $n=2$ (half turn) and $n=3$ (one turn). For $N=4$, $n=2$ (half turn) , $n=3$ (one turn) and $n=4$ (one and a half turns) etc.
\end{itemize}

Note that the results of distributions for $N = 10$ and $\theta = 10$ are illustrated in figures S1-S3.
\section{How to detect helical orbitals - algebraic approach}

Following S. Gunasekaran and L. Venkataraman in (\cite{guna}, p.5-6), it is possible to detect helical orbitals in a pure {\bf algebraic} way.\\

The idea is to use the {\bf Löwdin partitioning technique} \cite{lowdin}, which is a simple technique allowing focus on a particular part of the molecule by "constructing an {\bf effective Hamiltonian} which acts only on the target model space but gives the same result as the original Hamiltonian acting on the complete space" (see \cite{jin}). \\

The {\bf Hamiltonian} $H$ of the complete system can be written as:

\begin{equation}
H= 
\left ( 
\begin{array}{ccc}
H_{aa} & H_{aC} & 0 \\
H_{Ca} & H_{CC} & H_{Cb} \\
0 & H_{bC} & H_{bb} 
\end{array}
\right )
\end{equation}

The left group is associated to a set $\boldmath{p}_a =(p_{1},\dots ,p_{a} )$ of basis orbitals and the right group to $\boldmath{p}_b =(p^{\star}_1,\dots ,p^{\star}_{b} )$. The central chain consists of $2N$ orbitals given by $p_L$, $p_R$ and $2(N-2 )$ orbitals $p_{x,i}$, $p_{y,i}$, $i=1,\dots ,N-1$. We denote by $\boldmath{p}_C$ this vector.\\

The matrix $H$ is then a $a+2n+b$ square matrix. By construction of the Hamiltonian, we have $H_{Ca}^t = H_{aC}$ where $t$ denotes the transpose. The associated {\bf Schrödinger equation} $H\cdot \psi = E\psi$ gives for $\psi = \di\sum_{i=1}^a c_i p_i +c_L p_L +\di\sum_{i=1}^{n} c_{x,i} p_{x,i} +c_{y,i} p_{y,i} +c_R p_R +\di\sum_{j=1}^b c_j p^{\star}_j$ a linear system of the form:
\begin{equation}
\left \{
\begin{array}{l}
H_{aa} \boldmath{p}_a +H_{aC} \boldmath{p}_C = E \boldmath{p}_a ,\\
H_{Ca} \boldmath{p}_a +H_{CC} \boldmath{p}_C + H_{Cb}  \boldmath{p}_b =E \boldmath{p}_C ,\\
H_{bC} \boldmath{p}_C +H_{bb} \boldmath{p}_b = E \boldmath{p}_b ,\\
\end{array}
\right .
\end{equation}

For any integer $d>0$, let us denotes by $I_d$, the identity matrix of size $d$. Assuming that for a given $E$ the matrices $E I_a - H_{aa}$ and $EI_b - H_{bb}$ are {\bf invertible}, one can express $\boldmath{p}_a$ and $\boldmath{p}_b$ using $\boldmath{p}_C$, namely: 
\begin{equation}
\boldmath{p}_a = (EI_a -H_{aa} )^{-1} H_{aC} \boldmath{p}_C,\ \ \
\boldmath{p}_b = (EI_b -H_{bb} )^{-1} H_{bC} \boldmath{p}_C 
\end{equation}
Replacing in the second equation $\boldmath{p}_a$ and $\boldmath{p}_b$ by their expressions, we obtain the following equation:
\begin{equation}
H_{Ca} (EI_a -H_{aa} )^{-1} H_{aC} \boldmath{p}_C +H_{CC} \boldmath{p}_C + H_{Cb}  (EI_b -H_{bb} )^{-1} H_{bC} \boldmath{p}_C =E \boldmath{p}_C 
\end{equation}
This last equation can be understood as the action of an {\bf effective Hamiltonian} $H_{eff}$ defined over $\boldmath{p}_C$ by: 
\begin{equation}
H_{eff} =\Sigma_L + H_C +\Sigma_R ,
\end{equation}
where the matrix $\Sigma_L$ and $\Sigma_R$ are defined by:
\begin{equation}
\Sigma_L = H_{Ca} (EI_a -H_{aa} )^{-1} H_{aC} \ \ \mbox{and}\ \ \ \Sigma_R =H_{Cb}  (EI_b -H_{bb} )^{-1} H_{bC} ,
\end{equation}
leading to a $2N \times 2N$ Hamiltonian.\\

The two matrices $\Sigma_L (E)$ and $\Sigma_R (E)$ are $2N \times 2N$ and depend only on the coupling between the end groups and the terminal orbital of the chain $p_L$ and $p_R$.\\

The main result of S. Gunasekaran and L. Venkataraman in \cite{guna} can be stated as follows:\\

{\bf Existence criterion}: {\it A necessary condition for $H$ to yield helical states is that $\Sigma_L$ and $\Sigma_R$ do not commute, i.e. $[\Sigma_L ,\Sigma_R ]\not= 0$, where the bracket $[\dot ,\dot ]$ is defined by $[A,B]=AB-BA$.}\\

The main difficulty in applying this result is due to the computation of the two matrix $\Sigma_L$ and $\Sigma_R$ which is {\bf in general not tractable} and only possible in simple configurations. We refer to \cite{jin} for more details. However, we can deduce from the previous result the following statement:\\

{\bf Genericity of helical orbitals}: {\it A linear chain satisfying the previous assumptions admits generically helical orbitals}.\\

The proof of this statement relies on the existence criterion. Indeed, let $A$ and $B$ be two arbitrary square matrices. The condition of non-commutation is open, i.e. that sufficiently small perturbations of $A$ and $B$ do not commute. Another way of saying that arbitrary close perturbations of commutative matrices are in general non-commutative.\\

As a consequence, helical orbitals are not special; on the contrary they are the most common behavior of orbitals for linear chains.\\

Despite its interest, the previous criterion is {\bf not effective to detect helical orbitals} for a specific molecule as it demands the explicit computation of the matrices $\Sigma_L$ and $\Sigma_R$. In the following, we rely on an approach suggested by Garner and co-workers in \cite{garner} making an essential use of the symmetry properties of the molecule. 
\section{How to detect helical orbitals - symmetries approach}

\subsection{Basic structural properties}
\label{basic}
We restrict ourself to linear molecules made of $N$ atoms $A_i$, $i=1,\dots ,N$. We denote by $\mathbb{A}=\{ A_i \}_{i=1,\dots ,N}$ the family of atoms and by $L_{\A}$ the corresponding linear molecule. We assume that a coordinates system is given such that $z$ is in the direction of the molecule.\\

In order to observe helical MOs in a linear molecule $L_{\A}$, one needs to deal with atoms supporting a $\pi$-system which is two dimensional, i.e. with atoms having as a basis set $p_x$ and $p_y$, $p_z$ supporting the $\sigma$-bond. We assume that we have $n$ such atoms.\\

As a consequence, along the molecule we have two planes $\mathscr{P}_x$ and $\mathscr{P}_y$ containing the family of $p$ orbitals $p_x (z)$ and $p_y (z)$ respectively where $z$ denotes the integer position of each atom along the chain.\\   

The previous restriction implies that one has to {\bf restrict attention to atoms belonging to the bloc $p$ of the periodic table of the elements}. Moreover, in order to have a structure which can be understood, it seems reasonable to restrict our attention to elements of the bloc $p$ of the {\bf second and third line of the periodic table}.\\

The boundary of such a chain can be generated by a single MO. For more simplicity, we assume that the left boundary MO is $p_y =p_0$ and the right one is an MO denoted by $p_N$ which can be arbitrarily oriented in the $x-y$ plane. We have then $N=n+2$ atoms in the linear chain.\\

The left and right atoms of the linear chain belong to some structures denoted by $S_L$ and $S_R$.\\

The H\"uckel matrix of such a molecule takes the form, taking as a basis set $p_0=p_x$, $p_x^i$, $p_y^i$, $i=1,\dots ,n$, $p_N$:

\begin{figure}[H]
	\centering
	\includegraphics[scale=0.3]{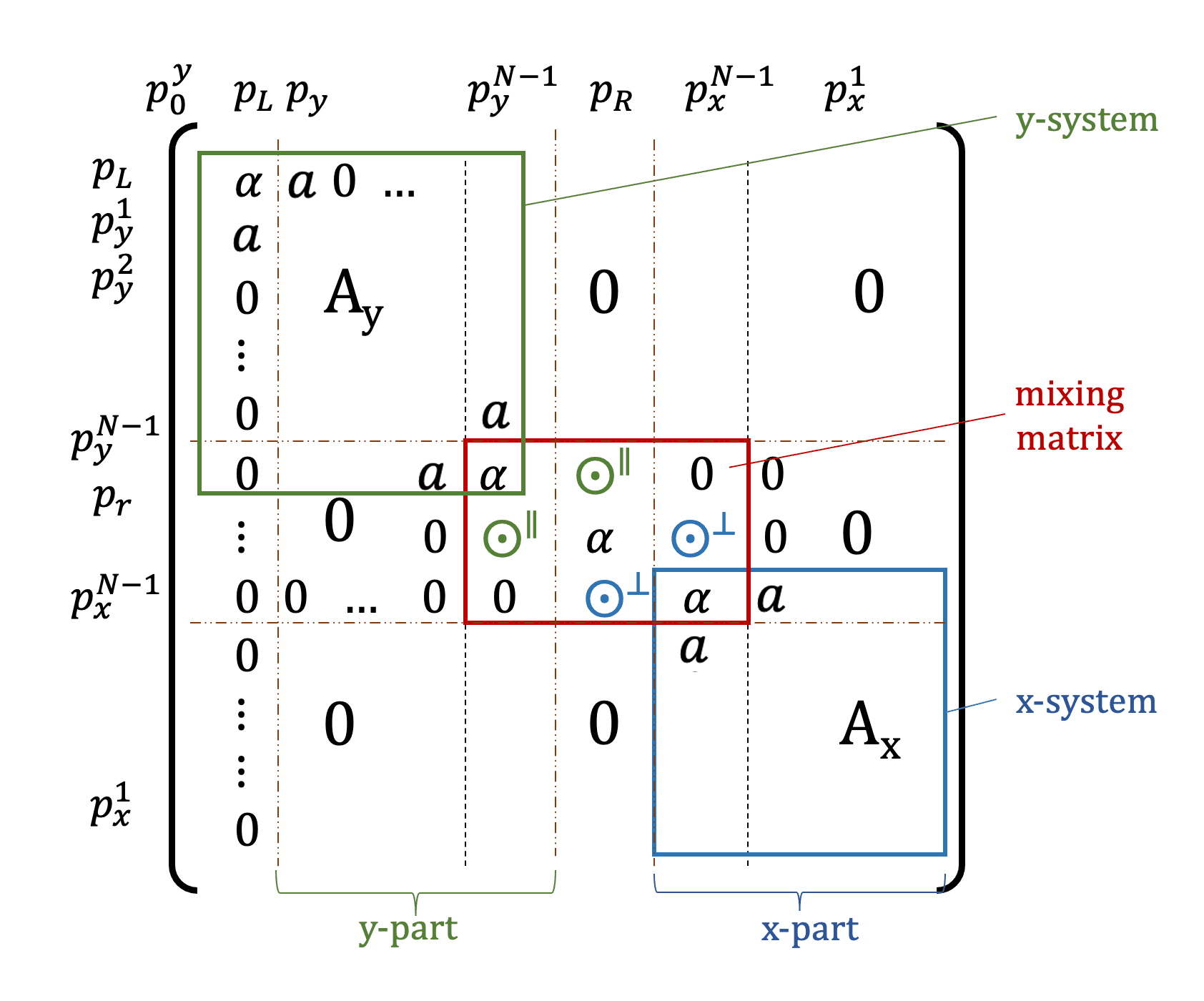}
	\caption{General matrix}
	\label{fig11-general-matrix}
\end{figure}

where the $A$ matrix is given by:

\begin{figure}[H]
	\centering
	\includegraphics[scale=0.2]{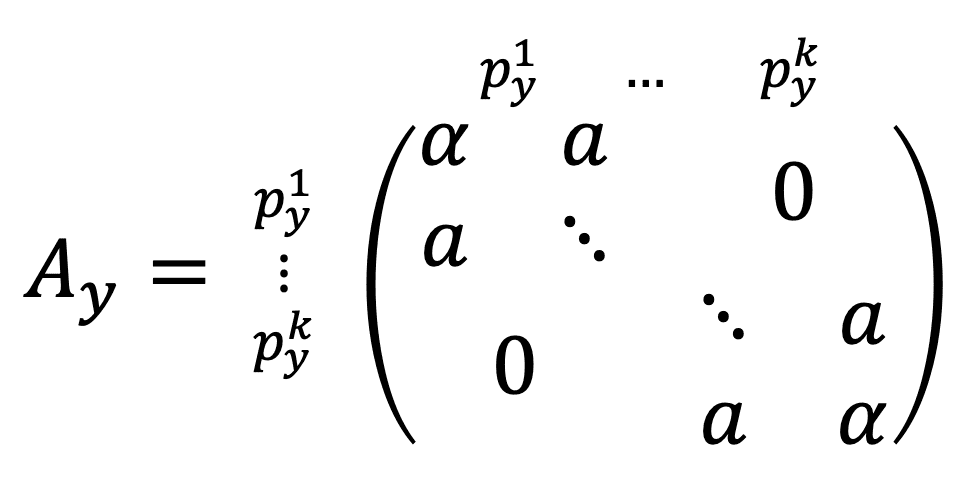}	
	\caption{A Matrix}
	\label{fig12-matrix-A}
\end{figure}

The characteristic polynomial of $H$ can be decoupled in two cases: 

\begin{enumerate}
\item when $\odot^{\parallel}=0$, corresponding to the case where $p_N$ is orthogonal to $p_x^n$, i.e. $p_N = p_y$. The mixing matrix reduces to:

\begin{figure}[H]
	\centering
	\includegraphics[scale=0.2]{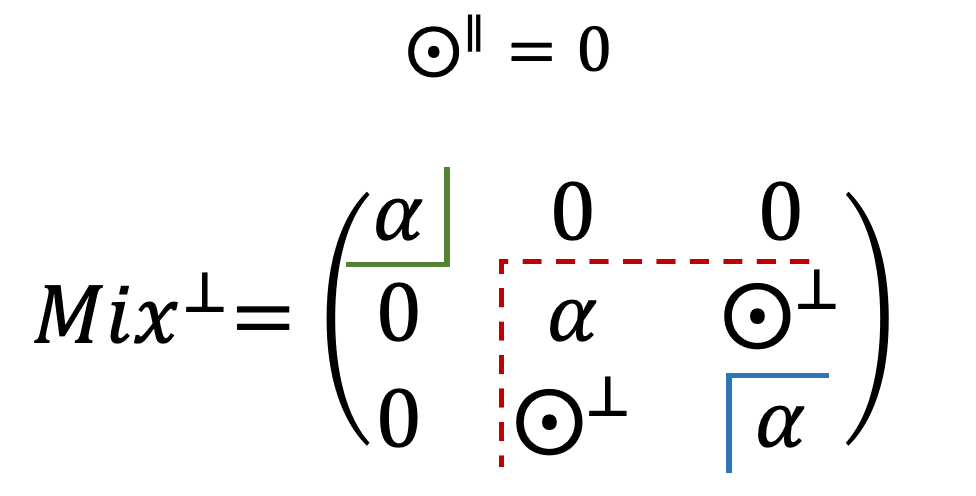}	
	\caption{Orthogonal matrix}
	\label{fig13-orthogonal-matrix}
\end{figure}

and the characteristic polynomial has the form:
\begin{equation}
P(x)=P_A (x) P_{A^{\perp} ,\odot^{\perp}} (x) .
\end{equation}

\item when $\odot^{\perp} =0$, corresponding to the case where $p_N$ is orthogonal to $p_y^n$, i.e. that $p_N =p_x$. The mixing matrix reduces to: 

\begin{figure}[H]
	\centering
	\includegraphics[scale=0.2]{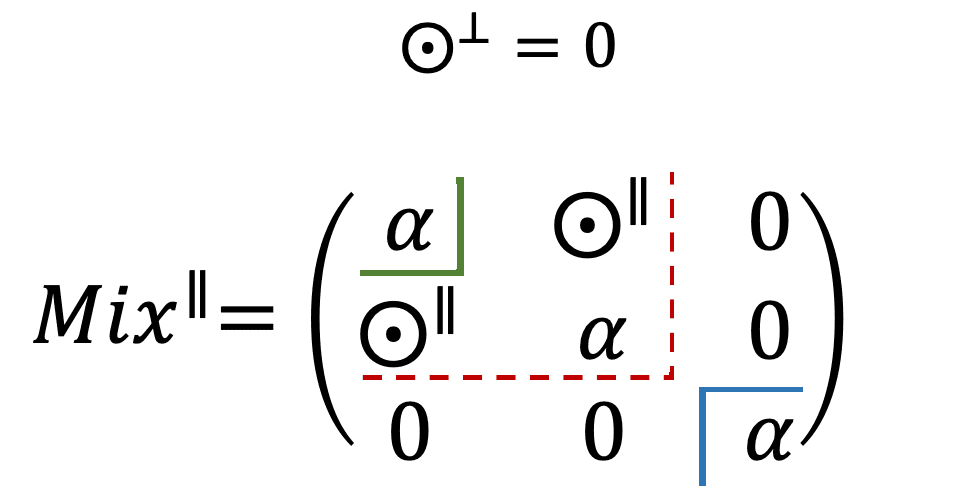}
	\caption{Mixing matrix}
	\label{fig14-mixing-matrix}
\end{figure}
and the characteristic polynomial is given by:  
\begin{equation}
P(x)=P_{A^{\perp}} (x) P_{A,\odot^{\parallel}} (x) .
\end{equation}
\end{enumerate}

The first case corresponds to the orthogonal configuration of the ethylene and the second one to the planar configuration.\\

In all the other configurations, we have a mixing between the components of the $p_x$ and $p_y$ orbitals which is a necessary condition for the existence of helical orbitals. 

\subsection{Symmetry properties}

The existence of helical states implies that we have a mixing between MOs $p_x$ and $p_y$ in a very special way. In particular, we assume that there exists what is called an {\bf helicogenic $C_2$ symmetry axis} (see \cite{garner}), i.e. a $C_2$ axis such that $C_2 (p_x)=p_y$.\\

As an example, for the ethylene we have:

\begin{itemize}
\item the planar configuration of the ethylene possesses three $C_2$ axes: one directed along the chain and two passing through the centre directed along $x$ and $y$. None of these $C_2$ axes are helicogenic.

\item the orthogonal configuration also possesses three $C_2$ axes: one along the chain and two which are dihedral. These last two axes are helicogenic.
\end{itemize}

However, as already discussed in the previous section, none of these two examples possess explicit helical MOs. In the first case, no mixing is induced by the symmetries so that one cannot wait for helical MOs. For the orthogonal case however, this is due to the fact that despite the mixing generated by the symmetry axes, the characteristic polynomial factorizes and the behaviour of the $p_x$ and $p_y$ family is disconnected. This last property can be related to the existence of another symmetry of the molecule, namely that the orthogonal configuration of the ethylene possesses a {\bf mirror-plane symmetry}. As a consequence, we have to assume that no mirror-plane symmetry must be present in order to generate helical MOs. 
\subsection{The helical orbitals criterion - symmetries}

Following the previous discussion, we are led to the following statement, which was originally made by M.H. Garner, R. Hoffmann, S. Rettrup and G.C. Solomon in \cite{garner}, p. J) for cumulenes:\\

{\bf The helical orbitals criterion}: {\it Linear chain $L_{\mathbb{A}}$ satisfying the structural assumption, without mirror-plane symmetry, admitting a $C_2$ helicogenic axis, has explicit helical MOs.}\\
 
The proof of such a theorem can be deduced from the properties of the H\"uckel matrix associated to such kind of molecules. \\

The absences of a mirror-plane symmetry implies that the molecule is {\bf chiral}, i.e. that the image of the molecule by a mirror plane is not invariant. This remark can be used to give an alternative statement of the helical orbitals theorem:\\

{\bf Helical orbitals criterion}: {\it linear chain $L_{\mathbb{A}}$ satisfying the structural assumption and chiral admitting a $C_2$ helicogenic axis, has explicit helical MOs.}\\

This statement can be related to the {\bf Curie's principle} in which he states informally in his paper "On Symmetry in Physical Phenomena" \cite{curie} in 1894 an intuitively plausible relationship between the symmetry of an effect and its cause, namely that "when certain causes produce certain effects, the elements of symmetry of the causes must be found in the effects produced" (p. 401). Indeed, we are waiting for helical molecular orbitals. These objects are naturally chiral so that following Curie's principle, one must find chirality in the initial geometry of the molecule which induces the orbitals structure. As a consequence, chirality is unavoidable. The Curie's principle gives a physical support to the sentence "the formation of helical symmetry-adapted MOs requires chirality; not surprising considering a helix is a chiral object" made by M.H. Garner et al. in (\cite{garner}, page J): this is not surprising indeed and is a consequence of the Curie's principle.\\

We refer to Chalmers \cite{chalmers} as well as Ismael \cite{ismael} for more details and discussion about the validity of the Curies's principle. 


\section{Induced Helical MOs for non helical linear chains}

The previous characterization of helical MOs can be used to induce helical states from a molecule which does not admit it at a first glance. Several strategies can be tested.
\subsection{Breaking of symmetries and chirality}

Let us consider a linear chain which does not satisfy the helical orbital criterion but satisfying the structural assumptions.\\ 

By definition, the core chain possesses two $C_2$ helicogenic axes and one $C_2$ axis along the chain. Moreover, the core chain also possesses also three orthogonal mirror planes $\sigma_{x,y}$, $\sigma_{x,z}$ and $\sigma_{y,z}$ (see figures \ref{fig: 3cum-0} and \ref{fig: 4cum-0}). \\

In order to satisfy the helical orbital criterion, we have to break the mirror-plane symmetries. This will depend on the fragment we put on the boundaries. This can be done in several steps:\\

\begin{figure}[H]
    \centering
    \includegraphics[width=0.7\linewidth]{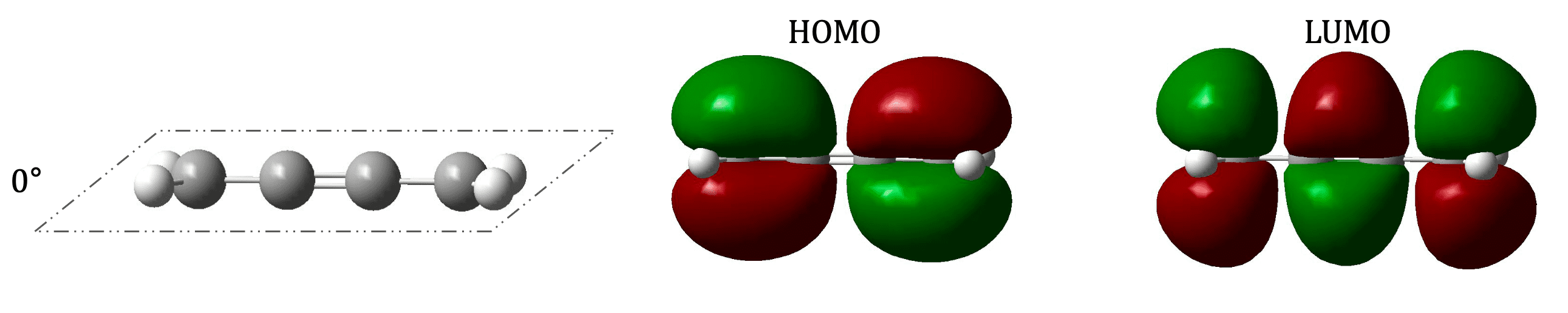}
    \includegraphics[width=0.7\linewidth]{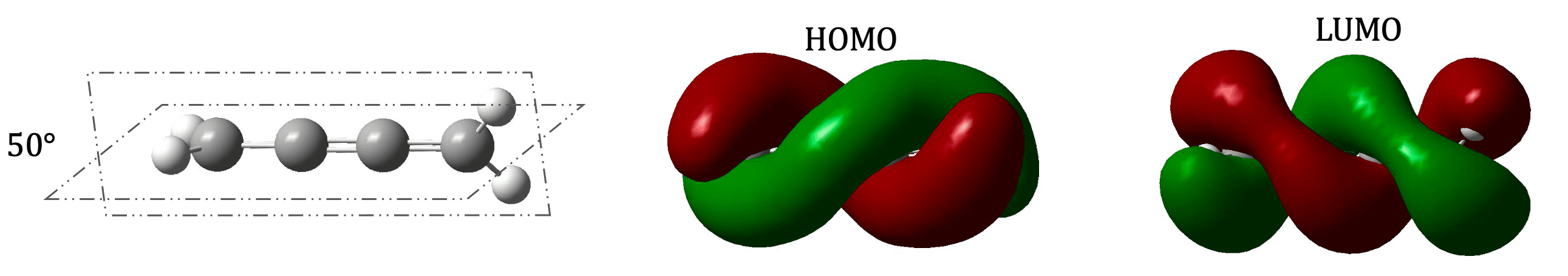}
    \includegraphics[width=0.7\linewidth]{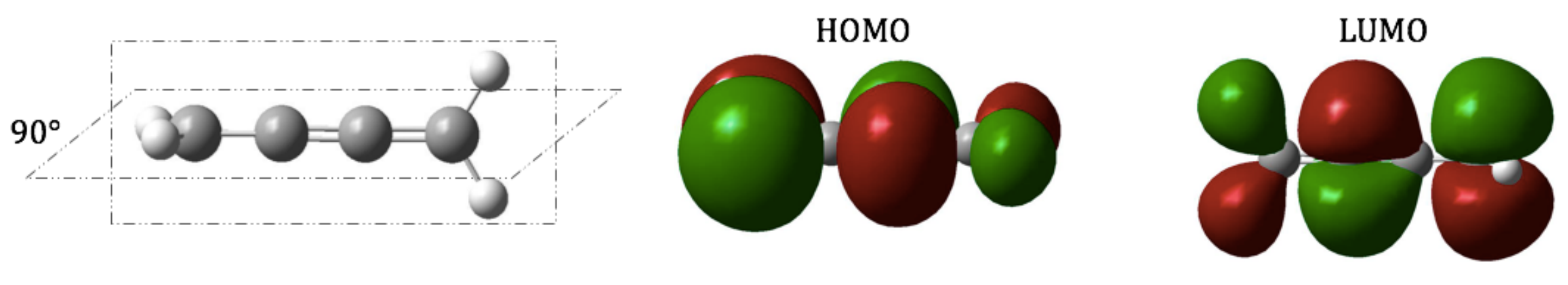}
    \caption{Representation of the HOMO and LUMO MOs obtained at the B3LYP/6-311G(d.p) level of theory for the $N=3$ cumulene in its ground singlet state  at a 0°, 50 ° and 90 °rotation (see Table S1 for details).}
    \label{fig15-3cum-0}
\end{figure}

 \begin{figure}[H]
    \centering
    \includegraphics[width=0.8\linewidth]{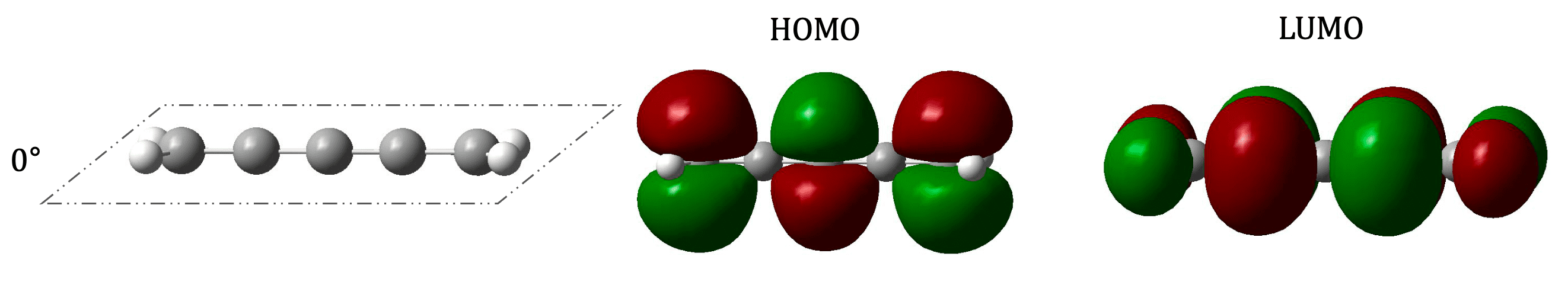}
    \includegraphics[width=0.8\linewidth]{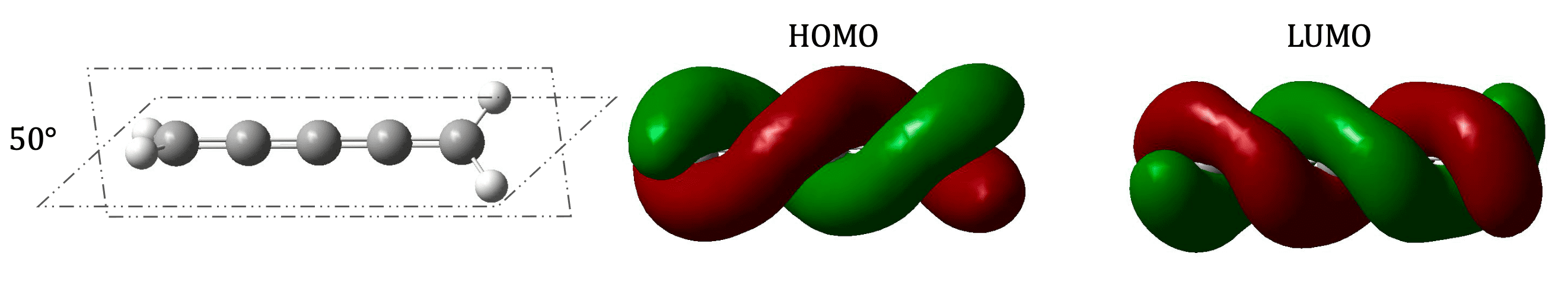}
    \includegraphics[width=0.8\linewidth]{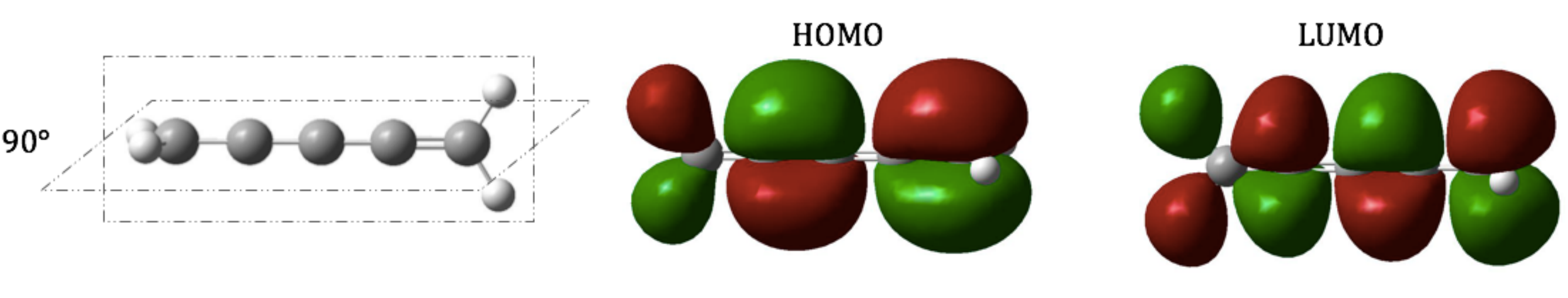}
    \caption{Representation of the HOMO and LUMO MOs obtained at the B3LYP/6-311G(d.p) level of theory for the $N=4$ cumulene in its ground singlet state  at a 0°, 50 ° and 90 °rotation (see Table S3 for details).}
    \label{fig16-4cum-0}
\end{figure}

\begin{itemize}
\item Breaking the $\sigma_{x,z}$ mirror-plane symmetry means that the molecule is not planar. 
\end{itemize}
\vskip 3mm
Let us now consider a non-planar linear chain. This condition is not sufficient as the orthogonal configuration of the [N]-cumulenes in their ground states does not support helical MOs for example. \\

\begin{itemize}
\item Taking the left fragment $L$ to be contained in the plane $\sigma_{x,z}$, the breaking of the mirror-plane $\sigma_{x,z}$ will be effective with a {\bf right fragment $R$ which is not contained in $\mathscr{P}_y$}. 
\end{itemize}
\vskip 3mm
Such a choice, will naturally induce the breaking of the $\sigma_{x,y}$ mirror-plane symmetry.\\

The previous remarks imply that an {\bf axis torsion} of the end group of a linear chain will {\bf generically induce helical MOs}.
\subsection{Using exited states of molecules}
 
The previous result is valid as long as one considers a stable minimal configuration, i.e. the fundamental configuration of a given molecule.\\

 \begin{figure}[H]
    \centering
    \includegraphics[width=0.7\linewidth]{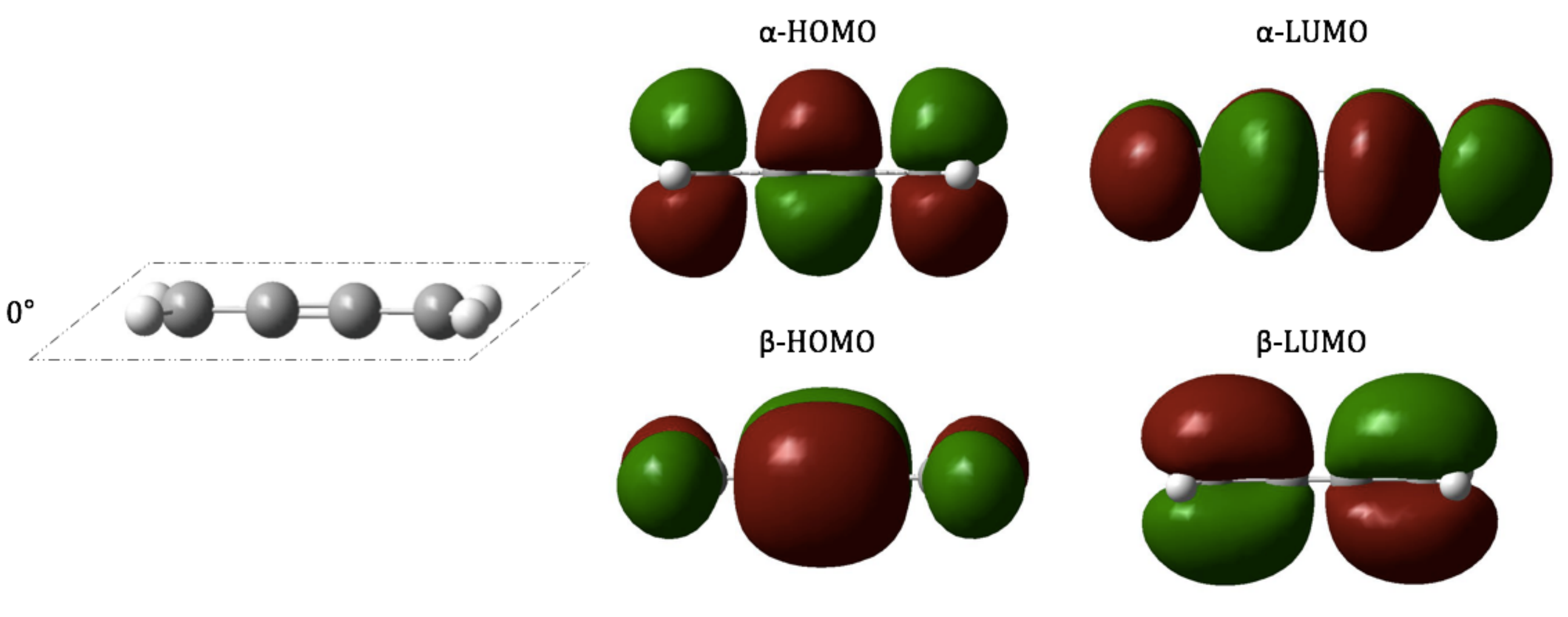}
    \includegraphics[width=0.7\linewidth]{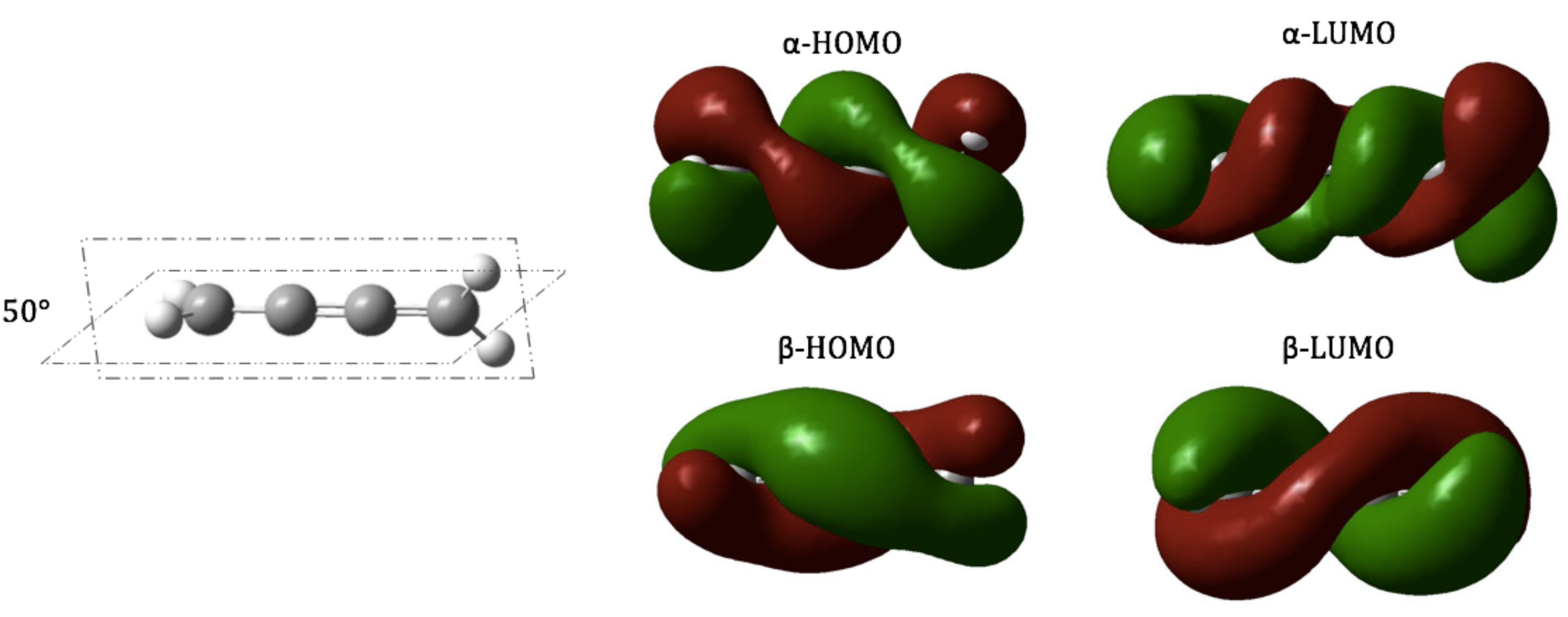}
    \includegraphics[width=0.7\linewidth]{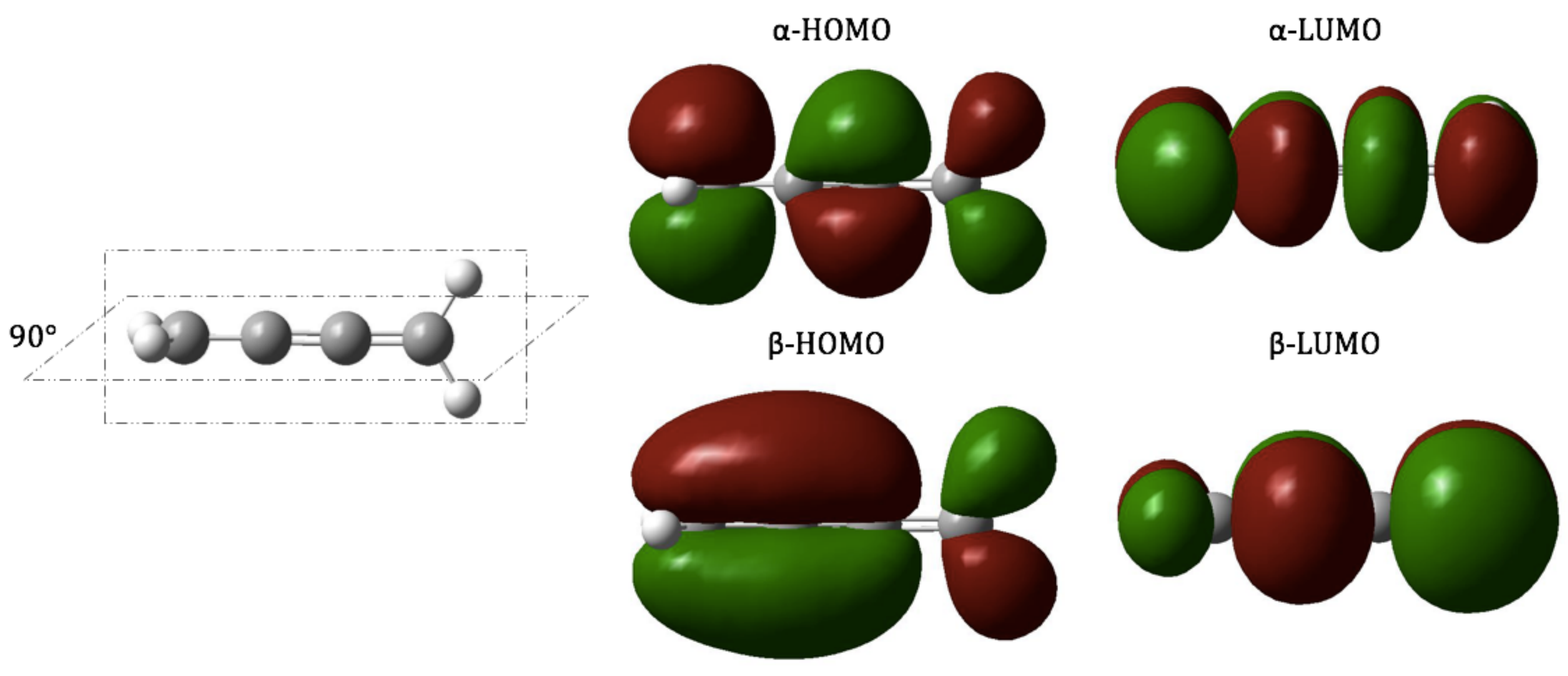}
    \caption{Representation of the HOMO and LUMO MOs obtained at the B3LYP/6-311G(d.p) level of theory for the $N=3$ cumulene in its first excited triplet state at a 0°, 50 ° and 90 °rotation (see Table S2 for details).}
    \label{fig17-3cum-trip}
\end{figure}

  \begin{figure}[H]
    \centering
    \includegraphics[width=0.7\linewidth]{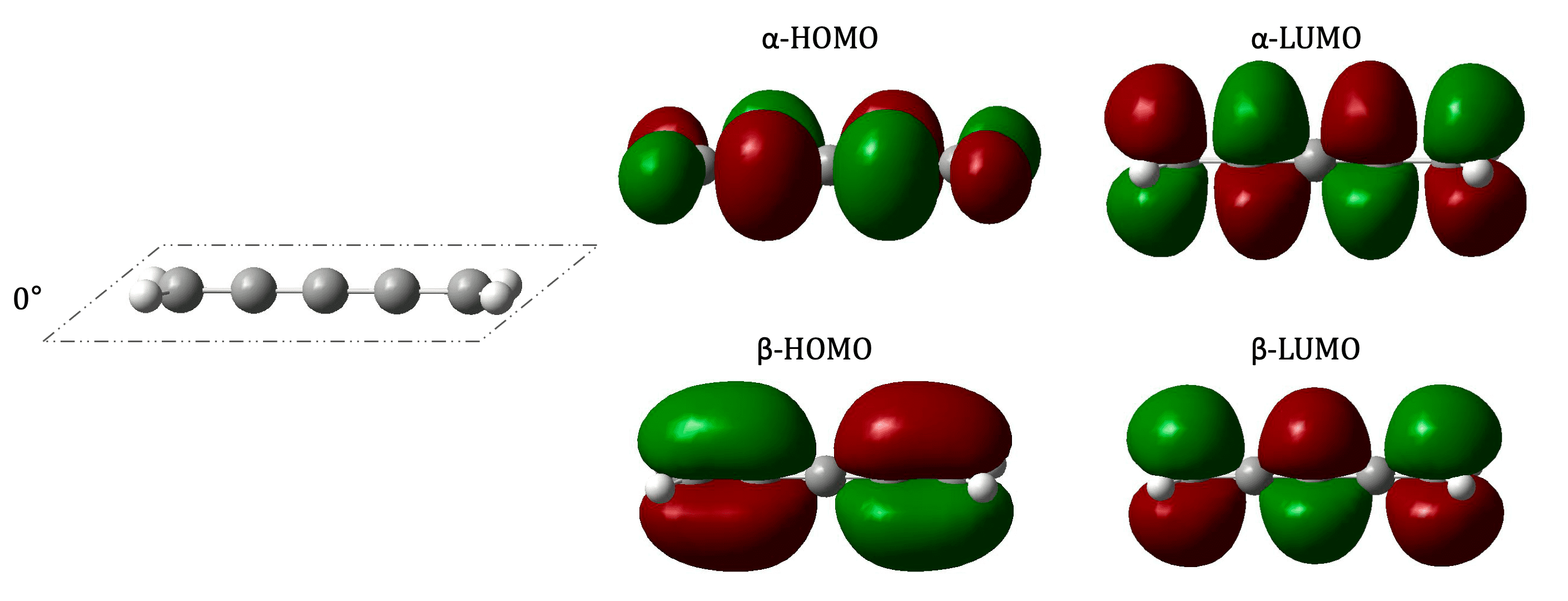}
    \includegraphics[width=0.7\linewidth]{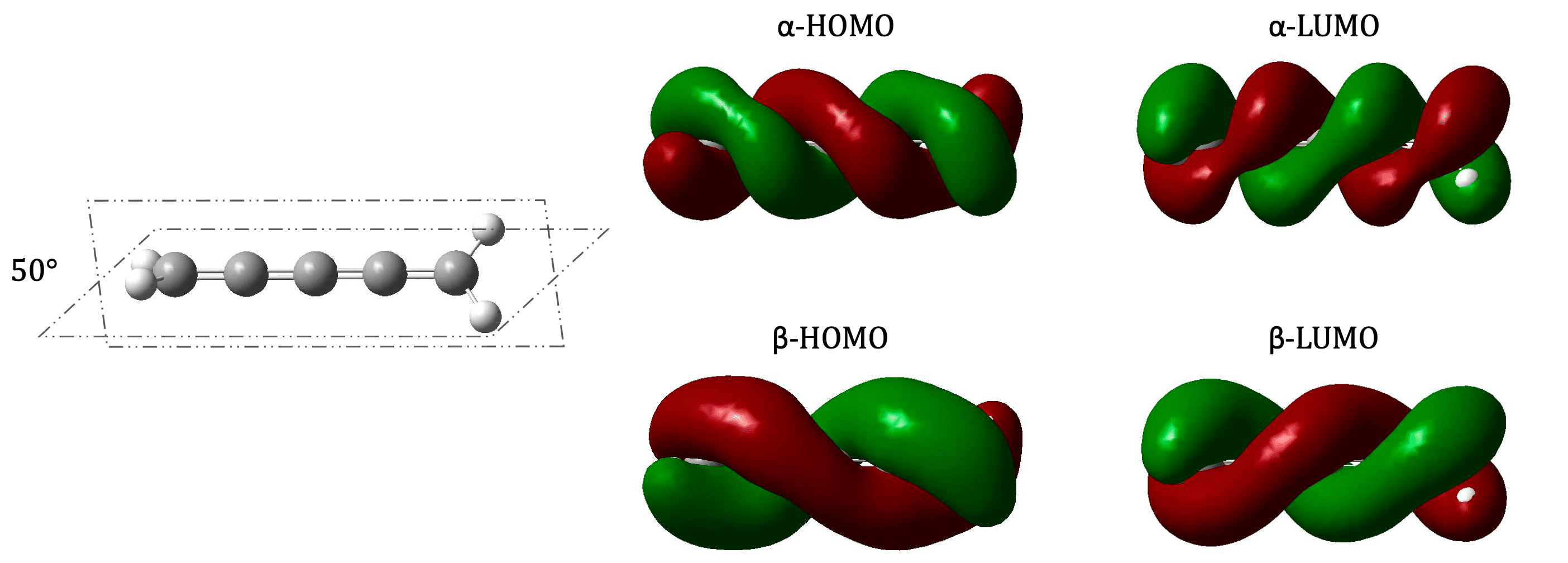}
    \includegraphics[width=0.7\linewidth]{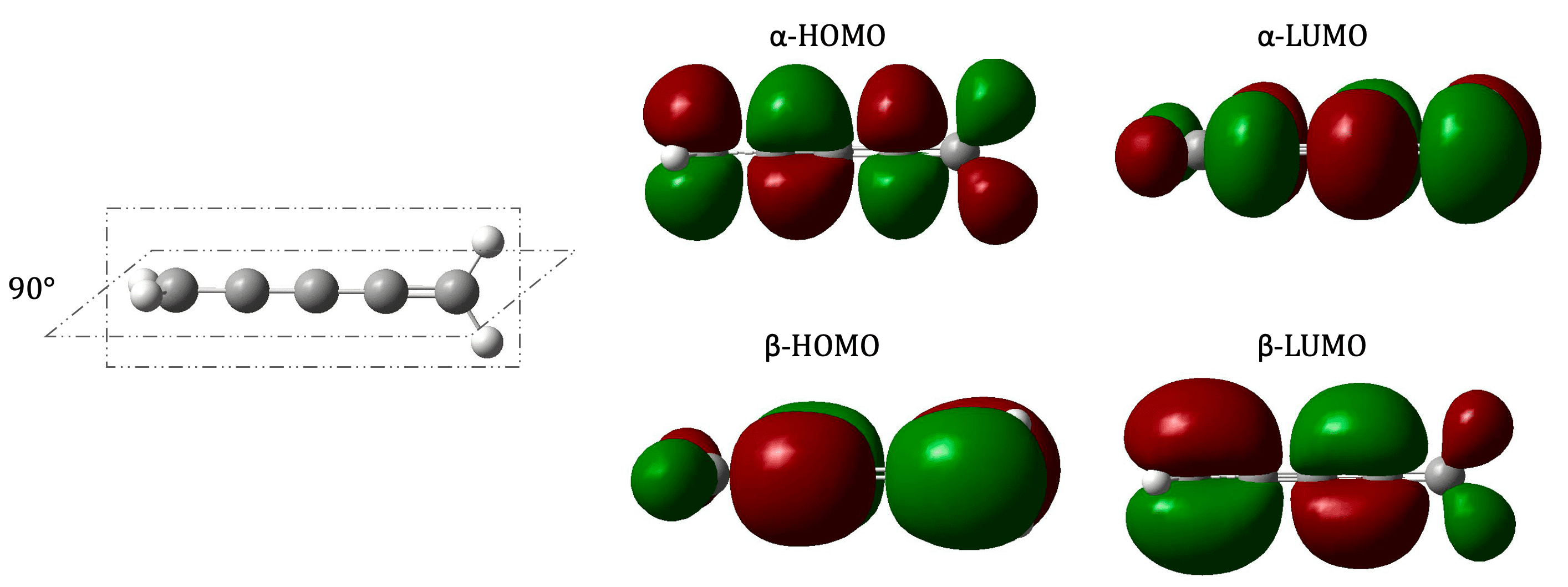}
    \caption{Representation of the HOMO and LUMO MOs obtained at the B3LYP/6-311G(d.p) level of theory for the $N=4$ cumulene in its first excited triplet  state at a 0°, 50 ° and 90 °rotation (see Table S4 for details).}
    \label{fig18-4cum-trip}
\end{figure}

However, the {\bf symmetry group of the MOs of a given molecule depends on its energy state}. As a consequence, if the symmetry group of a given molecule is such that the molecule does not admit helical OMs but contains symmetry elements which alone satisfy the helical state theorem, one can find an {\bf excited versions} of it, which can nevertheless {\bf exhibit helical MOs} thanks to {\bf symmetry adapted MOs}.\\ 

As we see, the previous remark also applies also for {\bf different electronic multiplicities}. \\

In the next section, we study more complex examples illustrating the previous discussion. In particular, we observe the following results:

\begin{itemize}
\item There exists generically excited version of the molecule leading to stable configuration admitting helical states. 

\item Symmetry adapted MOs can be used to predict the characteristics of the helixes. 
\end{itemize}
\newpage
\section{Induced helical MOs - Examples} 

\subsection{Cumulenes containing double-bonded heavy elements}

Cumulenes are a highly varied class of compounds, including such species as ketenes, allenes, ketenimines, and isocyanates, as well as analogues where carbons $C_0$ and $C_N$ are replaced by silicon, germanium, oxygen, sulfur, nitrogen phosphorus and/or arsenic. In agreement with the review of Escudié et al. \cite{escudie} fifteen families of compounds were studied in order to scan a major part of the experimentally known cumulenes containing double-bonded heavy elements. Examples of molecules (see table \ref{table6-heavy}) were tested at the B3LYP/6-311G(d.p) level of theory in order to study the presence of helical orbitals.  
  
\begin{table}[H]
    \caption{Cumulenes containing doubly bonded heavy elements} 
    \label{table6-heavy}
    \centering
    \begin{tabular}{cccccc}
    \midrule
                                  & $A_0$/$A_N$ & $L_1$ & $L_2$ & $R_1$ & $R_2$ \\
                                  \midrule
    Cumulene                      & C/C         & H     & H     & H     & H     \\
    \midrule
    $N$=3 :                       &             &       &       &       &       \\ 
    Phosphabutatriene             & P/C         & H     & -     & H     & H     \\ 
    Diphosphabutatriene           & P/P         & H     & -     & H     & -     \\ 
    Arsabutatriene                & As/C        & H     & -     & H     & H     \\ 
    Silabutatriene                & Si/C        & H     & H     & H     & H     \\ 
    \midrule                                
    $N$=2 :                       &             &       &       &       &       \\ 
    Phosphaallene                 & P/C         & H     & -     & H     & H     \\ 
    Diphosphaallene               & P/P         & H     & -     & H     & -     \\ 
    Phosphaazaallene              & P/N         & H     & -     & H     & -     \\ 
    Arasaallene                   & As/C        & H     & -     & H     & H     \\ 
    Arsaphosphaallene             & As/P        & H     & -     & H     & -     \\ 
    Diarsaallene                  & As/As       & H     & -     & H     & -     \\ 
    Silaallene                    & Si/C        & H     & H     & H     & H     \\ 
    Phosphasilaallene             & Si/P        & H     & H     & H     & -     \\ 
    Silaketene                    & Si/O        & H     & H     & -     & -     \\ 
    Germaallene                   & Ge/C        & H     & H     & H     & H     \\ 
    Germaphosphaallene            & Ge/P        & H     & H     & H     & -     \\
    \bottomrule
    \end{tabular}
\end{table}  

    \begin{table}[H]
    \caption{$N=3$. Molecular orbitals obtained at the B3LYP/6-311G(d.p) level of theory for a rotation of 25\Degre.} 
    \label{table7-mo-3-cum-25}
    \centering
    \begin{tabular}{ccccc}
        \midrule
           & \textbf{HOMO-1} & \textbf{HOMO} & \textbf{LUMO} & \textbf{LUMO+1} \\
             \bottomrule
    Phosphabutatrienes &
    {\includegraphics[height=1.40cm]{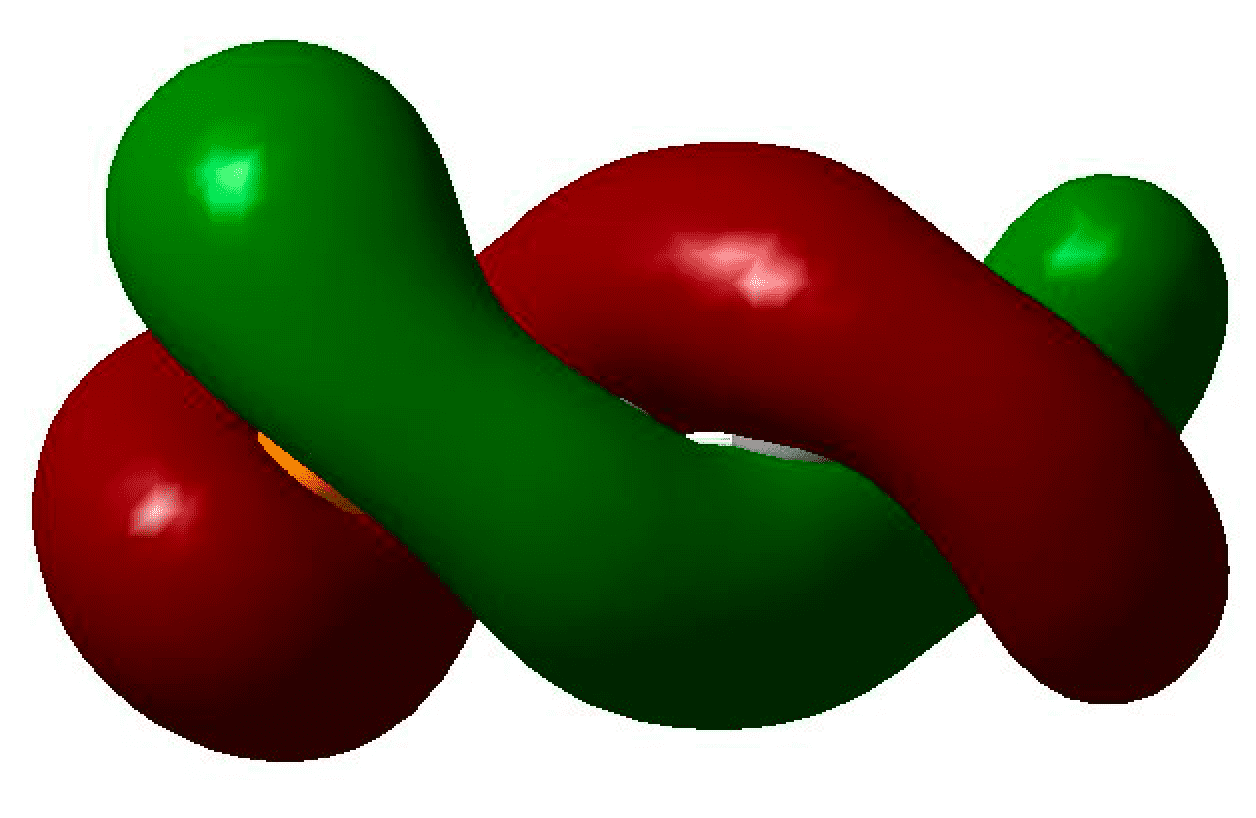}} &
    {\includegraphics[height=1.40cm]{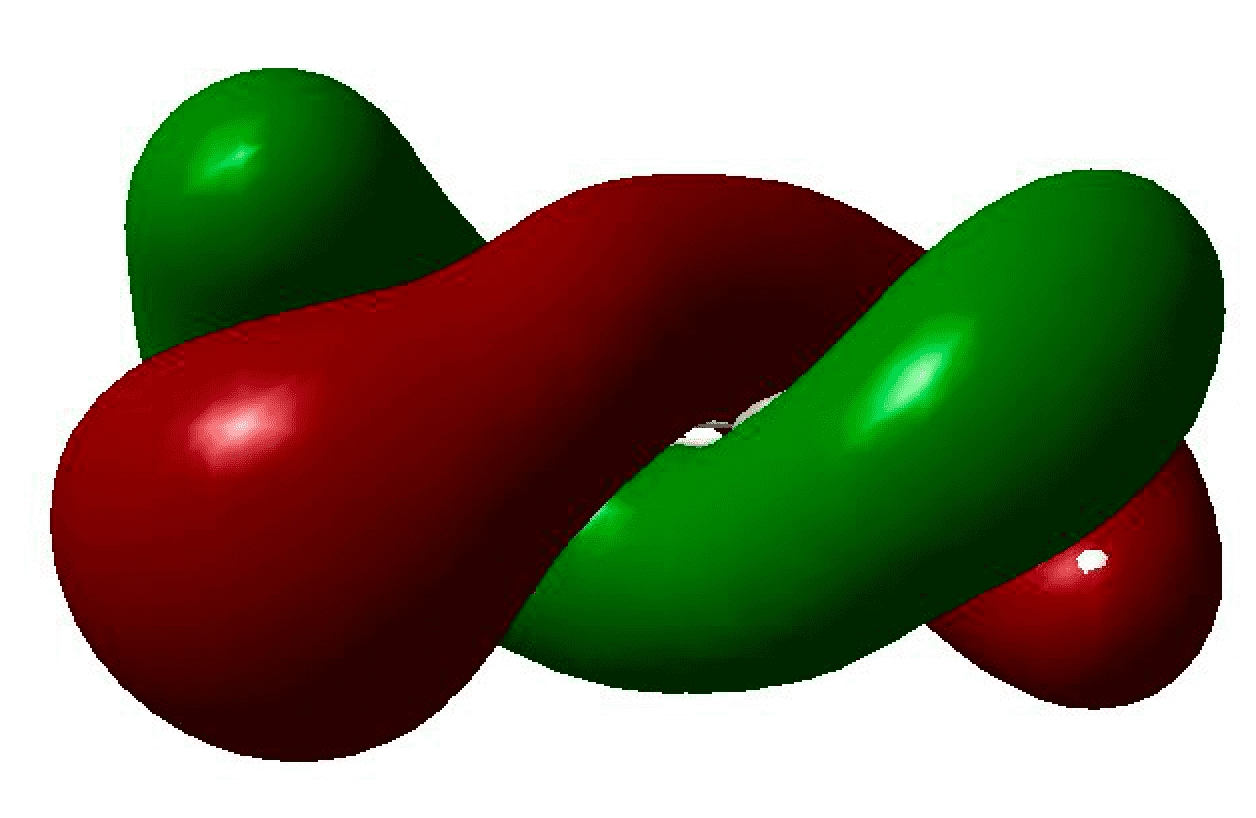}} &
    {\includegraphics[height=1.40 cm]{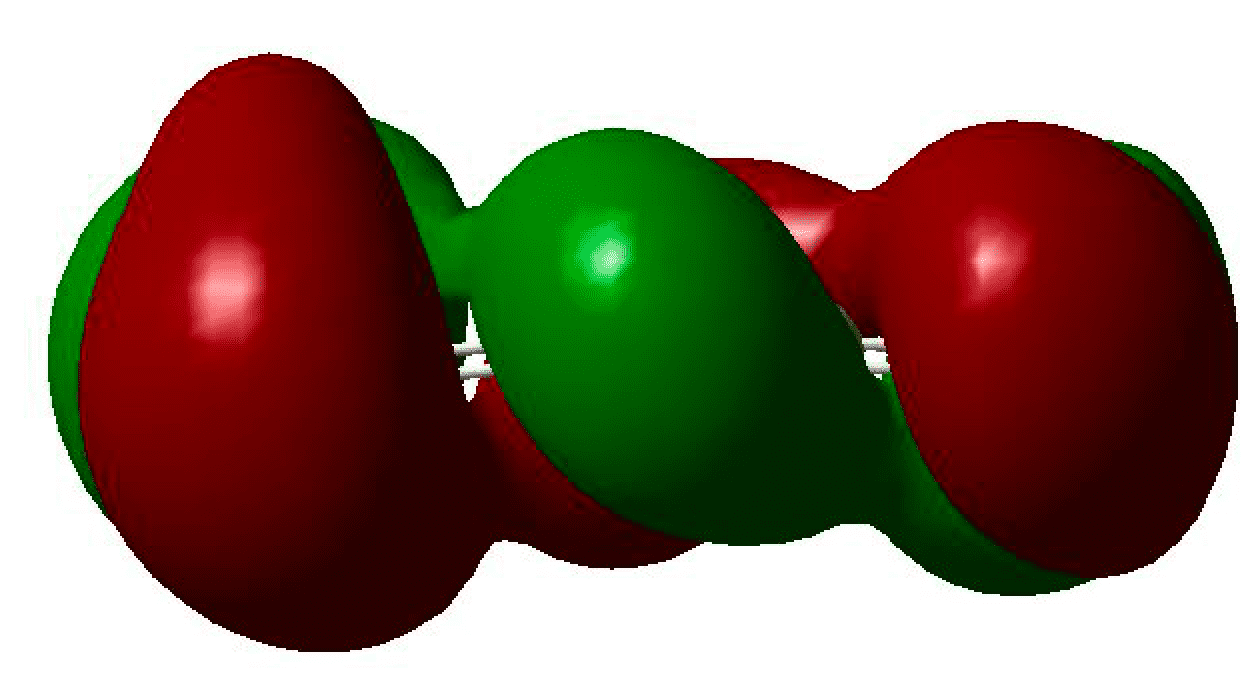}} & 
    {\includegraphics[height=1.40cm]{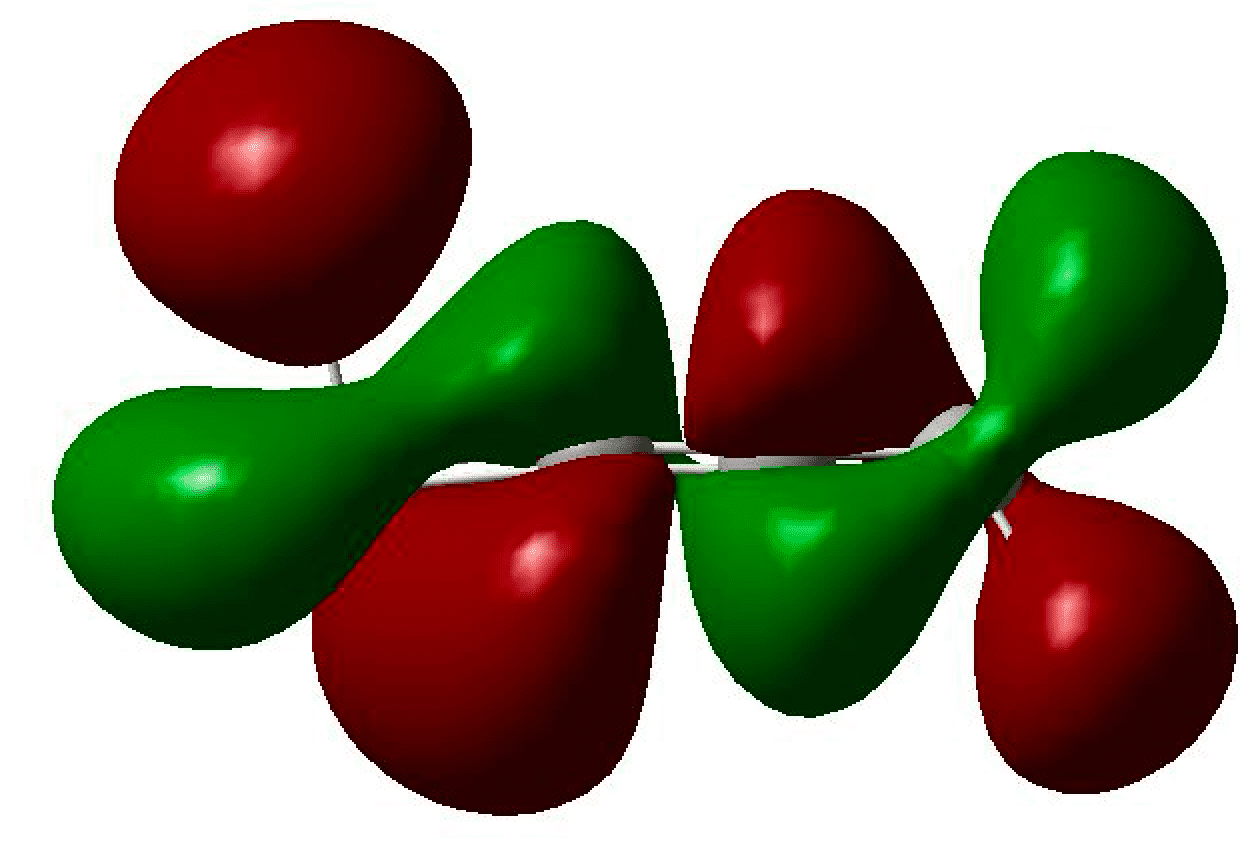}} \\                 
     \hdashline[1pt/1pt]                
    Diphosphabutatrienes &
    {\includegraphics[height=1.40cm]{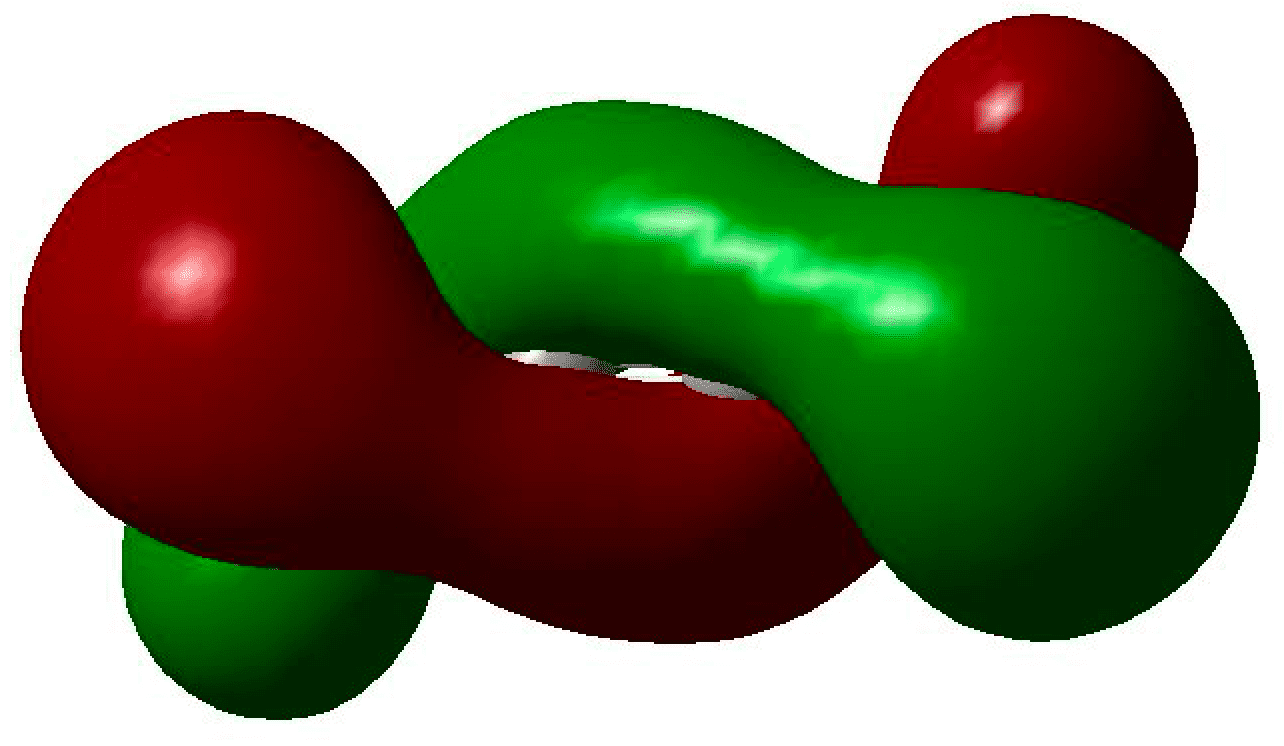}} &
    {\includegraphics[height=1.40cm]{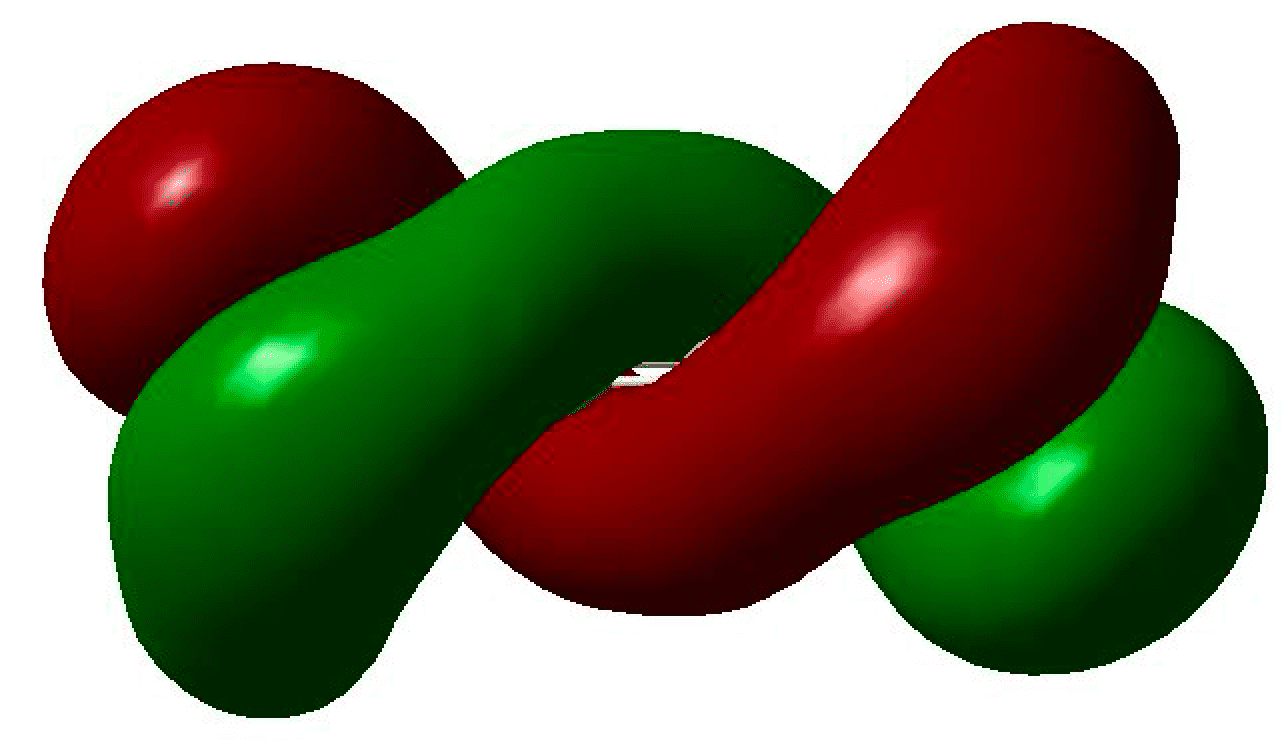}} &
    {\includegraphics[height=1.40 cm]{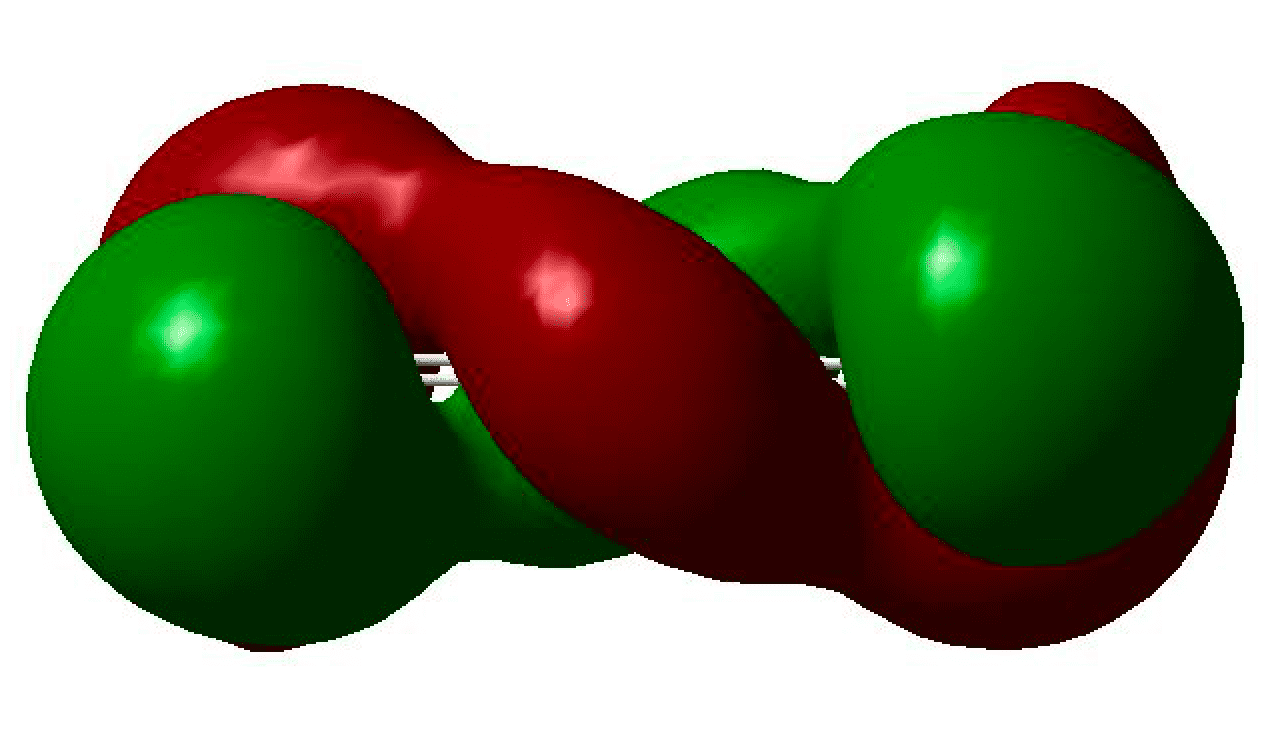}} & 
    {\includegraphics[height=1.40cm]{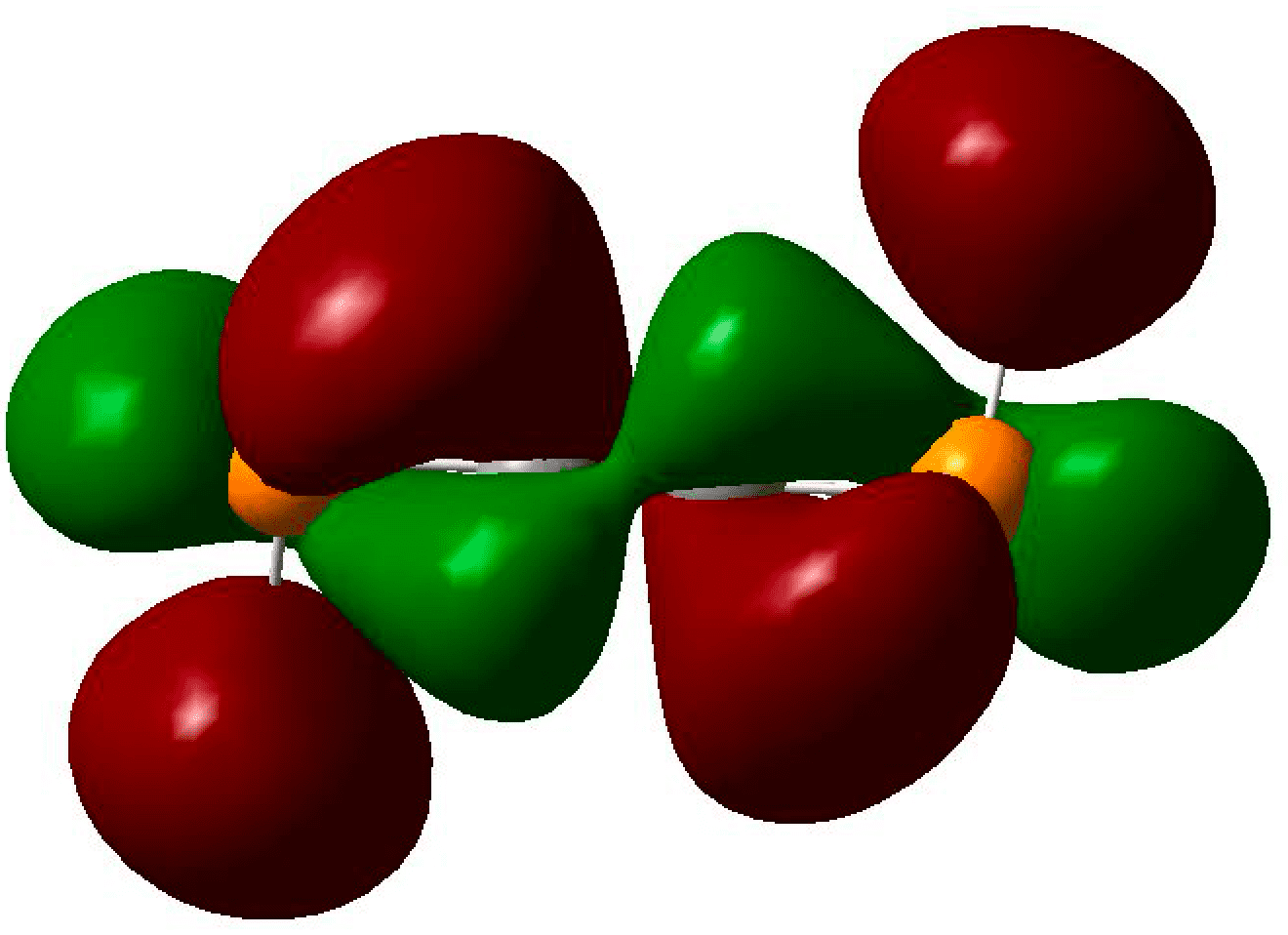}} \\
    \hdashline[1pt/1pt]
    Arsabutatrienes &
    {\includegraphics[height=1.40cm]{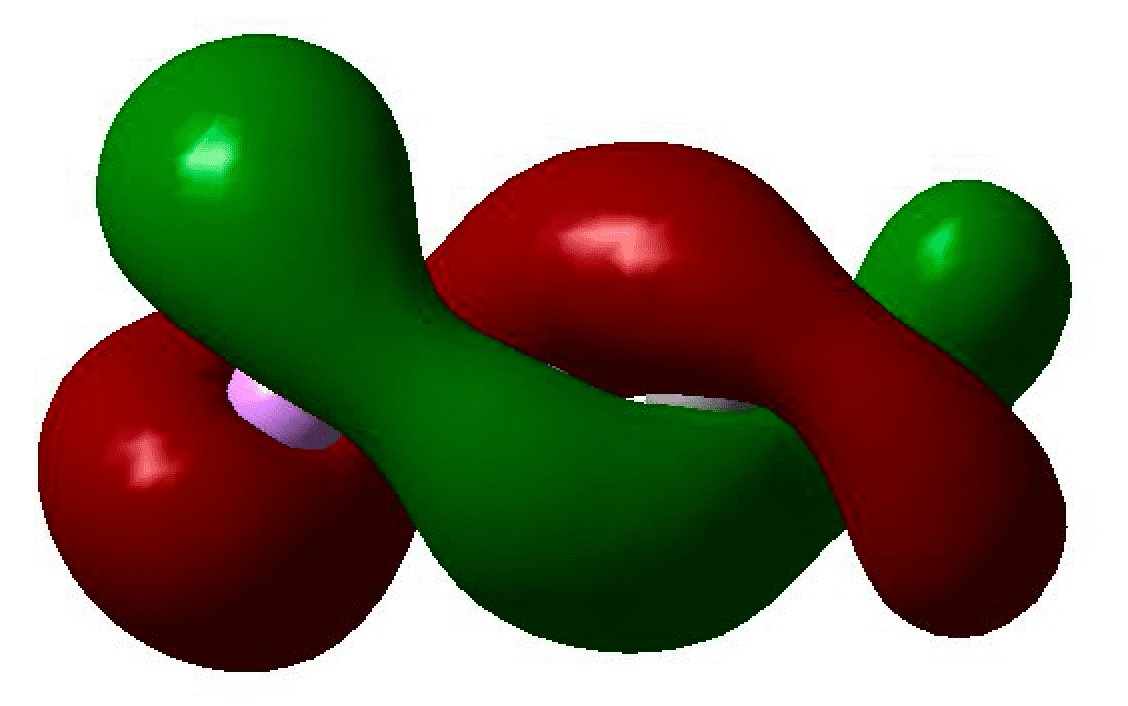}} &
    {\includegraphics[height=1.40cm]{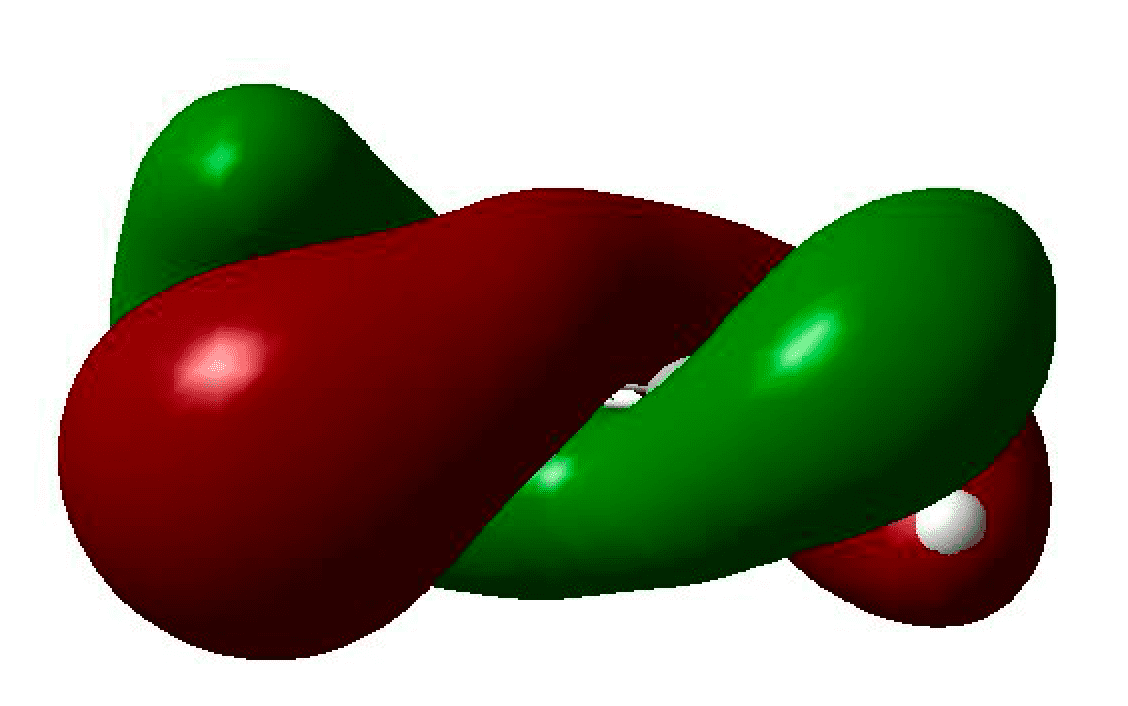}} &
    {\includegraphics[height=1.40 cm]{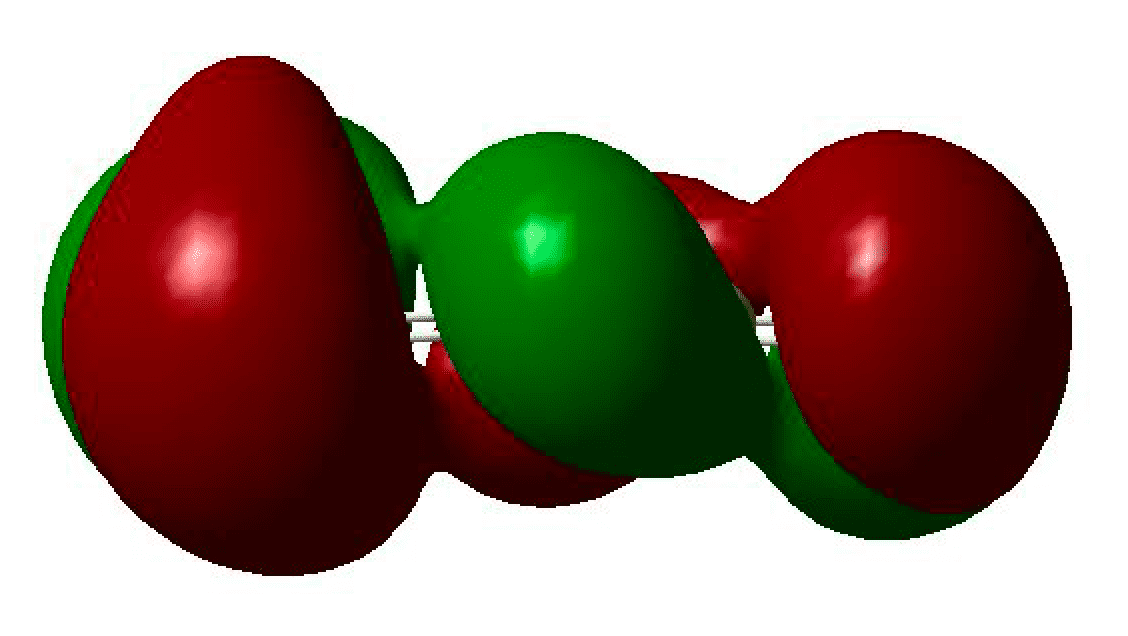}} & 
    {\includegraphics[height=1.40cm]{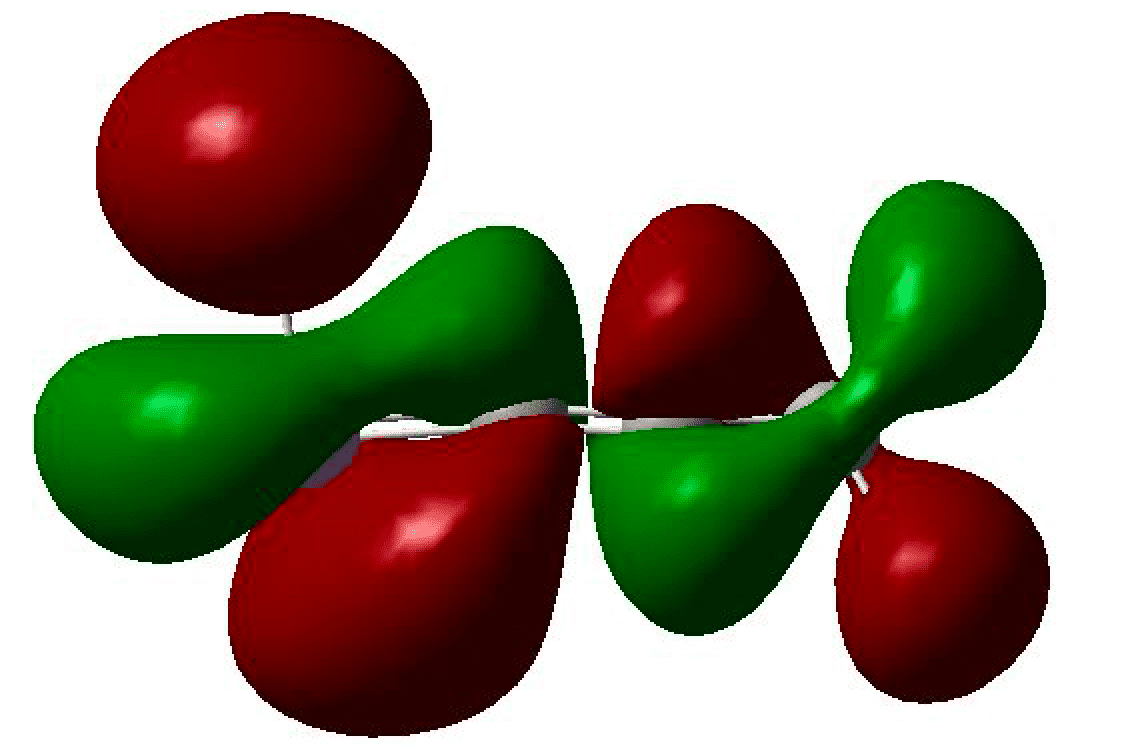}} \\
    \hdashline[1pt/1pt]
    Silabutatrienes &
    {\includegraphics[height=1.70cm]{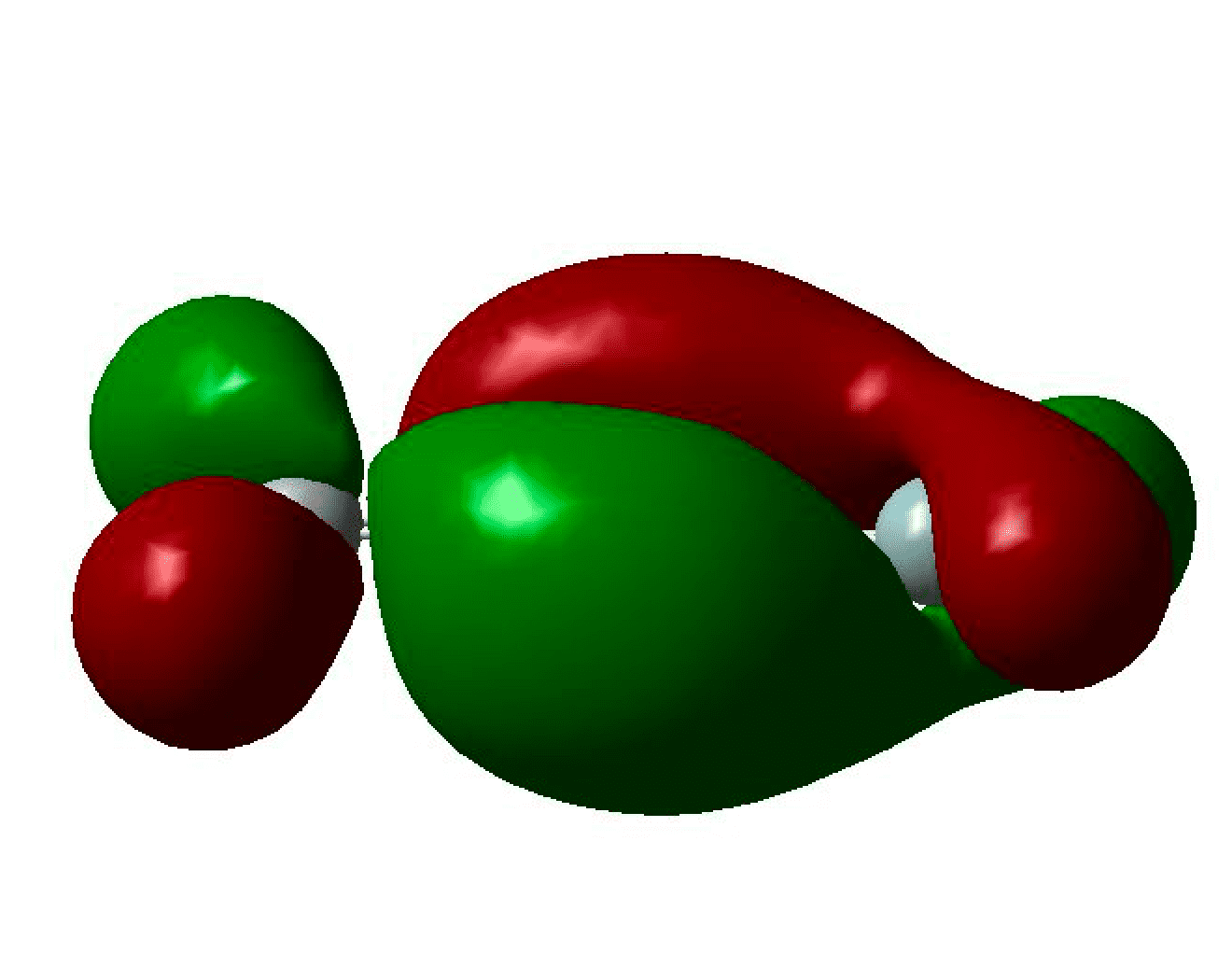}} &
    {\includegraphics[height=1.40cm]{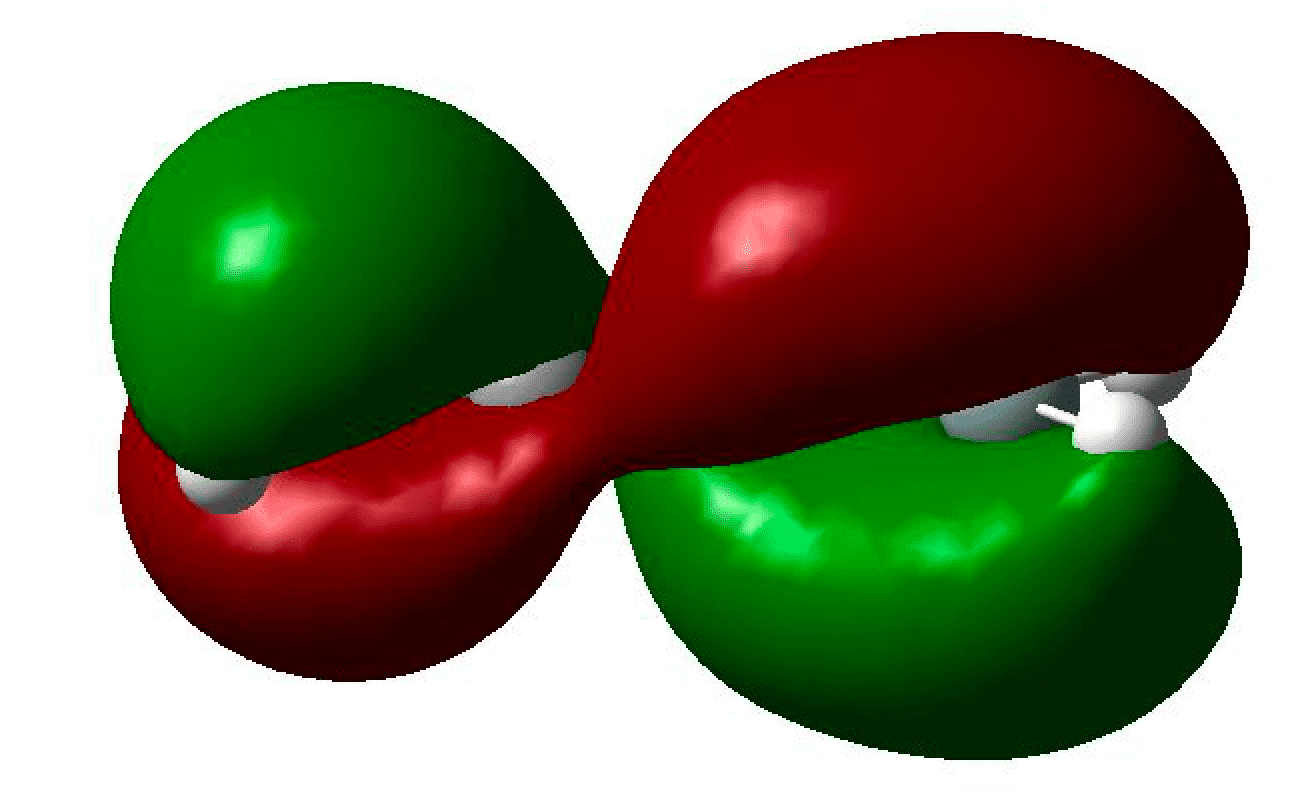}} &
    {\includegraphics[height=1.40cm]{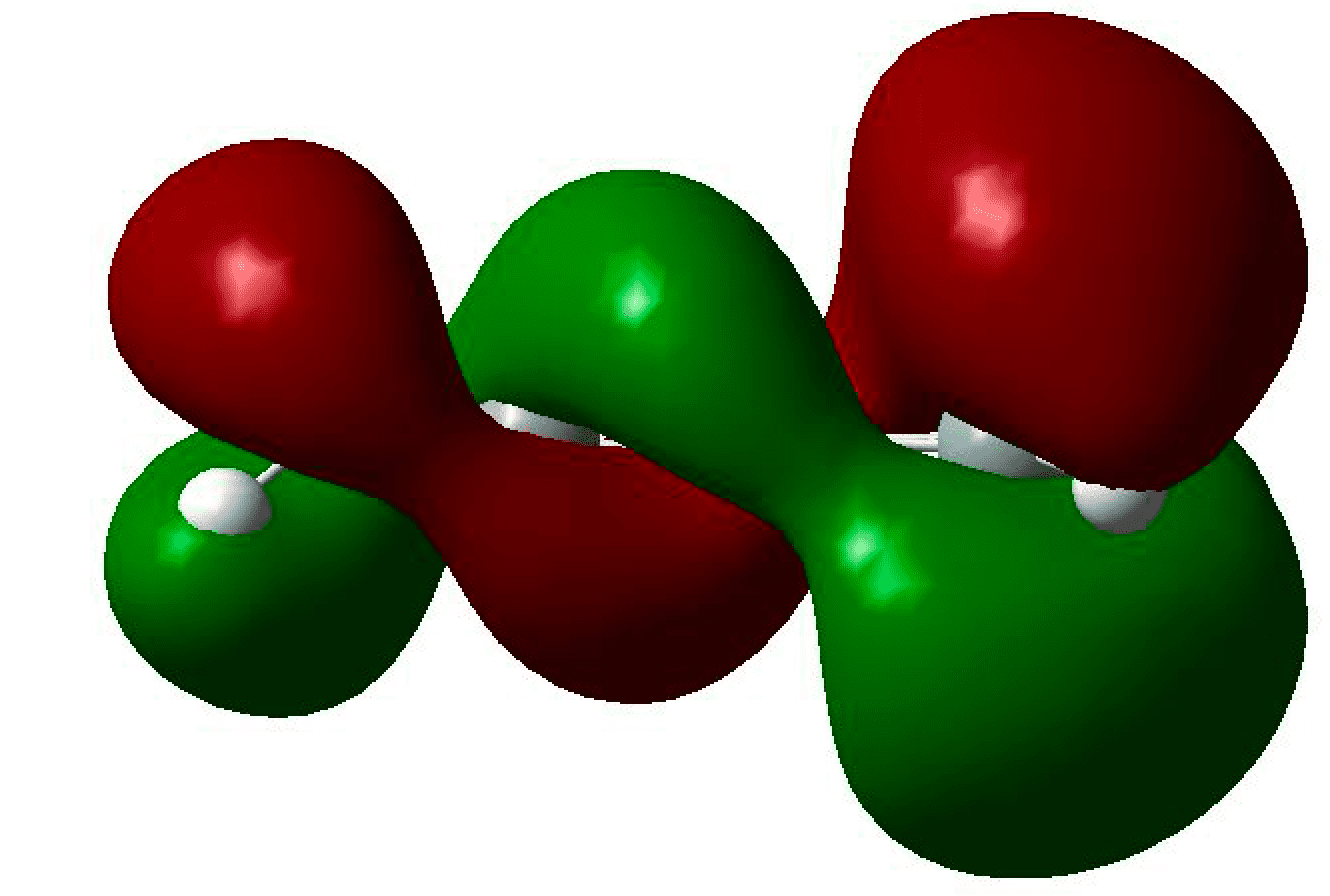}} & 
    {\includegraphics[height=1.40cm]{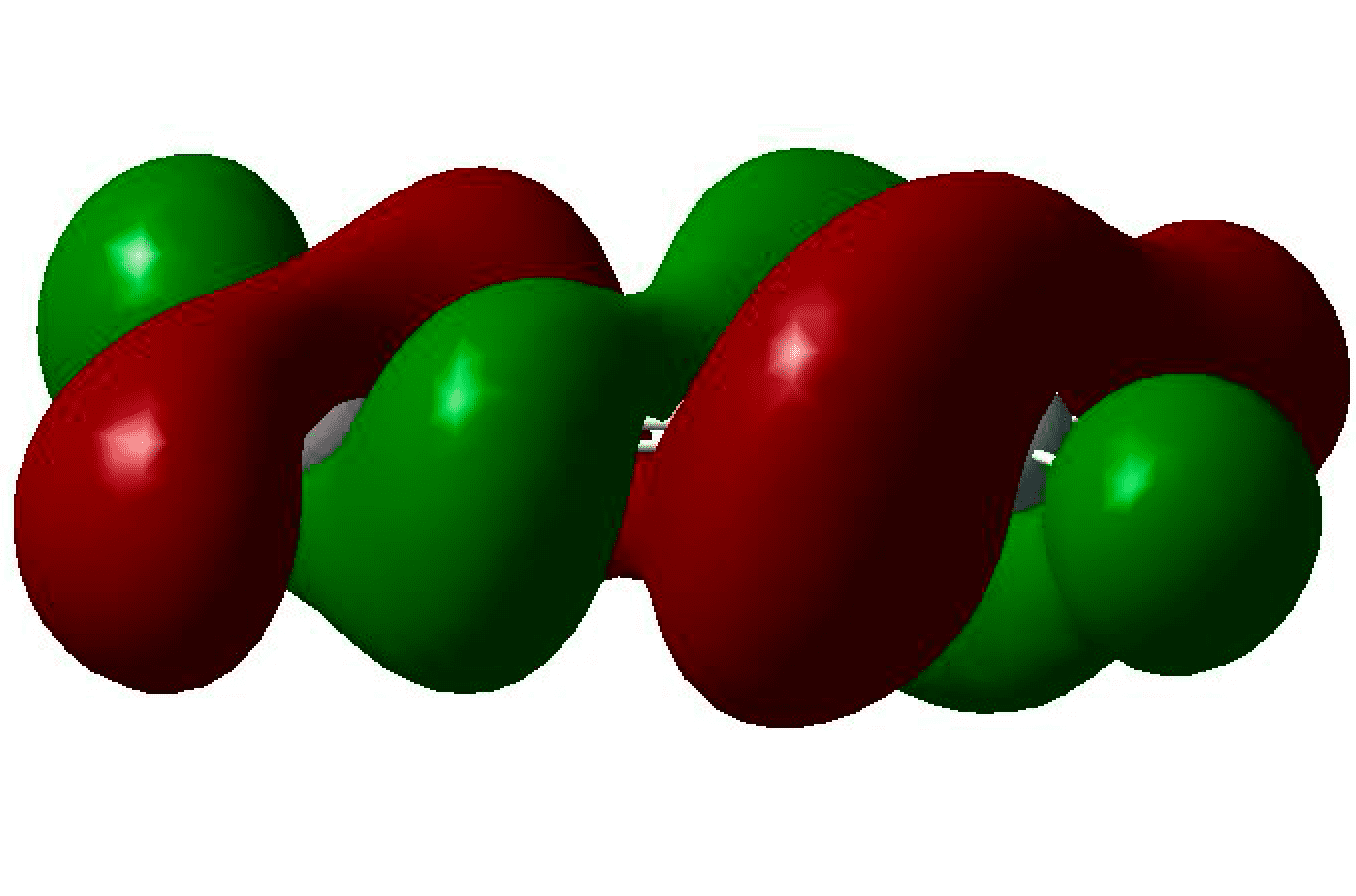}} \\
\bottomrule
        \end{tabular}
        \end{table}

\begin{table}[H]
    \caption{$N=2$. Molecular orbitals obtained at the B3LYP/6-311G(d.p) level of theory for a rotation of 25\Degre.} 
    \label{table8-2-cum-25}
    \centering
    \begin{tabular}{ccccc}
        \midrule
        & \textbf{HOMO-1} & \textbf{HOMO} & \textbf{LUMO} & \textbf{LUMO+1} \\
          \bottomrule
    Phosphaallenes &
    {\includegraphics[height=1.40cm]{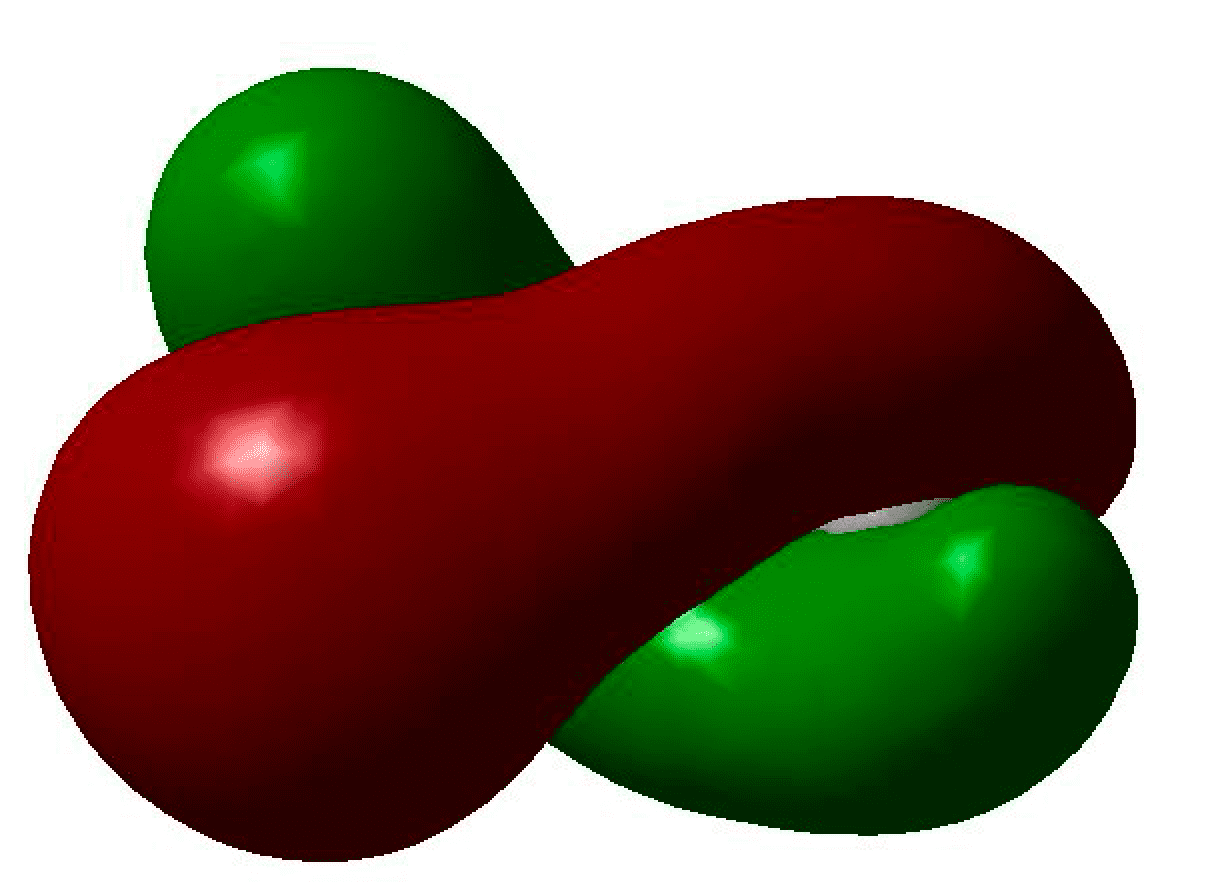}} &
    {\includegraphics[height=1.40cm]{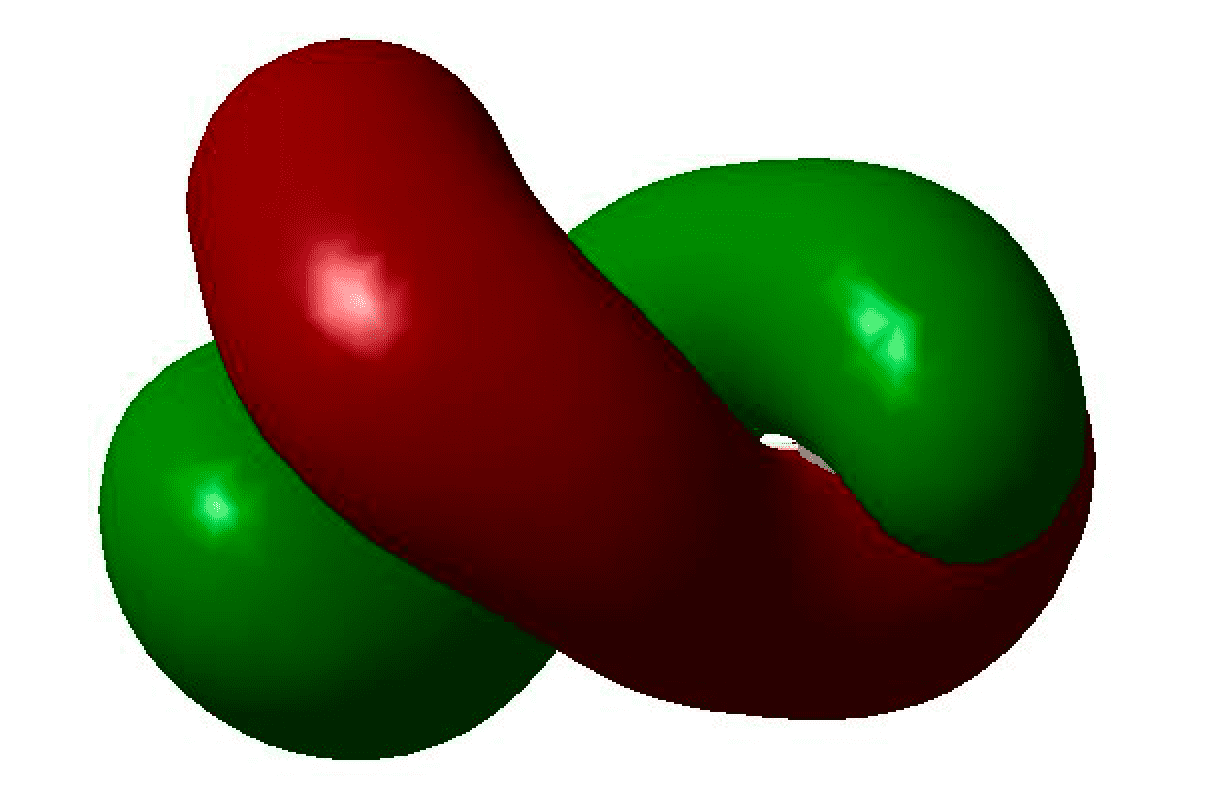}} &
    {\includegraphics[height=1.40 cm]{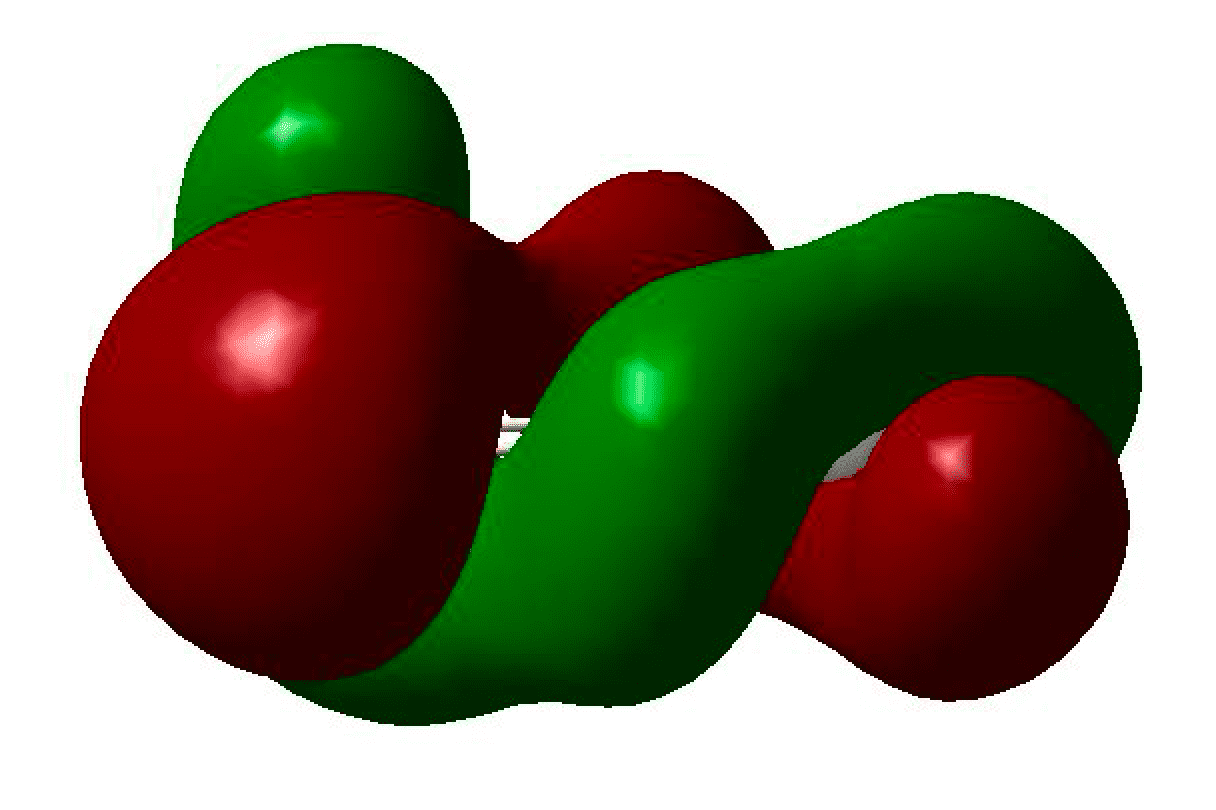}} & 
    {\includegraphics[height=1.40cm]{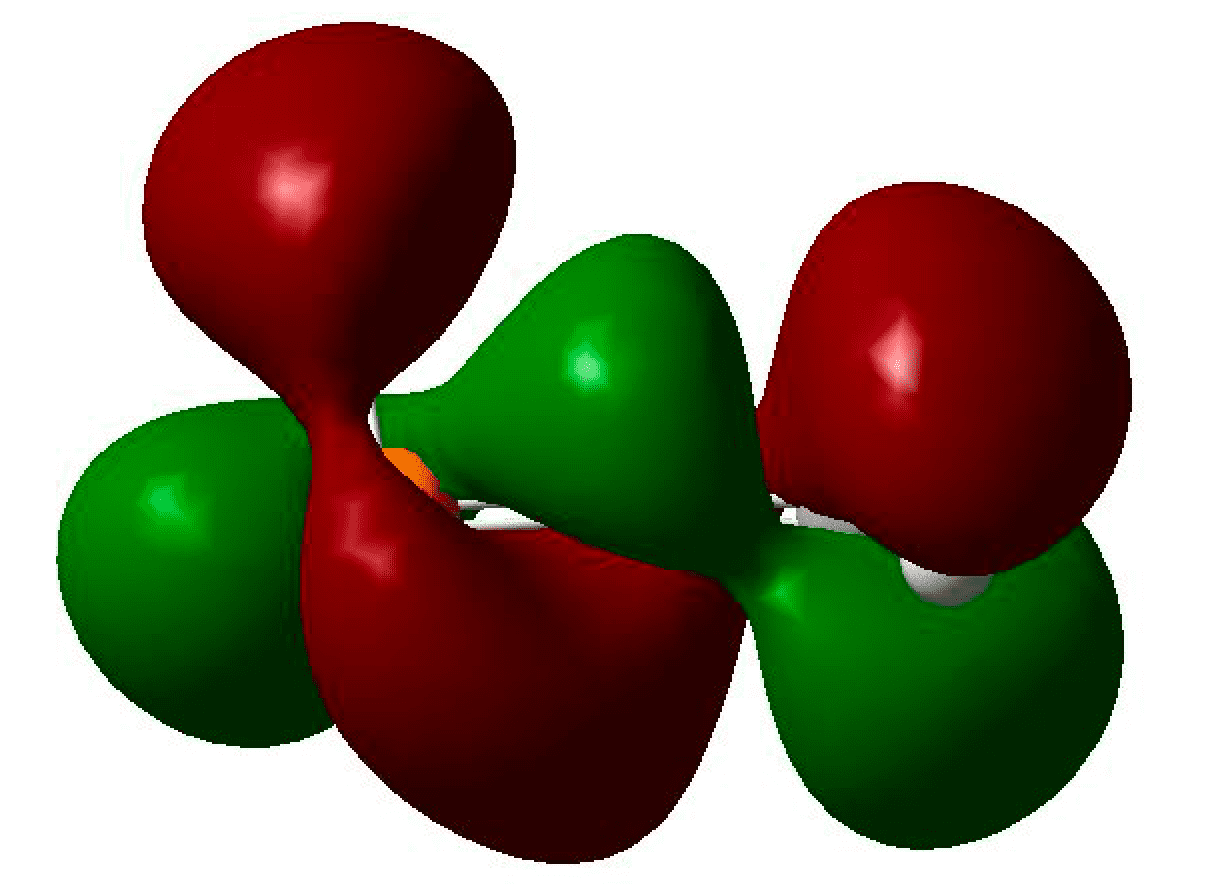}} \\
     \hdashline[1pt/1pt]
     Diphosphoallenes  &
    {\includegraphics[height=1.40cm]{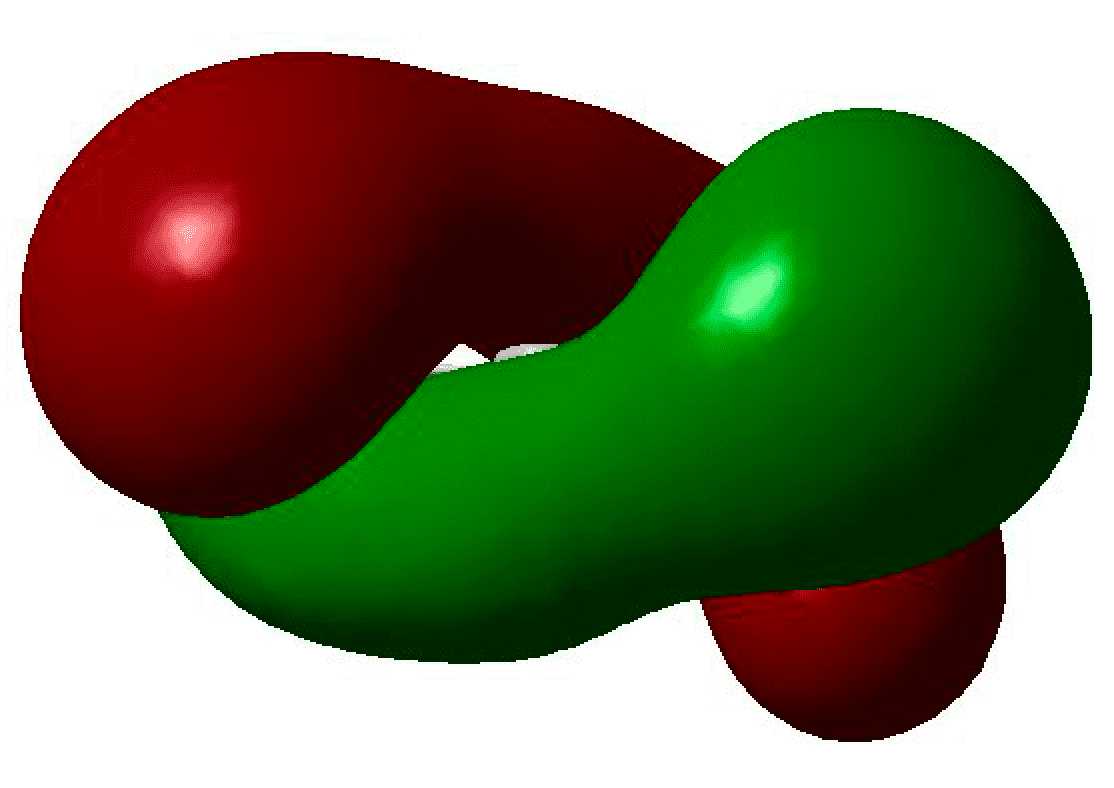}} &
    {\includegraphics[height=1.40cm]{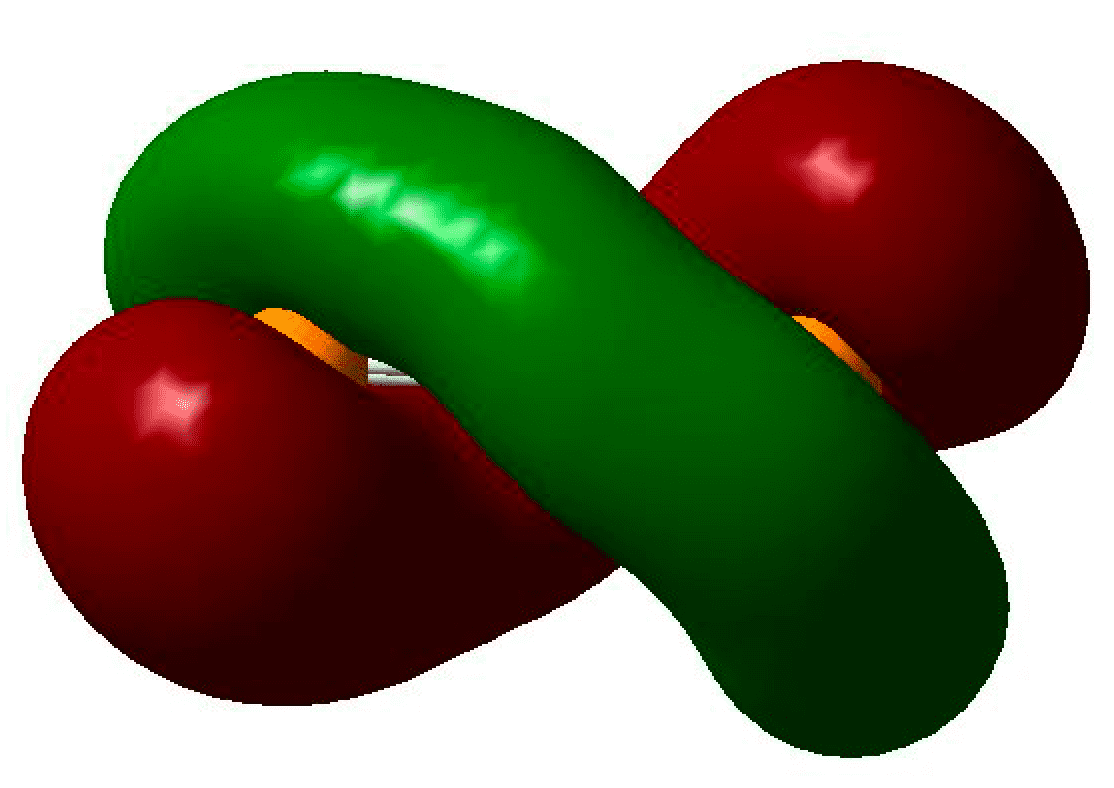}} &
    {\includegraphics[height=1.40 cm]{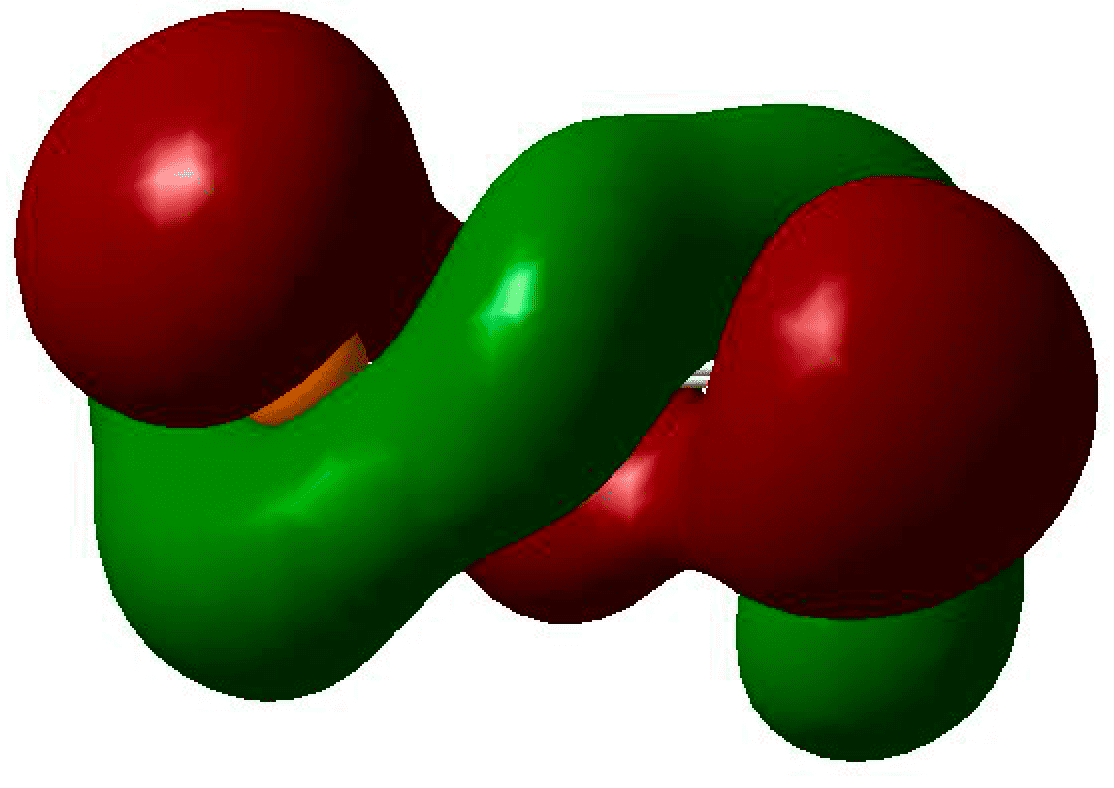}} & 
    {\includegraphics[height=1.40cm]{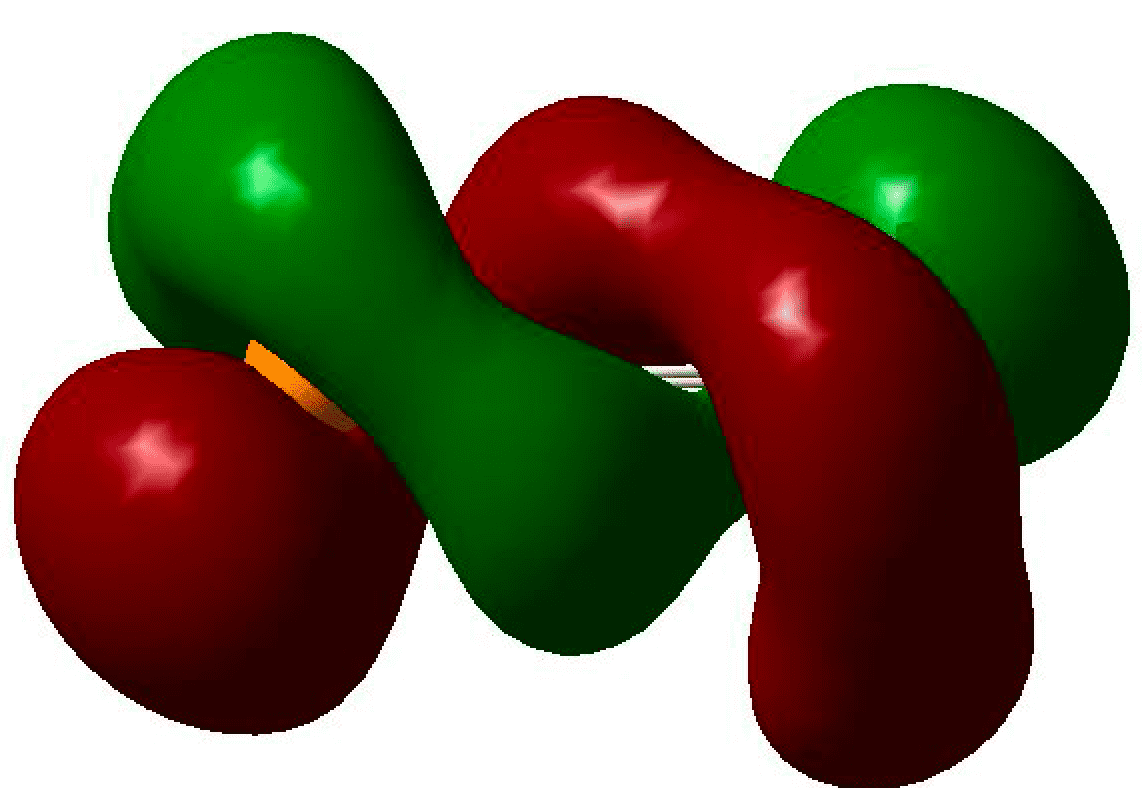}} \\
     \hdashline[1pt/1pt]
     Phosphaazaallenes  &
    {\includegraphics[height=1.40cm]{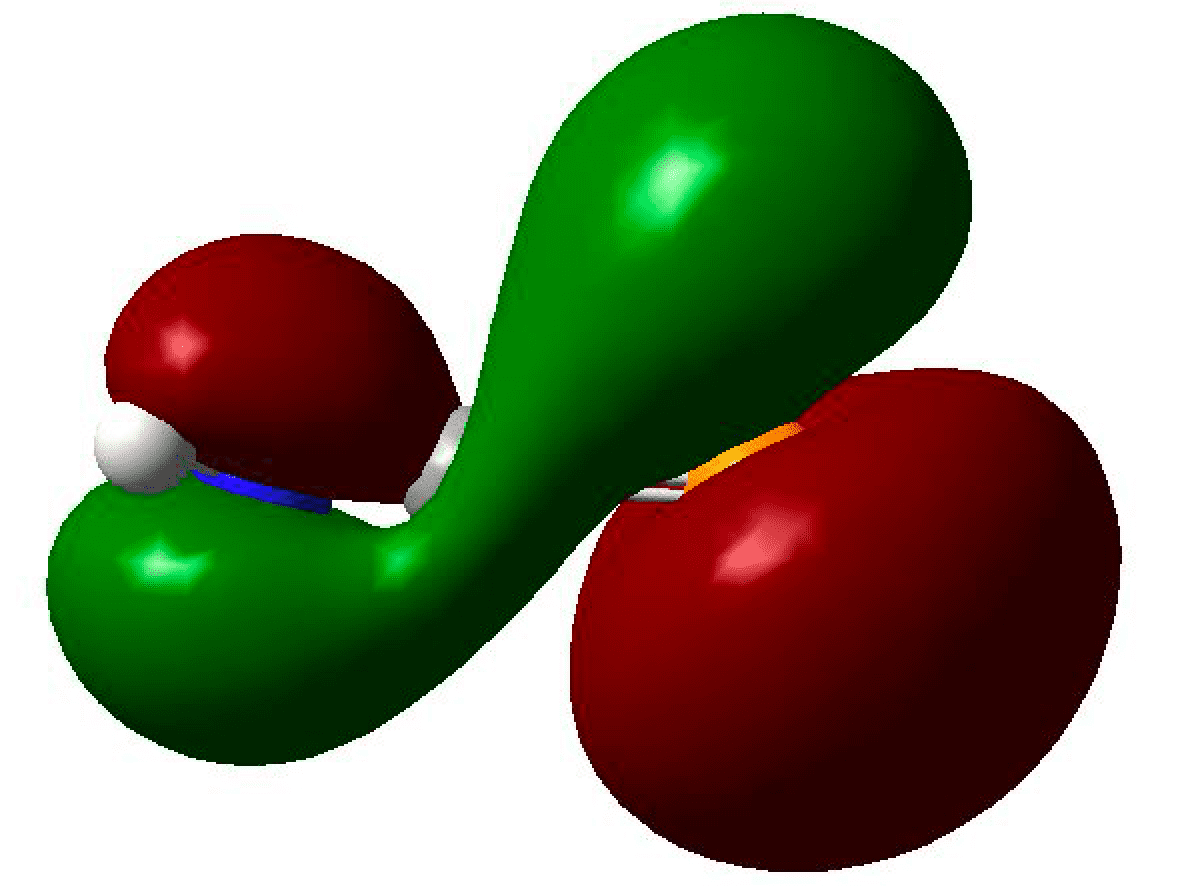}} &
    {\includegraphics[height=1.40cm]{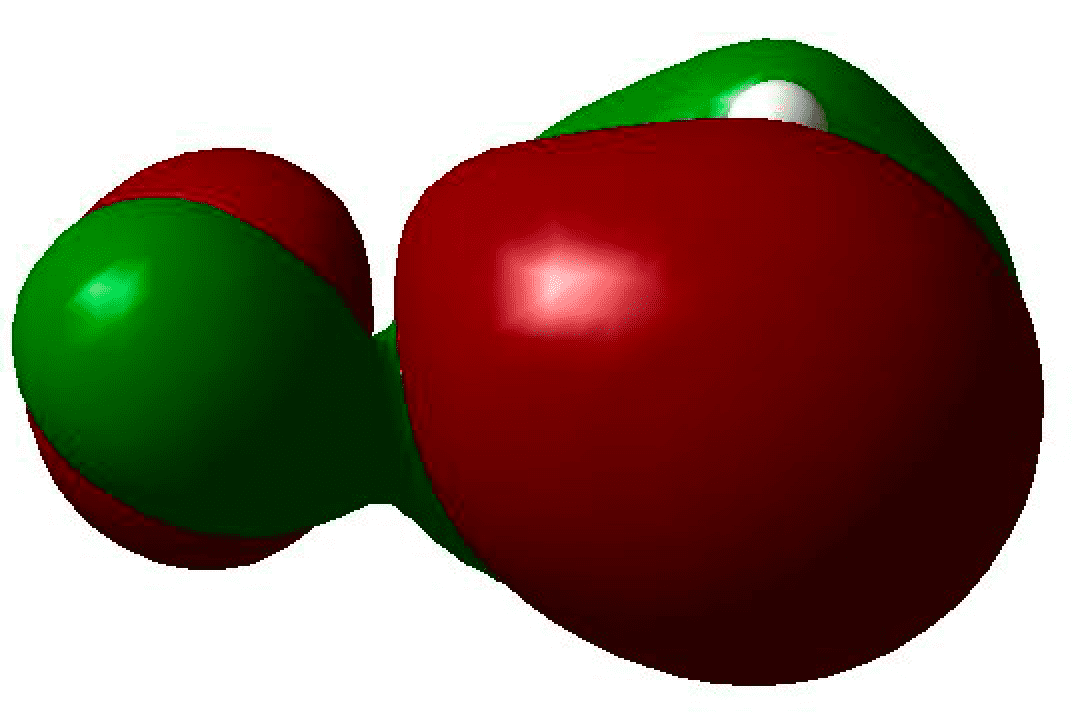}} &
    {\includegraphics[height=1.40 cm]{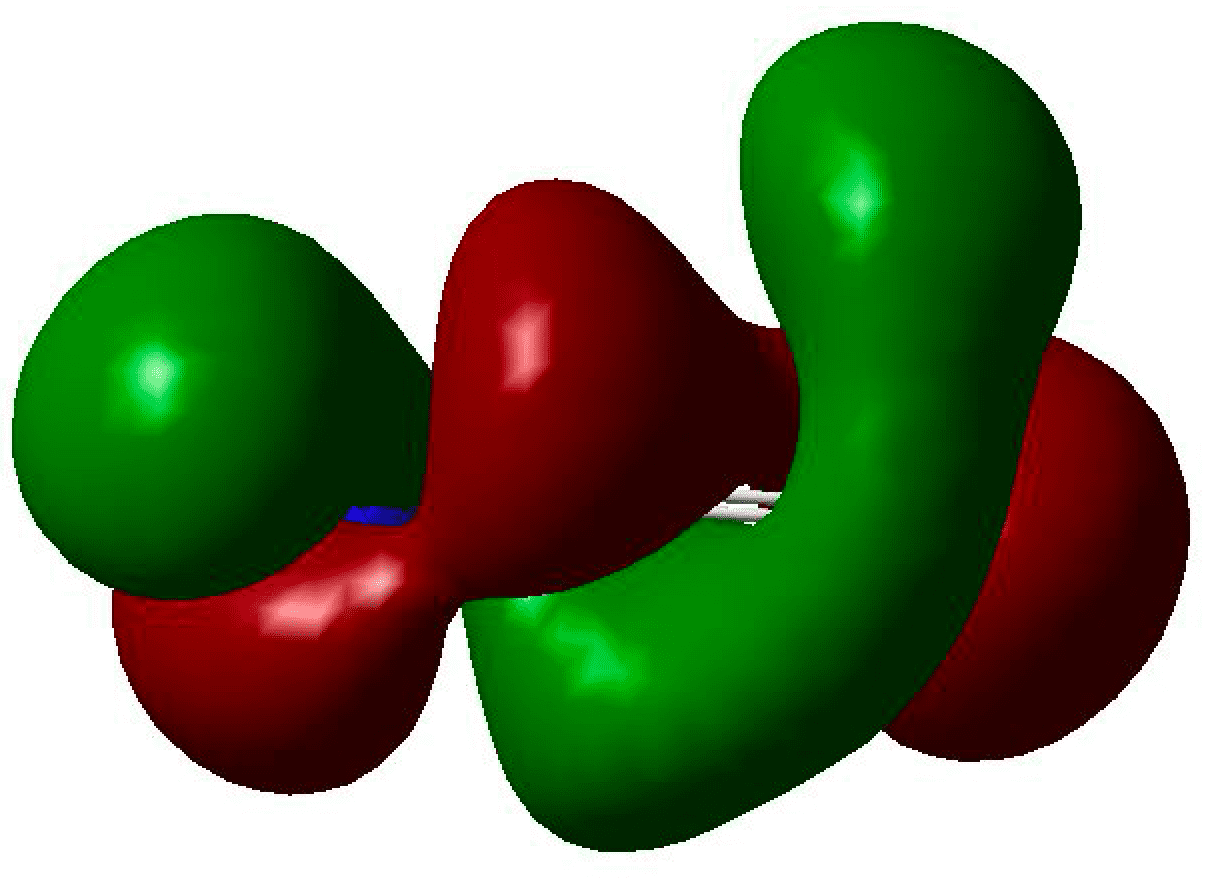}} & 
    {\includegraphics[height=1.40cm]{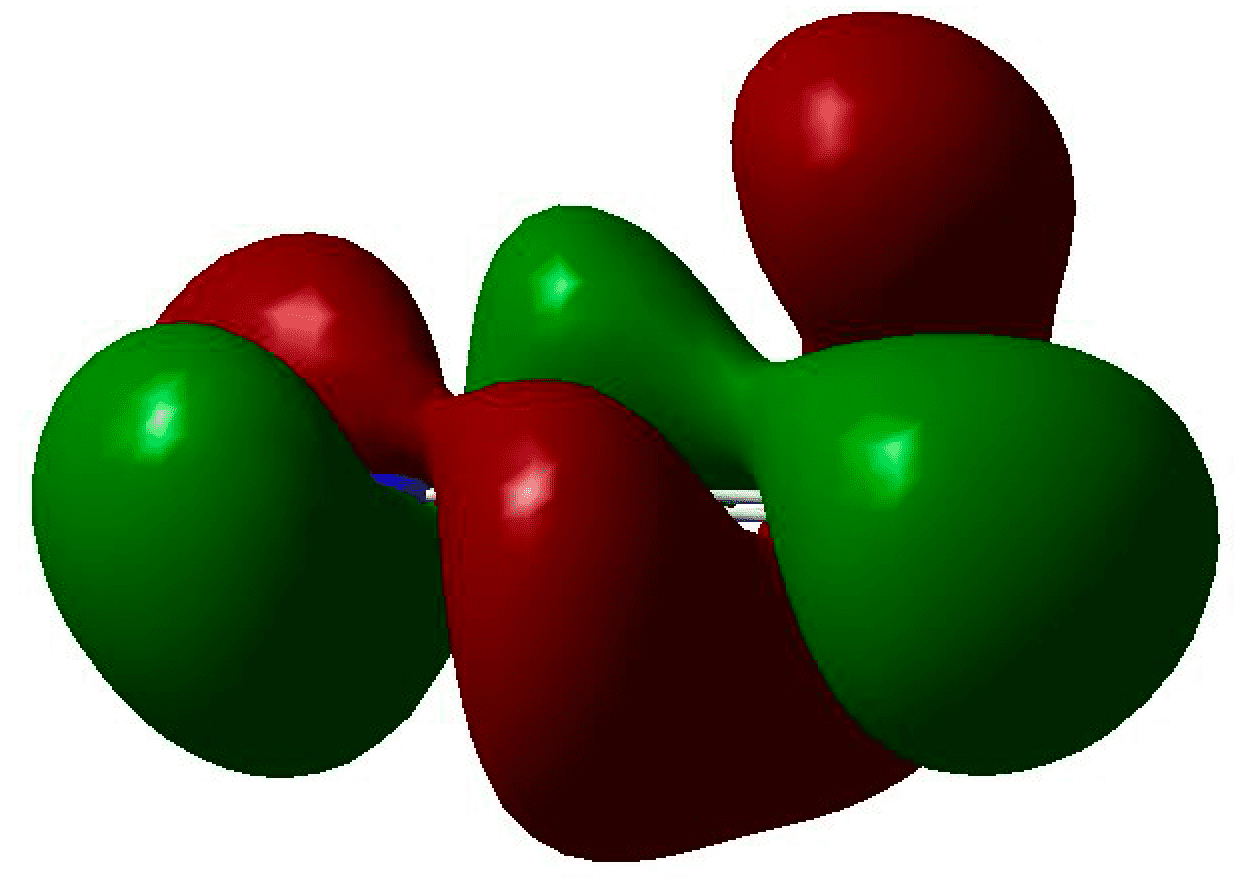}} \\
    \hdashline[1pt/1pt]
     Arsaallenes  &
    {\includegraphics[height=1.40cm]{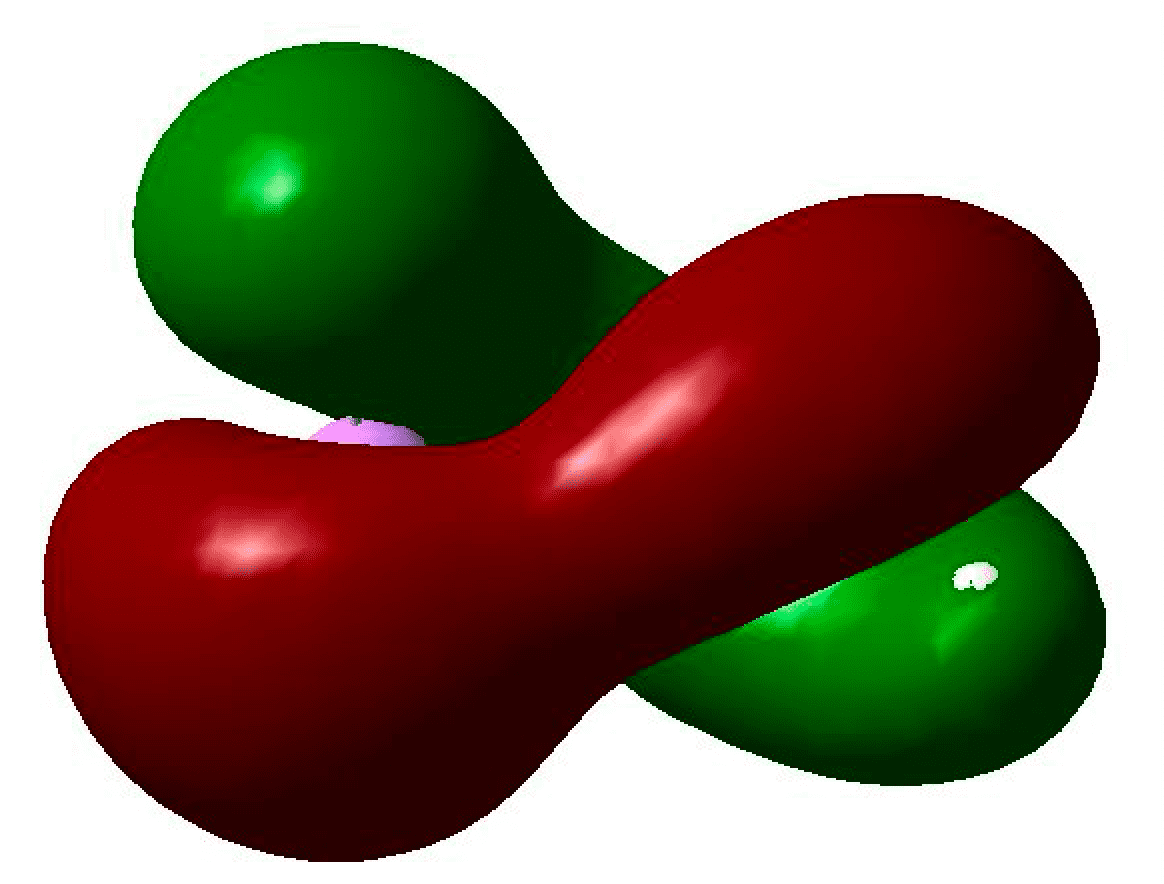}} &
    {\includegraphics[height=1.40cm]{HOMO-1-arsaallene-4.png}} &
    {\includegraphics[height=1.40 cm]{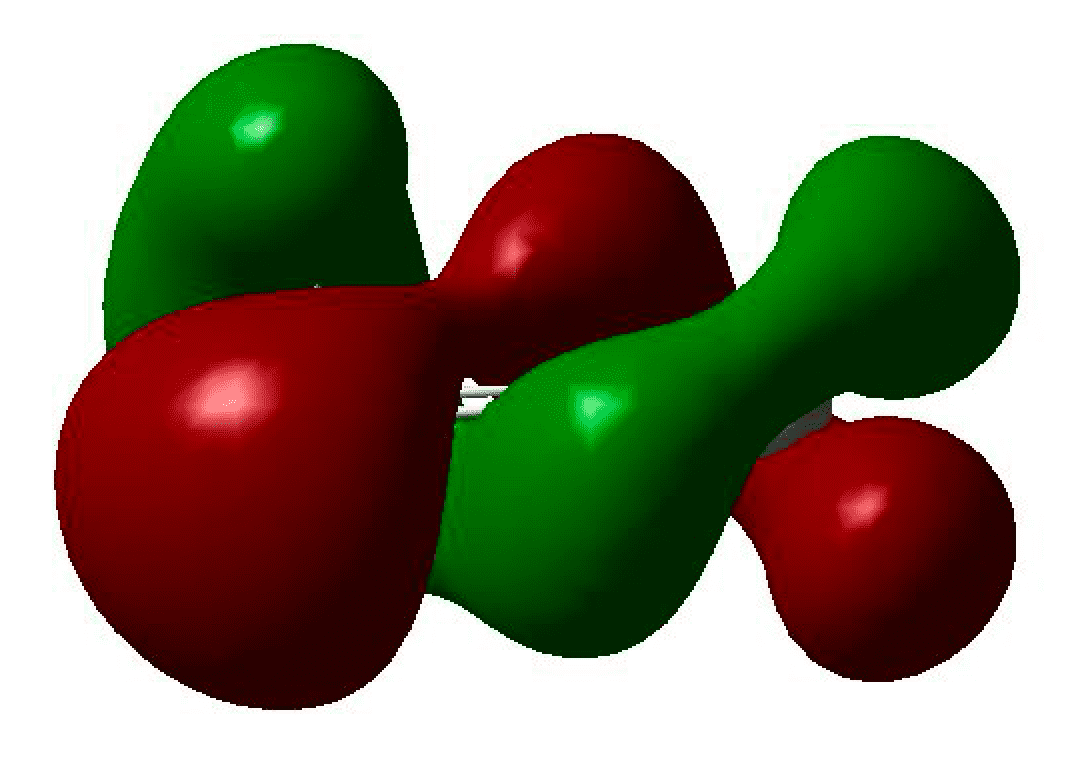}} & 
    {\includegraphics[height=1.40cm]{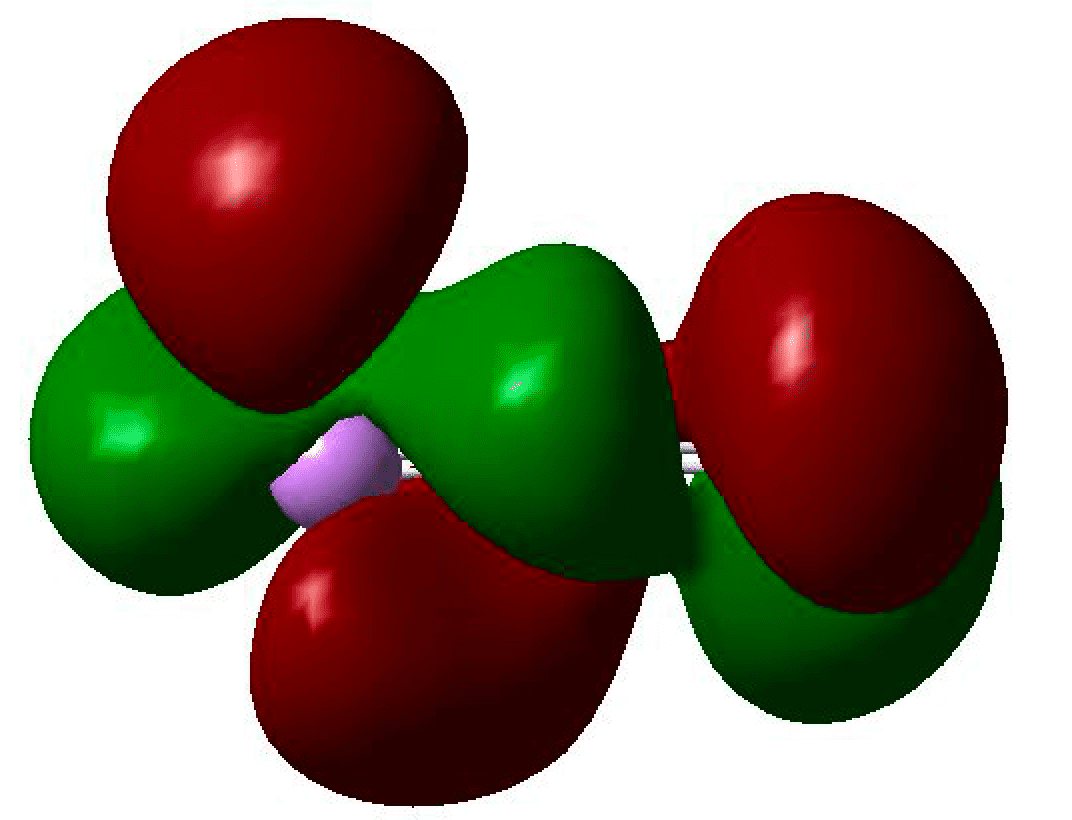}} \\
     \hdashline[1pt/1pt]
    Arsaphosphaallenes &
    {\includegraphics[height=1.40cm]{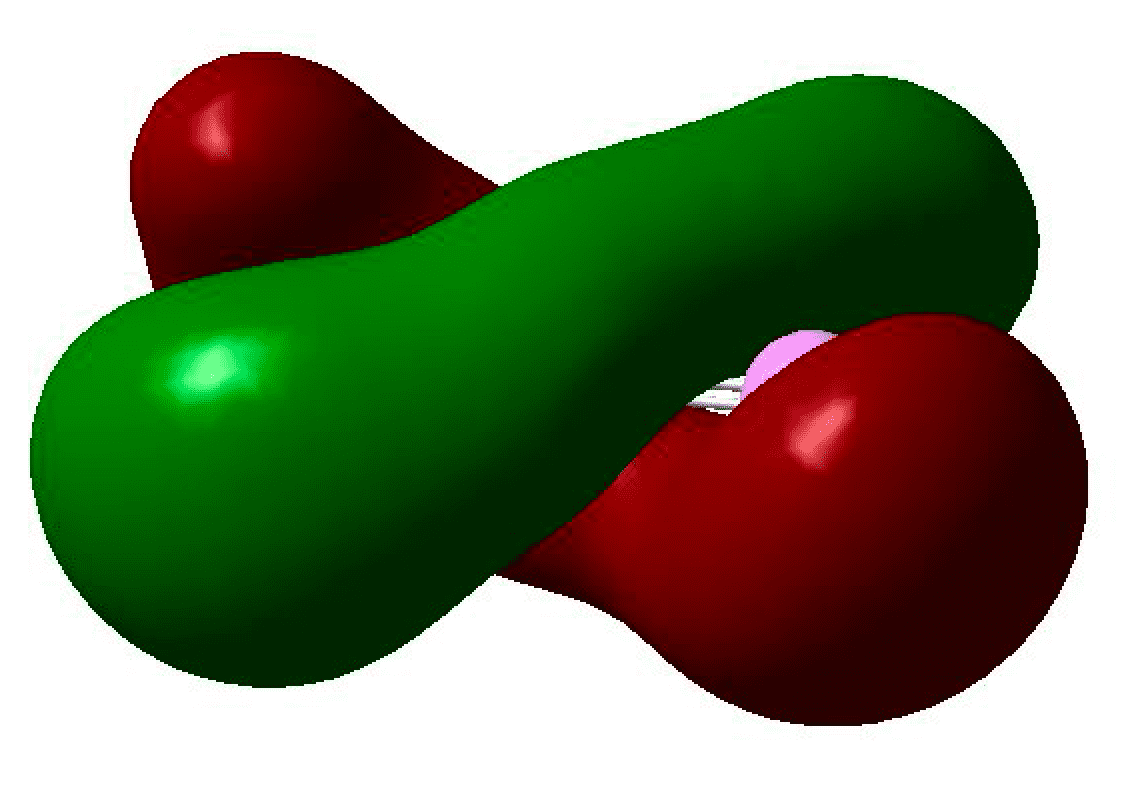}} &
    {\includegraphics[height=1.40cm]{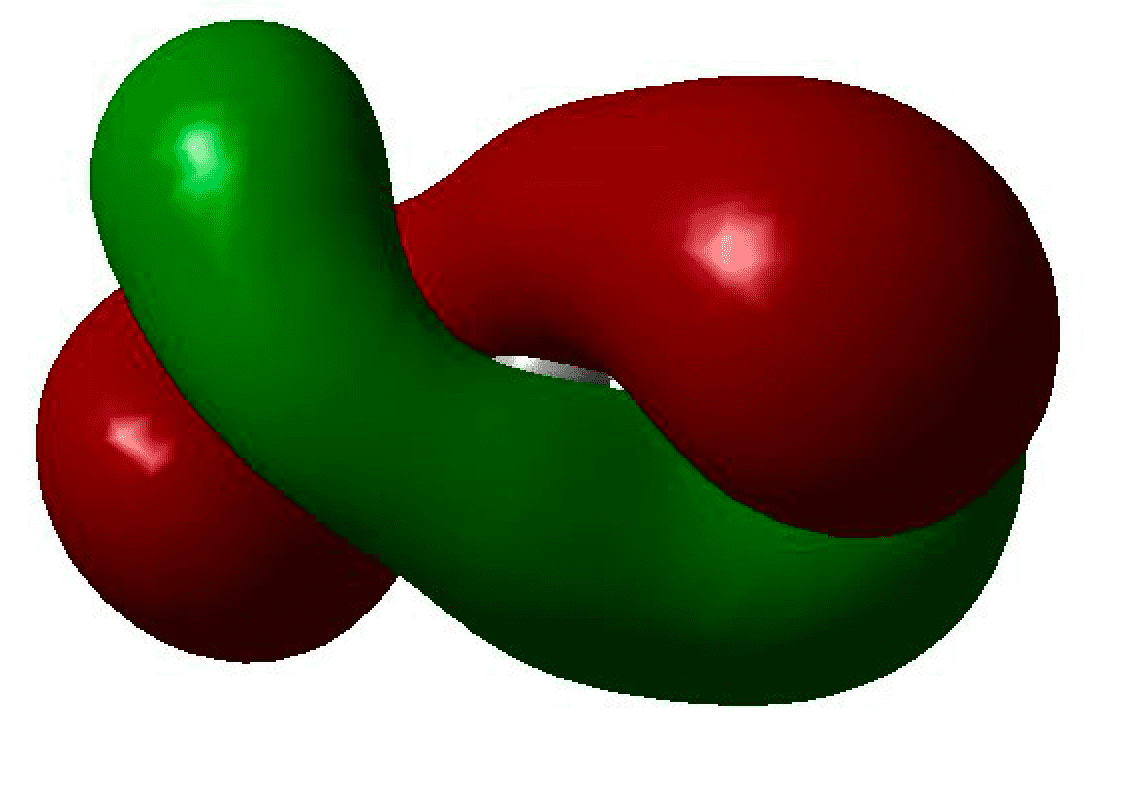}} &
    {\includegraphics[height=1.40 cm]{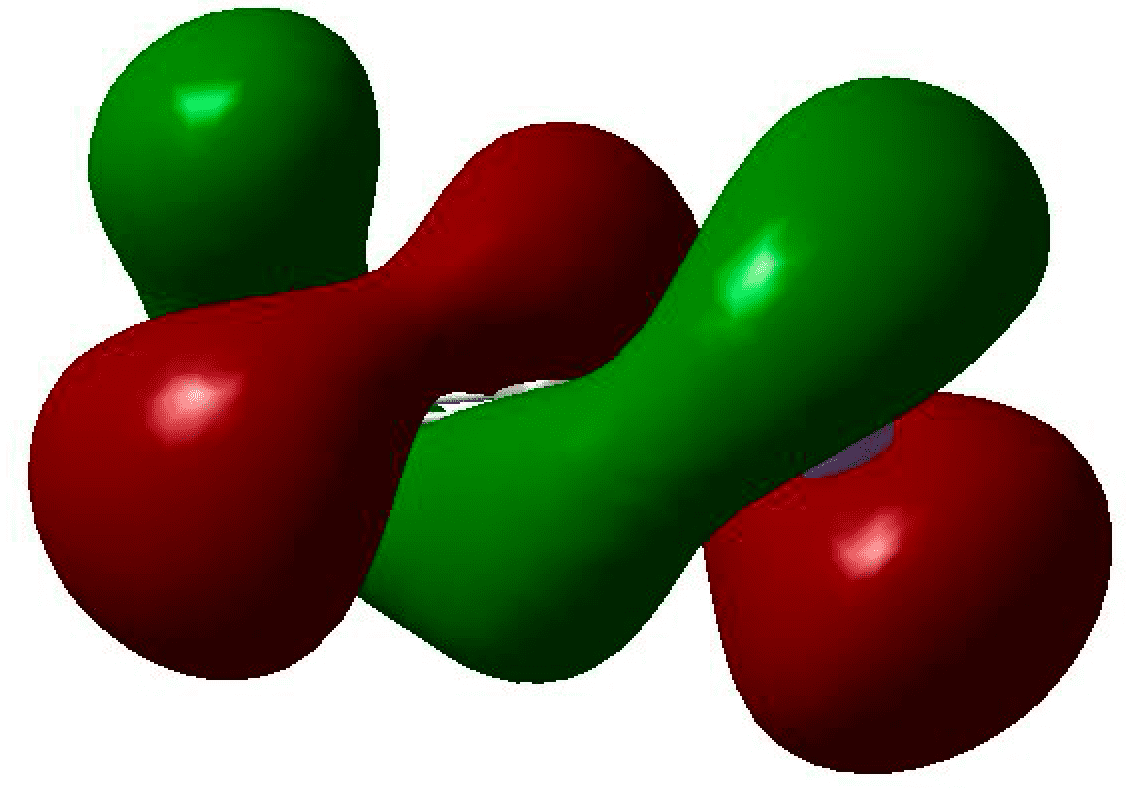}} & 
    {\includegraphics[height=1.40cm]{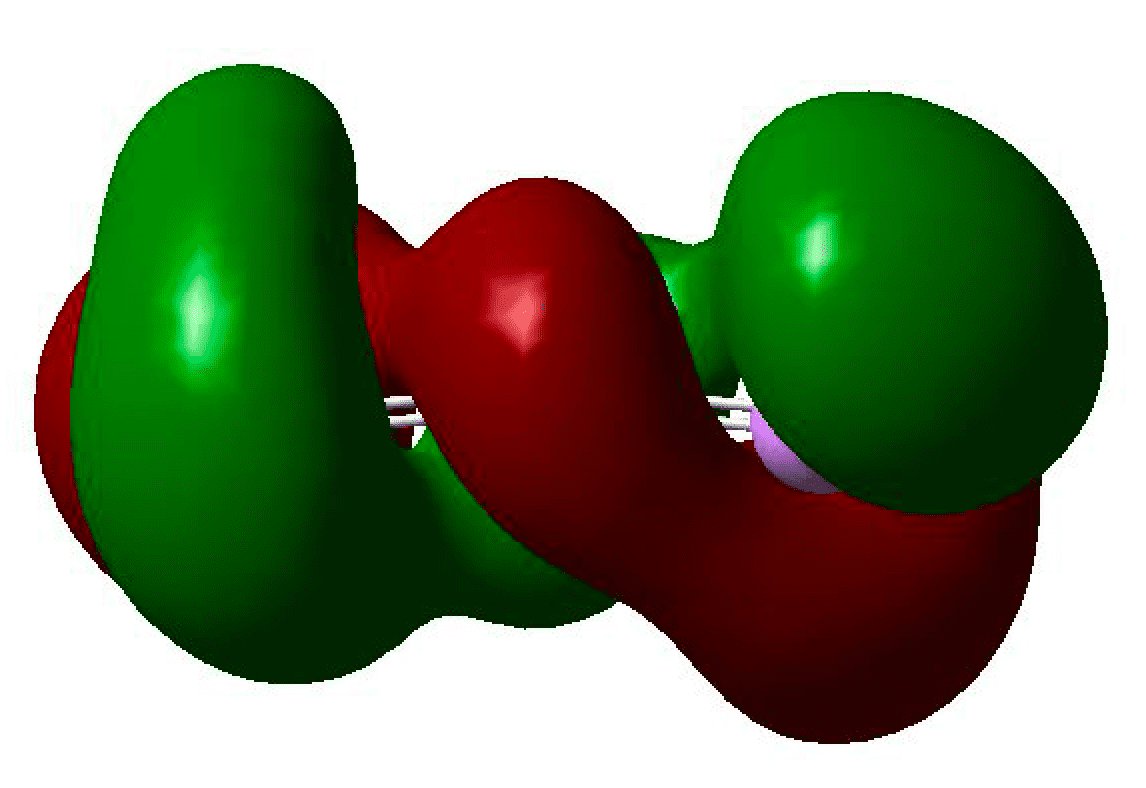}} \\
    \hdashline[1pt/1pt]
     Diarsaallenes &
    {\includegraphics[height=1.40cm]{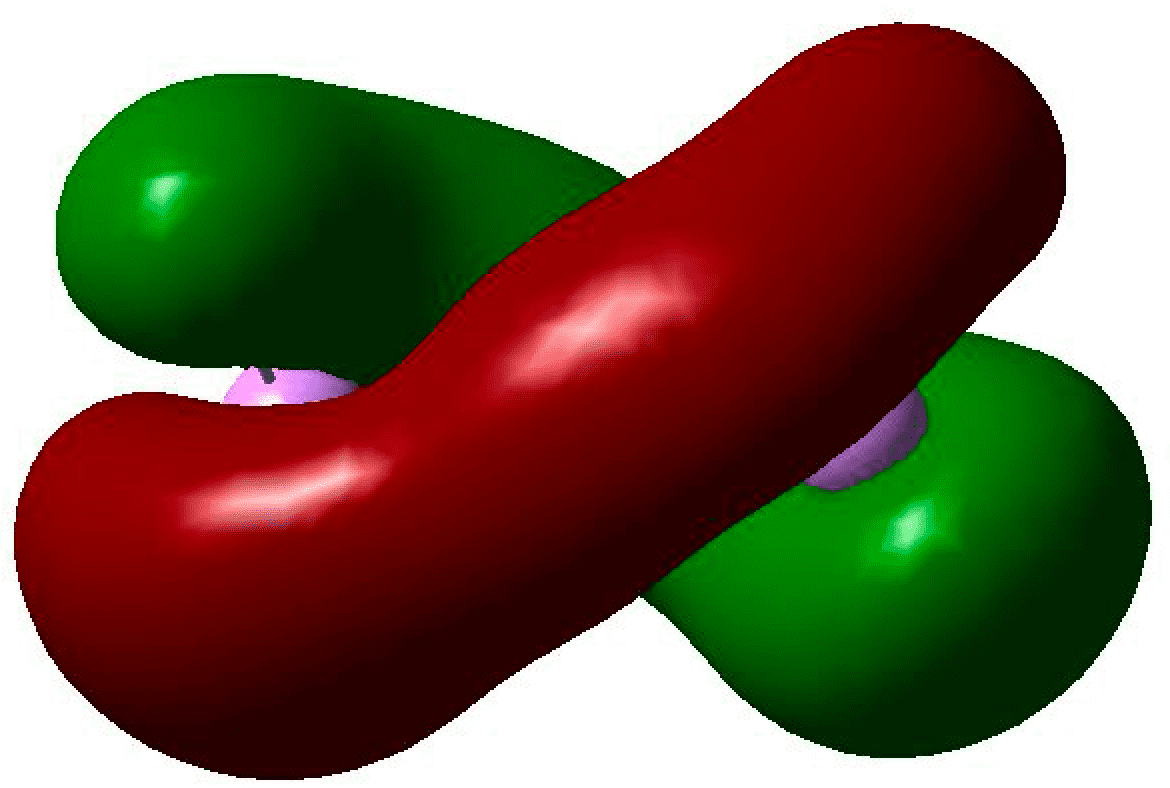}} &
    {\includegraphics[height=1.40cm]{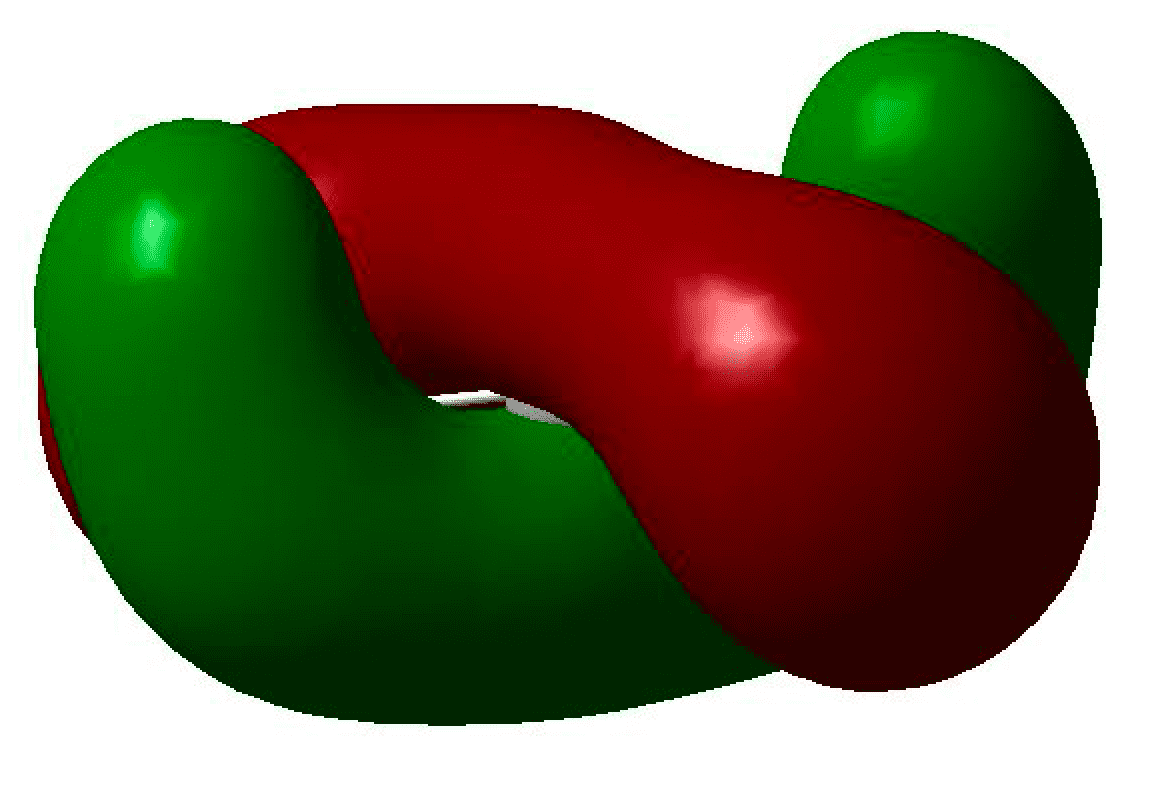}} &
    {\includegraphics[height=1.40 cm]{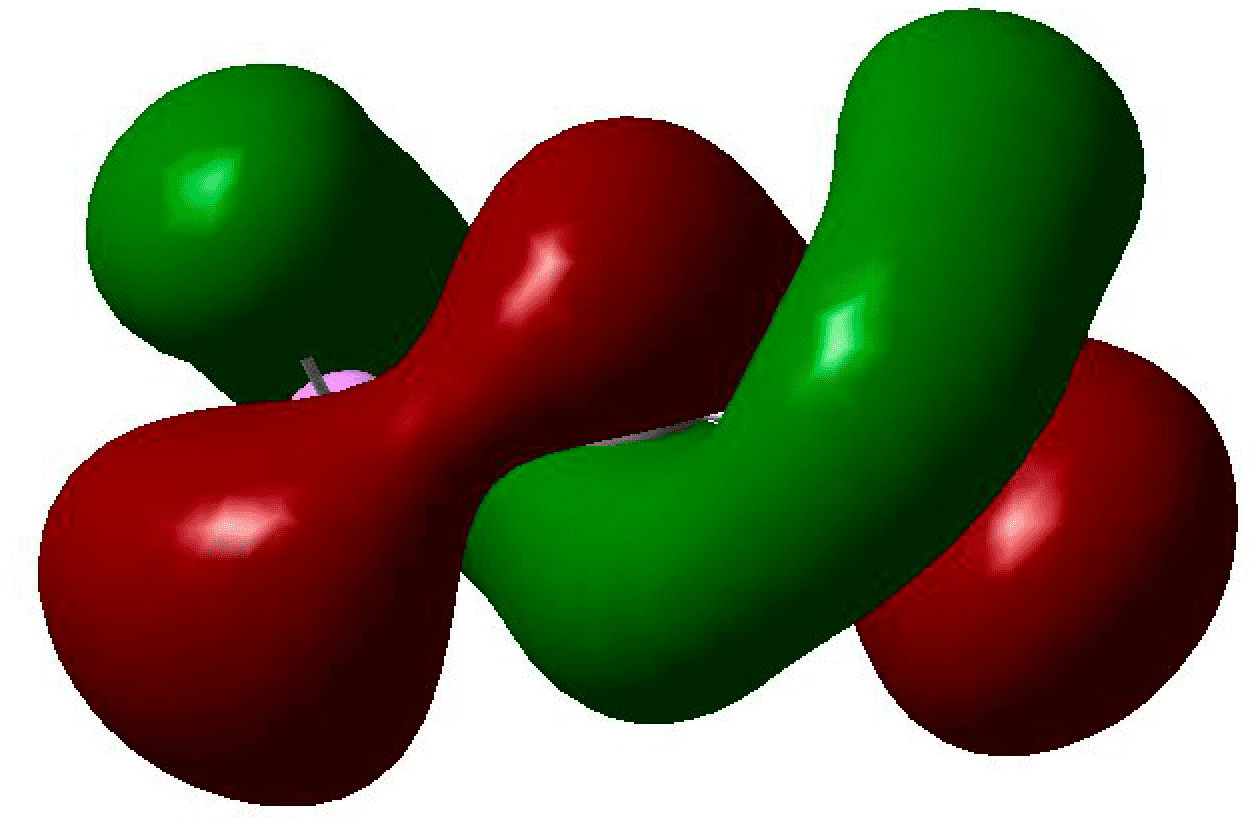}} & 
    {\includegraphics[height=1.40cm]{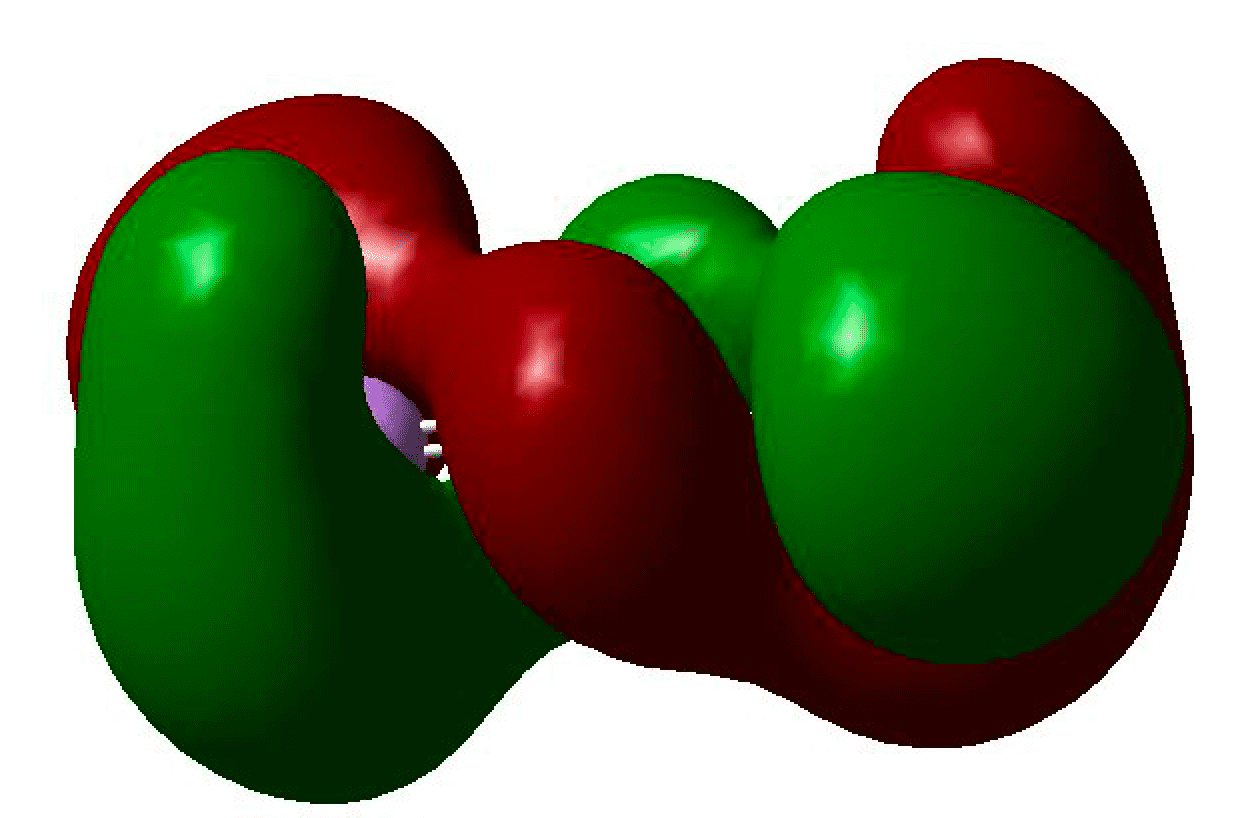}} \\
     \hdashline[1pt/1pt]
     Silaallenes &
    {\includegraphics[height=1.40cm]{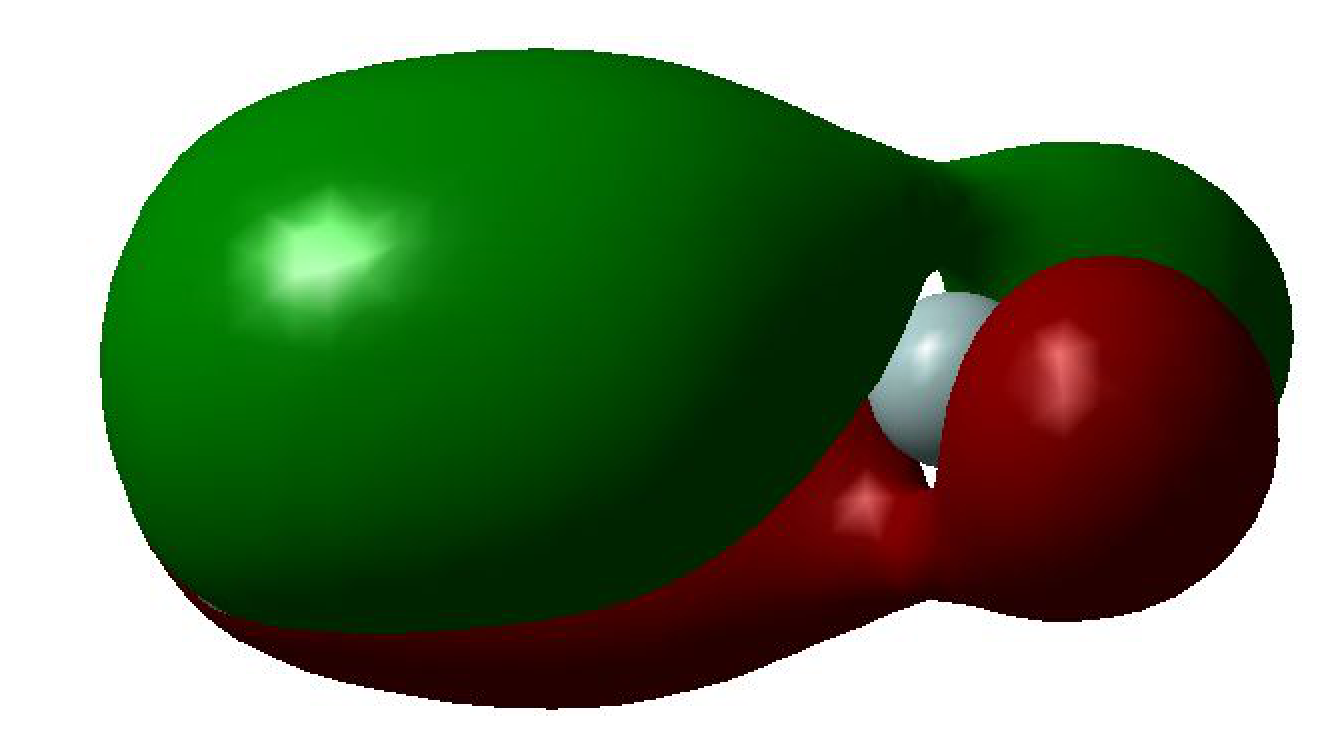}} &
    {\includegraphics[height=1.40cm]{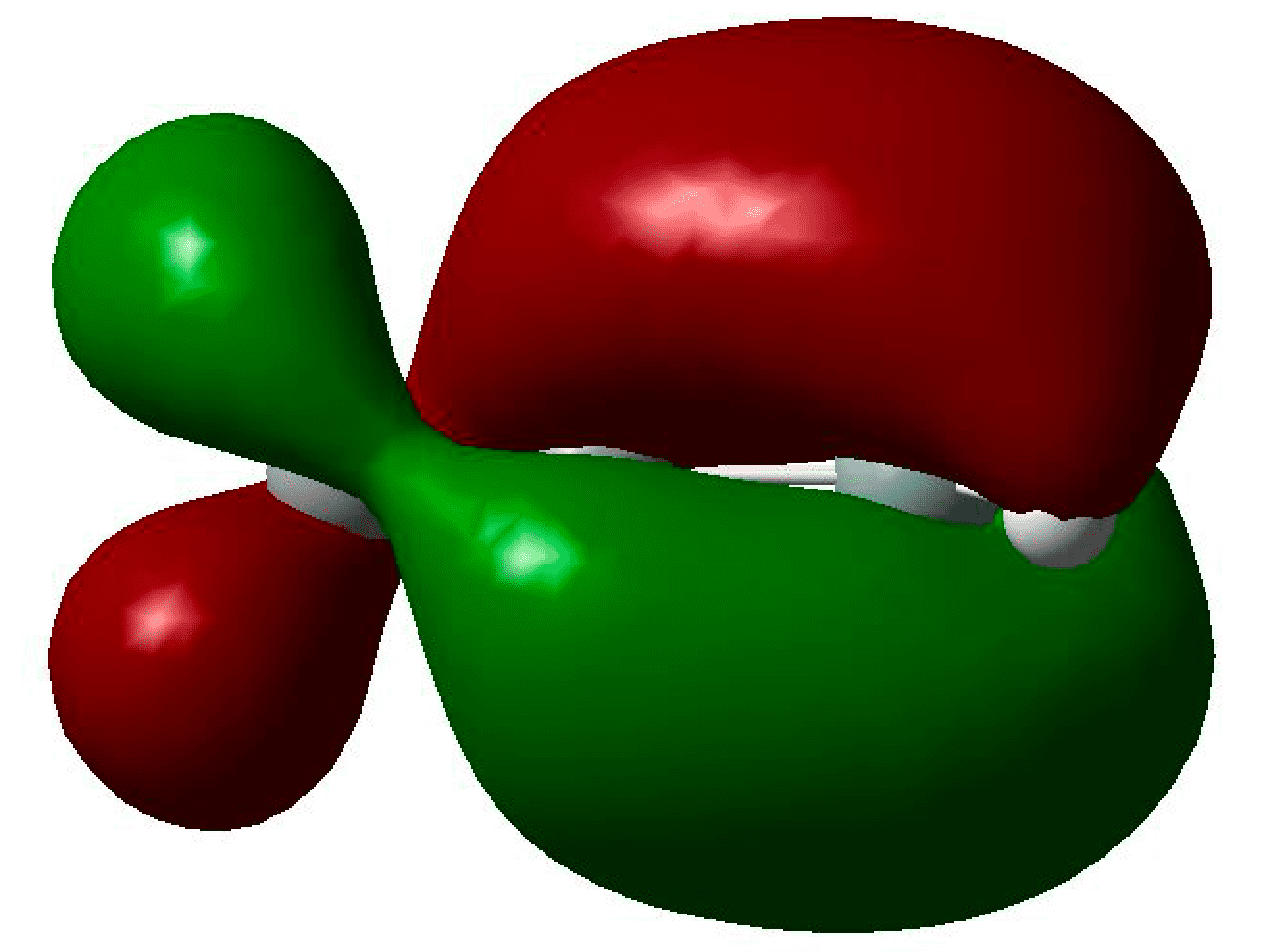}} &
    {\includegraphics[height=1.40cm]{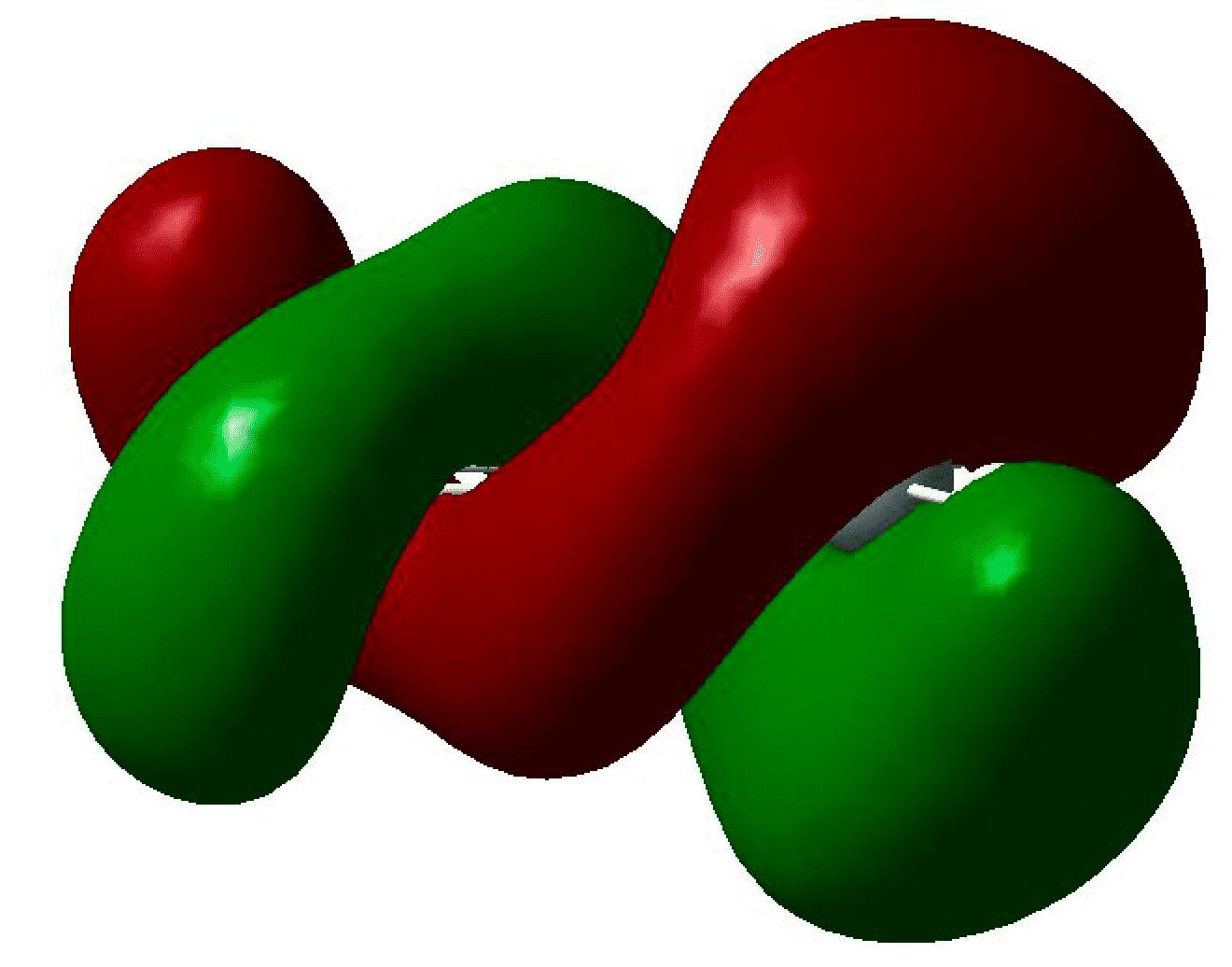}} & 
    {\includegraphics[height=1.40cm]{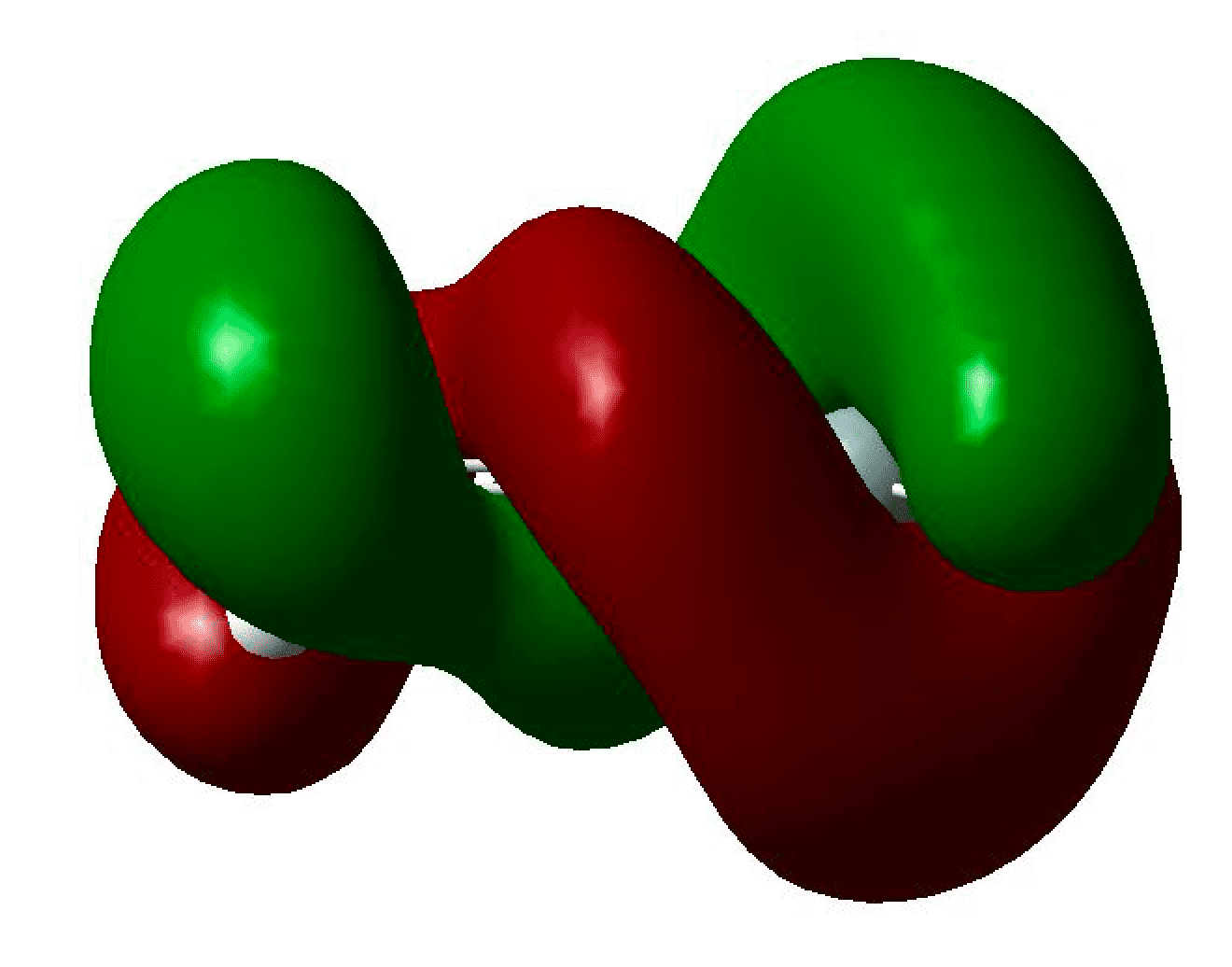}} \\
     \hdashline[1pt/1pt]
    Phosphasilaallenes &
    {\includegraphics[height=1.60cm]{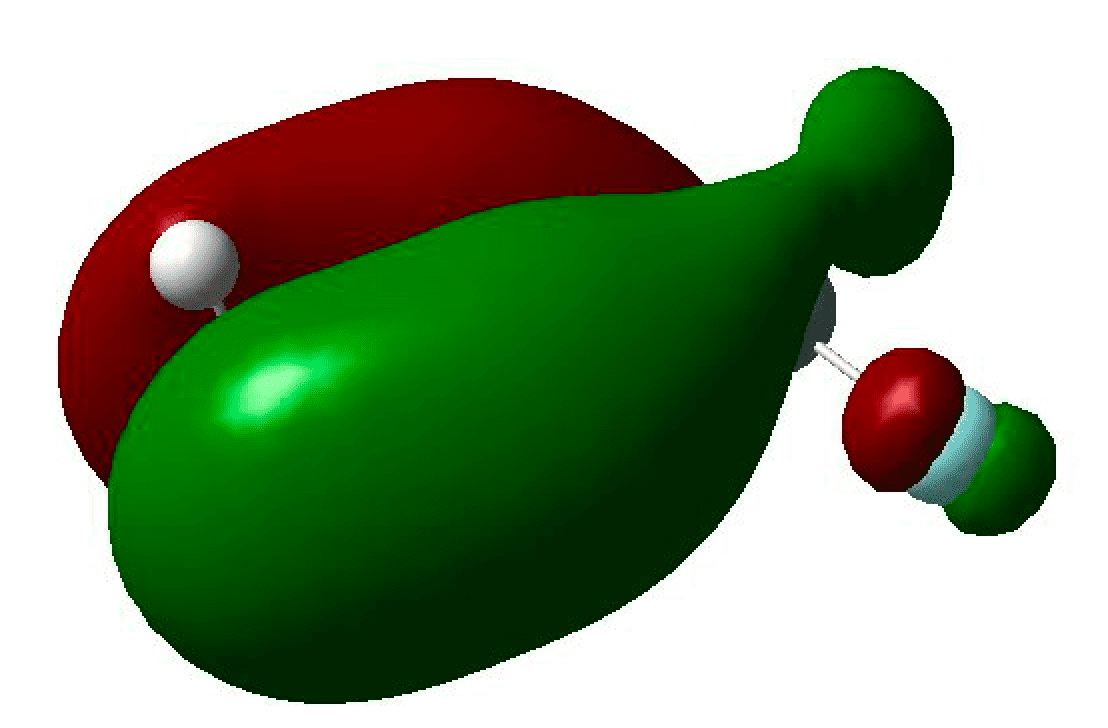}} &
    {\includegraphics[height=1.40cm]{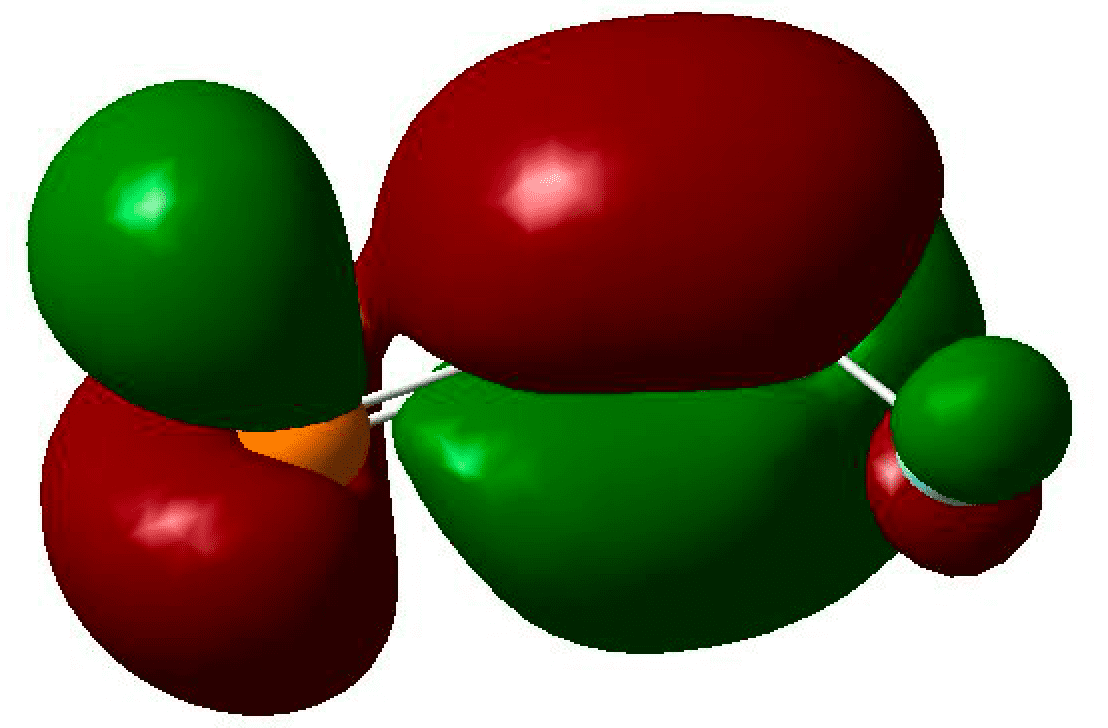}} &
    {\includegraphics[height=1.40cm]{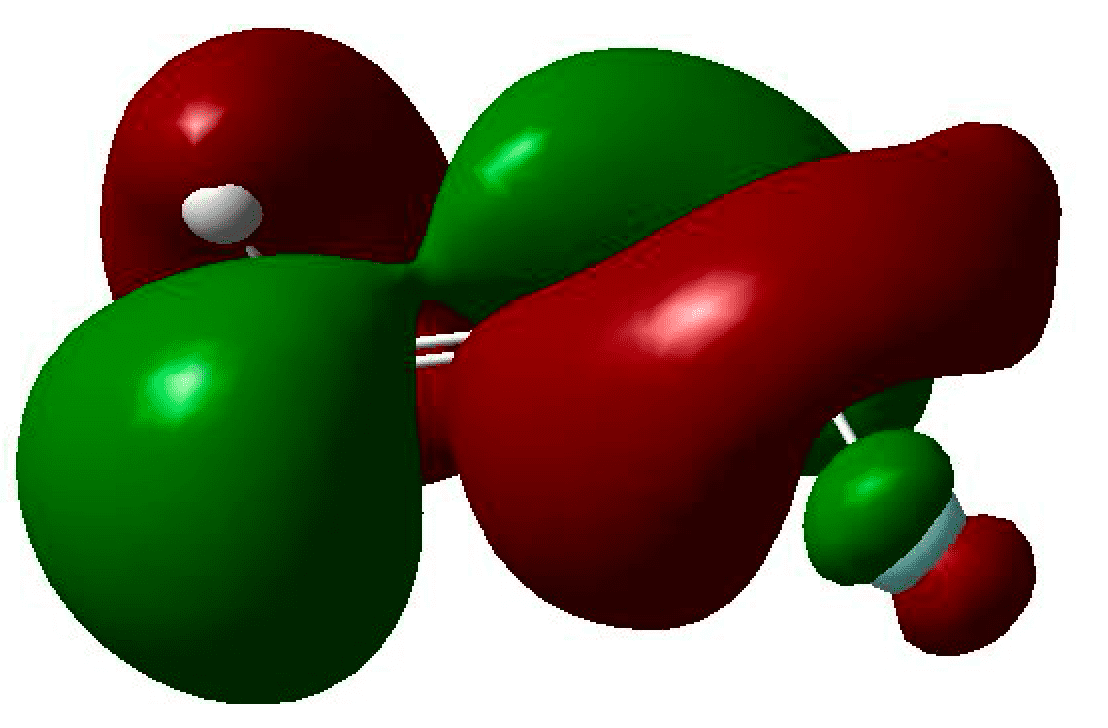}} & 
    {\includegraphics[height=1.40cm]{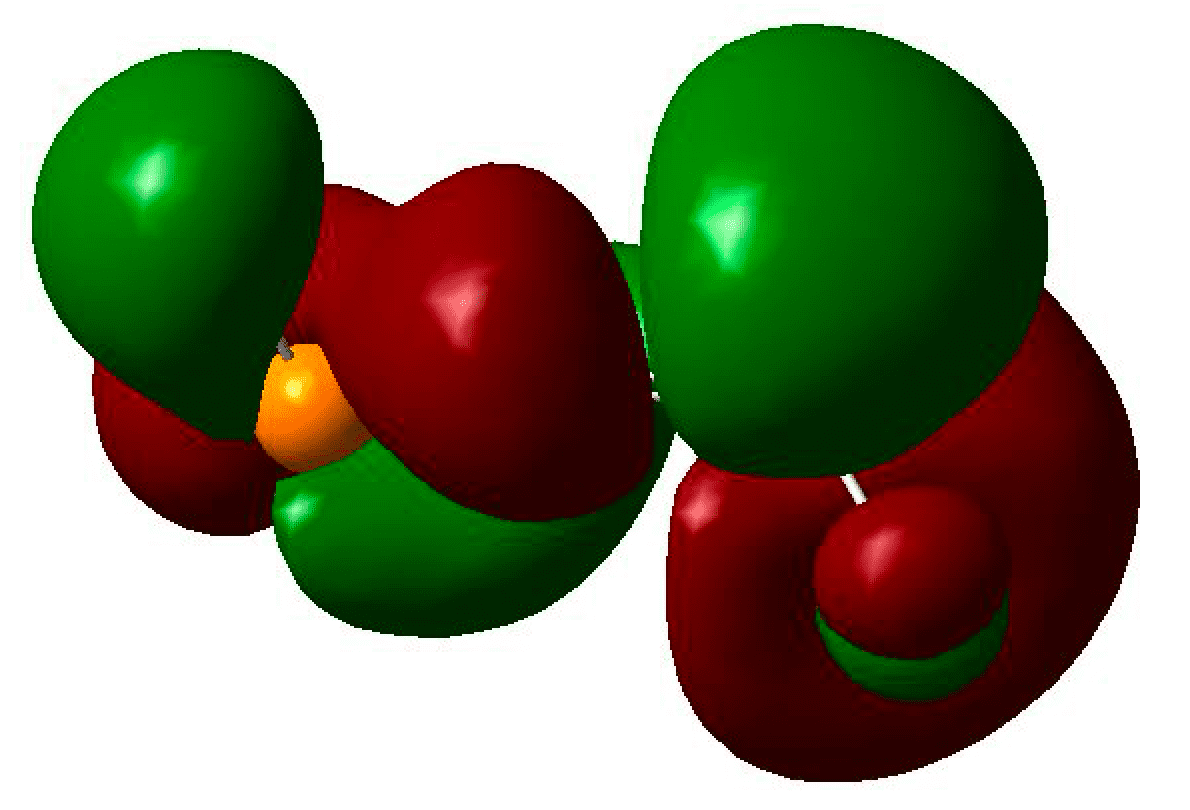}} \\
     \hdashline[1pt/1pt]
    Silaketenes &
    {\includegraphics[height=1.40cm]{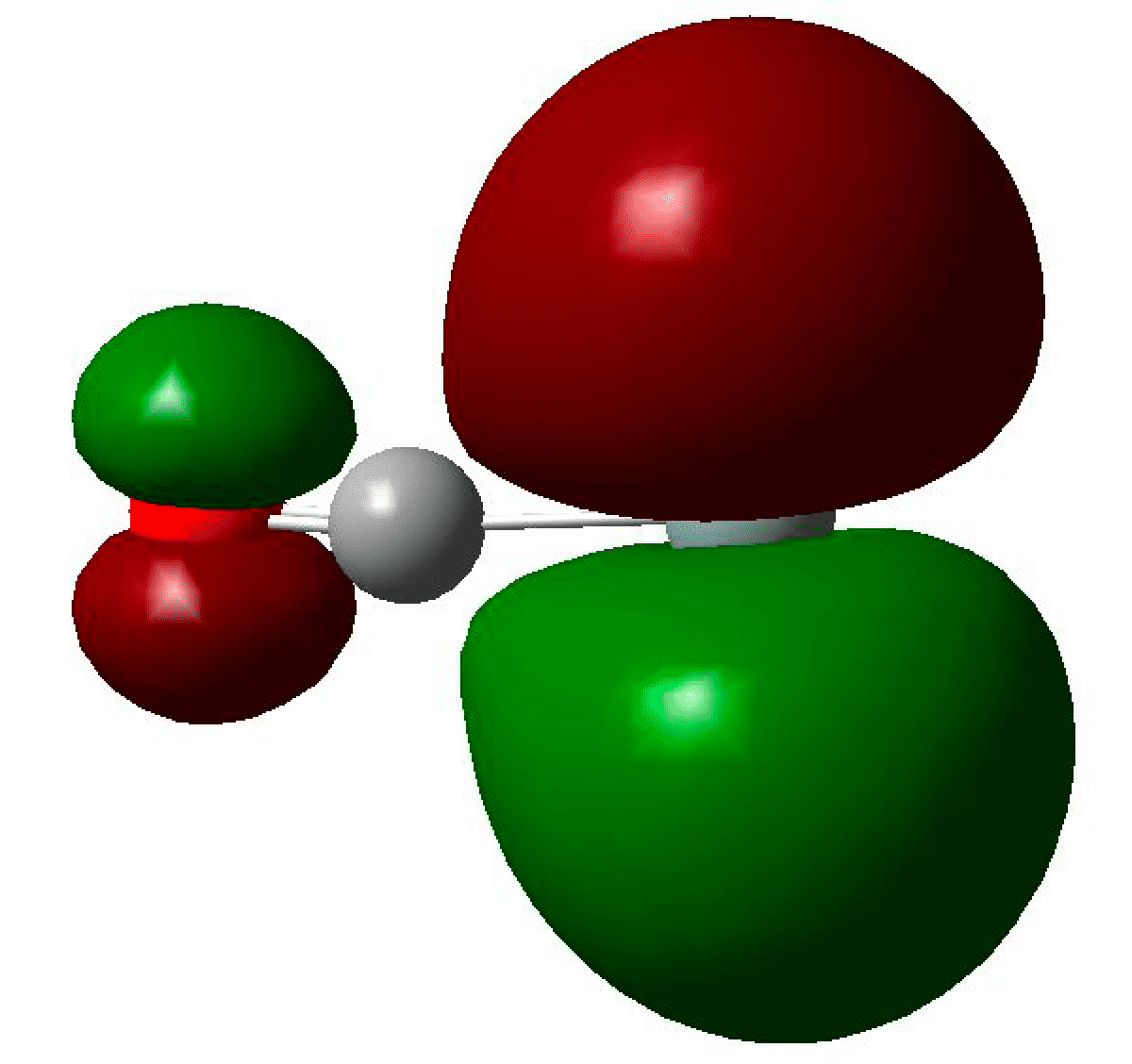}} &
    {\includegraphics[height=1.40cm]{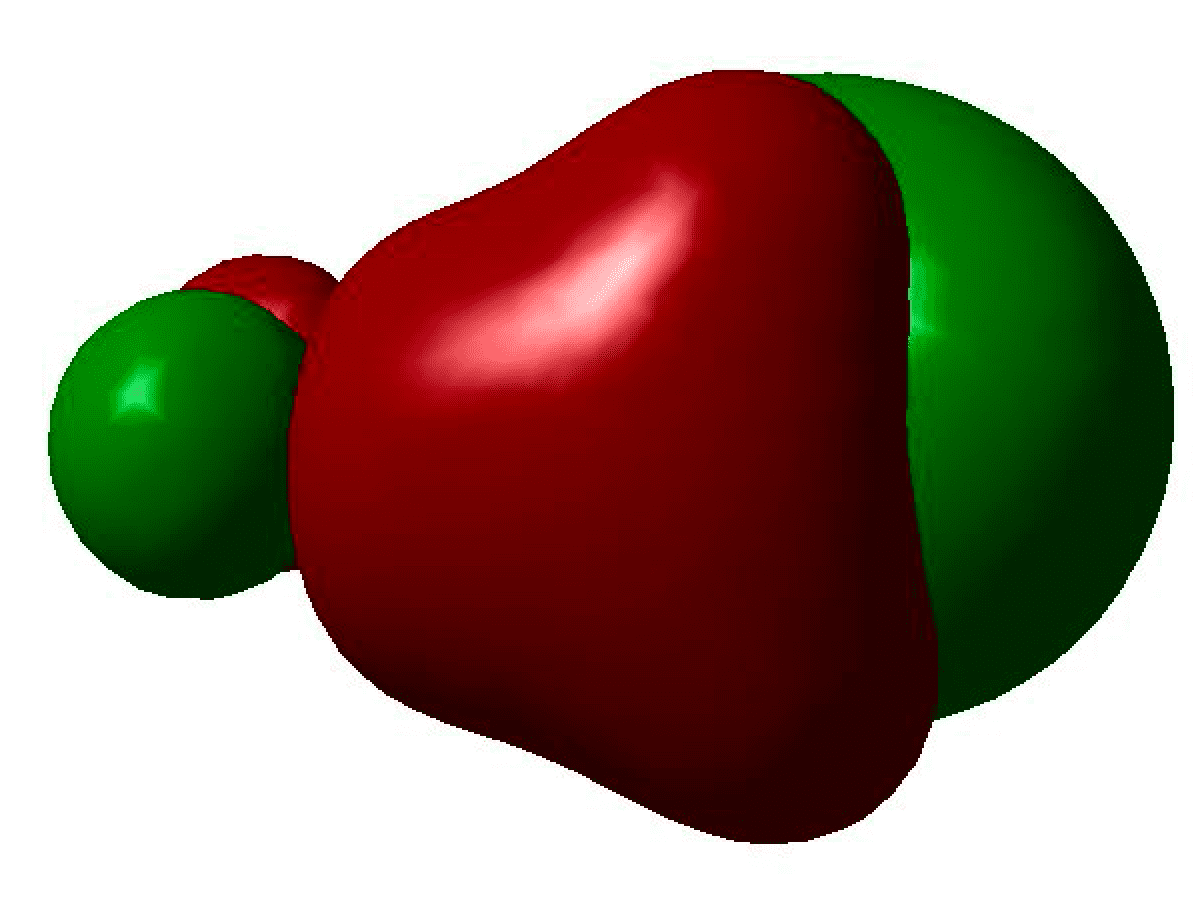}} &
    {\includegraphics[height=1.40cm]{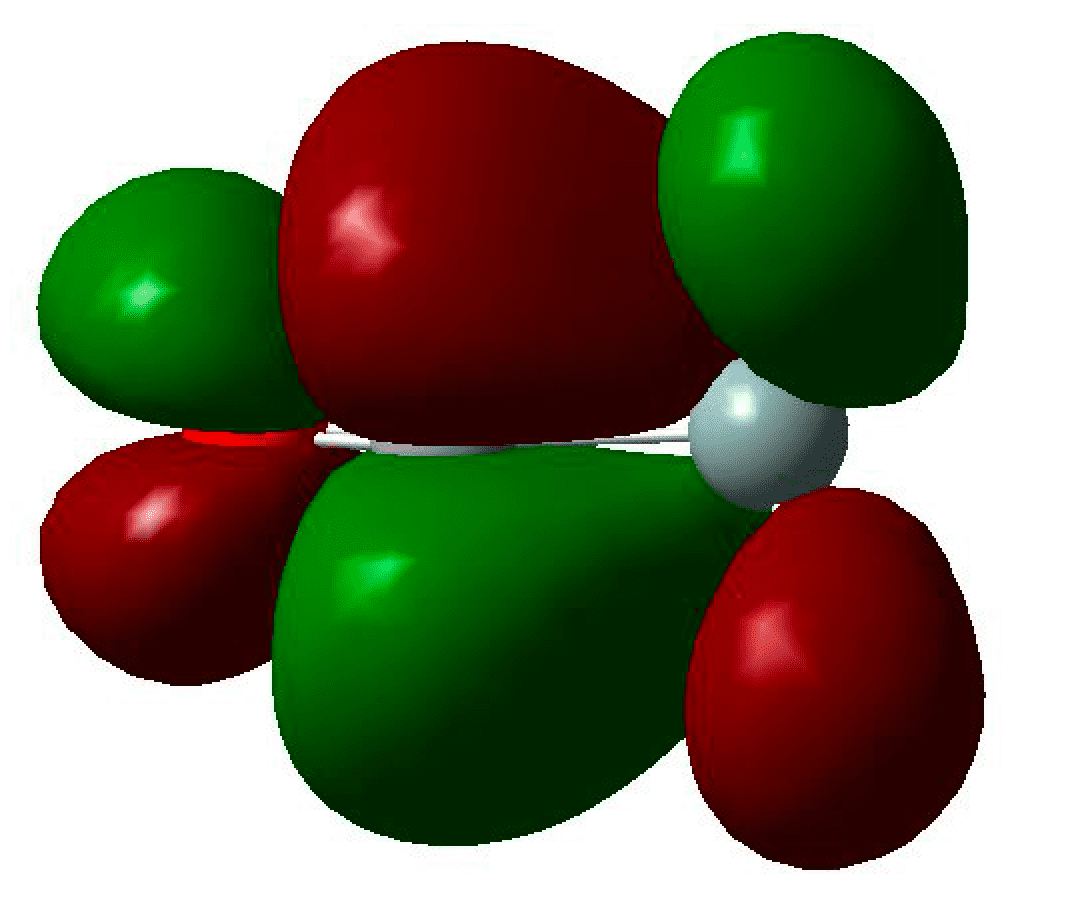}} & 
    {\includegraphics[height=1.40cm]{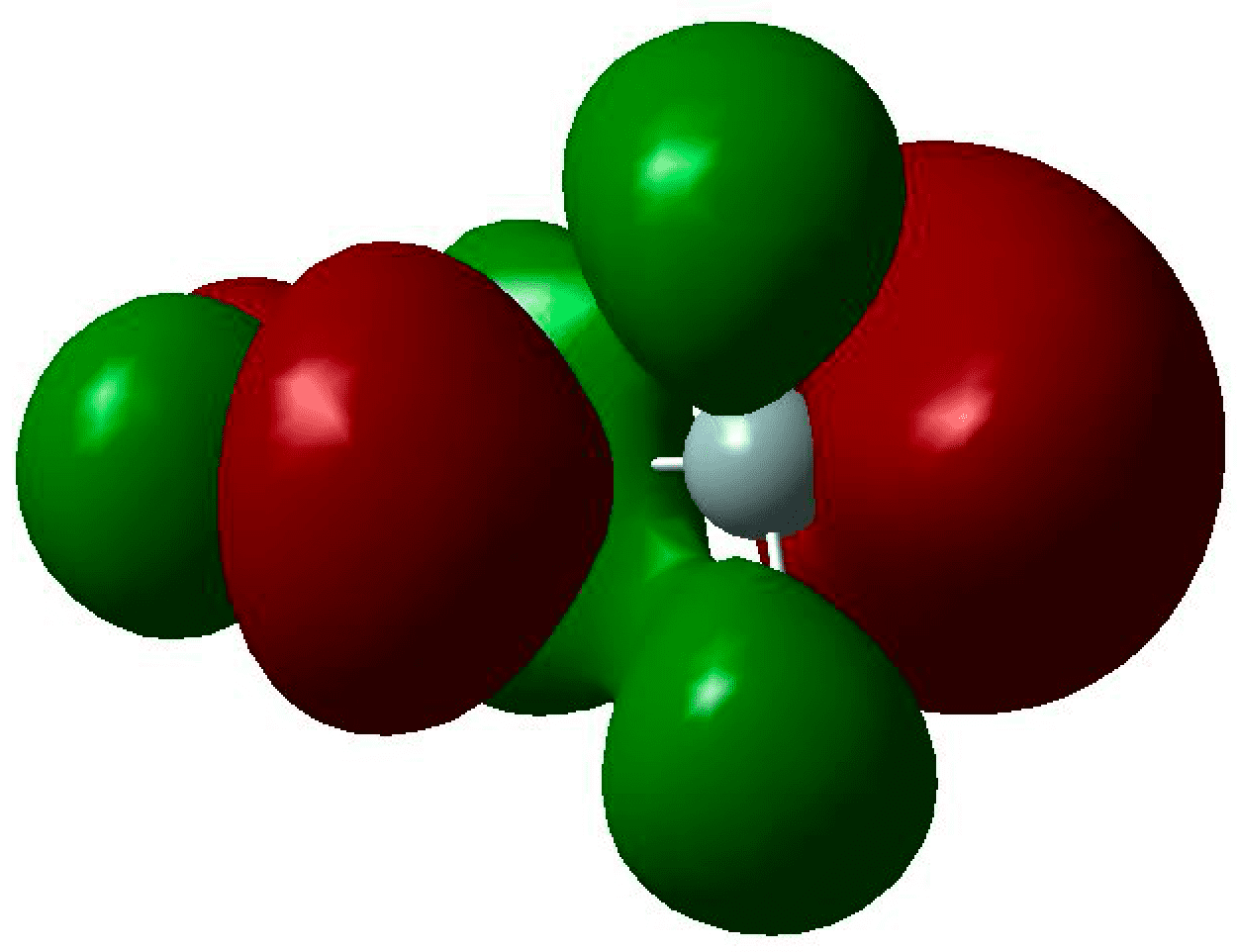}} \\
     \hdashline[1pt/1pt]
    Germaallenes &
    {\includegraphics[height=1.40cm]{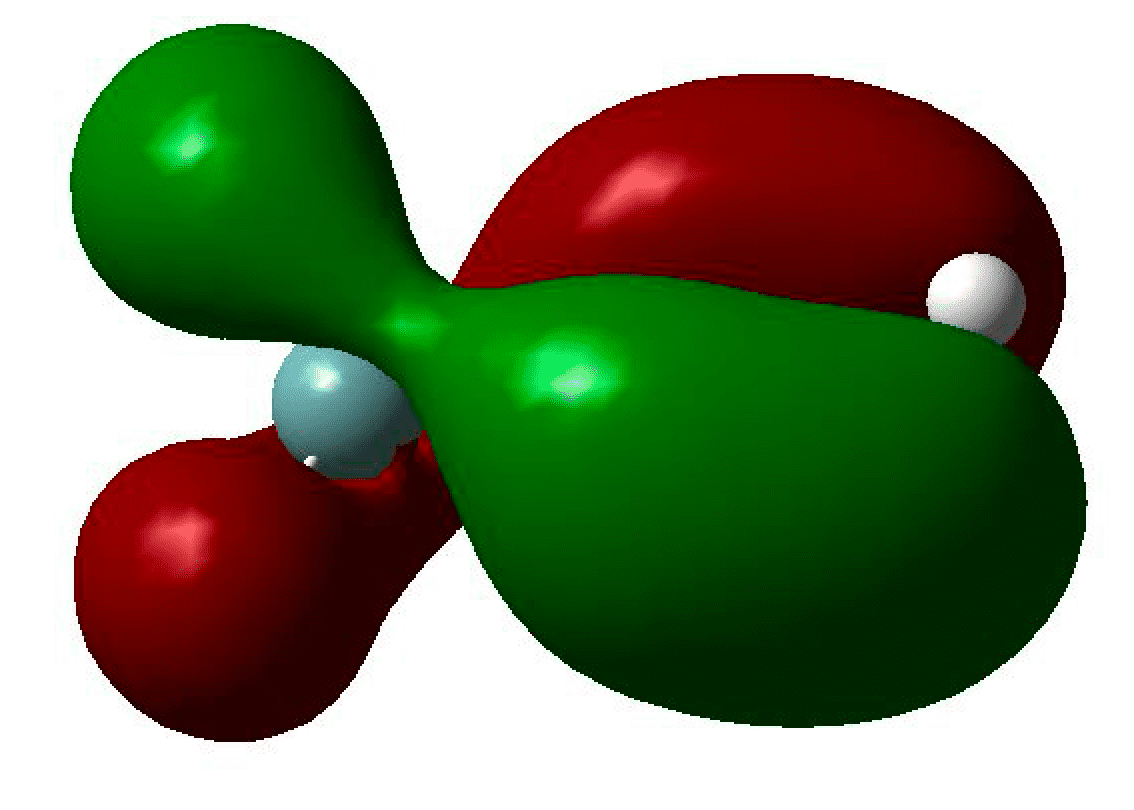}} &
    {\includegraphics[height=1.40cm]{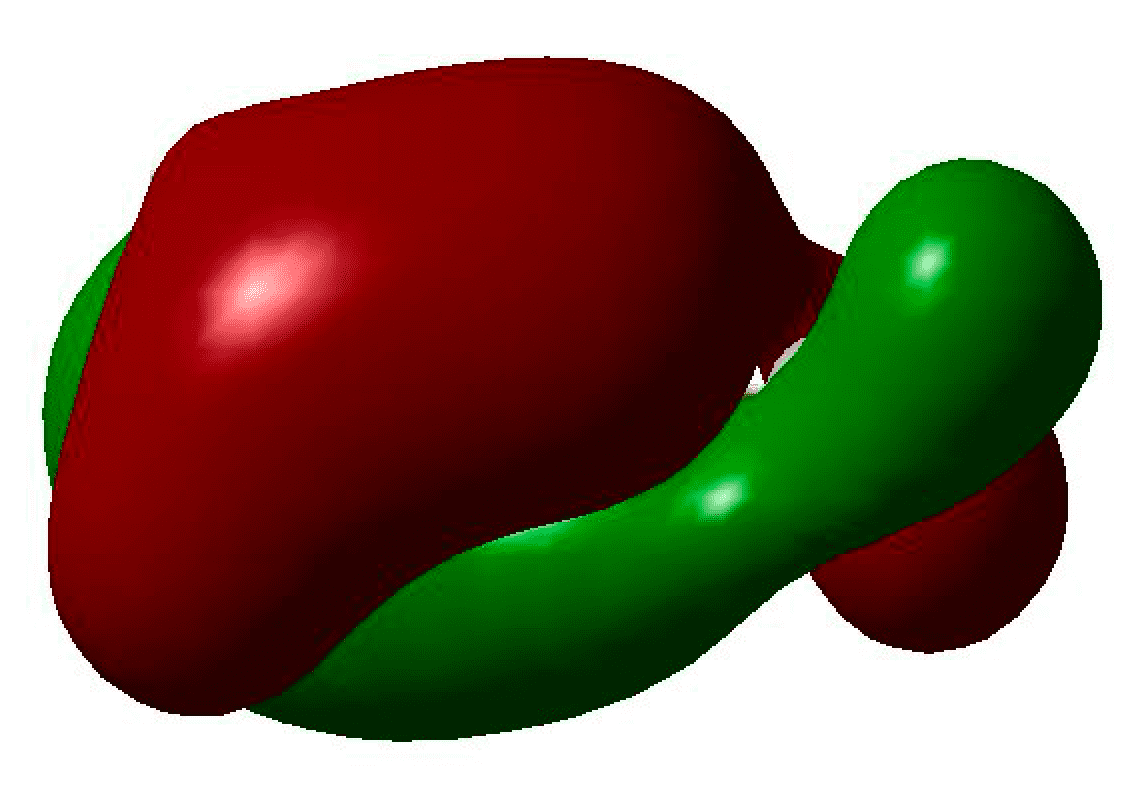}} &
    {\includegraphics[height=1.40cm]{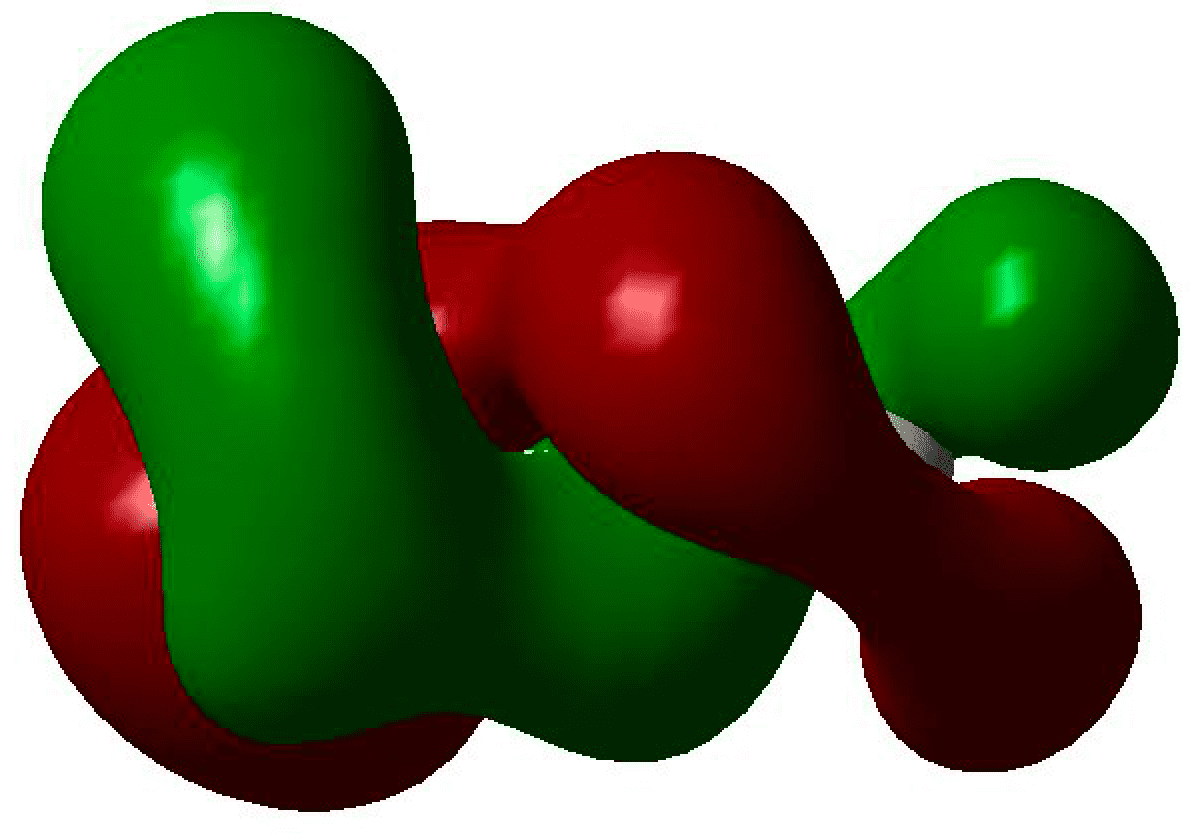}} & 
    {\includegraphics[height=1.40cm]{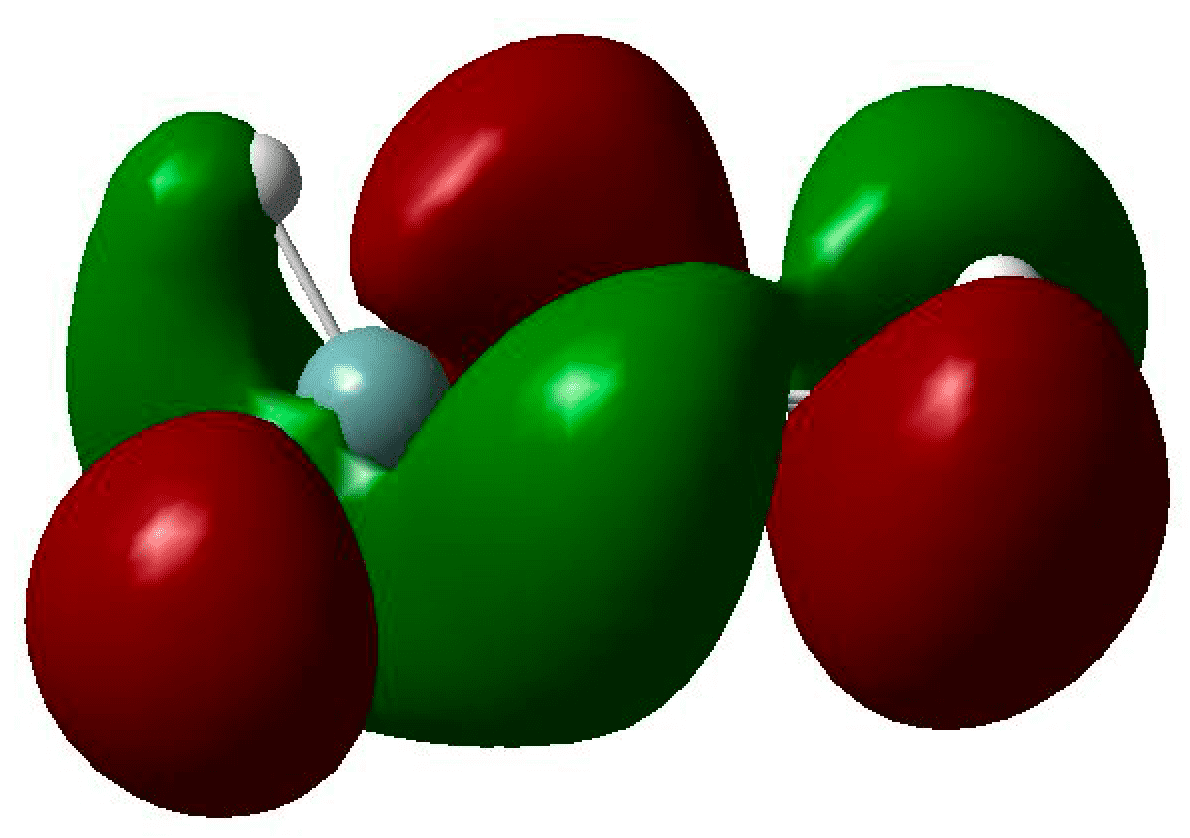}} \\
     \hdashline[1pt/1pt]
    Germaphosphaallenes &
    {\includegraphics[height=1.40cm]{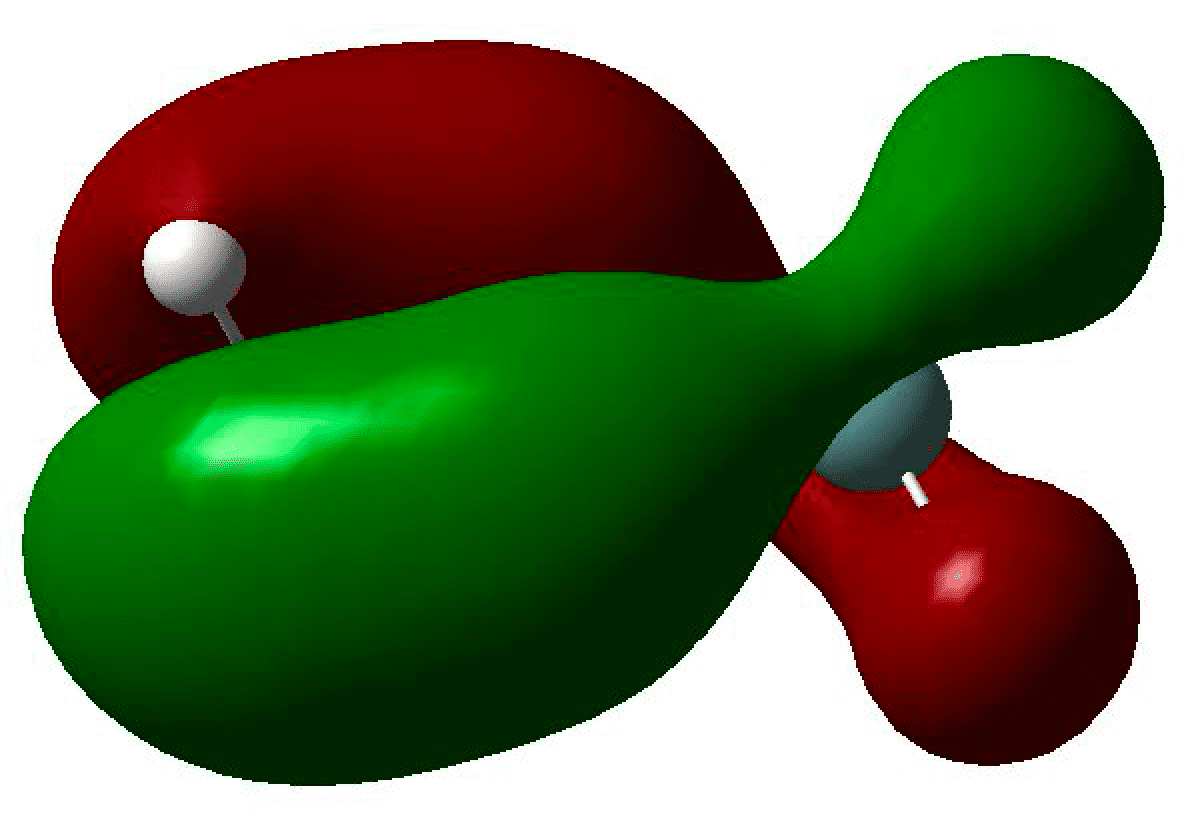}} &
    {\includegraphics[height=1.40cm]{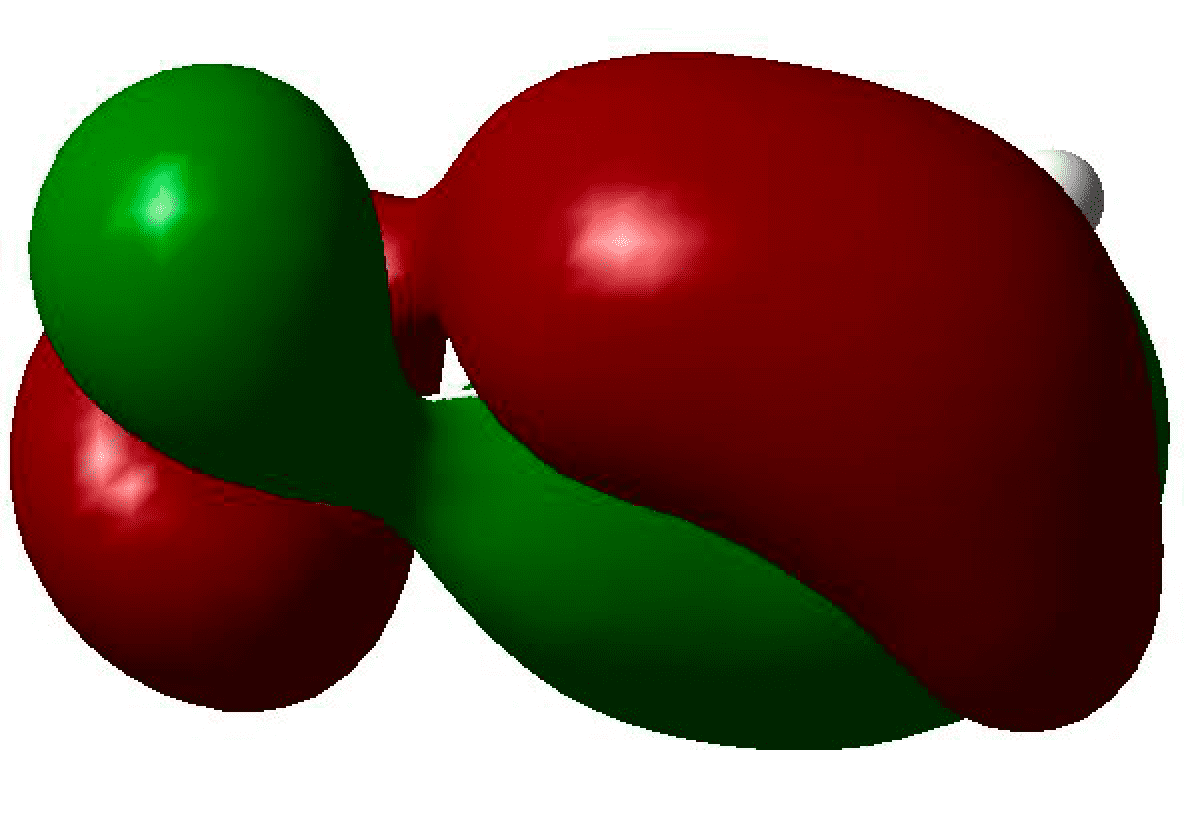}} &
    {\includegraphics[height=1.40cm]{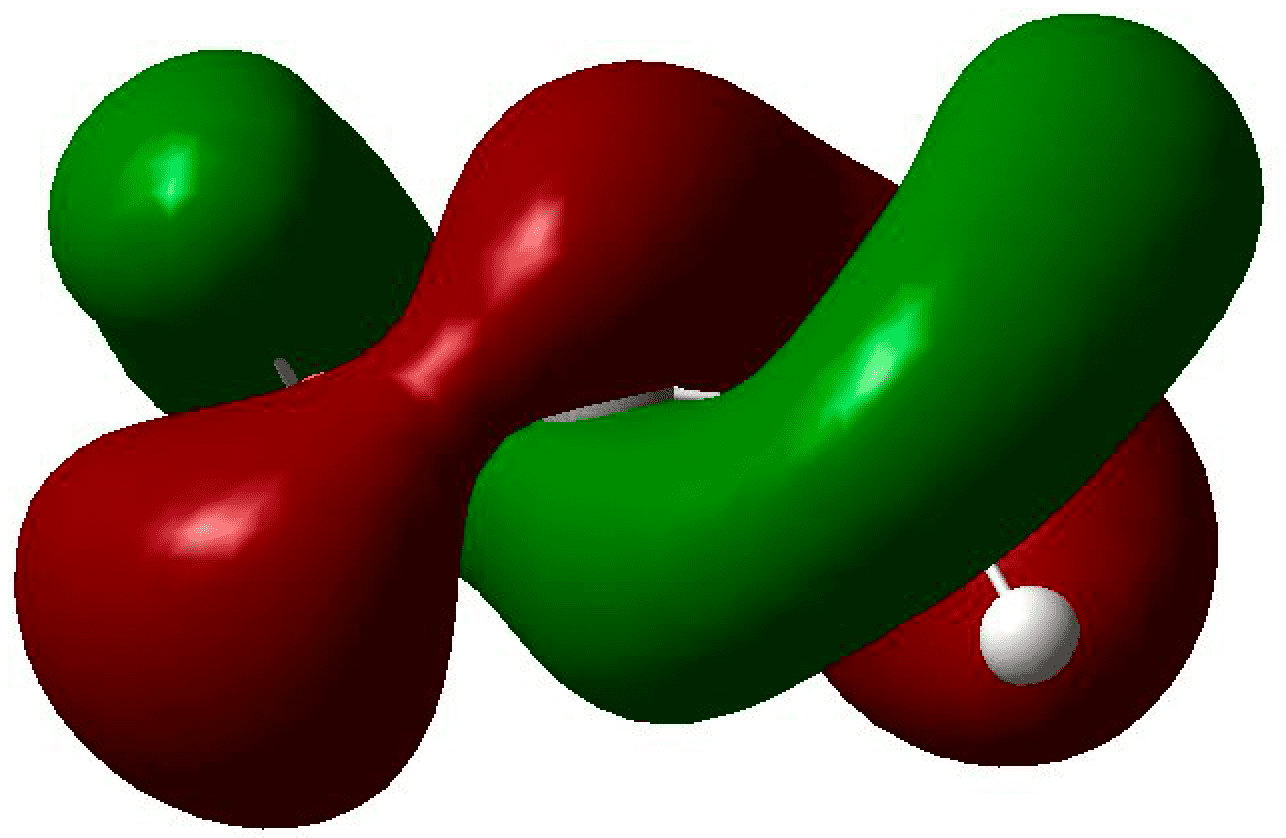}} & 
    {\includegraphics[height=1.40cm]{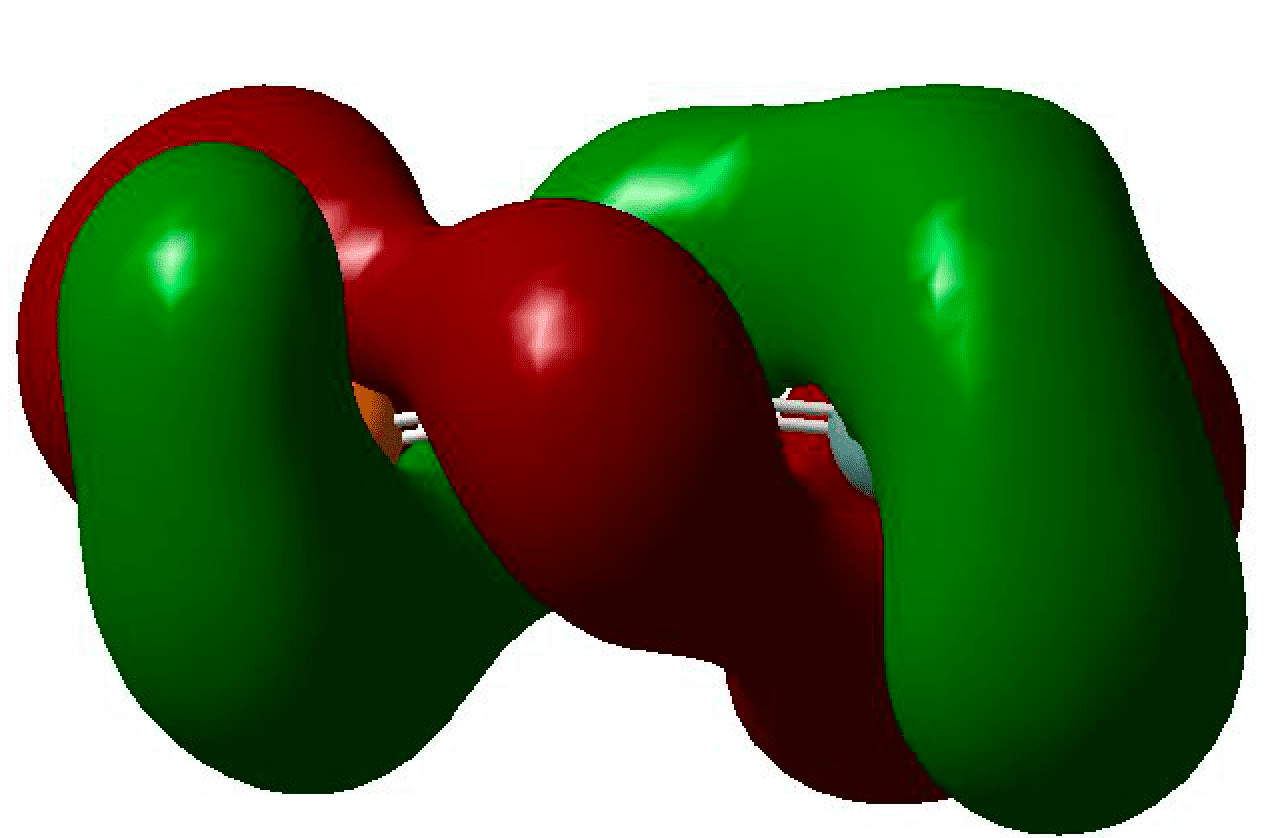}} \\
    \bottomrule
        \end{tabular}
        \end{table}  

\newpage

\subsection{Model DPBD (Diphenylbutadiyne)}

Our interest was also focused on the DPBD molecule for its particularity to have a linear chain containing an alternation of triple and simple bonds like carbyne systems \cite{ozcelik}.

 \begin{figure}[H]
    \centering
    \includegraphics[width=0.4\linewidth]{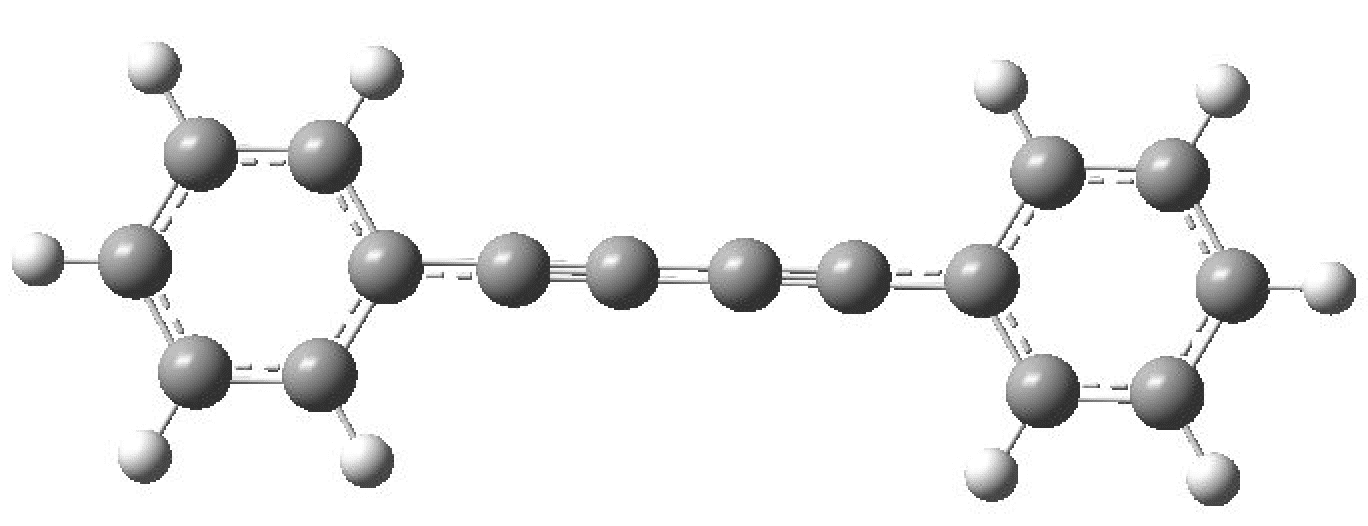}
    \caption{Representation of the molecule of Diphenylbutadiyne}
    \label{fig19-mol-dpbd}
\end{figure}

The B3LYP/6-311G(d.p) calculations were done with a $C_2$ symmetry. We have observed the shape of the orbitals for rotations from 0° to 90°. 
We have listed in the following tables the angles of rotation for which we observed changes in the shape of the orbitals.

\begin{table}[H]
    \caption{Molecular orbitals and energies obtained at the B3LYP/6-311G(d.p) level of theory for the DPBD system for rotations of 0\Degre, 20\Degre, 40\Degre, 60\Degre, 80\Degre and 90\Degre (see Table S5 for details).}
    \label{table9-mo-dpbd}
    \centering
    \begin{tabular}{ccccc}
        \midrule
  0° &
    {\includegraphics[height=1.40cm]{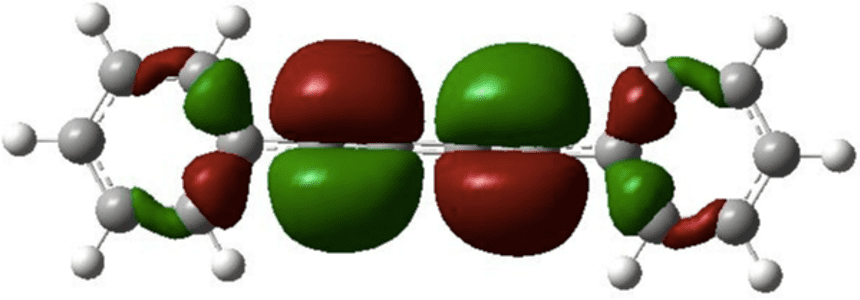}} &
    {\includegraphics[height=1.40cm]{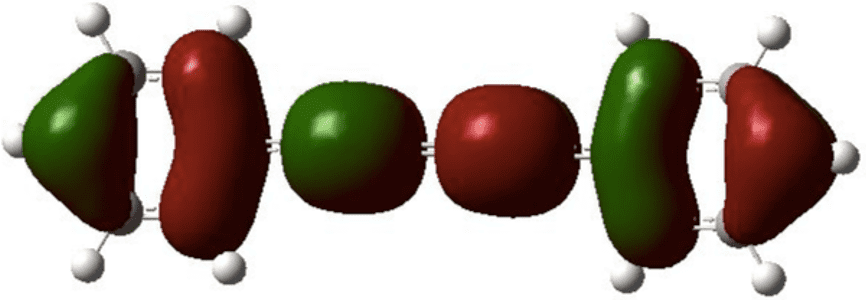}} &
    {\includegraphics[height=1.40 cm]{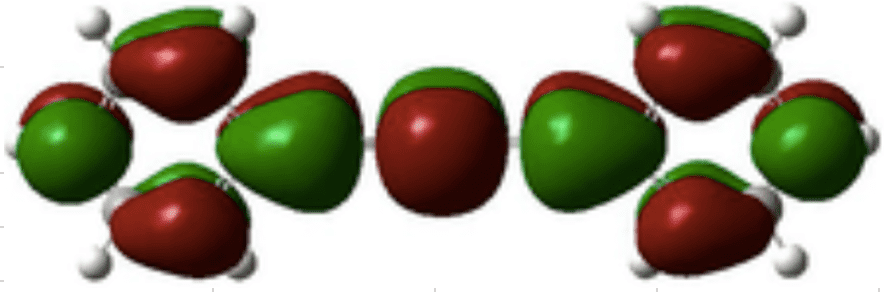} } & 
    {\includegraphics[height=1.40cm]{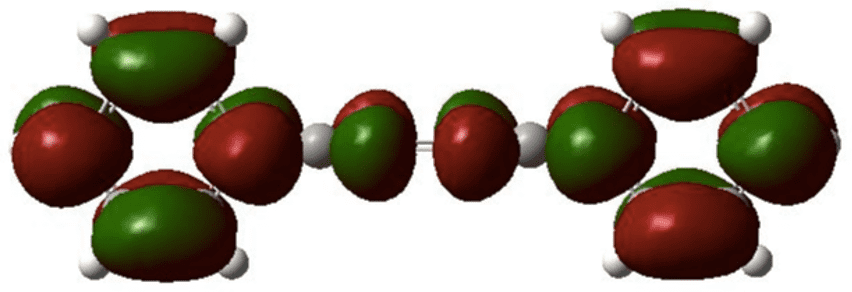} } \\
    &HOMO-1 (-0.29609 u.a) & HOMO (-0.25606 u.a) & LUMO (-0.01465 u.a) & LUMO+1 (0.03893 u.a) \\
  20° &
    {\includegraphics[height=1.40cm]{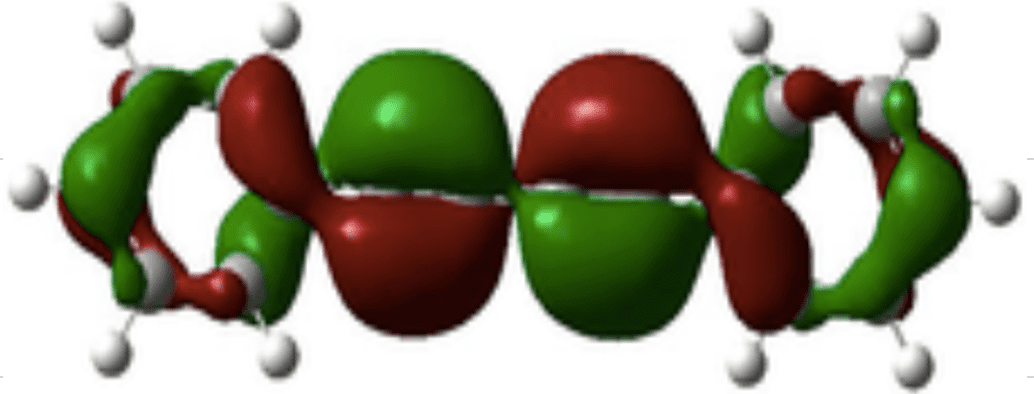}} &
    {\includegraphics[height=1.40cm]{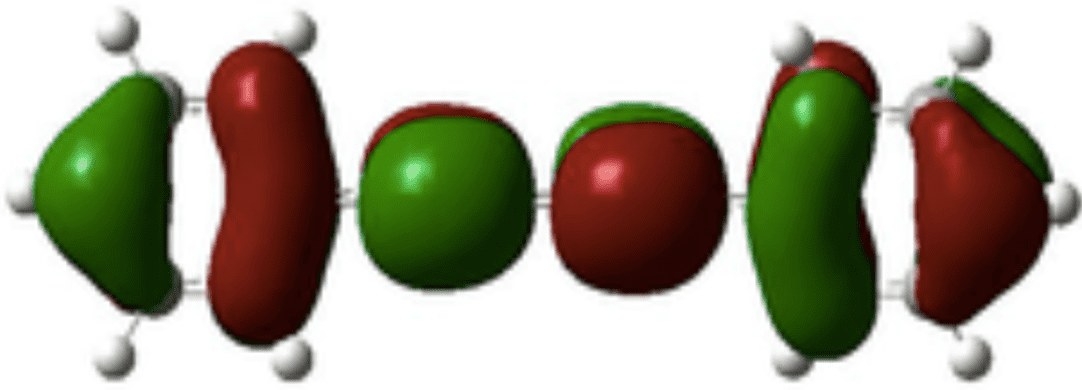}} &
    {\includegraphics[height=1.40 cm]{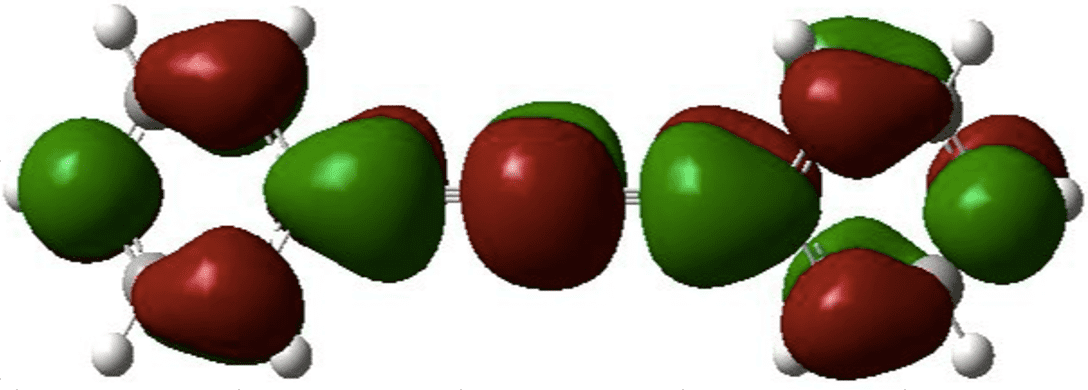} } & 
    {\includegraphics[height=1.40cm]{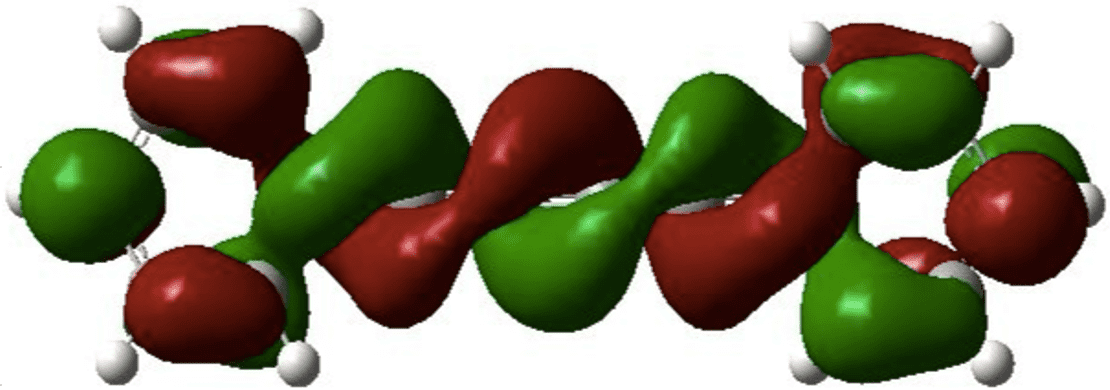} } \\
    & HOMO-1 (-0.29259 u.a) & HOMO (-0.25677 u.a) & LUMO (-0.01385 u.a) & LUMO+1 (0.03008 u.a) \\
   40°&
    {\includegraphics[height=1.40cm]{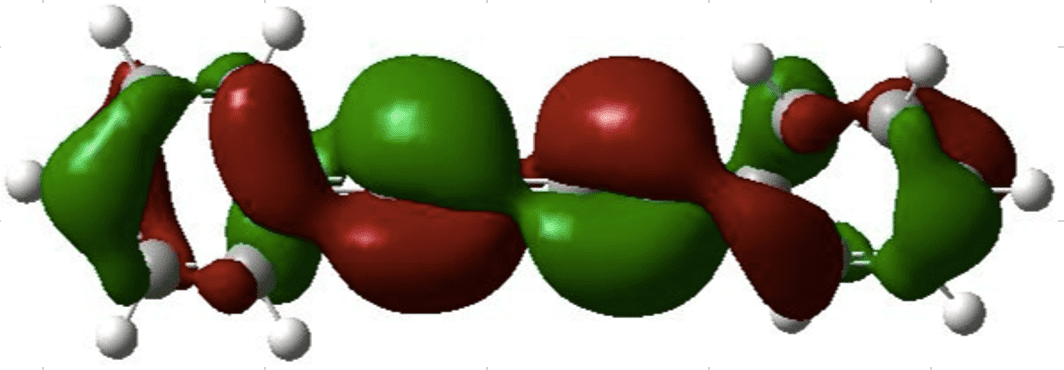}} &
    {\includegraphics[height=1.40cm]{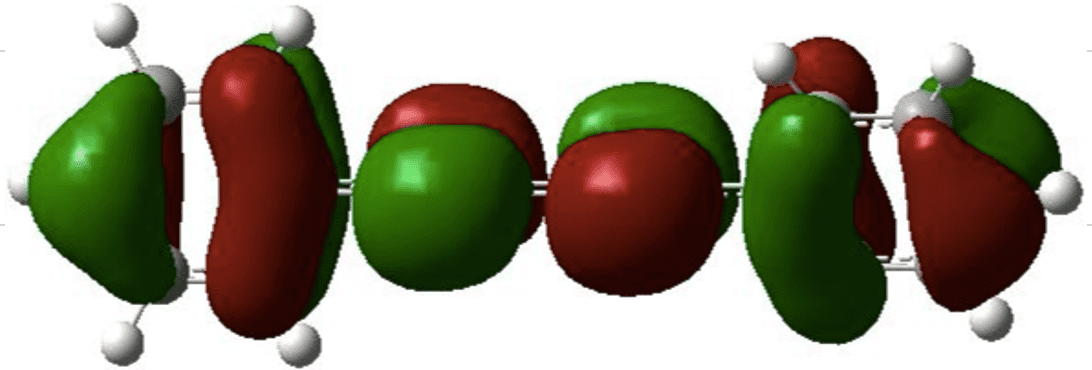}} &
     {\includegraphics[height=1.40 cm]{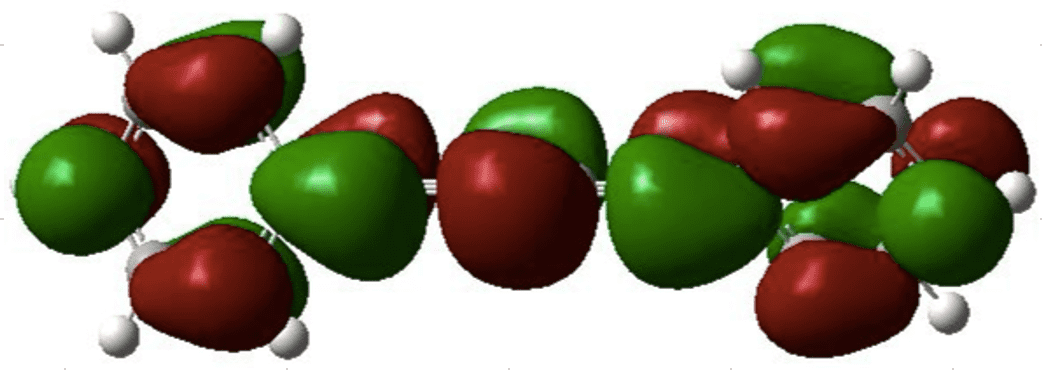} } & 
    {\includegraphics[height=1.40cm]{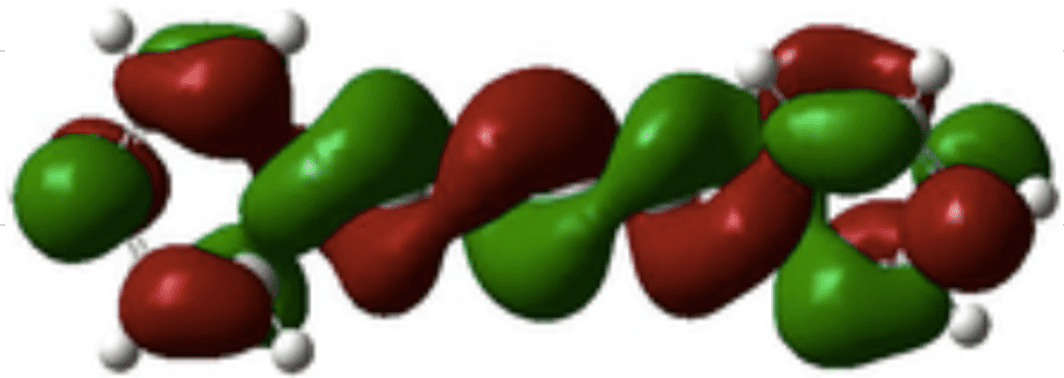} } \\
   & HOMO-1 (-0.28601 u.a) & HOMO (-0.25889 u.a) & LUMO (-0.01152 u.a) & LUMO+1 (0.02042 u.a)  \\
  60°&
    {\includegraphics[height=1.40cm]{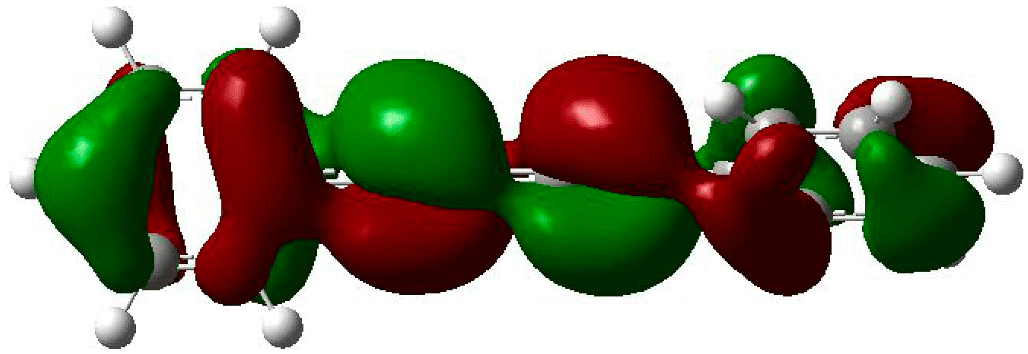}} &
    {\includegraphics[height=1.40cm]{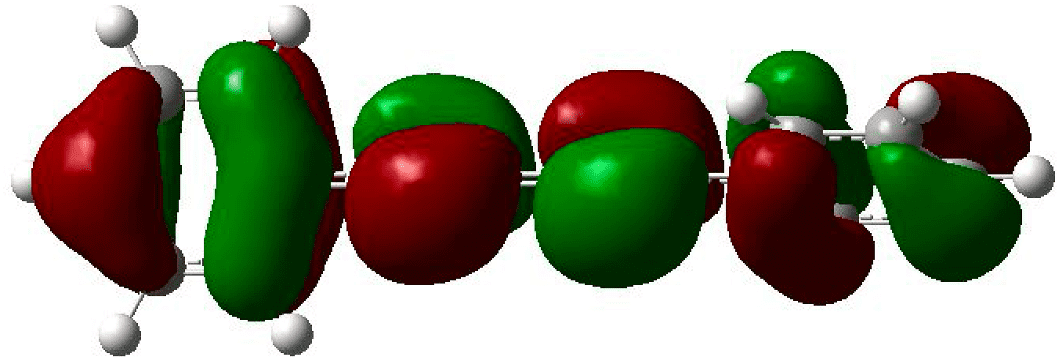}} &
      {\includegraphics[height=1.40 cm]{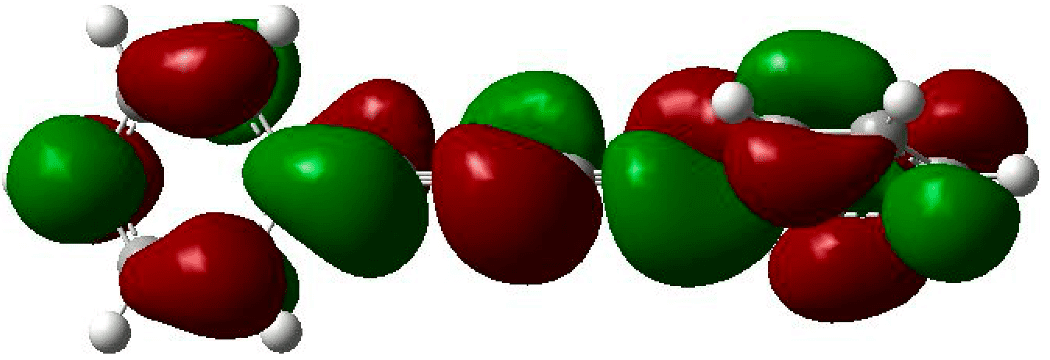} } & 
    {\includegraphics[height=1.40cm]{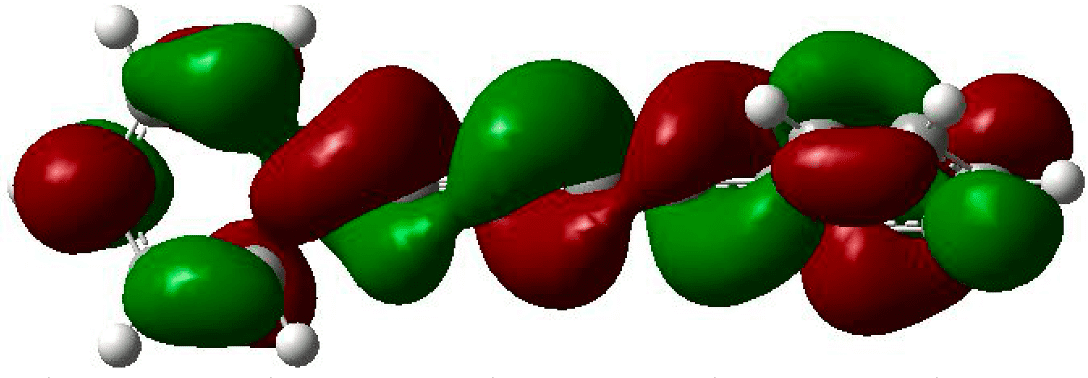} } \\
    & HOMO-1 (-0.27911 u.a) & HOMO (-0.26234 u.a) & LUMO (-0.00771 u.a) & LUMO+1 (0.01166 u.a)\\
  80°&
    {\includegraphics[height=1.40cm]{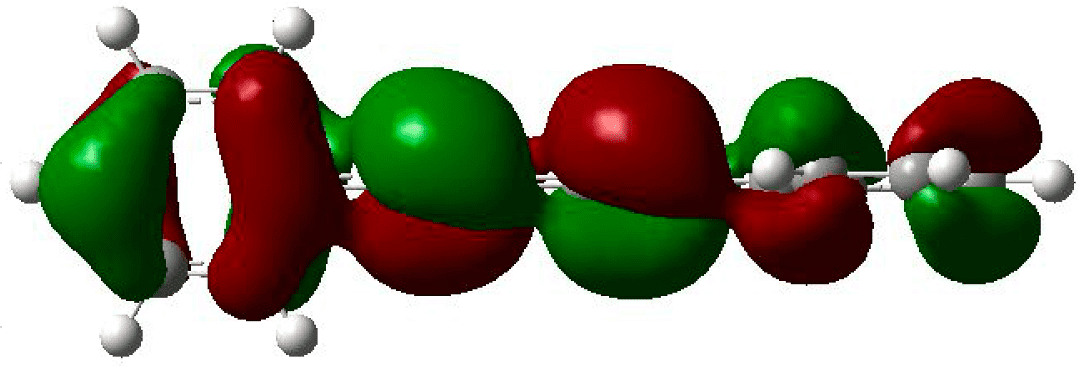}} &
    {\includegraphics[height=1.40cm]{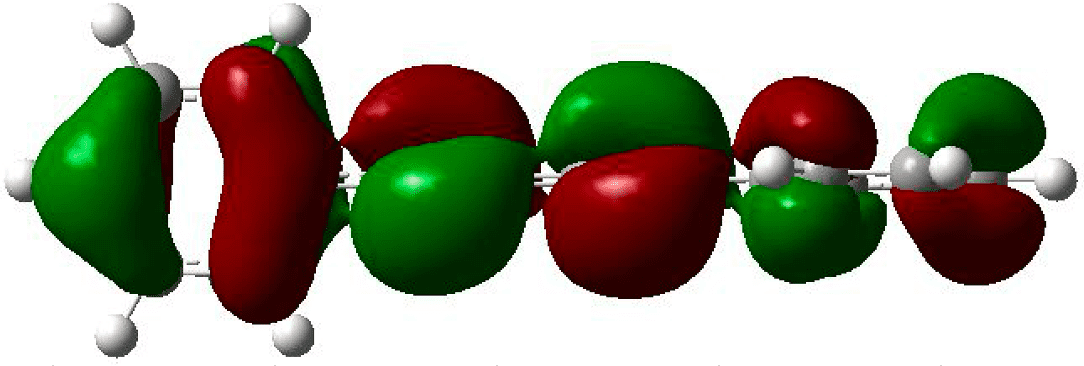}} &
    {\includegraphics[height=1.40 cm]{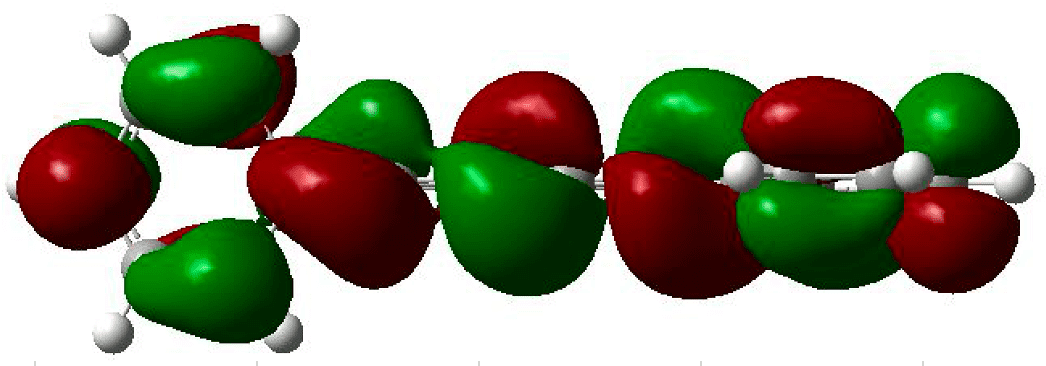} } & 
    {\includegraphics[height=1.40cm]{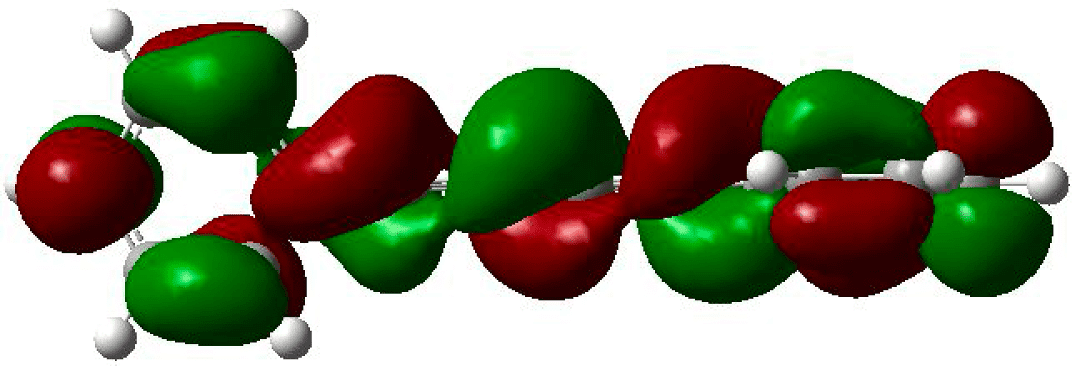} } \\
   & HOMO-1 (-0.27265 u.a) & HOMO (0.26698 u.a) & LUMO (-0.00251 u.a) & LUMO+1 (0.00398 u.a) \\
  90° &
    {\includegraphics[height=1.40cm]{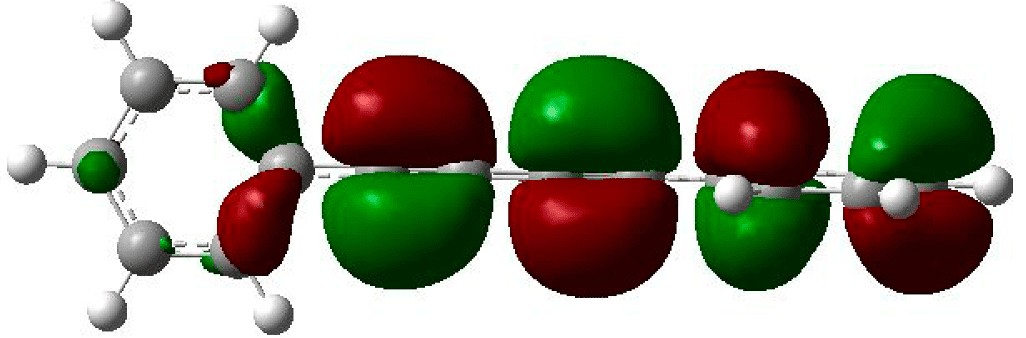}} &
    {\includegraphics[height=1.40cm]{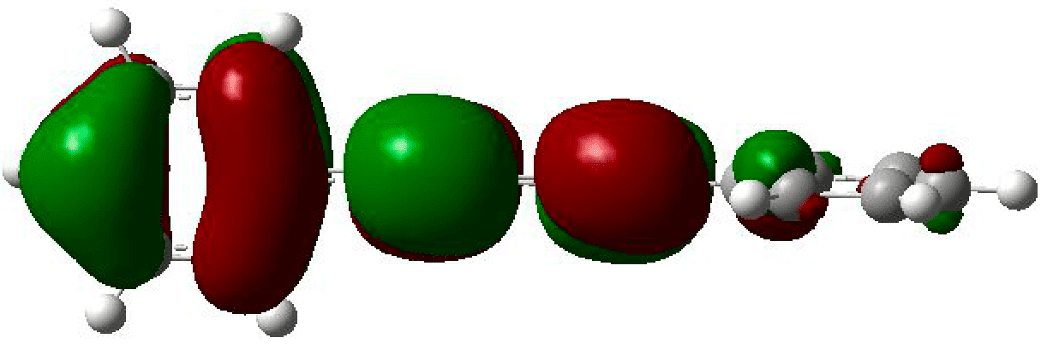}} &
    {\includegraphics[height=1.40 cm]{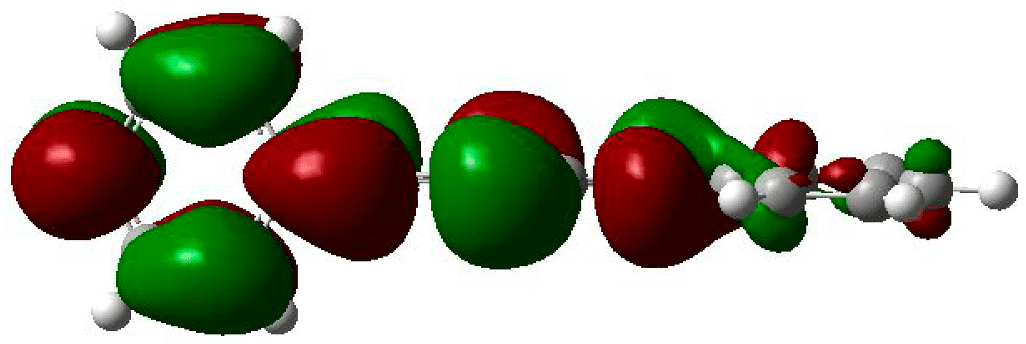} } & 
    {\includegraphics[height=1.40cm]{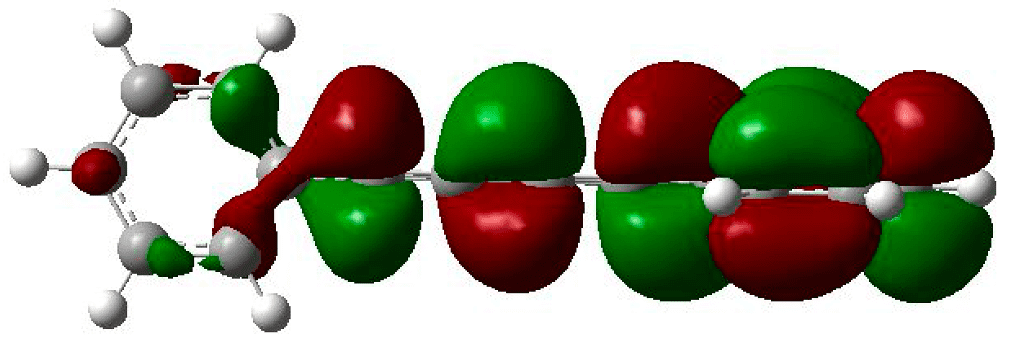} }\\
    &HOMO-1 (-0.26970 u.a) & HOMO (-0.26970 u.a) & LUMO (-0.00058 u.a) & LUMO+1 (-0.00058 u.a) \\
        \bottomrule
        \end{tabular}
        \end{table}  
        
\subsection{Tolanophane}

\begin{figure}[H]
    \centering
    \includegraphics[width=0.3\linewidth]{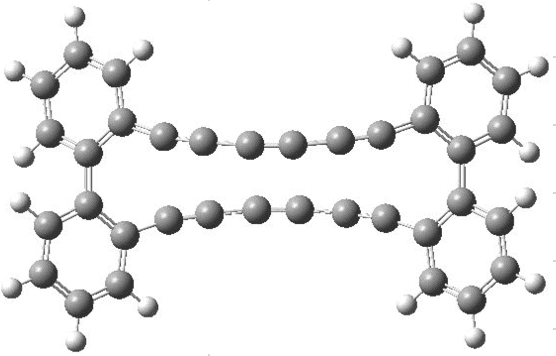}
    \caption{Representation of the molecule's tolanophane}
    \label{fig20-mol-tolanophane}
\end{figure}

Carbyne molecules are very promising for electronic and mechanical applications. They would offer, in their chain version, a "specific resistance surpassing that of all known materials to date" \cite{liu}. Flexible and very resistant, these materials also have singular electrical properties. Carbyne-carbyne cross-linking are equally promising. The tolanophane molecule is a representative example of these systems \cite{toyota}. It consists of two tolane (diphenylethyne) units (see fig.\ref{fig20-mol-tolanophane}). X-ray analysis revealed that this molecule has a twisted structure with $D_2$ symmetry in agreement with the modelling. Our study shows that these systems generate helical states. These helical states appear for a twist of the molecule in the form of a braid.

Note here the presentation of the LUMO+3 because it is the only energy level for which we obtain a helix.

\begin{table}[H]
    \caption{LUMO orbitals and energies obtained at the B3LYP/6-311G(d.p) level of theory for the tolanophane system for various selected rotations (see Table S6 for details).} 
    \label{table10-OM-tolano-DFT}
    \centering
    \begin{tabular}{ccc}
        \midrule
    {\includegraphics[height=2.10cm]{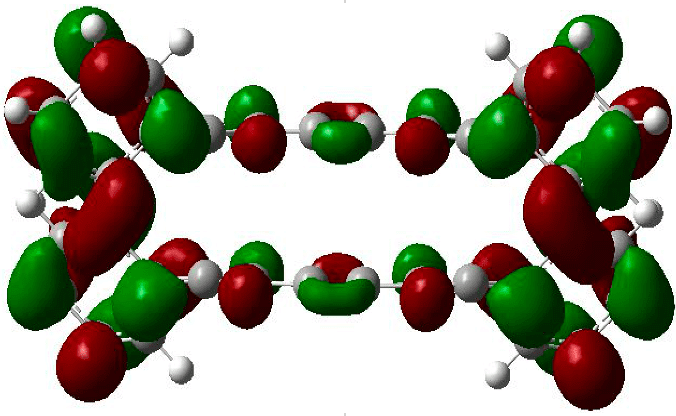} } &
    {\includegraphics[height=2.10cm]{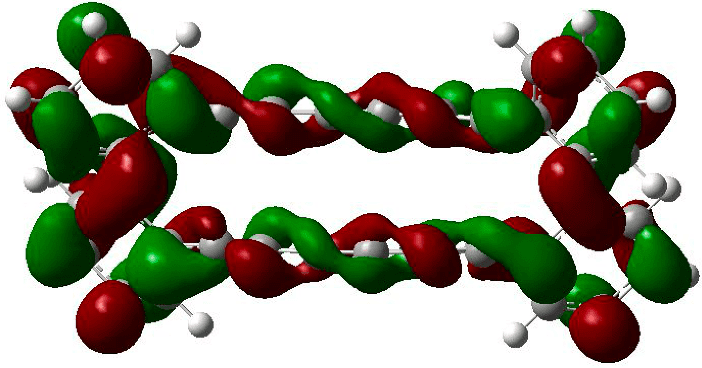} } & 
    {\includegraphics[height=2.10cm]{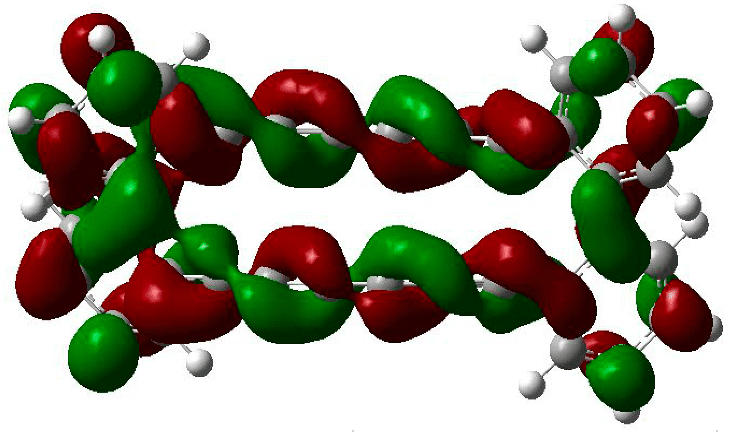} } \\
    0\Degre (E = 0.01055 u.a) & 5\Degre (E = 0.01011 u.a) & 15\Degre (E = 0.00778 u.a) \\
    
    {\includegraphics[height=2.10cm]{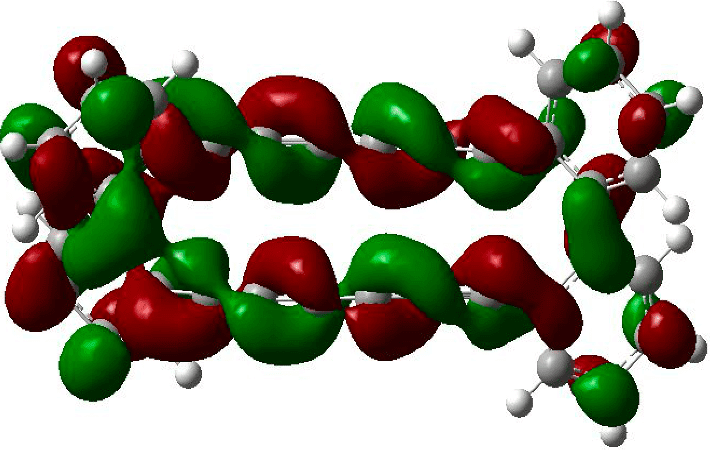} } &
     {\includegraphics[height=2.10cm]{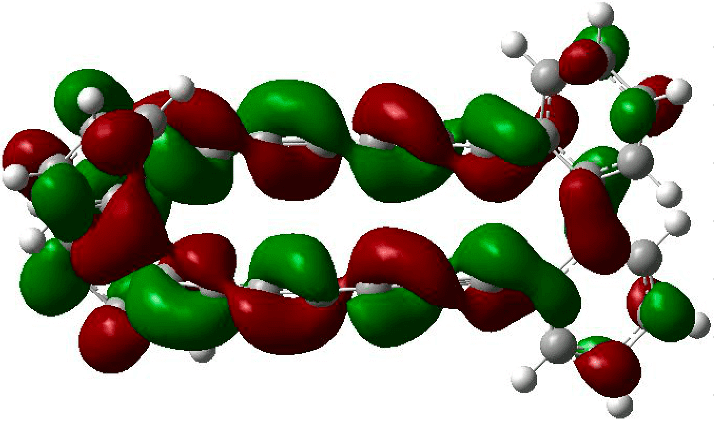} } &
    {\includegraphics[height=2.10cm]{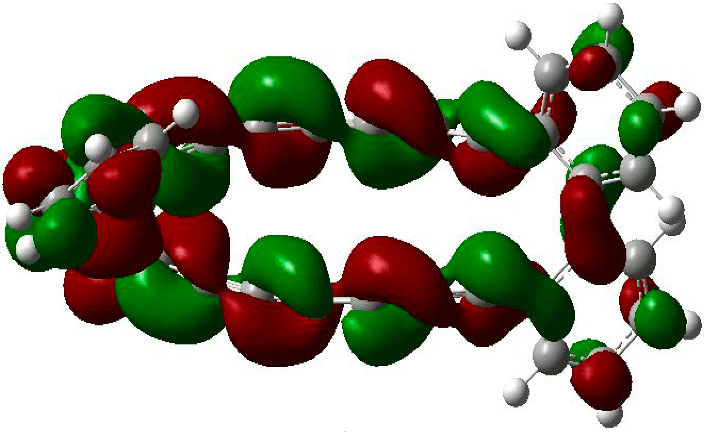} } \\
    25\Degre (0.00495 u.a)&  35\Degre (E = 0.00211 u.a) & 45\Degre (E = -0.00087 u.a)\\

    {\includegraphics[height=2.30cm]{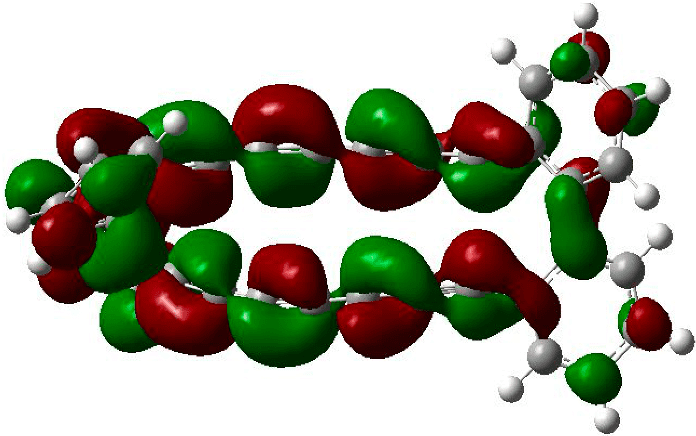} } &
    {\includegraphics[height=2.30cm]{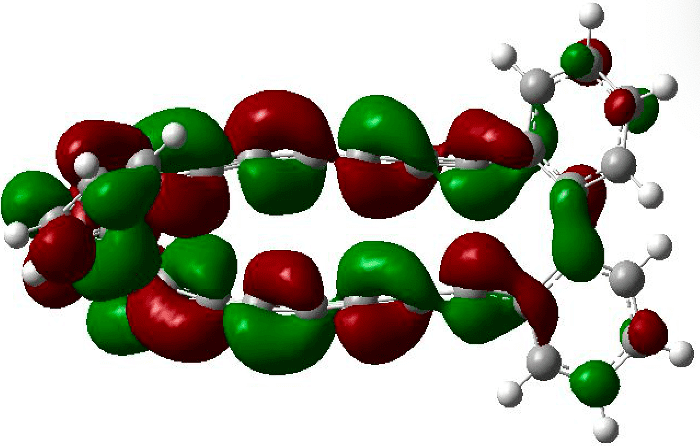} }& 
    {\includegraphics[height=2.30cm]{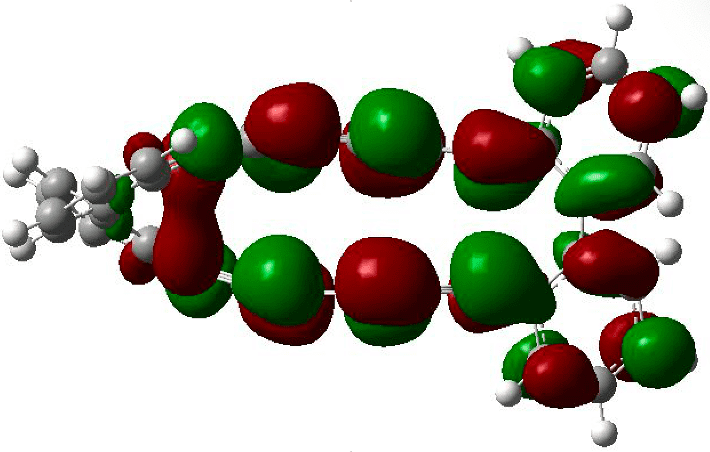} } \\
    55\Degre (E = -0.00369 u.a)  & 65\Degre (E = -0.00623 u.a) & 75\Degre (E = -0.00813 u.a) \\

    {\includegraphics[height=2.10cm]{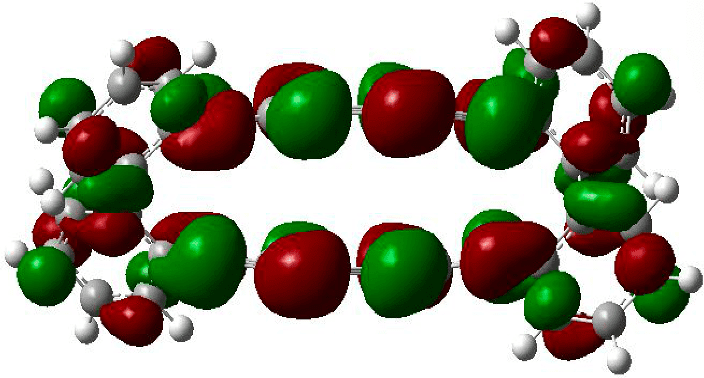} } & 
    {\includegraphics[height=2.10cm]{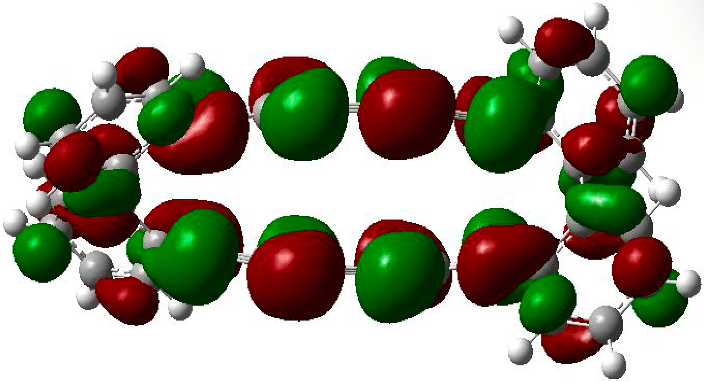} } &
    {\includegraphics[height=2.10cm]{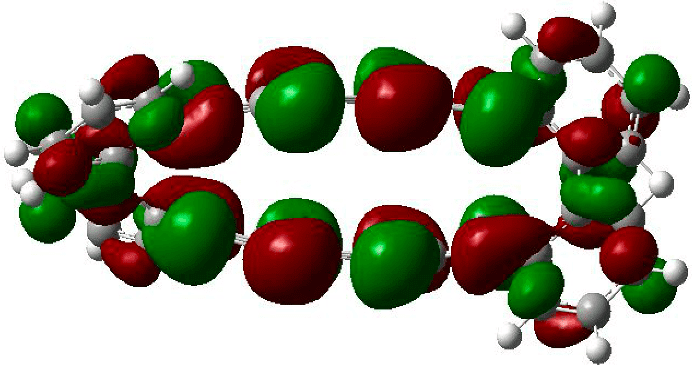} } \\
    85\Degre (E = -0.01097 u.a)  & 95\Degre (E = -0.00960 u.a) & 105 \Degre (E = -0.00819 u.a) \\
   
   {\includegraphics[height=1.60cm]{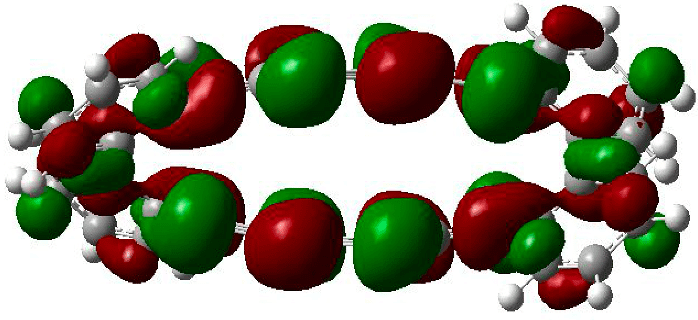} } &
    {\includegraphics[height=1.60cm]{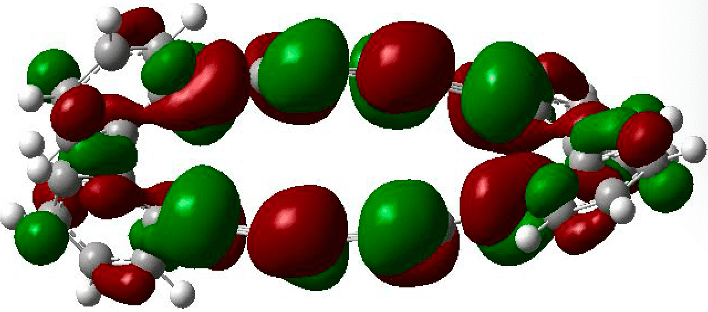} } & 
    {\includegraphics[height=1.55cm]{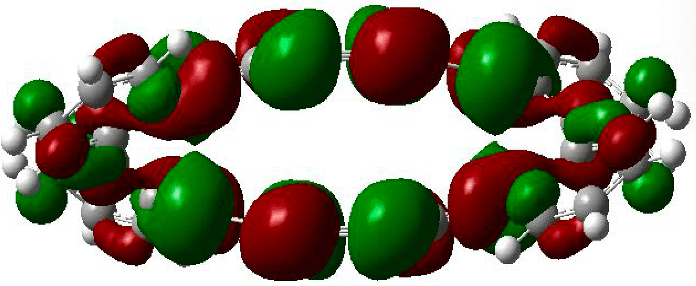} } \\
   115\Degre (E = -0.00659 u.a)  & 125\Degre (E = -0.00659 u.a) & 135\Degre (E = -0.00326 u.a) \\
   
   {\includegraphics[height=1.60cm]{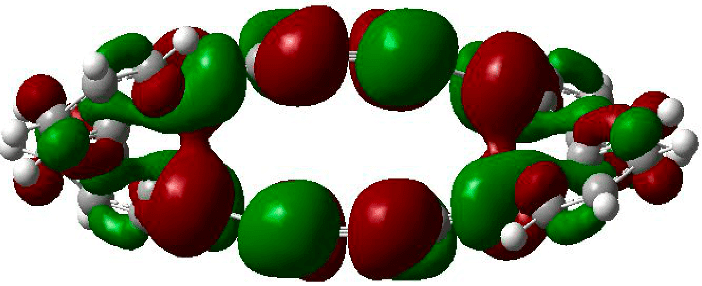} } &
   {\includegraphics[height=1.60cm]{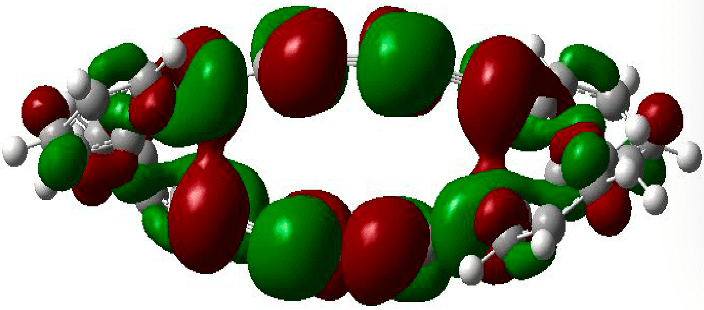} } &
    {\includegraphics[height=1.60cm]{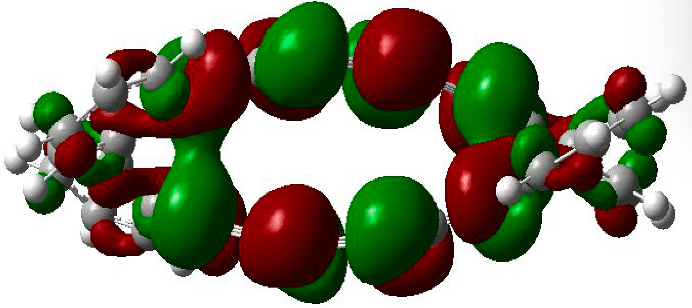} } \\
   145\Degre (E = -0.00149 u.a) & 155\Degre (E = -0.00149 u.a)  & 165\Degre (E = -0.00192 u.a)  \\
        \bottomrule
        \end{tabular}
        \end{table}  
\section{Beyond $p$-orbitals - helical states using $d$-orbitals and metallacumulenes}

As recently reported by Garner and coll. \cite{bro} the prospective of helical states using $d$-orbitals has started. In particular, the metallacumulenes are potential classes of molecules that may exhibit helical MOs in the linear fragments of the molecules. 
If some MOs of the trans-$[EtC=(C=)_4 C=Ru=(C=)_4 C Me]^{2+}$ system are somewhat helical, Garner also reports that at the ruthenium centre there exists a jump not yet identified in the evolution of the helical MOs. This jump is assigned by the authors to the inherent sign-change in the metal $d$-orbitals that couple to the carbon $\pi$-orbitals.

The presence of variable helicity in different parts of a molecule is then clearly highlighted. However, the question that now arises is whether the existence of possible helical MOs of the metal part. In a last perspective we finally wanted to imagine the helical MOs of the metallic part through, for example, the study of a metal-metal bonds. Theoretical investigations  refer to the existence of multiple metal-metal bonds, mainly consisting of a combination of $\sigma$ and $\pi$ interactions in all the $[M_2X_8]^{2-}$ species investigated \cite{Cavi}. In addition, $\delta$-like interactions also occur in the complexes of rhenium (M=Re) in particular. The conformation of the $[Re_2H_8]^{2-}$ system where hydrogen ligands are eclipsed ($D_{4h}$) were studied in its lowest energy configuration represented as $[\sigma^2 \pi^4 \delta^2]$. The main reason of this decision was that conformation allows for maximum $\delta$-$\delta$ overlap between the Re(III) centre resulting in the formation of a potential helix when the ($D_{4h}$) symmetry is broken as shown in figure \ref{fig21- Re}.

 \begin{figure}[H]
    \centering
    \includegraphics[width=0.2\linewidth]{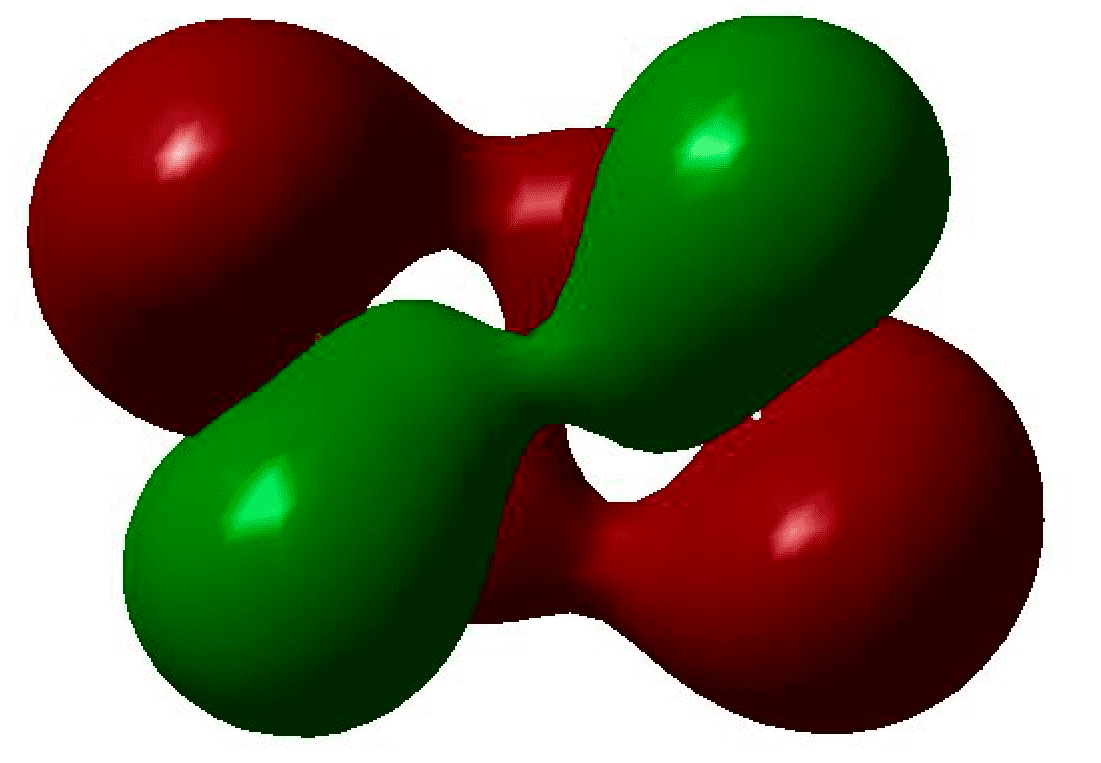}
    \includegraphics[width=0.18\linewidth]{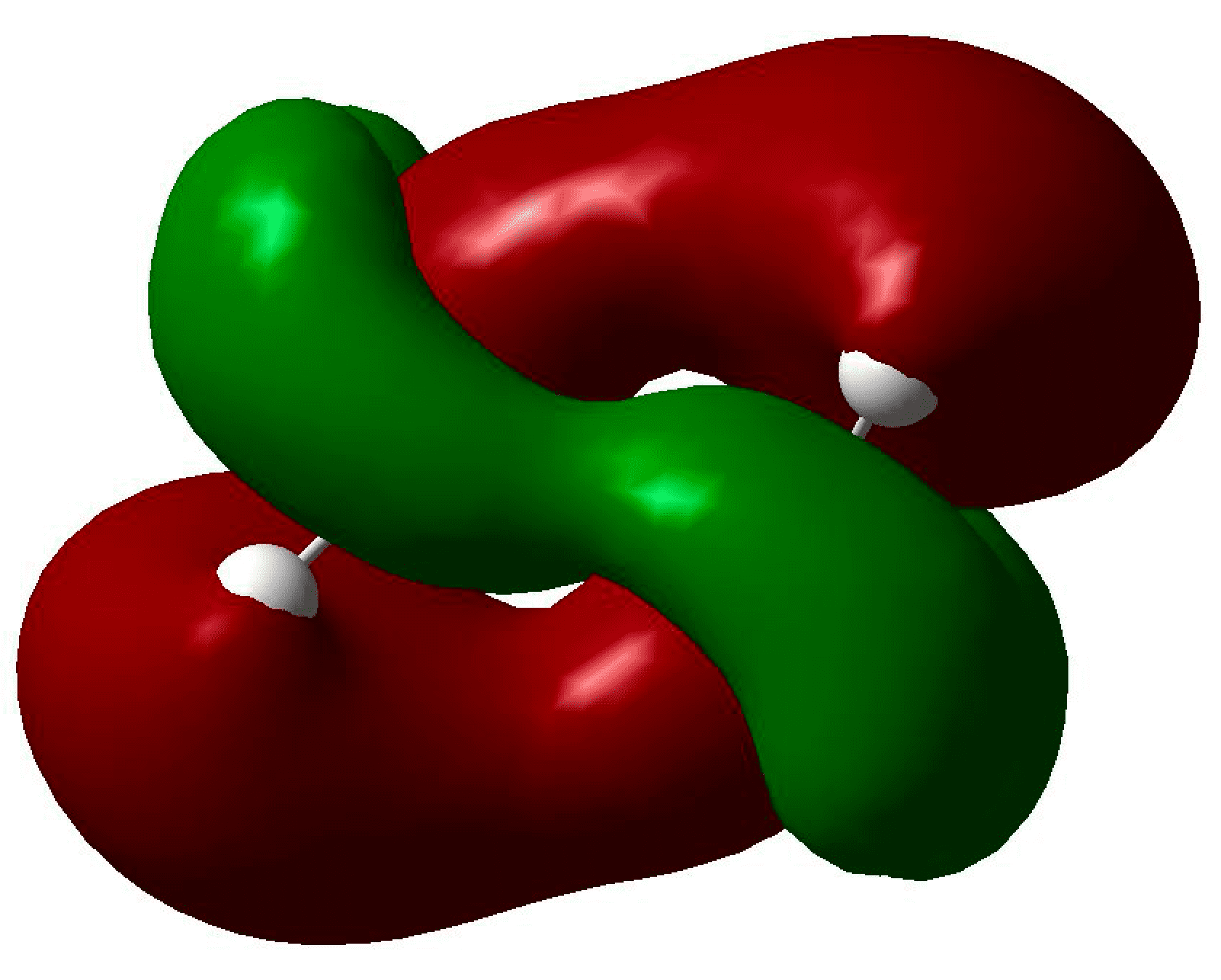}
    \caption{Representation of the HOMO-5 and the HOMO of the $[Re_2H_8]^{2-}$}
    \label{fig21-Re}
\end{figure}

\section{Conclusion}

Starting from the work of M.H. Garner et al. \cite{garner}, we studied in generality the generation of helical orbitals for particular chain of atoms which include in particular [N] cumulenes. 

We first discuss the definition of helical orbitals following \cite{garner} and \cite{guna}. In particular, we discuss the possibility to associate a perfect helix to a given distribution of angles representing the evolution of the $\pi$-system along the chain. We introduce as an index of helicity a correlation number obtained by linear regression between an hypothetical perfect helix and the exact feature of the $\pi$-system under consideration. This index is different from the MAD index of \cite{bro}. Taking as an example linear chains of boron nitride, we show that perfect helix are not the rule and a more general definition of helices are needed, as long as one is interested in characterizing as precisely as possible helical orbitals by a single parameterized function.\\

We then give a global description of distribution obtained using H\"uckel theory for twisted [N]-cumulenes generalizing and completing previous work of M.H. Garner et al. \cite{garner} which is limited to even numbered [N]-cumulenes. Explicit formula are given and several general properties of H\"uckel-distribution are proved. Simulations are provided in order to illustrate these properties. \\

We then discuss minimal assumptions under which a given linear chain admit helical orbitals. Several methods exist.\\ 

We first focus on an algebraic characterization of helical orbitals given by S. Gunasekaran et al. in \cite{guna} based on the Löwdin partitioning technique. Using this characterization, we deduce that helical orbitals are generic. Indeed, the existence criterion is based on the fact that two matrices do not commute which is a generic property for a set of matrices. Despite its interest, this criterion is not useful from a practical point of view because the computation of the matrices entering in the criterion are in general too complicated to compute explicitly.\\

We then discuss more precisely the structural properties of linear chain focusing on the role of symmetries of the molecule. As helical orbitals are associated to special geometrical properties of the molecule and are chiral objects, we deduce using Curie's principle, that a necessary condition for a molecule to possess helical orbitals is chirality. This assumption joins observation made by M.H. Garner et al. in \cite{garner}. Moreover, the existence of helicogenic axes of symmetries as defined by \cite{garner} is also necessary. The previous conditions are necessary when looking for helical orbitals at the ground state. If not, as the symmetry group of MOs for a given molecule depends on the energy state, one can look for excited version of a given molecule for which symmetry adapted MOs lead to helical orbitals. The previous remark opens many possibilities for a huge number of structures to exhibit helical orbitals, in particular when different electronic multiplicities are allowed. This point clearly proves again that helical orbitals must be more generic than usually believed. Several examples are given in order to illustrate these ideas. \\

Helical orbitals are constructed by focusing on $2p$ orbitals. However, the criterion for helicity based on helicogenic axes and chirality suggests that $d$ orbitals can be a good candidate to construct helical orbitals of a new kind. Following a recent result presented in \cite{bro} we discuss this possibility using metallacumulenes. The main observation is that thanks to $\delta$ interaction and breaking of symmetries, helical orbitals are again observed. \\

We hope that this work provides significant information and tools to study helical orbitals in various situations. The use of $d$-orbitals seems to be very promising as it opens the possibility to construct helical orbitals for new types of molecules with possibly different helical morphology.  
\begin{appendix}

\section{Computational Methods}
\label{computmet}

Ground and excited-states geometries were determined at the DFT level using the B3LYP exchange-correlation functional with the 6-311G(d.p) basis set. Calculations were performed using Gaussian 09 program package \cite{G09}.  In order to obtain reliable energies and OMs CASPT2(8:10)sp/6-311G(d.p) calculations were also carried out. Calculations were performed using Molpro program packages \cite{Molpro}.

\section{A technical result}
\label{techni}

For all $N\geq 1$, $z=0,\dots ,N$, the function $a_n (z)=\sin{\left( \frac{z n \pi}{N+1} \right)}$ satisfies the equality 
\begin{equation}
    a_n(N-z+1) = (-1)^{n+1} a_n(z)
\end{equation}

\begin{proof}
This is a simple computation. We have 
\begin{equation}
\left .
\begin{array}{lll}
     a_n(N-z+1) & = &  \sin{ \left ( (N-z+1) \frac{n\pi}{N+1} \right)} ,\\
     & = &  \sin{ \left (\frac{ (N+1) n \pi}{N+1}- \frac{z n \pi }{N+1} \right )} ,\\
     & = &  \sin{ \left( n \pi \right) } \cos{\left( \frac{z n \pi}{N+1} \right)} - \cos{\left( n \pi \right)} \sin{\left( \frac{z n \pi}{N+1} \right)} ,\\
    & = &  - \cos{ \left( n \pi \right) } \sin{\left( \frac{z n \pi}{N+1} \right)} = (-1)^{n+1} a_n(z) .
\end{array}
\right .
\end{equation}
\end{proof}

\section{Proof of the angle formula for cumulene}

\subsection{The case $\theta=\pi/2$}
\label{proofangle}

The aim of this section is to prove the following formula: 

\begin{equation}
    \mathscr{A}_{N,n,+,0,z} = \cos^{-1}  \left ( \frac{\psi_n(0) . \psi_n(z)}{\lVert\psi_n(0)\rVert \lVert \psi_n(z)\rVert} \right) = \cos^{-1} \left ( \frac{ \epsilon(a_n (1)) \epsilon (a_n (z+1))}{\sqrt{1 + (a_n(z) / a_n(z+1))^2 }} \right) ,
\end{equation}
where $\epsilon (x)$ is the function equal to $+1$ if $x>0$ and $-1$ if $x<0$.

\begin{proof}
The two vectors $\psi_n (0)$ and $\psi_n (z)$ are given by 
\begin{equation}
    \psi_0 (z)=\di\sqrt{\frac{2}{N+1}} \left ( 
    \begin{array}{c}
    0 \\
    a_n (1) 
    \end{array}
    \right ) ,
    \ \ \
    \mbox{\rm and}
    \ \ \ 
\psi_n (z)=\di\sqrt{\frac{2}{N+1}} \left ( 
    \begin{array}{c}
    a_n (z) \\
    a_n (z+1) 
    \end{array}
    \right ) ,
\end{equation}
whose scalar product $\psi_n (0)\cdot \psi_n (z)$ is given by 
\begin{equation}
\psi_n(0) \cdot \psi_n(z) = \frac{2}{N+1} a_n (1)  a_n (z+1) .
\end{equation}

The norm of each vector $\psi_n (0)$ and $\psi_n (z)$ is given by 
\begin{equation}
    \lVert\psi_n(0)\rVert = \sqrt{\frac{2}{N+1}} \mid a_n (1) \mid ,\ \ \ \mbox{\rm and}\ \ \
    \lVert\psi_n(z)\rVert = \sqrt{\frac{2}{N+1}} \sqrt{(a_n(z))^2 + (a_n(z+1))^2}
\end{equation}

We then obtain 
\begin{equation}
\mathscr{A}_{N,n,+,0,z} = \cos^{-1} \left ( \frac{a_n(1) a_n (z+1)}{ \mid a_n(1) \mid \sqrt{(a_n(z))^2 + (a_n (z+1) ^2}} \right) .
\end{equation}

As $a_n (z+1)\not= 0$ when $z=1,\dots ,N-1$, putting  $(a_n (z+1))^2$ in factor in the square root, we deduce  
\begin{equation}
    \mathscr{A}_{N,n,+,0,z} = \cos^{-1} \left ( \epsilon (a_n (1)) \epsilon (a_n (z+1)) \frac{1}{\sqrt{1+(a_n(z)/a_n (z+1))^2 }} \right ) .
\end{equation}

When $z=N$, we have $a_n (N+1)=\sin (n\pi)=0$ and $\mathscr{A}_{N,n,+,0,N}$ reduces to $\mathscr{A}_{N,n,+,0,z} = \cos^{-1} (0)=\pm \pi/2$. Due to the symmetry relation (\ref{symorb}), we deduce that the sign depends only on $n$ and is given by $(-1)^{n+1}$. This completes the proof.
\end{proof}

\subsection{The case $\theta=\pi/2$}
\label{proofangle0}

By definition, we have 
\begin{equation}
    \psi_n (0)=\left ( 
\begin{array}{c}
0 \\
\di\sqrt{\frac{2}{N+2}} \sin \left ( \frac{n\pi}{N+2} \right ) 
\end{array}
\right )
\ \ \mbox{\rm and}\ \  \psi_n (N)=\left ( 
\begin{array}{c}
0 \\
\di\sqrt{\frac{2}{N+2}} \sin \left ( \frac{n\pi(N+1)}{N+2} \right ) 
\end{array}
\right )
.
\end{equation}

As $\di\sin \left ( \di\frac{n\pi(N+1)}{N+2} \right ) = \sin \left ( 
n\pi -\di\frac{n\pi}{N+2} \right ) = (-1)^{n+1} \sin \left ( \di\frac{n\pi}{N+2} \right )$, we obtain 
\begin{equation}
    \psi_n (N)= (-1)^{n+1} \psi_n (0).
\end{equation}
We deduce directly that for $n$ odd, $\psi_n (N)=\psi_n (0)$ and the angle is null and for $n$ even $\psi_n (N)=-\psi_n (0)$ and the angle is $\pi$. This concludes the proof.

\section{Proof of the formula for the OM coefficients of $C_2$ Symmetry-adapted linear combinations for $\theta=0$ and $\theta=\pi/2$ twisted cumulene}
\label{proofcoef}

\subsection{The case $\theta=\pi/2$}
\label{proofcoefpisur2}

This is the simplest case. Indeed, denoting by $A_N (w)$ the $N\times N$ matrix given by 
\begin{equation}
    A_N (w)=\left (
    \begin{array}{ccccc}
    w & 1 & 0 & \dots & 0\\
    1 & w & \ddots & \ddots & \vdots\\
    0 & \ddots & \ddots & \ddots & 0 \\
    \vdots & & & & 1 \\
    0 & \dots 0 & & 1 & w
    \end{array}
    \right )
,
\end{equation}
and posing as usual $w=\di\frac{\alpha - \lambda}{\beta}$, the secular determinant $S_N (\lambda )$ of the H\"uckel matrix $H_N$ defined by $S_N (\lambda )=\det ( H_N -\lambda \mbox{\rm Id} )$ is such that 
\begin{equation}
\label{secular}
    S_N (\lambda ) = \di\beta^N \left ( P_N (w) \right ) ^2 ,
\end{equation}
where $P_N (w) =\det (A_N (w))$ and $A_N$ is the classical matrix associated with a $N$-linear chain of carbon atoms. \\

The $N$ roots of $S_N$ are doubly degenerate and are symmetrically distributed around the value $\alpha$ and are given by 
\begin{equation}
\lambda_n = \alpha +2\beta \cos \left ( k_{N,n} \right ) ,     
\end{equation}
with $k_{N,n} =n\pi /N+1$.\\

We refer to the work of C.A. Coulson (\cite{coulson}, Appendix p.393-394) for more details.

\subsection{The case $\theta=0$}
\label{proofcoefzero}

This case is more complicated. The secular determinant $S_N (\lambda)$ for the H\"uckel matrix reads in this case as 
\begin{equation}
    S_N (\lambda )=\di\beta^N P_{N+1} (w) P_{N-1} (w)
\end{equation}

We have $2N$ roots which are non-degenerate and obtained by intricating the symmetric roots of $P_{N+1} (w)$ and $P_{N-1} (w)$. The roots of $P_{N-1}$ corresponding to the $p_x$ system and those of $P_{N+1}$ to the $p_y$ system. \\

Symmetry-adapted linear combinations in $C_2$ of atomic orbitals coming from the $p_x$ and $p_y$ systems are possible only for atomic orbitals which are not too far in energy. As a consequence, we must understand how the two spectrums are intricated.  Denoting by \color{blue} $\lambda_{n,y}$ \color{black} the eigenvalues induced by the $y$ part and \color{red} $\lambda_{n,x}$ \color{black} the eigenvalues induced by the $x$ part we obtain an intrication of the following form
\begin{equation}
    \color{blue} \lambda_{N+1,y} \color{black} < \color{red} \lambda_{N-1,x} \color{black} < \color{blue} \lambda_{N,y} \color{black} <\dots < \color{blue} \lambda_{2,y} \color{black} < \color{red} \lambda_{1,x} \color{black} < \color{blue} \lambda_{1,y} .  
\end{equation}
\color{black} Taking explicit values (see just below), one can observe that taking $n$ for a particular energy level corresponding to \color{blue} $\lambda_{n,y}$ \color{black}, $n=2,\dots ,N$, the closest values of \color{red} $\lambda_{k,x}$ \color{black} is obtained for $k=n-1$.\\

As an example, taking $N=3,4$ and $5$ representing only the value $2\cos (k_{N+1,n} )$ for $\lambda_{n,y}$ and $2\cos (k_{N-1 ,n})$ for $\lambda_{n,x}$, we obtain:\\

for $N=3$ (see figures 19 and 20) : \color{blue} $\lambda_{4,y}=\alpha +1.618 \beta \color{black}  < \color{red} \lambda_{2,x}=\alpha +1.000 \beta \color{black}  < \color{blue} \lambda_{3,y}=\alpha +0.618a \color{black} < \color{blue} \lambda_{2,y}=\alpha -0.618\beta \color{black} < \color{red} \lambda_{1,x}=\alpha -1.000 \beta \color{black} < \color{blue} \lambda_{1,y}= \lambda_1 = \alpha -1.618\beta$ \\

\color{black} for $N=4$ : \color{blue} $\lambda_{5,y}=\alpha +1.732\beta \color{black} < \color{red} \lambda_{3,x}=\alpha +1.414\beta \color{black} < \color{blue} \lambda_{4,y}=\alpha +1.000\beta \color{black} < \color{red} \lambda_{2,x}=\alpha +0.000\beta \color{black} < \color{blue} \lambda_{3,y}=\alpha +0.0000\beta \color{black} < \color{blue} \lambda_{2,y}=\alpha -1.000\beta \color{black} < \color{red} \lambda_{1,x}=\alpha -1.414\beta \color{black} < \color{blue} \lambda_{1,y} = \alpha -1.732\beta$ \\

\color{black} for $N=5$ : \color{blue} $\lambda_{6,y}=\alpha +1.802\beta \color{black} < \color{red} \lambda_{4,x}=\alpha +1.618\beta \color{black} < \color{blue} \lambda_{5,y}=\alpha +1.247\beta \color{black} < \color{red} \lambda_{3,x}=\alpha +0.618\beta \color{black} \color{blue} < \lambda_{4,y}=\alpha +0.445\beta \color{black} < \color{blue} \lambda_{3,y}=\alpha -0.445\beta \color{black} < \color{red} \lambda_{2,x}=\alpha -0.618\beta \color{black} < \color{blue} \lambda_{2,y}=\alpha -1.247 \color{black} < \color{red} \lambda_{1,x}=\alpha -1.618\beta \color{black} < \color{blue} \lambda_{1,y} = \alpha -1.802\beta$ \\

\color{black}

The intrication of these values is resumed in the following table for each of these three cases:

 \begin{figure}[H]
    \centering
    \includegraphics[width=0.6\linewidth]{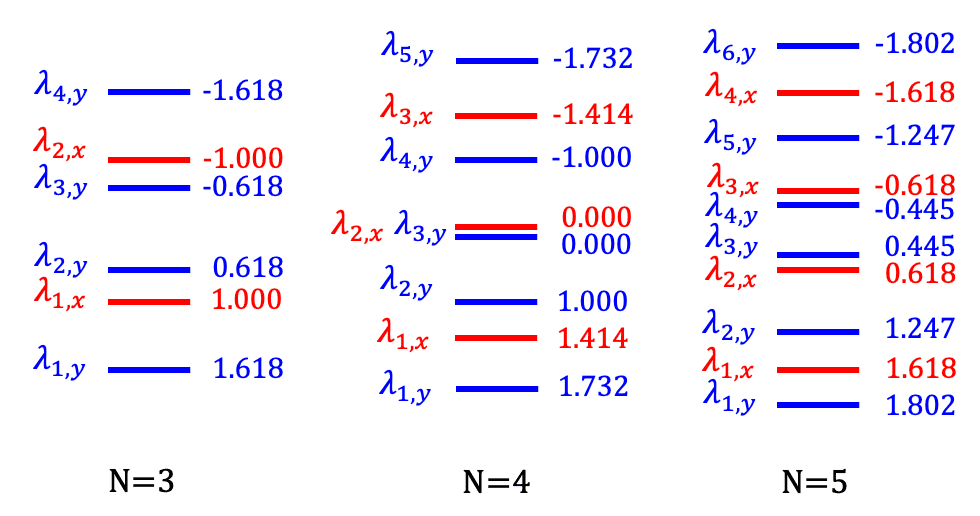}
    \caption{Representation in diagram form}
    \label{fig22-diagram}
\end{figure}

\color{black} As a consequence, taking the general form of the coefficients $c_{N,n} (z)$ for a linear chain given in \eqref{general}, we obtain for a given $N$ and $n=2,\dots ,N$
\begin{equation}
\label{formcoeff0}
    c_{y,n,N}^{\theta=0} (z) =  c_{N+1,n} (z+1),\ \ \mbox{\rm and}\ \ \ 
    c_{x,n,N}^{\theta=0} (z) =  c_{N-1,n-1} (z) .
\end{equation}
This concludes the proof. \\

A working example is given in the next Section for $N=3$ and $N=4$.

\section{Explicit computations of $C_2$-adapted linear combination of MOs for the $\theta=0$ twisted $[N]$-cumulene, $N=3,4$.}

As an example of the computations and quantities manipulated in the previous section, we derive explicit values for the case $N=3,4$ in the $\theta=0$ case, where the plane of the molecule is $(z,x)$ and the orthogonal one corresponds to $y$:

\begin{tabular}{p{9cm}|p{9cm}p{9cm}}
\centering \textbf{$\theta$ = 0 - N odd}  &  \centering \textbf{$\theta$ = 0 - N even}   \\
    &  \\
\centering{\includegraphics[height=2.0cm]{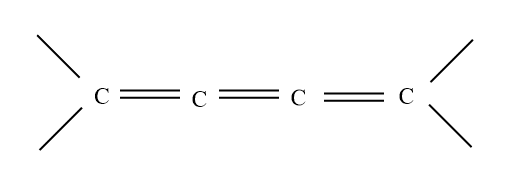}}  & 
\centering{\includegraphics[height=2.1cm]{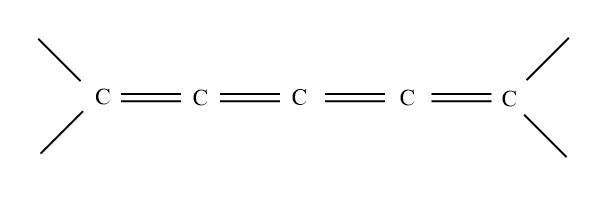}} \\
    &   \\
\centering $N = 3$ (6$\pi$ electrons) & \centering $N=4$ (8$\pi$ electrons)  
\end{tabular}

As already observed, the molecular orbitals are made of two perpendicular $\pi$ systems contained in the $(x,z)$ plane and $(z,y)$ plane respectively corresponding to a $N+1$ linear chain in the $(z,y)$ plane represented in blue and a $N-1$ linear chain in the $(x,z)$ plane. 

\begin{tabular}{p{9cm}|p{9cm}p{9cm}}
\center $y$ part :&  \center $y$ part :\\   
 & \\
\centering {\includegraphics[height=2.0cm]{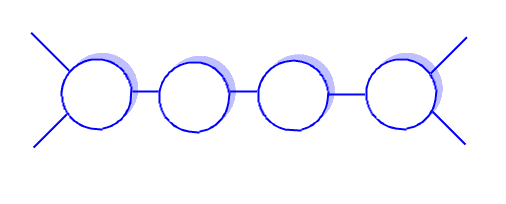}}  & 
\centering {\includegraphics[height=2.0cm]{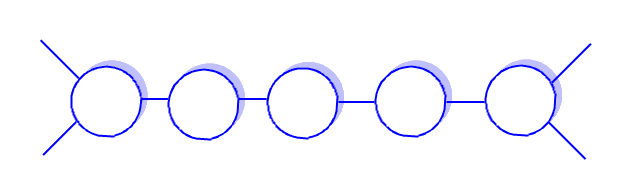}} \\
 & \\
\center $x$ part : & \center $x$ part :\\
 & \\
\centering {\includegraphics[height=2.0cm]{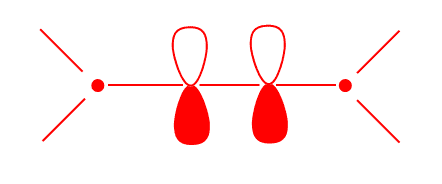}}  & 
\centering {\includegraphics[height=2.1cm]{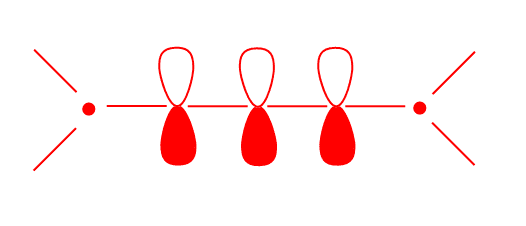}} \\
  & 
\end{tabular}

Computing the coefficients $c_{y,N,n}^{\theta=0} (z)$ for all value of $n$ and $z$ using formula \eqref{formcoeff0}  and finally the corresponding eigenvalues corresponding to $2\cos (\pi n/N+2 )$ gives:

\begin{tabular}{p{9cm}|p{9cm}p{9cm}}
    & \\    
\centerline{$c_{y,3,n}^{\theta =0}$(z)}  
\centering
      \color{blue}
   \begin{tabular}{c|cccc}
        \midrule
   z / n & 1 & 2 & 3 & 4 \\
   \hline
   0 & 0.37 & 0.60 & 0.60 & 0.37 \\
   1 & 0.60 & 0.37 & -0.37 & -0.60 \\
   2 & 0.60 & -0.37 & -0.37 & 0.60 \\
   3 & 0.37 & -0.60 & 0.60 & -0.37 \\
   \hline
   $\lambda_{n,y}$ & 1.618 & 0.618 & -0.618 & -1.618 \\
        \bottomrule
        \end{tabular} 
        & 
\centerline{$c_{y,4,n}^{\theta=0} (z)$}     
  \centering
     \color{blue}
   \begin{tabular}{c|ccccc}
        \midrule
   z / n & 1 & 2 & 3 & 4 & 5 \\
   \hline
   0 & 0.289 & 0.50  & 0.577  & 0.50  & 0.289\\
   1 & 0.50  & 0.50  & 0      & -0.50 & -0.5 \\
   2 & 0.577 & 0     & -0.577 & 0     & 0.577\\
   3 & 0.50  & -0.50 & 0      & 0.50  & -0.5 \\
   4 & 0.289 & -0.50 & 0.577  & -0.50 & 0.289 \\
   \hline
   $\lambda_{n,y}$ & 1.732 & 1 & 0 & -1  & -1.732\\ 
        \bottomrule
        \end{tabular}     
\end{tabular}

These values are corresponding for each $n$ to the following OMs configurations:

\begin{tabular}{p{9cm}|p{9cm}p{9cm}}
    & \\    
     \centering
      \color{blue}
   \begin{tabular}{c|cccc}
        \midrule
    n & 1 & 2 & 3 & 4 \\
   \hline
    & {\includegraphics[height=3.0cm]{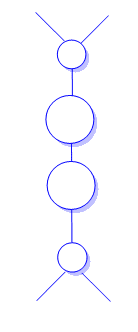}}  &
   {\includegraphics[height=3.0cm]{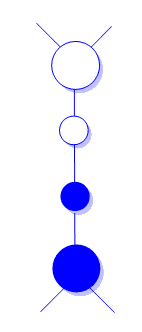}}& {\includegraphics[height=3.0cm]{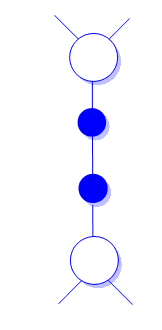}} &
   {\includegraphics[height=3.2cm]{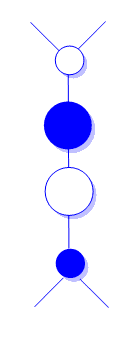}}\\
        \bottomrule
        \end{tabular} 
    &
   \centering
      \color{blue}
   \begin{tabular}{c|ccccc}
        \midrule
   n & 1 & 2 & 3 & 4 & 5\\
   \hline
    &{\includegraphics[height=3.0cm]{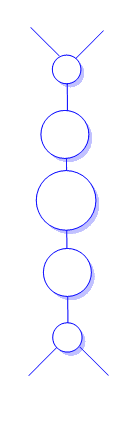}}  &  
    {\includegraphics[height=3.0cm]{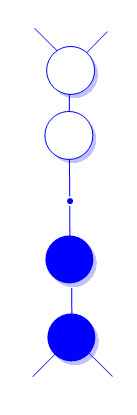}} & 
    {\includegraphics[height=3.0cm]{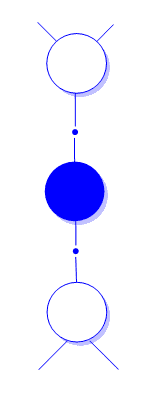}} & 
    {\includegraphics[height=3.0cm]{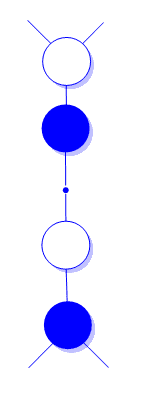}} &
    {\includegraphics[height=3.1cm]{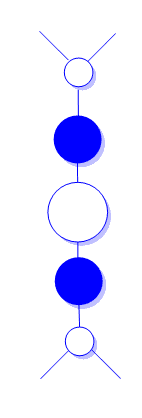}}\\
    \bottomrule
        \end{tabular} 
\end{tabular}
     
In the same way, the coefficients $c_{x,N,n}^{\theta=0} (z)$ are given by:  

\begin{tabular}{p{9cm}|p{9cm}p{9cm}}
    & \\    
\centerline{$c_{x,3,n}^{\theta=0} (z)$}  
\centering
      \color{red}
   \begin{tabular}{c|cccc}
        \midrule
   z / n & 1 & 2 & 3 & 4 \\
   \hline
   0 & . & 0 & 0 & . \\
   1 & . & 0.707 & 0.707 & . \\
   2 & . & 0.707 & -0.707 & . \\
   3 & . & 0 & 0 & . \\
   \hline
   $\lambda_{n,x}$ & . & 1.000 & -1.000 & . \\
        \bottomrule
        \end{tabular} 
        & 
\centerline{$c_{x,4,n}^{\theta=0} (z)$}     
  \centering
     \color{red}
   \begin{tabular}{c|ccccc}
        \midrule
   z / n & 1 & 2 & 3 & 4 & 5 \\
   \hline
   0 & . & 0  & 0  & 0  & .\\
   1 & .  & 0.50  & 0.707      & 0.50 & . \\
   2 & . & 0.707     & 0 & -0.707     & .\\
   3 & .  & 0.50 & -0.707      & 0.50  & . \\
   4 & . & 0 & 0  & 0 & . \\
   \hline
   $\lambda_{n,x}$ & . & 1.414 & 0 & -1.414  & .\\ 
        \bottomrule
        \end{tabular}     
\end{tabular}

which corresponds for each $n$ to the following MOs configurations:

\begin{tabular}{p{9cm}|p{9cm}p{9cm}}
    & \\    
     \centering
      \color{red}
   \begin{tabular}{c|cccc}
        \midrule
    n & 1 & 2 & 3 & 4 \\
   \hline
    & .  &
    {\includegraphics[height=3.0cm]{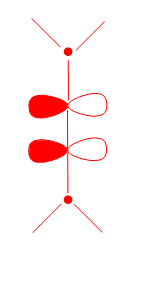}} &  {\includegraphics[height=3.0cm]{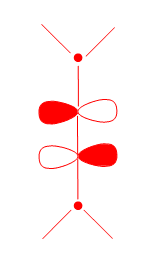}} &
   .\\
        \bottomrule
        \end{tabular} 
    &
   \centering
      \color{red}
   \begin{tabular}{c|ccccc}
        \midrule
   n & 1 & 2 & 3 & 4 & 5\\
   \hline
    & .  &  
    {\includegraphics[height=3.0cm]{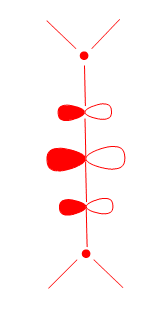}} &  {\includegraphics[height=3.0cm]{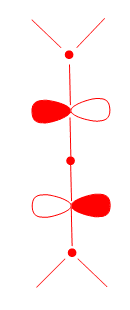}} &  {\includegraphics[height=3.0cm]{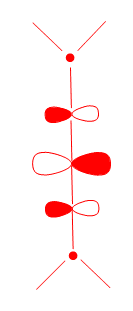}} &
    . \\
    \bottomrule
        \end{tabular} 
\end{tabular}

We then obtain the following table for $C_2$-adapted linear combinations of MOs using equation \eqref{psi_n(z)}:
\\

\begin{tabular}{p{9cm}|p{9cm}p{9cm}}
\centerline{$\psi_{3,n} (z)$}
\centering
   \begin{tabular}{c|cccc}
        \midrule
    z / n &  & 2 & 3 &  \\
   \hline
   $0$ &  & $ \begin{pmatrix} \textcolor{red}{0} \\ \textcolor{blue}{0.60} \end{pmatrix}$ & $ \begin{pmatrix} \textcolor{red}{0} \\ \textcolor{blue}{0.60} \end{pmatrix}$ & \\
   $1$ &  & $ \begin{pmatrix} \textcolor{red}{0.707} \\ \textcolor{blue}{0.37} \end{pmatrix}$ & $ \begin{pmatrix} \textcolor{red}{0.707} \\ \textcolor{blue}{-0.37} \end{pmatrix}$ &  \\
   $2$ &  & $ \begin{pmatrix} \textcolor{red}{0.707} \\ \textcolor{blue}{-0.37} \end{pmatrix}$ & $ \begin{pmatrix} \textcolor{red}{-0.707} \\ \textcolor{blue}{-0.37} \end{pmatrix}$ & \\
   $3$ &  & $ \begin{pmatrix} \textcolor{red}{0} \\ \textcolor{blue}{-0.60} \end{pmatrix}$ & $ \begin{pmatrix} \textcolor{red}{0} \\ \textcolor{blue}{0.60} \end{pmatrix}$ &  \\
        \bottomrule
        \end{tabular}
 
        & 
\centerline{$\psi_{4,n}(z)$}
\centering
    \begin{tabular}{p{1cm}|p{1.2cm}p{1.2cm}p{1.3cm}p{1.3cm}p{1.3cm}}  
        \midrule
    \centerline{z / n} &  \centerline{1} &  \centerline{2} &  \centerline{3} &  \centerline{4} &  \centerline{5}  \\
   \hline
   \centerline{$1$} & $ \begin{pmatrix} \textcolor{red}{0} \\ \textcolor{blue}{0.289} \end{pmatrix}$ & $ \begin{pmatrix} \textcolor{red}{0} \\ \textcolor{blue}{0.5} \end{pmatrix}$ & $ \begin{pmatrix} \textcolor{red}{0} \\ \textcolor{blue}{0.577} \end{pmatrix}$ & $ \begin{pmatrix} \textcolor{red}{0} \\ \textcolor{blue}{0.5} \end{pmatrix}$ & $ \begin{pmatrix} \textcolor{red}{0} \\ \textcolor{blue}{0.289} \end{pmatrix}$\\
   \centerline{$2$} & $ \begin{pmatrix} \textcolor{red}{0} \\ \textcolor{blue}{0.5} \end{pmatrix}$ & $ \begin{pmatrix} \textcolor{red}{0.5} \\ \textcolor{blue}{0.5} \end{pmatrix}$ & $ \begin{pmatrix} \textcolor{red}{0.707} \\ \textcolor{blue}{0} \end{pmatrix}$ & $ \begin{pmatrix} \textcolor{red}{0.5} \\ \textcolor{blue}{-0.5} \end{pmatrix}$ & $ \begin{pmatrix} \textcolor{red}{0} \\ \textcolor{blue}{-0.5} \end{pmatrix}$ \\
   \centerline{$3$} & $ \begin{pmatrix} \textcolor{red}{0} \\ \textcolor{blue}{0.577} \end{pmatrix}$ & $ \begin{pmatrix} \textcolor{red}{0.707} \\ \textcolor{blue}{0} \end{pmatrix}$ & $ \begin{pmatrix} \textcolor{red}{0} \\ \textcolor{blue}{-0.577} \end{pmatrix}$ & $ \begin{pmatrix} \textcolor{red}{-0.707} \\ \textcolor{blue}{0} \end{pmatrix}$ & $ \begin{pmatrix} \textcolor{red}{0} \\ \textcolor{blue}{0.577} \end{pmatrix}$\\
  \centerline{$4$} & $ \begin{pmatrix} \textcolor{red}{0} \\ \textcolor{blue}{0.5} \end{pmatrix}$ & $ \begin{pmatrix} \textcolor{red}{0.5} \\ \textcolor{blue}{-0.5} \end{pmatrix}$ & $ \begin{pmatrix} \textcolor{red}{-0.707} \\ \textcolor{blue}{0} \end{pmatrix}$ & $ \begin{pmatrix} \textcolor{red}{0.5} \\ \textcolor{blue}{0.5} \end{pmatrix}$ & $ \begin{pmatrix} \textcolor{red}{0} \\ \textcolor{blue}{-0.5} \end{pmatrix}$\\
   \centerline{$5$} & $ \begin{pmatrix} \textcolor{red}{0} \\ \textcolor{blue}{0.289} \end{pmatrix}$ & $ \begin{pmatrix} \textcolor{red}{0} \\ \textcolor{blue}{-0.5} \end{pmatrix}$ & $ \begin{pmatrix} \textcolor{red}{0} \\ \textcolor{blue}{0.577} \end{pmatrix}$ & $ \begin{pmatrix} \textcolor{red}{0} \\ \textcolor{blue}{-0.5} \end{pmatrix}$ & $ \begin{pmatrix} \textcolor{red}{0} \\ \textcolor{blue}{0.289} \end{pmatrix}$ \\
        \bottomrule
      \end{tabular}
        \\    
\end{tabular}

\end{appendix}

\section*{Associated content}
\noindent {\textbf{Supporting Information}} \\

\noindent The Supporting Information is available for free. 

\bibliographystyle{unsrt}

\end{document}